\documentclass[12pt]{book}
\usepackage{epsfig}
\usepackage{mathrsfs}
\usepackage[utf8]{inputenc}
\usepackage{color}

\definecolor{grey}{rgb}{0.4,0.4,0.4}

\begin{document}


\title{
{\bf The Higgs boson and the physics of $WW$ scattering 
before and after Higgs discovery}
}

\author{ {\Large Micha{\l} Szleper} \\
  \\
  {\sl National Center for Nuclear Research,} \\[-4pt] 
  {\sl Ho{\.z}a 69, 00-681 Warszawa, Poland} \\
\\
\\
\\
\\
\\
{\bf Abstract} \\
\\
This work presents a comprehensive overview of the physics of vector boson\\
scattering (VBS) in the dawn of Run 2 of the Large Hadron Collider (LHC).\\
Recalled~ here~ are~ some of~ its~ most basic physics principles,~ the historical\\
relation between vector boson scattering and the Higgs boson, then discussed\\
is the physics~ of VBS processes~ after Higgs discovery,~ and the prospects for\\
future VBS measurements at the LHC and beyond.  This monograph reviews \\
the work of many people, including previously published theoretical work as\\
well as~ experimental results,~ but also contains a portion of~ original simula-\\
tion-based studies that have not been published before.~~~~~~~~~~~~~~~~~~~~~~~~~~~~
}

\date{}

\maketitle


\tableofcontents

\chapter{Introduction}

{\it ``In the beginning there was symmetry."} - Werner Heisenberg

\section{Preamble - a little philosophical digression}

In the beginning there was symmetry and it was spontaneously broken.
It is time for this ingeneous and revolutionary idea to find its way to the 
recognition and acceptance of the wide public.
A recent google search for the phrase ``in the beginning there was" produced
several suggested endings for the query, including: the word, chaos,
light, death, darkness, and even more improbable ideas.
Only no symmetry.  But unlike all these other ideas, this one has by now a very
solid scientific basis.

In his popular book ``The Part and the Whole" (1969), Werner Heisenberg wrote:
``In the beginning there was symmetry".  He expressed the opinion that basic
symmetries of the world define the existing particle spectrum rather than the
other way around, and contrasted this view with ``in the beginning there was the
particle" for which he credited Democritus.
While Heisenberg most probably did not mean electroweak gauge symmetry, a concept being
back then in its early development and in which he admittedly did not
even show much interest, his point of view
is even more actual today than it was back in the 1960's.
Today we can tell his assertion was indeed correct.  As
of 2012, it even seems that we exactly know how this happens.
And yes, it was that symmetry breaking act that
defined what particles we have today.  Not the other way around.

\section{Spontaneous symmetry breaking}

The concept of spontaneous symmetry breaking is a milestone in physics.
The realization that certain key features of our physical world are
not explicitly determined by any fundamental laws of nature, but rather are the result
of a spontaneous choice of a single solution that happened possibly just once, very
early on in
the history of the Universe, has a profound impact on our understanding of the world.
Technically, spontaneous symmetry breaking is a mode of realization of symmetry breaking
in a physical
system, where the underlying laws are invariant under a symmetry transformation, but the
system as a whole changes under such transformations.  It thus describes systems where
the equations of motion or the Lagrangian obey certain symmetries, but the ground state
of the system, i.e., the lowest energy
solutions, do not exhibit that symmetry.  Rather than being reflected in the individual
solutions, the symmetry of the equations is reflected in the range of possible a priori
solutions, even if not observed in the physical world.
According to the Goldstone theorem, spontaneous breakdown of a symmetry is necessarily
associated with the appearance of new spinless particles, the so called Goldstone
bosons, one for each generator of the symmetry that is broken.
Unless the underlying symmetry is further broken explicitly, the Goldstone bosons are
massless.  Conversely, if the symmetry is not exact, the bosons acquire mass, albeit
are typically expected to be light.  In the latter case we talk of pseudo-Goldstone
bosons.
A well known example of this kind are the pions,
which can be identified with pseudo-Goldstone bosons related to the spontaneous breakdown
of the chiral-flavor symmetries of QCD as a result of the strong interaction.  The fact
that pions are not entirely massless is related to the approximative character of the
symmetry, which is due to the different quark masses.  Still, it is well known
that pions are much lighter than all the rest of the hadron spectrum.  Chiral
symmetry breaking is an important example of spontaneous symmetry breaking affecting
the chiral symmetry of strong interactions.  It is responsible for over 99\% of the mass
of the nucleons, as it allows to comprise heavy baryons out of nearly massless quarks,
and thus is in fact responsible for the bulk of the Universe mass.  Of course,
by saying this we implicitly take for granted the existence of that remaining 1\%.
Which is in fact far from trivial.  The whole clue is that quarks, and leptons for that
matter, are only {\it nearly} massless.  Without it, no mass would be there whatever.
It is perhaps not so paradoxical as it may seem at first glance that over 99\%
of the particle physicists' effort in the last few decades, in particular
in the field of experiment, has been put to reveal the origin of that
remaining 1\%.  
Finally, on July 4, 2012, the tiger broke free.
By announcing the discovery of the Higgs boson \cite{higgsdiscovery},
the ATLAS and CMS collaborations at CERN
strongly suggested that in fact this mass too originates from
spontaneous symmetry breaking - this time affecting the SU(2)$\times$U(1) symmetry of
electroweak interactions.  The effect, as understood from the Standard Model
perspective, is intrinsically connected with the so called Higgs
mechanism, which was originally proposed to explain how the weak gauge bosons $W$ and
$Z$ acquire masses.  Nevertheless, both historically and conceptually,
the idea of the Higgs mechanism as a theoretical means to explain the weak boson masses,
and that of the Higgs particle as its possible experimental consequence are two
autonomous entities.  And this distinction in a limited sense holds still today,
i.e., after the Higgs boson discovery.  A technical clarification is in place here.
By the term "Higgs mechanism" we will understand a mechanism of providing masses
to vector bosons that is based on a particular method developed within the
framework of Quantum Field Theory, namely by absorption of scalar fields that thus
become the missing longitudinal degrees of freedom of the initially massless vectors.
It works regardless of the rest of the model and what triggers 
electroweak symmetry breaking in particular.  As a matter of fact, the idea was
initially introduced to particle physics in a somewhat different (and obsolete
by now) context.
Existence of a physical
scalar particle, the Higgs boson, can be regarded as a sufficient proof of
correctness of the Higgs mechanism to provide masses to $W$ and $Z$ bosons, and indeed
is the only proof of it available at hand and
within reach of contemporary particle accelerators.  
Still, a Higgs boson is far from being a {\it necessary} condition for the Higgs 
mechanism be true.  Rather, it is
a phenomenological consequence of a certain way of realization of
this mechanism.  Moreover, it does not definitely settle the question of whether this 
realization is the one and only, let alone to all its intrinsic details.
In particular, we do not know if the Higgs is really an elementary particle or rather
a composite object.  This alone may have important phenomenological implications
to be discovered at some higher energy.
The reasons for this will be elucidated further on in the next chapters.
Furthermore, a physical Higgs boson per se is a necessary, yet not a sufficient condition
to prove that the very same mechanism is responsible for the generation of all
the fermion
masses.  For that one needs to show that it indeed couples to fermions with a
strength that is proportional to the fermion mass, such as is
assumed in the Standard Model - which is nonetheless another autonomous issue to
explore.
Cut a long story short, this is the main theme of this work: the success is not yet
complete and if we want
to understand fully the details of the mechanism of electroweak symmetry breaking,
there is still a long way ahead of us.
But before we can proceed along these lines, let us first back
up to the very origins of the problem: the concept of intermediate vector bosons and
the issues they brought up.

This work is organized as follows.  In the next chapter, a brief
quasi-historical overview of electroweak unification is given, with
special emphasis on the Higgs boson as the long-missing ingredient that
completes the Standard Model (SM) of elementary particles.  The twofold role of the
Higgs boson in the
Standard Model is discussed.  Chapter 3 summarizes the current experimental
status of the Higgs boson and discusses the importance of the $VV$ scattering
processes, where $V = W, Z$, in the aftermath of Higgs discovery, particularly in 
relation to existing scenarios of physics beyond the Standard Model (BSM).
We also briefly overview other existing LHC results that directly or indirectly 
relate to the physics
of $VV$ scattering.  In chapter 4 a detailed study of the $VV$ scattering
process at the LHC from a phenomenological point of view is presented.
Formal signal and background definitions are given, and computational
methods and problems in the evaluation of the signal are discussed.  Also
discussed are detector-related backgrounds
and the relevant detector-specific capabilities and limiting factors that define
the size of such backgrounds in a real experiment such as at the LHC.  Our present
knowledge of these effects, derived from
various available detector-specific analyses, is summarized.
Finally, it is argued that same-sign $WW$ scattering in particular deserves special
attention.  Chapter 5 presents a selective review of existing literature on the
subject, from the early pre-LHC works which discussed $VV$ scattering mainly in
the context of Higgsless models, up to the most recent papers and post-Higgs
discovery developments.  Ultimately, all this knowledge is gathered to sketch
a tentative simulation-based analysis and present the up to date prospects for the
observarion of physics beyond the Standard Model via $VV$ scattering in the
LHC at $\sqrt{s}$ = 13 TeV and beyond it.  

Any studies of detector-related effects affecting
the evaluation of signal and backgrounds contained in this work are based solely on
those results which have been officially presented by
the relevant collaborations.
This work reviews a lot of earlier work published by many people, including the work of
theorists as well as results of experiments, but contains also an amount of
original studies and results that have not been published before.  For the latter,
nothing else than publicly available simulation and analysis tools have been used.
Conceptually, this work is a continuation of the analysis presented in Ref.\cite{doroba},
with substantial updates and improvements, and with a much extended scope.

\chapter{The Higgs boson in the Standard Model}

This chapter sketches a vaguely historical derivation of the Standard
Model and the Higgs boson as its key component.

\section{Issues of electroweak unification}

Four-fermion contact interactions cannot exist in any complete theory
of elementary particles.  This is because the corresponding coupling
constant is forced to have the dimension of cross section (i.e., inverse
energy squared, in units where Planck's constant is set to 1)
\footnote{In the framework of Quantum Field Theory, everything is measured in units
of some power of
energy.  The Lagrangian density has dimension 4, fermionic fields have
dimension 3/2, bosonic fields have dimension 1.},
while the cross section in the lowest order of
perturbative expansion must in turn be proportional
to the square of the coupling constant.  From simple dimensional analysis
it immediately follows that asymptotically, at energies much larger than the
masses of the particles involved, the total cross section is bound to be
quadratically divergent with energy.  Unbound amplitude growth inevitably leads
to unitarity violation at some energy.
Unitarity is an imperative property of quantum systems which ensures the sum of
probabilities of all possible final states evolving from a particular initial
state is always equal to 1, and so it must hold in any acceptable theory.
It is somewhat less straightforward to show that four-fermion interactions also
inevitably lead to non-renormalizable perturbation expansion, meaning that
calculations of decay rates and cross sections suffer of an increasing number
of divergences arising from Feynman diagrams involving loops, and ultimately making
the theory lose predictive power.
These facts were well known even while the Fermi theory of weak
interactions, governed by a coupling constant $G_F$ expressed in GeV$^{-2}$,
was the only existing one and indeed provided a good
description of existing data in the low energy region.
An improvement of Fermi's
model, inescapable from a purely theoretical point of view,
required the introduction of a force carrier, which necessarily
had to be a vector boson, in analogy to the photon of QED.  
Note that in this case
the lowest order weak interaction process is a second order process in
the coupling
constant which now governs the coupling of fermions to the intermediate
vector bosons.  It is easy to see that due to this the coupling constant itself
now becomes dimensionless, pretty much as the fine structure
constant of QED, thus eliminating the theoretical shortcomings associated
to the Fermi theory.  Indeed, the similarity of weak and electromagnetic
interactions suggested the possibility of a unified theory, where the photon
and the $W$ bosons were part of the same SU(2) multiplet.
However, the prerequisite of correspondence of the
intermediate vector
boson model with the Fermi theory at low energy has an important consequence:
the weak force carriers must have a non-zero mass in order to
describe a short-range interaction.
The success is only partial since non-zero mass is a source of two paramount
problems and these provide the two in principle independent ways to derive the
full Standard Model as we know it today.

For the sake of completeness one should mention here also 
another potential problem with electroweak unification, which resided in the
parity violating character of the weak interactions.  Parity violation effects
were first observed experimentally in 1957.  The solution 
consisted in enlarging the gauge group to SU(2)$\times$U(1) to involve parity violating
interactions.  This had to be followed by the introduction of another neutral gauge
boson, the $Z$.
We know today that {\it three} weak force carriers, $W^+$, $W^-$ and $Z$,
are necessary to describe accurately all the observed phenomenology of weak interactions.

\section{The Higgs boson from the principle of gauge invariance}

One problem with non-zero mass is related to the fact that SU(2)$\times$U(1) gauge
invariance forbids explicit mass terms for
gauge bosons.  Gauge boson masses must be introduced to the theory in a dodgy
way and here is where spontaneous symmetry breaking comes in.  The theory must
be written so that the Lagrangian exhibits required gauge invariance, but 
its lowest energy solution cannot.  This, however, poses another problem:
how to avoid massless spin-zero particles, excluded by experiment, which
according to the Goldstone theorem are bound to appear as a result of
spontaneous symmetry breaking?  In a paper of 1964, Higgs showed \cite{higgs} that the
Goldstone bosons need not physically appear in a relativistic theory when
a local symmetry is spontaneously broken.  Instead, they may turn a massless
vector field into a massive vector field.
Regardless of how the relevant concepts
actually evolved from the historical point of view, it seems in principle clear that
the simplest implementation of this requires adding extra scalar fields
to the theory.  Namely, at least {\it three} scalar fields, playing the role of
the would-be Goldstone bosons of the broken symmetry, are needed to endow the
three gauge bosons, $W^+$, $W^-$ and $Z$, with mass.  Expressed the idea
in simple words, each apparently
massless gauge field and an apparently massless scalar field need to combine to form a
massive vector field, while the total number of helicity states remains unchanged.
This is the essence of the so called Higgs mechanism, which more adequately is
also referred to as the Englert-Brout-Higgs-Guralnik-Hagen-Kibble mechanism
\cite{englertbrout} \cite{ghk}, to
properly honor the many contributors to the idea in its presently known shape.
There are no massless scalar fields left, as would be predicted by the Goldstone theorem.
In particle physicists'
jargon it is often said that those fields are ``eaten up".
Exactly why longitudinal polarization is intrinsically
connected with the emergence of non-zero mass will be explained further below.
For the time being let us stick now to the most fundamental question: how does
this happen?

\subsection{The Electroweak Chiral Lagrangian formalism}

Inclusion of three scalar fields is the minimum required to provide masses to three gauge
bosons.
This in itself carries no clues as to the origins of
symmetry breaking and moreover, it inherently demands some additional terms
in the Lagrangian to couple the three Goldstone bosons to the gauge bosons and
known fermions in a gauge invariant way.
Once again now, real history put aside,
the most general formalism that can be developed at this point,
derived from the basic principles of Effective Field Theory,
is known as the Electroweak Chiral Lagrangian (EWChL)
approach.  Without getting here into too much detail of the EWChL (more information
on the subject can be found, e.g., in Refs.~\cite{ewchl}), let us only recall
its main principles and basic features.  The main idea is one of having a low-energy
effective parameterization of a full theory expressed in a model independent way.
The general leading order (LO) Lagrangian in a practically useful form must
be SU(2)$_L\times$U(1)$_Y$-invariant and contain all
the Lorentz-, C- and P-invariant operators up to dimension 4
(in theorists' jargon this means the dimension of the fourth power of energy).
In such effective formulation the full Lagrangian
can be symbolically written down in the form:

\begin{equation}
\mathscr{L} = \mathscr{L}_{SM} + \mathscr{L}_{EWChL} =
\mathscr{L}_{SM} + \sum_i a_i \mathscr{O}_i.
\end{equation}

\noindent
where $\mathscr{L}_{SM}$ are
the familiar pieces that emerge from the Standard Model Lagrangian in the
infinite Higgs mass limit and $\mathscr{L}_{EWChL}$ is a collection of additional
dimension-4 operators expressed in terms of a 2$\times$2 unitary matrix $U$

\begin{equation}
U = exp(i \frac{\vec{\sigma}\vec{\pi}}{v}).
\end{equation}

\noindent
In the above, $\vec{\pi}$ is a triplet of scalar fields, $\vec{\sigma}$ are 
Pauli matrices and $v = (\sqrt{2} G_F)^{-1/2} \approx$ 246 GeV
\footnote{Note that this is exactly the quantity known as the Higgs vacuum expectation
value, but it does not have such interpretation in this framework}.
Numerical coefficients $a_i$ play the role of effective new couplings.  
There is no explicit fundamental Higgs field included and,
in the general case, the matrix is {\it non-linearly} parameterized with the three
fields $\vec{\pi}$.

Each specific set of coefficients $a_i$ reproduces the full
phenomenology associated to a given physical scenario.  
It can be shown that only 5 independent operators
account for SU(2)$_{L+R}$-conserving contributions.
These coefficients
may contribute to gauge boson self-energies ($a_1$), triboson couplings ($a_2$,
$a_3$) and effective four-boson couplings ($a_4$, $a_5$).  If the $a_i$'s are 
understood as
originating solely from new physics at a TeV scale, then their typical sizes are
expected to be of the order $10^{-3}-10^{-2}$.
Precision electroweak data prior to the Higgs discovery defined more
stringent experimental limits on their respective values.  Once data from LEP
became available, it was noticed that
there were actually only two dimension-4 operators left in this Lagrangian that could
modify the phenomenology related to the mechanism of electroweak symmetry breaking at
some higher energy without contradicting any of the existing low energy data
from the electroweak sector.  The numerical coefficients for these
operators are the ones traditionally denoted as $a_4$ and $a_5$.  
Thus, the relevant part of
the Lagrangian was

\begin{equation}
\mathscr{L}_{EWChL} = a_4 [{\tt Tr}(V_\mu V_\nu)]^2 + a_5 [{\tt Tr}(V_\mu V^\mu)]^2
\end{equation}

\noindent
where we have defined $V_\mu = U(D_\mu U)^\dagger$ and $D_\mu$ is the
electroweak gauge covariant derivative.
For reasons that will become
completely transparent in the next section, an effective modification of
the four-boson couplings could be realized in terms of Higgs exchange (if we do
not assume explicitly its existence in the model) or exchange of new, heavy particles.
Thus, our ignorance
of the electroweak symmetry breaking mechanism could be effectively shown in
terms of a two-dimensional ($a_4$, $a_5$) plane of which parts had been already excluded
on theoretical grounds and other parts remained unexplored.
The effective Lagrangians by themselves could only describe accurately the
electroweak physics at low energy.
They necessarily invoked some new physics to tackle the issues of 
renormalizability and unitarity.  Perturbative EWChL predictions can be extended
to higher energies using known techniques of unitarization, the two most commonly
known classes of them are called Pad{\'e} unitarization (a.k.a.~Inverse Amplitude
Method) and K-matrix unitarization (a.k.a.~N/D Protocol).
Typically this procedure leads to predictions of new resonances in the particle spectrum.
The entire nature: masses, widths, couplings and spins of those
resonances are in principle determined by the choice of ($a_4$, $a_5$), but in
practice some theoretical uncertainty related to the use of Chiral Perturbation
Theory is bound to be present and manifest in that quantitative predictions depend
on the unitarization method that had been chosen.  Here is where model-independence
ends, because unitarization scheme is part of a model.
This uncertainty becomes the larger the lighter the predicted resonances,
which certifies that the entire approach is for technical reasons mostly suited for
Higgsless scenarios (the term Higgsless here should be understood as anything not
involving a physical scalar lighter than, say, 700 GeV).
The entire formalism makes no a priori assumptions as to
the nature or dynamics of the gauge symmetry breaking mechanism.
It is interesting to notice that for a specific choice of parameters, the
Standard Model phenomenology could also be reproduced {\it in principle}. 
Correspondence of the EWChL with the Standard Model has been in fact
demonstrated, albeit only in the heavy Higgs limit \cite{ewchl2}.  This correspondence
is given by setting $a_4=0$ and $a_5$ being inversely proportional to the Higgs mass
squared.  However, the resonance widths obtained by applying
e.g.~the Inverse Amplitude Method are not exactly the same as the uniquely
determined - for a given Higgs mass - Standard Model Higgs widths.
Full correspondence
between soft and hard electroweak symmetry breaking has not been demonstrated,
at least within known unitarization schemes.
In any case, existence of a light scalar resonance makes this kind of
description of little practical use.

The EWChL approach
used to be an important theoretical
framework to study the effective phenomenologies in different scenarios
of electroweak symmetry breaking.  It allowed to confront their predictions
with those of the Standard Model with a light Higgs without running into
strict model-dependence or into
unphysicalness (like in the Higgsless Standard Model).
With the Higgs discovery, the minimum list of operators up to dimension 4
has been completed.
In principle the SM Higgs can be added to the EWChL by hand
and the same formalism can still be applied with this modification.
This is the simplest possible upgrade of the EWChL formalism and indeed some
studies of physics beyond the Standard Model have been carried in this language.
Coefficients $a_4$ and $a_5$ can be reinterpreted as modifications to SM
couplings which potentially induce simultaneous existence of heavier resonances
\cite{espriu}.
However, a somewhat different approach has nowadays become more popular.
The SM is built from
operators of up to dimension 4.\footnote{Strictly speaking, as long as we do not
include Majorana neutrinos.  Those can be generated via a dimension 5 operator.}
Extensions to the SM can be
parameterized in terms of higher dimension operators.
On this we will elaborate in a bit more detail in the next chapter.  
But for now let us still back up to the Standard Model.

\subsection{The Standard Model solution}

All the above being said, history went actually a different way.
At this point in history, Higgs and
independently Englert and Brout \cite{englertbrout}, had already come up with a
somewhat arbitrary and yet elegant idea of the exact mechanism that triggers
the gauge symmetry breakdown, that could solve the problem in a 
remarkably economical way compared to the technical complicacy of the EWChL
formulation.  The idea was effectively incorporated into the theory of electroweak 
interactions by Weinberg \cite{weinberg} and ever since then it became the core of
the Standard Model of elementary particles.  The concept did not
call for any new phenomenology, with just one exception: the Higgs boson.
And only one parameter suffices here for a complete quantitative description: the 
Higgs mass.
To understand the whole mechanism, let us first consider a toy model.
The following explanation is modeled on the one from Ref.~\cite{scholar}.
In relativistic field theory the simplest Lagrangian that
can realize spontaneous symmetry breaking is given by the addition of a
complex scalar field $\phi$ such that

\begin{equation}
\mathscr{L} = (\partial_\mu\phi)^* \cdot \partial^\mu\phi - V(\phi),
\end{equation}

\noindent
where

\begin{equation}
V(\phi) = \mu^2 \phi^*\phi + \lambda (\phi^*\phi)^2
\end{equation}

\begin{figure}[htbp]
\vspace{-6cm}
\begin{center}
\epsfig{file=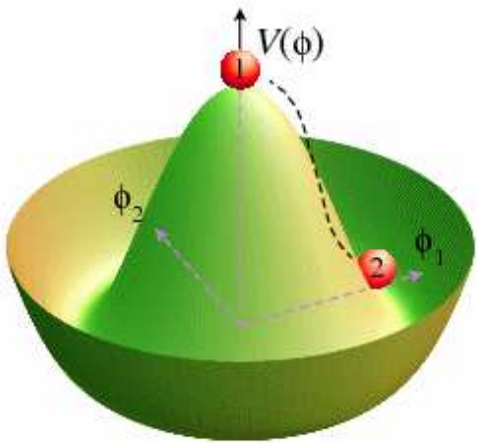,width=1.0\linewidth}
\end{center}
\vspace{-8cm}
\caption{The ``Mexican hat" potential $V(\phi)$ - the case of $\lambda>0$,
$\mu^2<0$.}
\end{figure}

\noindent
and $\mu$ and $\lambda$ are the mass and self-interaction coupling constant of the 
physical scalar particles related to the field.
The model is invariant under the global transformation
$\phi(x) \rightarrow \phi(x) e^{i\alpha}$.
If $\mu^2>0$, this model describes just a self-interacting scalar field and
nothing happens of special interest.
If however $\mu^2<0$, then $\phi=0$ is a local
maximum of the potential and therefore is bound to be an unstable state.
The minimum of potential $V$ has now the form of a circle defined by
$|\phi|^2=-\mu^2/2\lambda$.
In other words, in the ground state the value of $\phi$ is non-zero,
its magnitude being actually $v/\sqrt{2}$ with $v=\sqrt{-\mu^2/\lambda}$,
but with arbitrary phase.  
Thus, there will be a
degenerate family of vacuum states, accordingly to possible choices of
the phase angle $\alpha$.
By choosing a particular minimum, say the one where $\phi$ is real and positive,
one breaks the symmetry with respect to $\alpha$.
We can perform a Taylor series expansion around this location.  Defining two real
shifted fields $\phi_{1,2}$ such that

\begin{equation}
\phi = \frac{1}{\sqrt{2}} (v+\phi_1+i\phi_2),
\end{equation}

\noindent
the Lagrangian rewrites

\begin{equation}
\mathscr{L}=\frac{1}{2} [(\partial_\mu\phi_1)^2 + \partial_\mu\phi_2)^2] - V,
\end{equation}

\noindent
with

\begin{equation}
V = -\frac{1}{4} \lambda v^4 + \lambda v^2 \phi_1^2 +
\lambda v \phi_1 (\phi_1^2+\phi_2^2) +
\frac{1}{4} \lambda (\phi_1^2+\phi_2^2)^2.
\end{equation}

Because we have defined $\phi_1$ and $\phi_2$ so that the vacuum corresponds
to a non-zero value of $\phi_1$ only,
the model describes effectively two kinds of particles: $\phi_1$ of
mass $\sqrt{2\lambda}v$ and the massless $\phi_2$, along with 
their respective triple and quartic couplings.
Particle $\phi_2$ is then the Goldstone boson related to breaking the initial 
symmetry of the system as a result of
the spontaneous choice of a vacuum state, while
$\phi_1$
is an extra massive scalar particle, prototype of the Higgs boson.

Now comes the Higgs mechanism.
By adding a massless gauge field into the picture, e.g., the electromagnetic
field with potential $A_\mu$, the
Lagrangian of the model can be expressed as

\begin{equation}
\mathscr{L} = 
(D_\mu\phi)^*D^\mu\phi - \frac{1}{4} F_{\mu\nu}F^{\mu\nu} - V(\phi),
\end{equation}

\noindent
where we can explicitly define the covariant derivative as
$D_\mu\phi = \partial_\mu\phi - ieA_\mu\phi$, and
$F_{\mu\nu} = \partial_\mu A_\nu - \partial_\nu A_\mu$
is the electromagentic field tensor.
This Lagrangian is invariant under the local gauge transformations

\begin{equation}
\phi(x) \rightarrow \phi(x)e^{i\alpha(x)}, 
\end{equation}
\begin{equation}
A_\mu(x) \rightarrow A_\mu(x)+\frac{1}{e}\partial_\mu\alpha(x).
\end{equation}

\noindent
Expansion, as before, around the chosen vacuum, yields a term of the form

\begin{equation}
\mathscr{L} = ... + \frac{1}{2}(\partial_\mu\phi_2-evA_\mu)^2 + ...,
\end{equation}

\noindent
which is nothing but a mass term of an effective vector field $B_\mu$ defined as

\begin{equation}
B_\mu = A_\mu - \frac{1}{ev}\partial_\mu\phi_2,
\end{equation}

\noindent
with a mass equal to $ev$.  There is no Goldstone boson left.  Instead, the
gauge field acquired mass by interaction with the scalar field $\phi_2$.

So what does it all have to do with providing masses to $W$ and $Z$ bosons
while leaving the photon massless in a way that does not violate gauge symmetry
of the Standard Model?
The key feature behind implementing this idea within the context of the Higgs 
mechanism resided in postulating a fourth scalar field, in addition
to the three would-be Goldstone bosons discussed before, for the
formation of an SU(2) isospin doublet of complex scalar fields, usually
denoted as 

\begin{equation}
\Phi = \frac{1}{\sqrt{2}} \left( \begin{array}{cc} \phi^+ \\ \phi^0 \end{array} \right) =
\frac{1}{\sqrt{2}} \left( \begin{array}{cc} w_1 + i w_2 \\ h + i z \end{array} \right).
\end{equation}

\noindent
Contrary to the general EWChL case, here a {\it linear} parameterization in the
scalar fields is assumed.
The vacuum is chosen so that $\langle 0|\Phi|0\rangle = \frac{1}{\sqrt{2}} (0, v)^T$,
which means
that it carries a non-vanishing value of the neutral $h$ field.  Here $v$ is
exactly the same quantity we have introduced in the previous section.  Note that
consequently this vacuum carries a weak charge, but no electromagnetic charge.  In
this case
not the whole SU(2)$\times$U(1) symmetry is broken.  There is an unbroken subgroup 
related to the fact that $\Phi$ does not interact with the photon.  Like before,
massless fields $w_1$, $w_2$ and $z$ are absorbed to form mass terms for the 
apparenty massless weak bosons.  The masses are effectively given by:

\begin{equation}
M_W = \frac{gv}{2} = \sqrt{\frac{\pi\alpha}{\sqrt{2}G_F}}\frac{1}{sin\theta_W}, 
\end{equation}

\begin{equation}
M_Z = \sqrt{\frac{\pi\alpha}{\sqrt{2}G_F}}\frac{1}{sin\theta_Wcos\theta_W}.
\end{equation}

\noindent
Here we have introduced the ``electroweak mixing (Weinberg) angle" defined via

\begin{equation}
sin\theta_W = e/g,
\end{equation}

\noindent
the ratio of the original electromagnetic and the weak coupling constants.
The photon remains massless.  Since the value of $sin\theta_W$ may be determined
experimentally, e.g., from a measurement of fermion scattering processes,
the above formulae represent in fact a {\it prediction} for the $W$ and $Z$ masses.
The fourth field $h$
is needed to trigger spontaneous symmetry breaking by its non-vanishing
vacuum expectation value.  As a result, the
shifted field $H = h-v$ becomes a physical, massive, self-interacting scalar -
the Standard Model Higgs boson.  The mass of the Higgs boson is given by

\begin{equation}
M_H = \sqrt{2\lambda v^2}
\end{equation}

\noindent
and hence it is not known a priori without knowledge of $\lambda$.
However, everything else in the theory is completely determined or at least
calculable.

Rewriting the Lagrangian in terms of physical particles, we see that
the Higgs couples to the gauge bosons:

\begin{equation}
\mathscr{L} = ... + M_W^2 \cdot W^{+\mu}W^-_\mu \cdot (1+H/v)^2 + \frac{1}{2} M_Z^2 \cdot
Z^\mu Z_\mu \cdot (1+H/v)^2 + ...
\end{equation}

\noindent
with a coupling proportional to the mass squared.
As a byproduct, it also couples to fermions generating their mass terms:

\begin{equation}
\mathscr{L} = ... + \sum_f m_f f \bar{f} \cdot (1+H/v)^2 + ...
\end{equation}

\noindent
By construction, the coupling to fermions is proportional to the fermion masses.

This completes the Standard Model of elementary particles and fundamental
interactions, the most successful theory in modern physics, from the point
of view of gauge invariance.

\section{The Higgs boson from the principle of unitarity}

The second reason why non-zero mass is a problem concerns polarization and the
issue of unitarity.  As already mentioned, the
unitarity condition is equivalent to the requirement of the sum of
probabilities of all possible final states evolving from a particular initial
state be always equal to 1.  This
sum of probabilities must be in principle calculated to infinite order in
perturbative expansion, which is of course impossible to achieve.
For the technical issues regarding the concept of unitarity and
the connection between unitarity and
renormalizability, the reader is referred to more topical literature, e.g.,
Ref.~\cite{horejsi}.
In the following we will use the commonly accepted practical criterion of
{\it tree} unitarity which demands for any $2 \rightarrow 2$ process
predicted by the theory its tree level amplitude be asymptotically at most flat
with energy.  Discussion on the validity of this criterion can be found in
Ref.~\cite{horejsi}.

Let us derive the essence of the problem in detail, as this
constitutes the theoretical basis of our proper subject.  In relativistic Quantum
Field Theory, a massive vector boson
can be described in terms of a wave function $B^\mu$ whose form is a wave-plane solution
of the Klein-Gordon equation:

\begin{equation}
B^\mu (x) = C \epsilon^\mu(p) e^{-ipx}
\end{equation}

\noindent
with the so called Lorenz condition
\footnote{Not Lorentz condition, as is often erroneously called.}, 
$\partial_\mu B^\mu = 0$, imposed.
Here $p$ is the four-momentum of the particle, C is a normalization constant whose
value is inessential at the present moment and $\epsilon^\mu$ is the polarization
vector corresponding to the plane wave.  In the boson rest frame, $\epsilon^\mu$
can be decomposed into the individual Cartesian coordinates, where

\begin{equation}
\epsilon^\mu_x = (0, 1, 0, 0),
\end{equation}
\begin{equation}
\epsilon^\mu_y = (0, 0, 1, 0),
\end{equation}
\begin{equation}
\epsilon^\mu_z = (0, 0, 0, 1),
\end{equation}

\noindent
and these correspond to the three possible linear polarization states.
We have already taken advantage of the fact that the zeroth component is bound to
be zero by the Lorenz condtion.  Alternatively, we may define two linear combinations
of $\epsilon^\mu_x$ and $\epsilon^\mu_y$,

\begin{equation}
\epsilon^\mu_+ = \frac{1}{\sqrt{2}}(0, 1, i, 0),
\end{equation}
\begin{equation}
\epsilon^\mu_- = \frac{1}{\sqrt{2}}(0, 1, -i, 0),
\end{equation}

\noindent
which correspond to two possible circular polarization states.  Let us now suppose
the boson moves along the $z$ axis.  The quantities $\epsilon^\mu_+$ and
$\epsilon^\mu_-$ will now stand for the two degrees of freedom of polarization
transverse to the boson direction, while $\epsilon^\mu_z$ will become the
longitudinal polarization and be further on denoted as $\epsilon^\mu_L$.  
Translated into the language of helicity, i.e.,
the projection of the boson's spin onto its direction of motion,
$\epsilon^\mu_+$, $\epsilon^\mu_L$ and $\epsilon^\mu_-$ correspond to
helicities +1, 0 and -1, respectively.

The general expression for the three components of $\epsilon^\mu$ for a boson
with mass M, energy E and 3-momentum $p_z$ directed along the $z$ axis
can be simply found 
by applying Lorentz transformation.  It is however fully sufficient to
consider that the transverse polarizations need not change, while $\epsilon^\mu_L$,
by definition directed along the momentum 3-vector, must be of the form

\begin{equation}
\epsilon^\mu_L = (a \frac{|\vec{p}|}{E}, a \frac{\vec{p}}{|\vec{p}|}) =
(a \frac{p_z}{E}, 0, 0, a),
\end{equation}

\noindent
where $a>0$.  The normalization condition readily yields $a = E/M$, hence

\begin{equation}
\epsilon^\mu_+ = \frac{1}{\sqrt{2}}(0, 1, i, 0),
\end{equation}
\begin{equation}
\epsilon^\mu_- = \frac{1}{\sqrt{2}}(0, 1, -i, 0),
\end{equation}
\begin{equation}
\epsilon^\mu_L = \frac{1}{M}(p_z, 0, 0, E).
\end{equation}

\noindent
One can quickly verify that the above indeed satisfies the Lorenz condition
expressed as $p_\mu\epsilon^\mu = 0$.

In this moment we have arrived at a very important conclusion.
The requirement of $\epsilon_L \neq 0$ makes sense only if $M \neq 0$.
For a massless boson there is no solution satisfying the Lorenz condition
that would correspond to longitudinal polarization.  And this is why, in the
languauge of relativistic Quantum Field Theory, on-shell
photons are purely transverse.  More generally, for a massless boson of spin $J$,
Lorentz invariance forbids other helicities than $+J$ and $-J$.

\begin{figure}[htbp]
\vspace{1cm}
\begin{center}
\epsfig{file=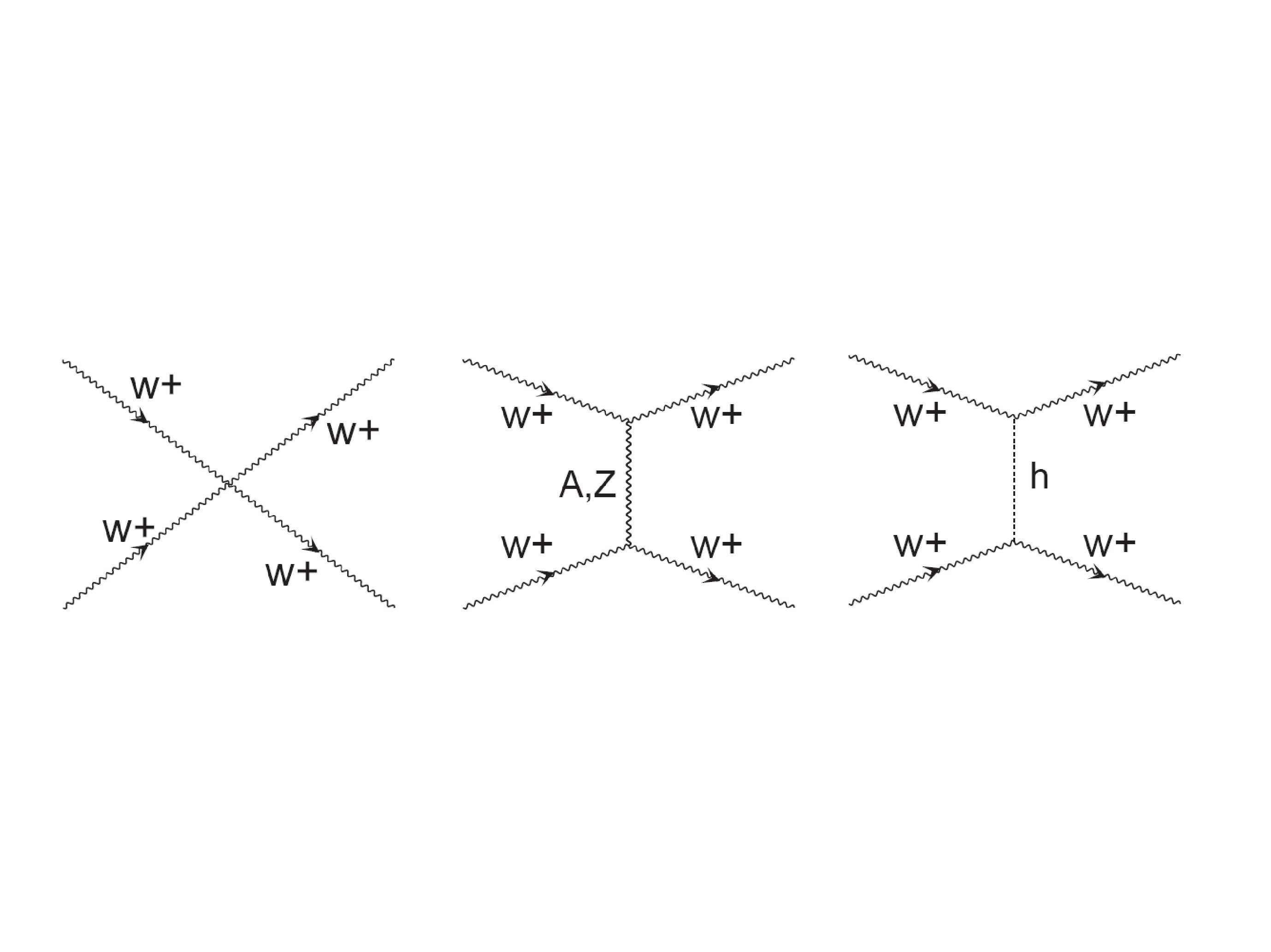,width=0.8\linewidth}
\end{center}
\caption{Feynman diagrams for the Standard Model process $W^+W^+ \rightarrow W^+W^+$:
the four-$W$ contact interaction graph, the $\gamma/Z$-exchange graph and the
Higgs exchange graph.}
\end{figure}
\vspace{0.5cm}

The form of $\epsilon_L$ defines its key feature which lies in its energy dependence.
It is clear that
at energies much larger than the boson mass, it grows indefinitely with energy,
like $\epsilon_L \sim E$, being a source of 
potentially fatal misbehavior
of the gauge boson sector.  To elucidate the problem, let us consider a simple
scattering process involving
two on-shell, same-sign, longitudinally polarized $W$ bosons:

\vspace{4mm}

\begin{center}
$W_L^+W_L^+ \rightarrow W_L^+W_L^+$.
\end{center}

\vspace{3mm}

In the lowest order, three subprocesses readily contribute to
this process: the four-$W$ contact interaction and $t$-channel (space-like)
photon and $Z$
exchange.  The amplitude of the contact interaction part must be proportional to

\begin{equation}
\mathscr{M} \sim \epsilon_L \epsilon_L \epsilon_L \epsilon_L \sim s^2,
\end{equation}

\noindent
where $s$ is the center of mass energy squared of the interacting bosons.  
Obviously it diverges like the fourth power of energy and so, paradoxically, the
electroweak theory leads to similar difficulties as did before the Fermi theory.
Even worse at first sight may look the diagram involving $t$-channel $Z$ exchange, as
one may expect in this case the leading divergence to be like $\sim E^6$ from
an appropriate combination of all the longitudinal $W$ and $Z$ components.
It can be shown, however,
that the longitudinal part of the $Z$ propagator vanishes and the full
contribution from the $t$-channel photon and $Z$ exchange in fact also goes like
$s^2$ in the leading term.  Moreover, by appropriate choice of the coupling
constant for the four-$W$ contact interaction, which in practice is secured by
the Standard Model gauge invariance, the two leading terms can be made to
cancel each other exactly.  It is worth to remember this point, since it will
come back to us in further considerations.  The triple gauge boson couplings,
$WWZ$ and $WW\gamma$, are well
constrained by experiment and we need not consider their variation at this point.
The same can hardly be told of the quartic couplings which remain largely unconstrained
from the experimental point of view.
Altogether, there are four quartic boson couplings
allowed in the Standard Model: $WWWW$, $WWZZ$, $WWZ\gamma$ and $WW\gamma\gamma$,
and their values within the model are completely specified.  
The $WWWW$ coupling in itself can be probed experimentally
at the LHC in an independent way, via measurements of triboson production.
Generally, it is expected that new physics may manifest
itself in changes of the effective quartic couplings relative to the Standard
Model and therefore alter the Standard Model predictions for triboson production,
{\it as well as} the high energy behavior of $WW$ scattering amplitudes.
For the sake of this chapter we will assume that
quartic couplings correspond exactly to their Standard Model values.
Consequently, we are left with

\begin{equation}
\mathscr{M}_{Gauge} = -g^2 \frac{s}{4M_W^2}+\mathscr{O}(s^0).
\end{equation}

Technically, this still implies unitarity violation and non-renormalizability.
As usual in particle physics, such problems are fixed by postulating new
particles and appropriate interactions to produce counter-terms that will
cancel the unwanted divergences.  Because of the form of the scalar propagator,
the inclusion of a neutral scalar particle $H$ that can be exchanged between the
two $W$ lines will result in an additional term

\begin{equation}
\mathscr{M}_{H} = g^2_{HWW} \frac{s}{M_W^4}+\mathscr{O}(s^0).
\end{equation}

From dimensional analysis it follows that the coupling constant that governs the
interaction of $H$ with the $W$ boson must have the dimension of energy.  By
looking at the expressions for $\mathscr{M}_{Gauge}$ and $\mathscr{M}_{H}$ one
easily notices that
the leading divergences will cancel out {\it exactly}
if and only if the condition $g_{HWW} = g M_W$ is {\it exactly} fulfilled.
Recalling that $g$ in itself is related to
the $W$ mass, this in particular means that the scalar $H$ must couple to
the $W$ proportionally to $M_W^2$.
We already have such candidate: it is the Standard Model Higgs boson.  Indeed, tedious
calculations within the framework of the Standard Model yield the asymptotic result

\begin{equation}
\mathscr{M}_{Gauge} + \mathscr{M}_{H} = g^2 \frac{M_H^2}{4M_W^2}
\end{equation}

\noindent
at energies much larger than the Higgs mass.

The same arguments apply to the opposite sign $W$ boson scattering process

\vspace{4mm}

\begin{center}
$W_L^+W_L^- \rightarrow W_L^+W_L^-$.
\end{center}

\vspace{3mm}

\noindent
In this case we have to take into account additional diagrams corresponding to
$s$-channel (time-like) photon
and $Z$ exchange, as well as an $s$-channel Higgs exchange diagram.
Without repeating the main points nor getting into detailed calculations
we can immediately write down the final results for the corresponding amplitudes:

\begin{equation}
\mathscr{M}_{Gauge} = -g^2 \frac{u}{4M_W^2}+\mathscr{O}(s^0),
\end{equation}

\begin{equation}
\mathscr{M}_{H} = g^2_{HWW} \frac{u}{M_W^4}+\mathscr{O}(s^0).
\end{equation}

\noindent
where $u$ is the familiar Mandelstam variable and we have used the
high energy approximation $s+t+u=0$.

Similarly, to the process of $W^\pm Z$ scattering

\vspace{4mm}

\begin{center}
$W_L^\pm Z_L \rightarrow W_L^\pm Z_L$,
\end{center}

\vspace{3mm}

\noindent
the lowest order diagrams that contribute are the $WWZZ$ contact interaction,
$s$- and $t$-channel $W^\pm$ exchange and $t$-channel Higgs exchange.
And likewise, the divergence resulting from the sum of the former three
is exactly canceled by the Higgs exchange diagram in the SM.

With the $ZZ$ scattering process the question is seemingly different, since in the SM
it can only occur via Higgs exchange (both $s$- and $t$-channel).  
However, in any real hadron-hadron experiment this
process cannot be separated from the dominant $W^+W^- \to ZZ$ process, where
three additional graphs contribute in the lowest order, including the
$WWZZ$ contact interaction, $t$-channel $W^\pm$ exchange and $s$-channel
Higgs exchange.  Once again here, Higgs exchange provides cancelation of
unwanted divergences.

By introducing a Higgs boson with appropriate couplings to other particles,
unitarity in the theory is established.  This in turn completes the Standard Model
from the point of view of the unitarity principle.  But there is more here.
A Higgs boson is necessary before the energy scale of unitarity violation.
A Higgs that is too heavy is useless in the SM.  From these considerations an upper
bound on the Higgs mass \cite{leequigg} was derived way before its actual observation.

\chapter{Standard Model experimental status and prospects for BSM}

The LHC has finished Run 1.
Both ATLAS and CMS have produced their preliminary (now every day closer to
being final) results based on combinations of the entire datasets from 7 TeV 
and 8 TeV.  Even if some of the results that have been published until now are not yet
to be considered final,
the most important findings are unlikely to change
significantly until the LHC is restarted again with a higher energy (13 TeV)
and collects enough new data.  To discuss physics of Run 2 of the LHC,
it is important to realize what exactly has
become known from Run 1 and within what uncertainty margins,
then how these uncertainty margins translate into the
potential of new discoveries in the forthcoming years.  This is of course
true not only for $VV$ scattering,
but for the entire LHC physics.  But the relation between the Higgs boson and $VV$
scattering is special and so this dependence is here even more strict.
This chapter will review our current, most up to date,
knowledge about the Higgs boson and summarize other measurements with
direct or indirect impact on the physics of $VV$ scattering in the next years.

\section{Higgs boson experimental status}

Four well known mechanisms of Higgs production at the LHC are: gluon-gluon fusion
via heavy quark loops, Vector Boson Fusion (VBF),
Higgsstrahlung off a gauge boson and heavy quark fusion (also called
$t\bar{t}$- or $b\bar{b}$-associated production).  Their relative importance
varies with the Higgs mass and the kind of physics we want to study, to a
lesser degree with the actual proton beam energies.  For a Higgs mass in the
vicinity of 125 GeV, gluon-gluon fusion is by far the dominant production mode, with VBF
contributing roughly an order of magnitude less and the other modes less still.
For Higgs-like resonance masses above the $t\bar{t}$ threshold, the relative
amounts of gluon-gluon fusion and VBF become gradually closer to unity, up to 
the point of the latter becoming over 1/3 of the total cross section at around
1 TeV.

On the other end of the Higgs boson, the relative importance of different decay
modes is driven by the respective mass thresholds for the decays into heavy
particles.  For $M_H < 2M_W$, as is indeed the
case for the Standard Model Higgs, decays to fermions like the $b$ or $c$ quarks or
to the $\tau$ leptons are strongly preferred
as far as raw branching fractions are concerned.  Background and event reconstruction
efficiency issues define
however $ZZ$ and $WW$ as being among the leading light Higgs decay modes to study,
the only other fully competitive channel being in fact $\gamma\gamma$ which occurs
solely via loop corrections.
In the Higgs-like resonance mass region above 150 GeV, decays into $WW$ and $ZZ$
become just about the only
relevant ones, and this assertion changes only marginally on the opening of
the $t\bar{t}$ channel for masses larger than 350 GeV.

Our process of interest is intrinsically connected with Vector Boson Fusion
followed by decay into a pair of vector bosons.  In the resonance region it is quite
identical with it and so the $W^+W^-$ and $ZZ$ scattering modes are naturally the most
widely studied to date.  Of course, in an experiment we only know the bosons in the
final state.  The process $ZZ \rightarrow ZZ$ is in principle the most direct probe
of the Higgs boson, as it only proceeds via Higgs exchange in the Standard Model,
but it cannot be separated from $W^+W^- \rightarrow ZZ$.
Specific VBF analyses have been performed in the low Higgs
mass range and will be the basis for future heavy resonance searches at 13 TeV.
These studies have come up with a typical experimental VBF signature to search for.
It consists
of two energetic forward jets and all the final Higgs decay products
usually well isolated in the barrel region of the detector, the two
direct decay products being typically reconstructed in opposite hemispheres.  The purely
electroweak character of the process means no QCD color flow occuring in the event
and reflects in a large rapidity gap between the two leading jets.  The typical
VBF signature used in Higgs searches does not explicitly discriminate between the gauge
boson polarizations.  Indeed such discrimination is impractical in a kinematic regime
where at least one of the gauge bosons must be off-shell.
The spin and parity of the resonance can be nonetheless
determined afterwards from the angular distributions of the decay products, where
naturally the $ZZ$ channel keeps the most complete information available in
the detector.

\begin{figure}[htbp]
\vspace{-7cm}
\begin{center}
\epsfig{file=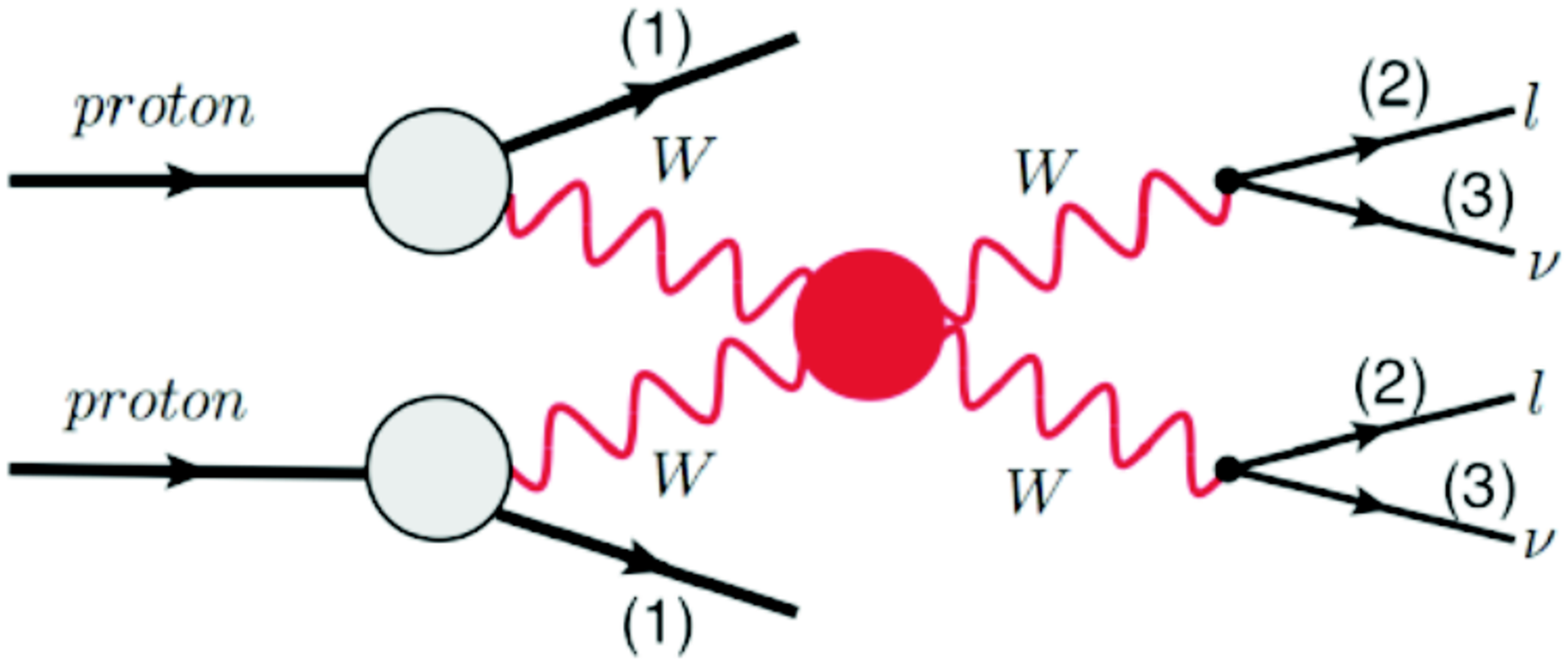,width=0.9\linewidth} \\
\vspace{-10cm}\epsfig{file=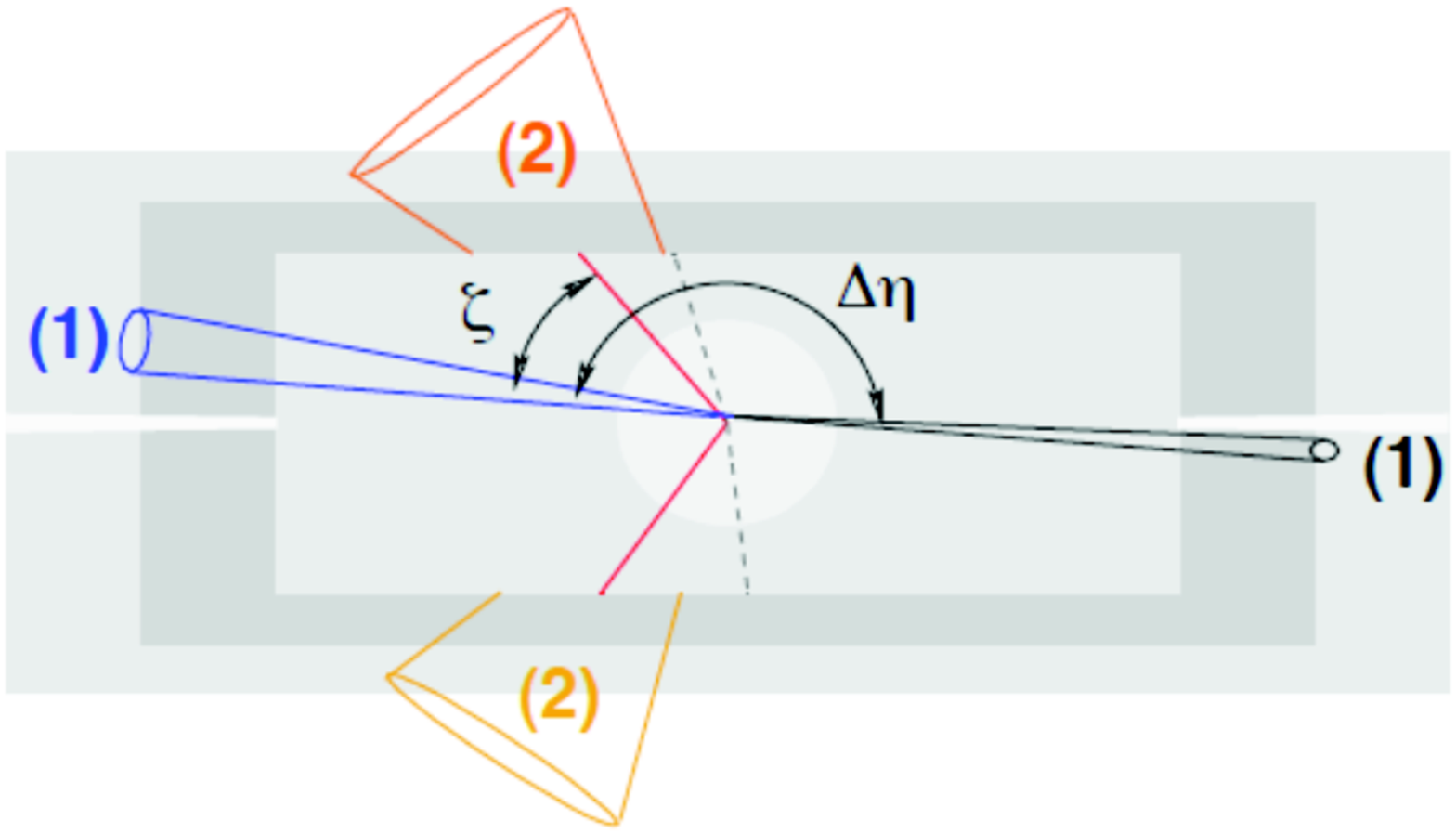,width=0.55\linewidth}
\epsfig{file=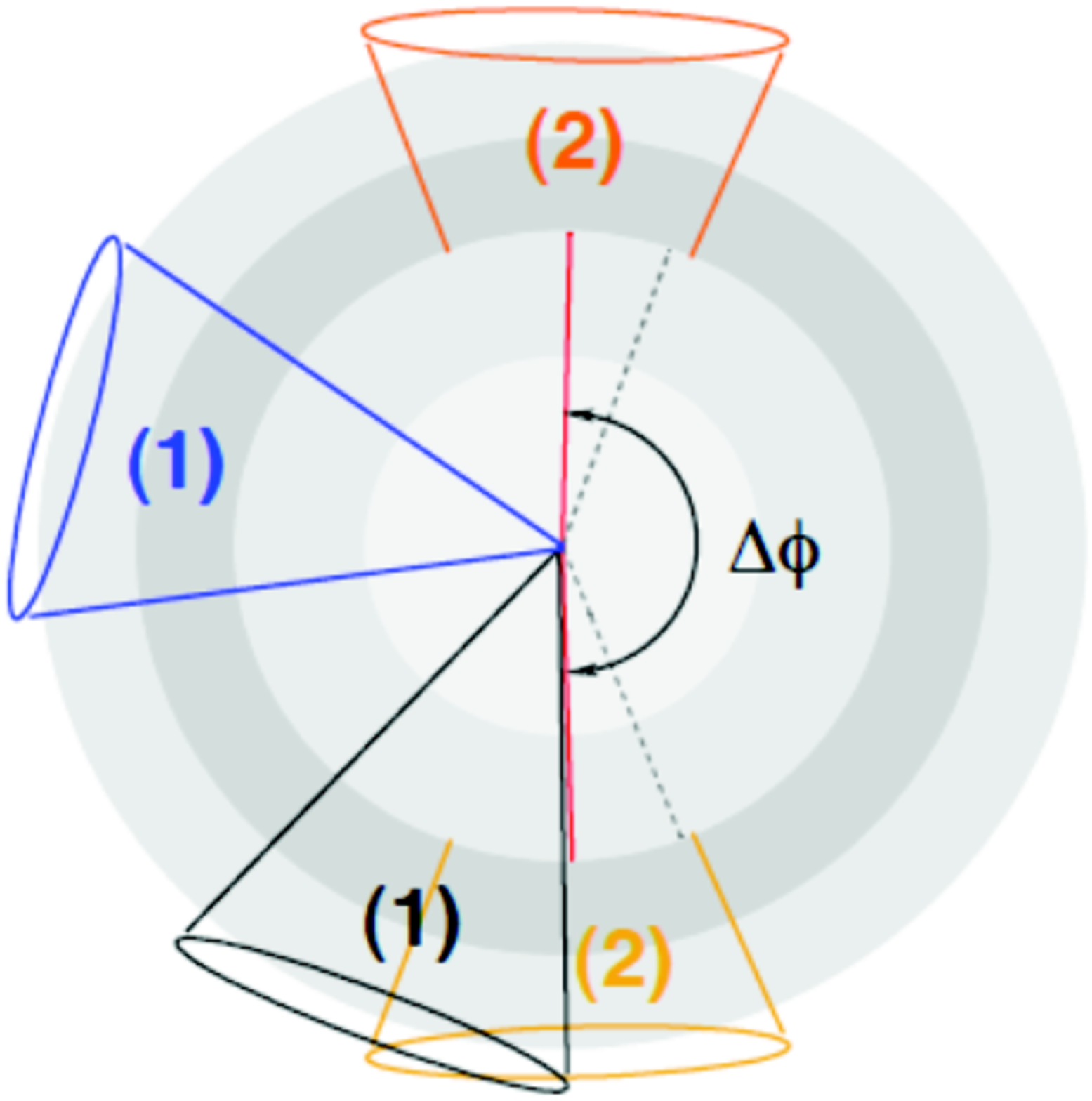,width=0.35\linewidth}
\end{center}
\vspace{-2cm}
\caption{The Vector Boson Scattering process at the LHC: a schematic drawing (top)
and two views of its basic signature in an LHC detector (bottom).  A purely
leptonic $WW$ decay channel is assumed.  A typical signature consists of
two forward high energy jets (labeled (1)) with a large pseudorapidity gap,
and two central leptons (labeled (2)) with a large gap in the azimuthal angle.
Quantities like $\Delta\eta$ and $\Delta\phi$ are instrumental in isolating
the process from the bulk of the background.}
\end{figure}

\subsection{Higgs mass, width and decay modes}

Higgs signal has been independently observed with more than 5$\sigma$ significance,
in both ATLAS \cite{atlashiggs} and CMS \cite{cmshiggs}, in 
two decay modes: $H \rightarrow ZZ^* \rightarrow l^+l^-l^+l^-$ (CMS: 6.5$\sigma$,
ATLAS: 6.6$\sigma$) and
$H \rightarrow \gamma\gamma$ (CMS: 5.6$\sigma$, ATLAS: 7.4$\sigma$).  A third 
bosonic decay mode, $H \rightarrow W^+W^-$,
comes close (CMS: 4.7$\sigma$, ATLAS: 4.1$\sigma$).
Observed significances agree with SM expectations.

The Higgs boson mass has now been precisely determined from a combination of data from
the two most sensitive decay modes which not unexpectedly also provide the best
mass resolution: $H \rightarrow 4l$ and $H \rightarrow \gamma\gamma$.
Its final values have been reported to be:

\vspace{3mm}

$M_H = 125.36 \pm 0.37 (stat) \pm 0.18 (syst)$ GeV (ATLAS) and

\vspace{2mm}

$M_H = 125.03 ~^{+~0.26}_{-~0.27} (stat) ~^{+~0.13}_{-~0.15} (syst)$ GeV (CMS).

\vspace{3mm}

\noindent
Higgs masses determined from the two channels separately are in satisfactory agreement
at CMS, with 
the final mass difference being quoted as
$M_H^{\gamma\gamma} - M_H^{4l} = -0.87 ~^{+~0.54}_{-~0.59}$ GeV.
ATLAS observed a marginally larger difference whose statistical significance is
likewise weak, $M_H^{4l} = 124.51 \pm 0.52$ GeV
vs.~$M_H^{\gamma\gamma} = 125.98 \pm 0.50$ GeV.
It should be noted that, as far as the central values are concerned,
$M_H^{4l} > M_H^{\gamma\gamma}$ for CMS, but
$M_H^{\gamma\gamma} > M_H^{4l}$ for ATLAS, which clearly favors statistical and
systematic uncertainties rather than physics as the most plausible interpretation
of any possible mass shifts.  There is no experimental support to the
idea of there actually being two nearly degenerate resonances, at least within
the present resolutions.

A study of Higgs decays into $\tau^+\tau^-$
revealed independent evidence of a Higgs signal at the 4.5$\sigma$ level in ATLAS
\cite{atlash2tautau} and at the 3.2$\sigma$ level in CMS \cite{cmsh2tautau}, 
both being compatible with
the expectations for a $\sim$125 GeV Standard Model Higgs boson and hence strongly
suggesting that the Higgs indeed does couple to fermions.  On the other hand,
no other fermionic decay has been firmly and directly established on its own.
The ones that have been directly searched for are the decays to
$b\bar{b}$, $\mu^+\mu^-$ and more recently $e^+e^-$ \cite{h2ff}.  
Analysis of the former does indeed reveal hints at a
roughly 2$\sigma$ level, in consistency with SM expectations.
A CMS combination of data from the two most important fermionic Higgs decays:
$\tau^+\tau^-$ and $b\bar{b}$, does not yet reach the
5$\sigma$ significance level \cite{cmsh2fermions}.
With the amount of data collected so far,
lack of signal observation in decays to lighter
fermions is fully consistent with the SM.
Of course, decays $H \rightarrow \gamma\gamma$ occurring at a rate roughly
consistent with the Standard Model indirectly suggest that 
Higgs couples to the top quark, too.  Moreover, theory predicts the main Higgs
production mechanism be gluon-gluon fusion via top quark loops and so the total
Higgs production rate
is driven predominantly by the Higgs coupling to the top.  In other words,
the simple observation of total Higgs production
occurring at a rate roughly consistent with the SM is another (and actually,
the strongest), albeit indirect, confirmation that the Higgs couples to fermions.

Certain rare Higgs decays predicted by the Standard Model have been
searched for as well.
Measurement of the rate of Higgs decaying into,
e.g., $Z\gamma$ would be a very interesting test of the Standard Model,
but so far data are of not enough statistical precision to do so \cite{h2zgamma}.

The width of the Higgs boson in the SM is fully determined by its mass.
For a 125 GeV Higgs, the expected width is close to 5 MeV, which is unfortunately
far beyond present experimental resolution.  From an analysis of data
in the resonance region of the 4-lepton
decay channel, CMS found the observed resonance width in agreement with the
detector resolution width and placed a 95\% CL upper bound on the intrinsic Higgs
width at 3.4 GeV.  A novel method has been proposed to constrain the Higgs
width by examination of the 4-lepton mass spectrum away from the Higgs peak
\cite{hwidth}.
In the dominant gluon fusion process, Higgs off-shell production and decay into
4 leptons gets enhanced due to the proximity of the $Z$ pair production threshold.
The ratio of cross sections for off-shell and on-shell Higgs production,
$\sigma (gg \to H^* \to ZZ)/\sigma (gg \to H \to ZZ^*)$
is directly proportional to the Higgs width.  Using this technique,
CMS placed a much better upper bound on it at 22 MeV (95\% CL) \cite{cmshwidth}.

\subsection{Higgs spin and parity}

Crucial to the identification of the 125 GeV resonance with the SM Higgs
is determination of its spin and parity.
The SM Higgs boson has spin-parity $J^P=0^+$.
Different spin-parity hypotheses of the observed Higgs-like resonance
have been severely constrained by the data.
The spin and parity of the Higgs resonance can be independently analyzed
in each decay mode, based on angular distributions of the respective decay products.
In CMS, this has been achieved so far using the three leading decay modes
\cite{cmshspin}:
$H \to ZZ^* \rightarrow l^+l^-l^+l^-$, $H \to W^+W^- \to l^+l^-\nu\nu$ and
$H \to \gamma\gamma$ \footnote{The possibility of providing additional evidence
based on the $\tau^+\tau^-$ decay mode is being studied}.  By far the most
sensitive of them is the 4-lepton channel.
Strictly speaking, $J^P$ is not {\it measured}, only
the likelihood of different hypotheses can be determined relative to each other
(one can of course argue that such procedure qualifies as being a measurement).
Each $J^P$ hypothesis translates into specific predictions of the angular
distributions that are computed directly from
the corresponding matrix elements.  For every pair of hypotheses their relative
likelihood of consistency with the data can then be quantified.
The procedure is more likely to end up in a conclusive result only as long as one
of the two selected hypotheses is the correct one (and the other incorrect).  Thus,
in practice, each non-standard $J^P$ hypothesis is tested against the $J^P=0^+$
hypothesis.  A $q$ value is then determined from data that is equal to the relative
likelihood of the tested hypothesis against the reference $J^P=0^+$ case.
Statistical significance of each result is determined by comparing
the single $q$ value obtained from data with its predicted probability distributions
that are calculated under the assumptions that one or the other $J^P$ hypothesis is
correct.  The respective probability distributions are obtained from a number
of simulated ``fake" experiments.  The hypotheses that have been tested
include $J^P=0^+_h$ (scalar with higher order couplings),
$0^-$ (pseudoscalar), $1^\pm$ (vector and pseudovector), $2^\pm_m$ (tensor
and pseudotensor with minimal couplings to SM particles - a graviton analogue)
and $2^\pm_h$ (tensor and pseudotensor with higher order couplings).
In addition, a maximum likelihood function can be defined in which a mixed
$J^P$ state is allowed, e.g., $0^+$ with $0^-$.

Analysis of the 4-lepton channel is based on a technique described in detail
in literature \cite{spindet}.
It exploits information on five angles that characterize the decay: two angles describe
the orientation of the decay plane of one $Z$ boson in the lab, a third angle
the relative azimuthal orientation of
both $Z$ decay planes and the last two angles describe the two $Z$ decays
in the respective $Z$ rest frames.  The 4-lepton channel alone allows to
reject all of the tested hypotheses at a confidence level (CL) greater than 95\%.
In the $H \to \gamma\gamma$ channel, Higgs spin correlates to
the polar angle of the $\gamma\gamma$ pair in the Higgs rest frame.
According to the Landau-Yang theorem, decays of a massive
vector into a couple of massless vectors are forbidden, so $J=1$ is here excluded.
All spin-zero scenarios
produce an identical isotropic $\gamma\gamma$ distribution and therefore
the $J^P=0^-$ hypothesis cannot be studied using this channel.  Results for the
tensor hypotheses depend on the production mechanism, but currently none can be fully
excluded at 95\% CL.
Finally, in the leptonic $H \to WW$ channel, analysis in a two-dimensional plane
spanned by the event transverse mass and the lepton-lepton mass was done.
The exclusion of the pseudoscalar hypothesis from this channel is marginal,
but $J^P=2^+_m$ can be excluded at 95\% CL or more in the cases where the
preferred production mechanism is quark-antiquark fusion.
From a combination of $H \to ZZ$, $WW$ and $\gamma\gamma$ results, the $J^P=2^+_m$
model is excluded at a 99.9\% CL regardless of the combination of the gluon-gluon
and quark-antiquark production modes and other spin-2 hypotheses are excluded
at 99\% CL or higher.  Likewise spin-1 hypotheses are excluded
at more than 99.99\% CL from the combination of decays $H \to ZZ$ and $WW$.
The pseudoscalar hypothesis is excluded at 99.5\% CL.  This of course refers
to pure $J^P$ hypotheses.  The 95\% CL limit on the fractional pseudoscalar
cross section in the Higgs resonance is 0.43 and so a significantly mixed parity
state is by all means allowed.

A combined spin-parity analysis from the three main decay modes was also published
by ATLAS \cite{atlashspin}.  
This analysis excluded the graviton-inspired $2^+$ hypothesis at a
more than 99.9\% CL, spin-1 hypotheses at 99.7\% CL and the pure $0^-$ hypothesis
at 97.8\% CL.  They do not quote numbers for the maximum allowed pseudoscalar
admixture.
Exclusion limits have been also set on the hypothesis
that the observed signal is shared between two nearly degenerate mass states.

\subsection{Higgs couplings}

Finally and most importantly for the sake of this work, Higgs couplings have been
probed via measurements of
branching fractions for the main decay modes: $W^+W^-$, $ZZ$, $\gamma\gamma$ and
$\tau^+\tau^-$ (and $b\bar{b}$, in principle), and the respective production
mechanisms.  All Higgs couplings can in principle
be inferred from fits
to the observed rates in different combinations of Higgs production mechanisms
and decay modes, where each full production $\times$ decay path can be
parameterized as a function of the relevant couplings.
However, data are not precise enough to determine independently all the
couplings with a reasonable accuracy.  For this reason, results are usually
presented in one of two forms.
In the first approach, events are categorized by final state, including
the contributions from all production mechanisms and a single parameter $\mu$
for each final state is fit to the observed signal yield.
The quantity $\mu$ is
the measured signal strength (cross section $\times$ branching fraction) relative to
the predicted SM signal strength.  The most recent results of the overall signal
strength relative to the SM that is obtained from
a simultaneous fit to all Higgs decay channels are \cite{cmshiggs} \cite{atlashiggs}:

\vspace{3mm}

$\mu = 1.00 \pm 0.09 (stat) \pm 0.07 (syst) ~^{+~0.13}_{-~0.15} (theo)$ (CMS), and

\vspace{2mm}

$\mu = 1.30 \pm 0.13 (stat) ~^{+~0.14}_{-~0.11} (syst)$ (ATLAS).

\vspace{3mm}

\noindent
In the channels of most interest for us here, CMS results were:

\vspace{3mm}

$\mu_{WW} = 0.83 ~^{+~0.22}_{-~0.20}$ (from $W^+W^-$) and

\vspace{2mm}

$\mu_{ZZ} = 1.00 ~^{+~0.32}_{-~0.26}$ (from $4l$).  

\vspace{3mm}

\noindent
ATLAS most recently published values are:

\vspace{3mm}

$\mu_{WW} = 1.08 ~^{+~0.16}_{-~0.15} (stat) ~^{+~0.16}_{-~0.13} (syst)$ and

\vspace{2mm}

$\mu_{ZZ} = 1.44 ~^{+~0.40}_{-~0.33}$.

\vspace{3mm}

\noindent
While consistent with the SM, these numbers still keep room for sizeable deviations.

Fits of $\mu$ were also done in separate categories where events were
tagged by production mode, exploiting the distinct kinematic and topological 
signatures of VBF,
Higgsstrahlung and $t\bar{t}$-associated production (the largest ``untagged"
sample corresponds mostly to gluon fusion).  They revealed consistency with
the SM within rather large errors.

\begin{figure}[htbp]
\begin{center}
\epsfig{file=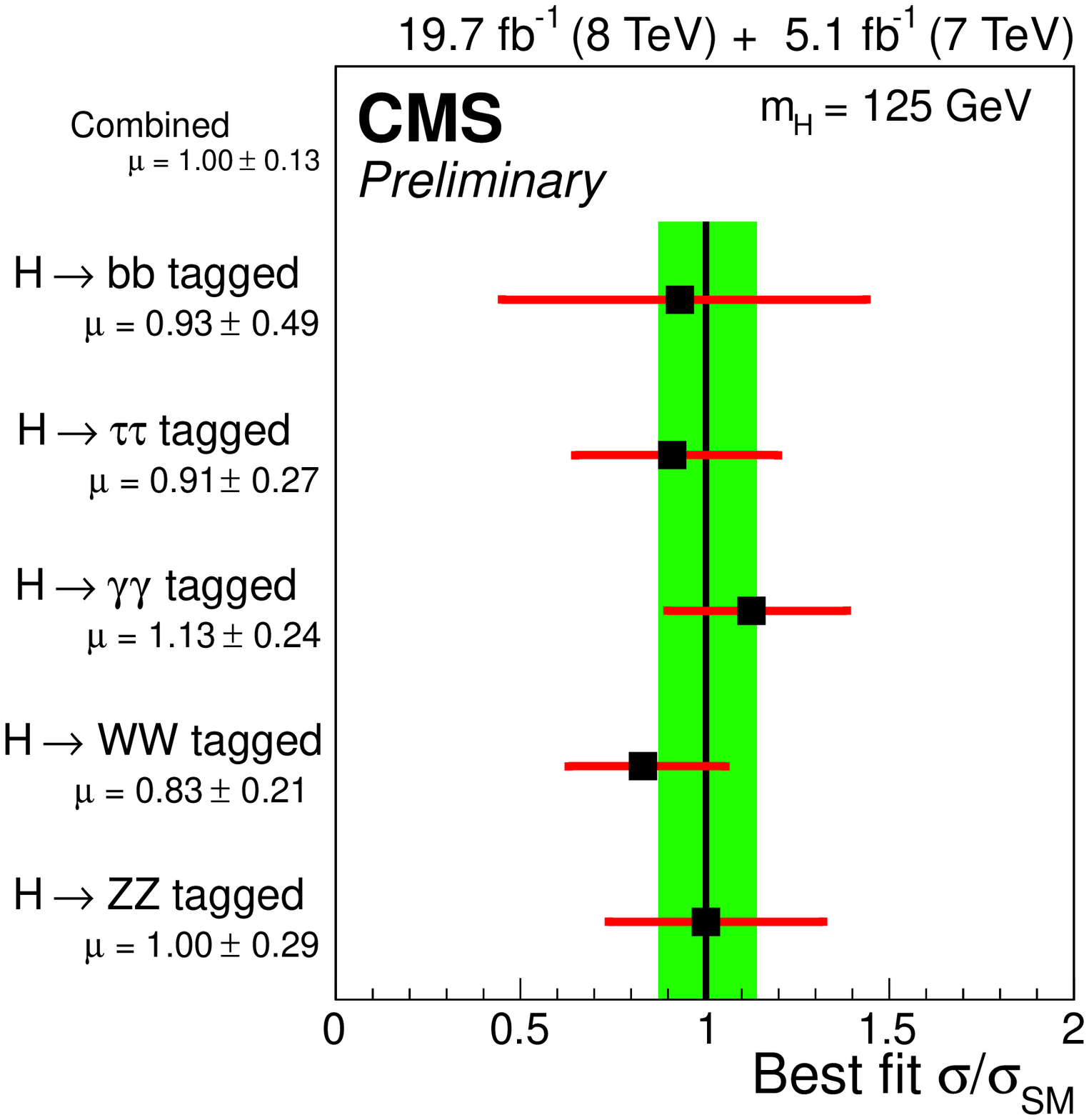,width=0.58\linewidth}\\
\epsfig{file=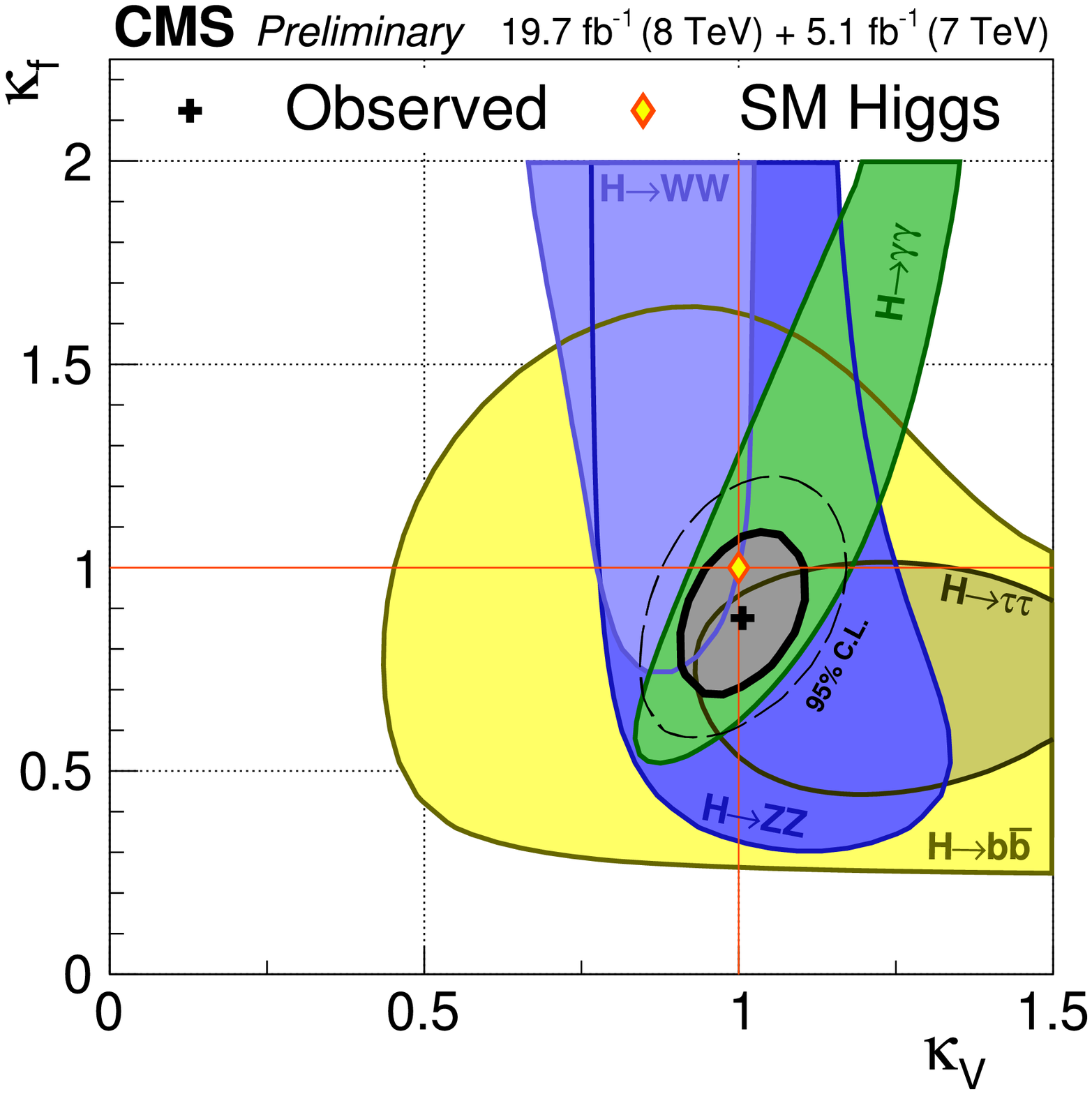,width=0.65\linewidth}
\end{center}
\caption{Upper plot: Values of the best fit $\sigma/\sigma_{SM}$ for the combination
(solid vertical line) and by predominant decay mode.  The
vertical band shows the overall $\sigma/\sigma_{SM}$ uncertainty.
Lower plot: 68\% CL contours for individual channels and for the overall
combination (thick curve) for the ($\kappa_V$, $\kappa_f$) parameters. 
The cross indicates the global best-fit values.  The dashed contour bounds the 95\% CL
region for the combination.  The yellow diamond represents the SM expectation.
Results from the CMS collaboration.  The shown parameter space was here restricted
to the first quadrant where the global minimum of the fit was found.  A second
minimum was also obtained for $\kappa_f<$0 (see also Fig.~\ref{hcoupatlas}).
Images reproduced from Ref.~\cite{wwwhiggscms}.
}
\label{hcoupcms}
\end{figure}

\begin{figure}[htbp]
\begin{center}
\epsfig{file=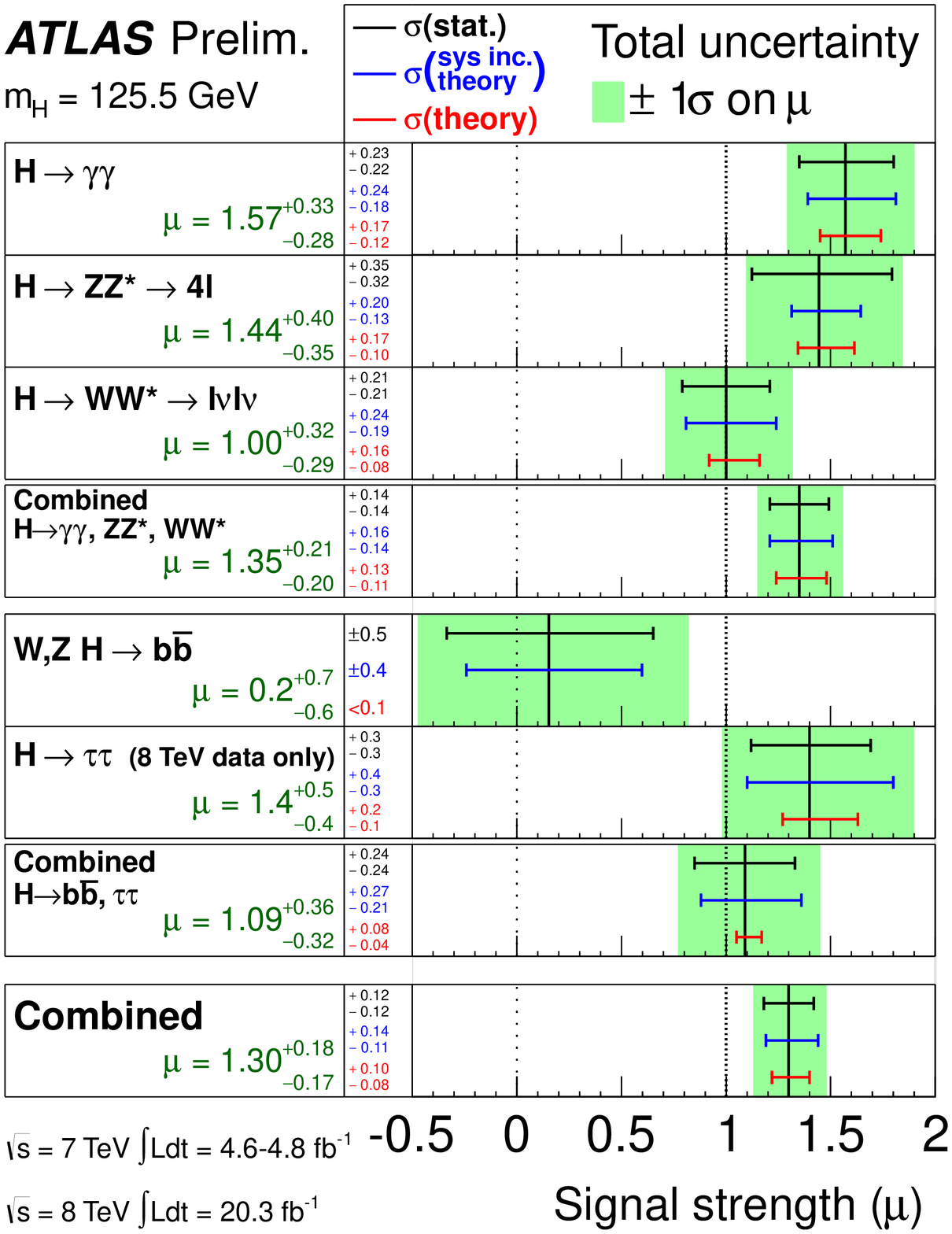,width=0.48\linewidth}\\
\vspace{5mm}
\epsfig{file=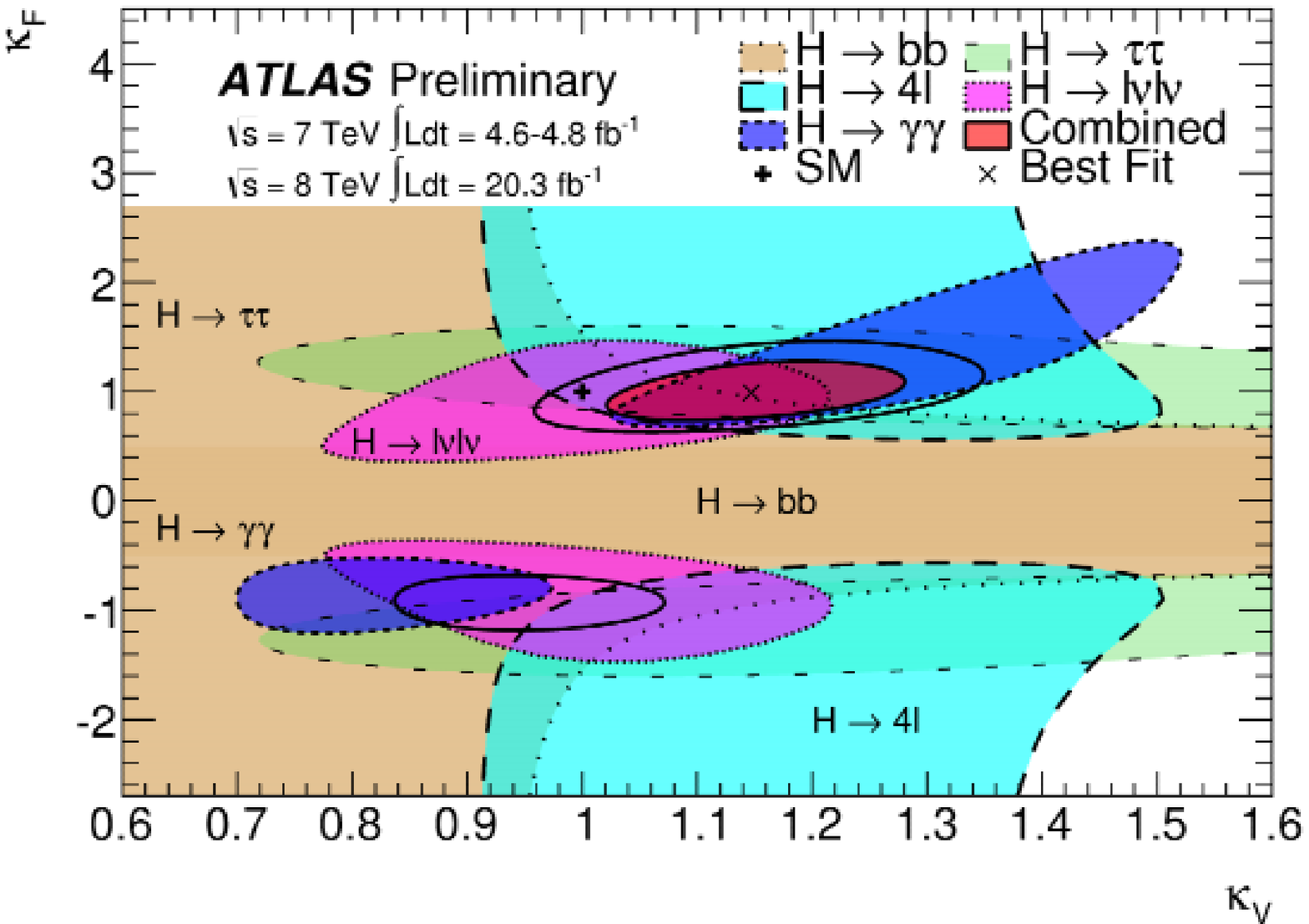,width=0.80\linewidth}
\end{center}
\vspace{-5mm}
\caption{Upper plot: The measured signal strengths normalized to SM expectations for the
individual final states and various combinations.  The best-fit values are shown by 
solid vertical lines.  The total $\pm1\sigma$ uncertainties are indicated by green shaded
bands, with the individual contributions from the statistical, systematic (including
theory) and theoretical uncertainties (from QCD scale, PDF, and branching ratios) 
are shown as superimposed error bars.
Lower plot: Results of fits that probe different coupling strength scale factors for
fermions
and vector bosons, assuming only SM contributions to the total width: 68\% CL contours
from individual decay channels and their combination.  Results from the ATLAS
collaboration.  Images reproduced from Ref.~\cite{wwwhiggsatlas}.
}
\label{hcoupatlas}
\end{figure}

The above are pure experimental results, with no model-dependence involved.
However, their relations to the genuine Higgs couplings are entangled.
In the other approach, events were categorized according to their full
production $\times$ decay chains and a simultaneous theoretical fit of the
corresponding cross sections $\times$ branching fractions was done to the data in
which only two parameters were allowed to vary freely: one to
globally modify the Higgs couplings to bosons, another to globally modify the
Higgs couplings to fermions.  From such fits, CMS obtained \cite{wwwhiggscms}
both scale factors,
$\kappa_V$ and $\kappa_f$, consistent with unity within 1$\sigma$; the accuracy
is roughly $\sim$10\% for $\kappa_V$ and $\sim$20\% for $\kappa_f$
(see Fig.~\ref{hcoupcms}).  From
one-dimensional parameter scans (in which the other coupling was set to its
SM value), one gets the following 95\% CL intervals:
$\kappa_V ~\epsilon$ [0.88, 1.15] and
$\kappa_f ~\epsilon$ [0.64, 1.16].
A similar analysis was done by ATLAS \cite{wwwhiggsatlas} (see Fig.~\ref{hcoupatlas}).

It should be stressed here that this procedure is not completely model-independent
because the content of the loops in gluon fusion and in $H \to \gamma\gamma$
decays must be explicitly assumed in
order to relate production mechanisms with decay modes via the same parameters: in
the SM, the loops are dominated by respective contributions from the top quark
and from the $W$ boson.
Agreement with the SM of the total Higgs signal
strength, in particular in the dominant ``untagged" category, as well as
that of the $H \to \gamma\gamma$ signal strength, justifies the approach.
More generally, the procedure can be regarded self-consistent for any model
that does not involve significant contributions from unknown heavy particles
within the present energy reach.  This may in fact be the case in
an interesting wide class of theories beyond the SM,
known as SILH models, that we will discuss further on.  
The procedure itself of scaling the Higgs couplings by only two
independent factors, $\kappa_V$ and $\kappa_f$, is likewise consistent with
the expected low-energy phenomenology of these models.
Therefore, the above result is of special interest from this point of view.
A dedicated test for the presence of BSM particles
was carried based on $\gamma\gamma$ data.
A fit to the data where all tree-level couplings were assumed equal to their
SM values, $\kappa_V=\kappa_f=1$, and varied freely were the effective Higgs
couplings to gluons
and photons, $\kappa_g$ and $\kappa_\gamma$, revealed consistency with the SM
within 1$\sigma$.

By reverting the procedure, the top coupling is probed by assigning a common
signal strength factor for the gluon fusion production mechanism,
with addition of the little $t\bar{t}H$ production mode, because they both
scale predominantly with the Yukawa coupling of the top quark in the SM.
The assumption that the Higgs couples proportionally to the fermion mass has
been indirectly supported by the data.

Other tests included modified up-type to down-type fermion couplings, motivated
by SUSY models, and modified independently top, bottom and $\tau$ couplings.
No deviations from the SM were observed.

\subsection{Searches for a non-SM Higgs}

By contrast, all dedicated searches for a non-Standard Higgs to date gave
negative results.  

Additional, heavy SM-like Higgses were excluded
at the 95\% CL or more up to the mass of 710 GeV from a combination of data
from $ZZ$ and $WW$ decays \cite{hheavyvv}.  
Likewise, no additional resonances have been observed
in the $\gamma\gamma$ spectrum between 150-850 GeV \cite{hheavygaga}.

Dedicated searches were carried for neutral and charged Higgses within
the framework of the Minimal Supersymmetric extension of the Standard Model (MSSM).
The most stringent exclusion limits come from the search for the MSSM decay
$h, H, A \rightarrow \tau^+\tau^-$ \cite{hmssm2tautau}
for which the standard $\tau^+\tau^-$ analysis was modified
so as to maximize the sensitivity to BSM effects.  An MSSM scalar Higgs
differs from the SM Higgs in terms of the relative contributions from different
production mechanisms and decay branching fractions.  
In particular, $b\bar{b}$-associated production followed
by decay into a $\tau^+\tau^-$ pair gets
enhanced because Higgs couplings to down-type fermions and third generation
fermions increase with $tan\beta$.
Additional exclusions were obtained from searches for the MSSM-specific effects
affecting the decays into
$b\bar{b}$ and $\mu^+\mu^-$.  Charged Higgses, predicted by the MSSM,
were searched for in the decay channels $H^\pm \rightarrow \tau^\pm\nu$,
$H^\pm \to cs$ and $H^\pm \to tb$ \cite{hmssmmisc}.  The combination of all these results
severely constrain the available MSSM parameter space in the Higgs sector,
although the hypothesis that the only discovered boson so far is in fact the lighter
of the two scalar Higgses of the MSSM cannot be ruled out completely.

Other, non-minimal supersymmetric models have been constrained as well.
This includes in particular the Next-to-Minimal Supersymmetry (NMSSM),
predicting a light Higgs scalar
decaying into a pair of light Higgs pseudoscalars, with a final state
consisting of 4 muons, $h \rightarrow aa \rightarrow 4\mu$.
Such decay chain, once thought
to be an alternative to the SM/MSSM scenario that should attract physicists'
main attention in case
the LHC fails to observe Higgs signal in one of its mainstream SM channels,
is inconsistent with the data \cite{cmshnmssm}.  An upper limit has been set on the cross
section for standalone light pseudoscalar Higgs production via gluon fusion
followed by decay into a muon pair, $\sigma(pp \to a) Br(a \to \mu^+\mu^-)$, which 
translates into further limits in the NMSSM
parameter space \cite{cmsa}.

Explicit searches for Higgs anomalous couplings have been carried.
Higgs production in association with a single top quark (and a light quark jet)
is particularly sensitive to the relative sign of the Higgs boson coupling to
fermions and bosons and to the value itself of the Higgs to top coupling.
Such studies were carried independently based on the $b\bar{b}$ and $\gamma\gamma$
decay modes,
but their results were inconclusive \cite{hcoupl}.

Inconsistency of the Higgs boson with models assuming the existence of a fourth
lepton generation, as well as fermiophobic Higgs models, was shown early on
\cite{hexo1}.
Other dedicated searches include heavy scalar and pseudoscalar Higgses in a general
two-doublet
model (2HDM), doubly charged Higgses, invisible decays of the SM-like Higgs
and lepton flavor violating decays and were translated into respective
exclusion limits \cite{hexo2}.

To summarize, consistency of all the data with the Standard Model holds invariably
in what regards Higgs physics.
Most key analyses have already been performed on the whole 7+8 TeV dataset and
so the main conclusions are unlikely to change significantly anytime before late 2015.
No hints of new physics have been observed, whether in the Higgs sector, or
for that matter in the many non-Higgs related searches carried at both ATLAS and CMS
(for a review of the latter the reader is referred elsewhere
\cite{cmssusy} \cite{cmsexo}).
On the other hand, plenty of room for new physics is still there to be unraveled
at some higher energy, or even possible to show up eventually at the currently
available energy
if only more LHC luminosity was available.  There are no clear indications
so far as to what this physics beyond the Standard Model might be.
Contrary to SUSY, which may not provide
any measurable hints of new physics unless by an increase of energy, SILH models in
general predict new physics in both ways.  The second phase
of the LHC, due to start in 2015, will increase both the energy and the
luminosity and has chances to solve the puzzle.

There is one more important thing to learn from the spin-parity analyses in
particular.  Since the $H \to W^+W^-$ channel offers relatively little sensitivity
to the Higgs spin-parity, the same weakness is bound to apply to measurements of $W$
helicity in the final state.  Even more difficult this will become
in the most interesting high mass region where the $W$'s are more boosted.
Separation of $W$
helicities in $WW$ scattering requires other techniques to be used.

\section{Electroweak physics results}
\label{lhcatgc}

Both ATLAS and CMS have produced a large number of results concerning gauge boson
production in general \cite{wwwewkcms}.  
The most directly relevant for us are those concerning diboson
and triboson production.  Their importance for the study of $VV$ scattering is
twofold.  Measurements of total cross sections for diboson production
cross check our calculations of
irreducible background.  More specific analyses of the respective
kinematic distributions allow to place limits on anomalous triple and quartic
vector boson couplings.

In what regards triboson production,
95\% CL limits were set at CMS on anomalous quartic couplings for
$WW\gamma\gamma$, $WWZ\gamma$ \cite{cmswvgamma}.  
These were based by searches for the $WW\gamma$
and $WZ\gamma$ final states, respectively.  There is no directly equivalent limit 
so far on the
$WWWW$ coupling, i.e., based on a measurement of triboson $WWW$ production,
either from ATLAS or CMS.

Results abound as far as diboson production is concerned.  Let us review the most
important of them.
The total inclusive $W^+W^-$ cross section at 7 TeV as measured by CMS was found to be
\cite{cmsw+w-7}

\vspace{3mm}
$\sigma (pp \rightarrow W^+W^-)|_{7TeV} =
52.4 \pm 2.0 (stat) \pm 4.5 (syst) \pm 1.2 (lumi)$\footnote{Errors labeled $lumi$
are those related to the LHC luminosity measurement.} pb,

\vspace{3mm}

\noindent 
which is
consistent within the errors with Standard Model predictions in the next-to-leading
order, including the two main production mechanisms of quark-antiquark annihilation
and gluon-gluon
fusion.  That the measured value is actually marginally higher than the prediction can
be ascribed to other production mechanisms such as: diffractive production, double
parton scattering, QED exclusive production, and Higgs boson production with decay
to $W^+W^-$, expected to yield additional contributions up to about 5\% altogether.
The ATLAS Collaboration measured \cite{atlasw+w-7}

\vspace{3mm}

$\sigma (pp \rightarrow W^+W^-)$ =
$51.9 \pm 2.0 (stat) \pm 3.9 (syst) \pm 2.0 (lumi)$ pb.

\vspace{3mm}

\noindent
The total $pp \rightarrow ZZ$ cross section at $\sqrt{s}$ = 7 TeV was measured to be

\vspace{3mm}

$\sigma (pp \rightarrow ZZ)|_{7TeV} = 6.24 \pm^{0.86}_{0.80} (stat)$
$\pm^{0.41}_{0.32} (syst) \pm 0.14 (lumi)$ pb (CMS) \cite{cmszz7}, and

\vspace{2mm}

$\sigma (pp \rightarrow ZZ)|_{7TeV} = 6.7 \pm 0.7 (stat)$
$\pm^{0.4}_{0.3} (syst) \pm 0.3 (lumi)$ pb (ATLAS) \cite{atlaszz7}.

\vspace{3mm}

\noindent
The inclusive $W^\pm Z$ production cross section from CMS was \cite{cmswz78}

\vspace{3mm}

$\sigma (pp \rightarrow WZ)|_{7TeV} = 20.76 \pm 1.32 (stat) \pm 1.13 (syst)$
$\pm 0.46 (lumi)$ pb,

\vspace{3mm}

\noindent
and from ATLAS it was \cite{atlaswz7}

\vspace{3mm}

$\sigma (pp \rightarrow WZ)|_{7TeV} = 19.0 \pm^{1.4}_{1.3} (stat) \pm 0.9 (syst)$
$\pm 0.4 (lumi)$ pb.

\vspace{3mm}

\noindent
All the $ZZ$ and $WZ$ cross sections are consistent with Standard Model NLO predictions.

Sadly, there is no dedicated measurement of the much lower same-sign $WW$ production,
although some bounds on it can be in principle indirectly inferred
using a combined measurement of $WW+WZ$ based on events with a $W$
decaying leptonically and two jets.  
%
%
%
%
%
%
Unfortunately, these measurements \cite{wwwz7}
are of not enough precision to extract the tiny same-sign $WW$ contribution.

Finally, the $W\gamma$ and $Z\gamma$ cross sections are \cite{cmsvgamma7}:

\vspace{3mm}

$\sigma (pp \rightarrow W\gamma)|_{7TeV} \times Br(W \to l\nu) = 37.0 \pm 0.8 (stat)$
$\pm 4.0 (syst) \pm 0.8 (lumi)$ pb (CMS), and

\vspace{2mm}
 
$\sigma (pp \rightarrow Z\gamma)|_{7TeV} \times Br(Z \to ll) = 5.33 \pm 0.08 (stat)$
$\pm 0.25 (syst) \pm 0.12 (lumi)$ pb (CMS).

\noindent
The ATLAS collaboration does not quote their total cross section values, but
restricts the measurements to a predefined fiducial region.  In any case, no
deviations from the SM were observed \cite{atlasvgamma7}.

At 8 TeV, CMS measured \cite{cmswwwzzz8}:

\vspace{3mm}

$\sigma (pp \rightarrow W^+W^-)|_{8TeV} = 69.9 \pm 2.8 (stat) \pm 5.6 (syst)$
$\pm 3.1 (lumi)$ pb,

\vspace{2mm}

$\sigma (pp \rightarrow ZZ)|_{8TeV} = 8.4 \pm 1.0 (stat) \pm 0.7 (syst)$
$\pm 0.4 (lumi)$ pb, and

\vspace{2mm}

$\sigma (pp \rightarrow WZ)|_{8TeV} = 24.61 \pm 0.76 (stat) \pm 1.13 (syst)$
$\pm 1.08 (lumi)$ pb \cite{cmswz78}.

\vspace{3mm}

\noindent
The $W^+W^-$ value is slightly higher than the Standard Model NLO prediction
of 57.3 $\pm^{2.3}_{1.6}$ pb, but again an extra 5\% increase of this value is expected
from the additional contributions calculated at the next-to-next-to-leading order,
chiefly from Higgs boson production.  Explanations in terms of new physics have
also been suggested, but are not quite convincing.
The $ZZ$ and $WZ$ values agree with Standard Model NLO predictions within the errors.
ATLAS showed:

\vspace{3mm}

$\sigma (pp \rightarrow W^+W^-)|_{8TeV} = 71.4 \pm 1.2 (stat) ^{+~5.0}_{-~4.4} (syst)$
$ ^{+~2.2}_{-~2.1} (lumi)$ pb \cite{atlasw+w-8},

\vspace{2mm}

$\sigma (pp \rightarrow ZZ)|_{8TeV} = 7.1 \pm^{0.5}_{0.4} (stat) \pm 0.3 (syst)$
$\pm 0.2 (lumi)$ pb \cite{atlaszz8}, and

\vspace{2mm}

$\sigma (pp \rightarrow WZ)|_{8TeV} = 20.3 \pm^{0.8}_{0.7} (stat) \pm^{1.2}_{1.1} (syst)$
$\pm^{0.7}_{0.6} (lumi)$ pb \cite{atlaswz8}.

\vspace{3mm}

Limits on anomalous triple gauge couplings, and in particular the ones of most direct
relevance for us, namely $WWZ$ and $WW\gamma$, have been derived so far from the 7 TeV
data.  An up-to-date summary of these measurements, together with a set of
references to the original papers, is available in Refs.~\cite{atgc}.
Also included in the summary are the respective results from the TeVatron and LEP.
In all these works, anomalous couplings were studied within the formalism known as the
effective Lagrangian approach, in which the most general form of the $WWZ/\gamma$ 
vertex is considered, including all terms that respect
Lorentz invariance and conserve $C$ and $P$.  The couplings
are taken to be constant parameters of the
Lagrangian and therefore independent of the boson momenta.
The conceptual basis of this approach is explained in detail in Ref.~\cite{tgbc}.
Accordingly, experimental limits are set on five quantities: $\Delta g_1^Z$,
$\Delta\kappa_Z$, $\lambda_Z$, $\Delta\kappa_\gamma$ and $\lambda_\gamma$.
The first three of these modify the $WWZ$ vertex, the following two modify the
$WW\gamma$ vertex.
In the SM, $\lambda_Z$=$\lambda_\gamma$=0 and $g_1^Z$=$\kappa_Z$=$\kappa_\gamma$=1.
Their actual values are determined from studies of diboson production processes
for which these vertices play a primary role, namely
$WW$, $WZ$ and $W\gamma$.  An anomalous triple gauge coupling would be manifest
in the rate of diboson production at high boson $p_T$ and invariant mass.  Typically,
it is ascertained in CMS by a one-dimensional evaluation of the $p_T$ spectrum
of the leading lepton or of the dijet
(both from $W/Z$ decay) or of the photon.  For the theoretical calculation of the
expected spectrum, either one or two anomalous parameters are varied at a time.
Correlations between couplings that contribute to the same vertex are rather weak
and so one-dimensional limits are generally sufficient.

\begin{figure}[htbp]
\begin{center}
\vspace{5mm}
\epsfig{file=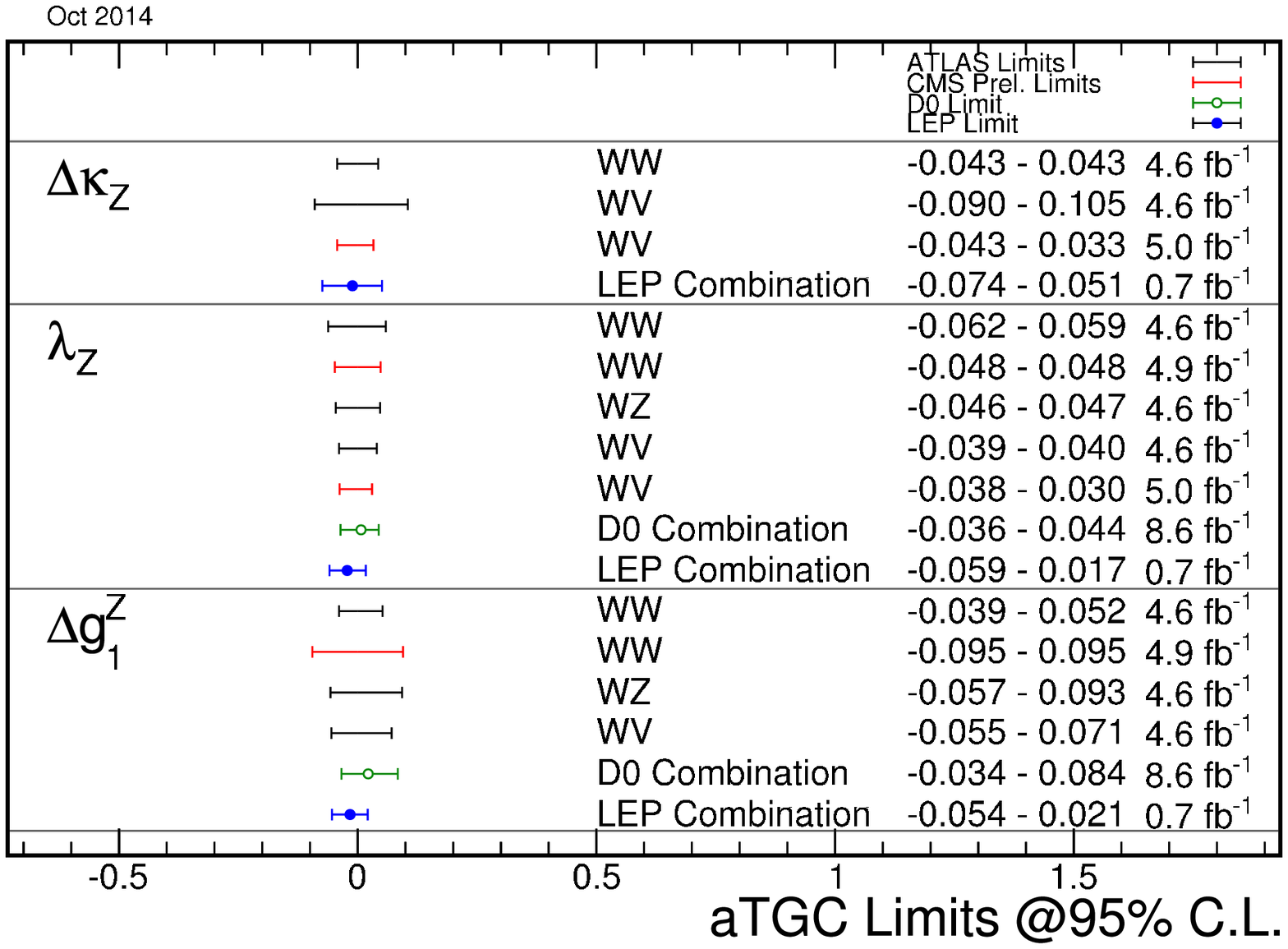,width=0.75\linewidth}
\end{center}
\vspace{-5mm}
\caption{Current limits on the anomalous couplings that contribute to the $WWZ$
vertex - compilation of results coming from LEP, TeVatron and LHC experiments.
Image reproduced from Ref.~\cite{atgc}.
}
\vspace{5mm}
\end{figure}

\begin{figure}[htbp]
\begin{center}
\vspace{5mm}
\epsfig{file=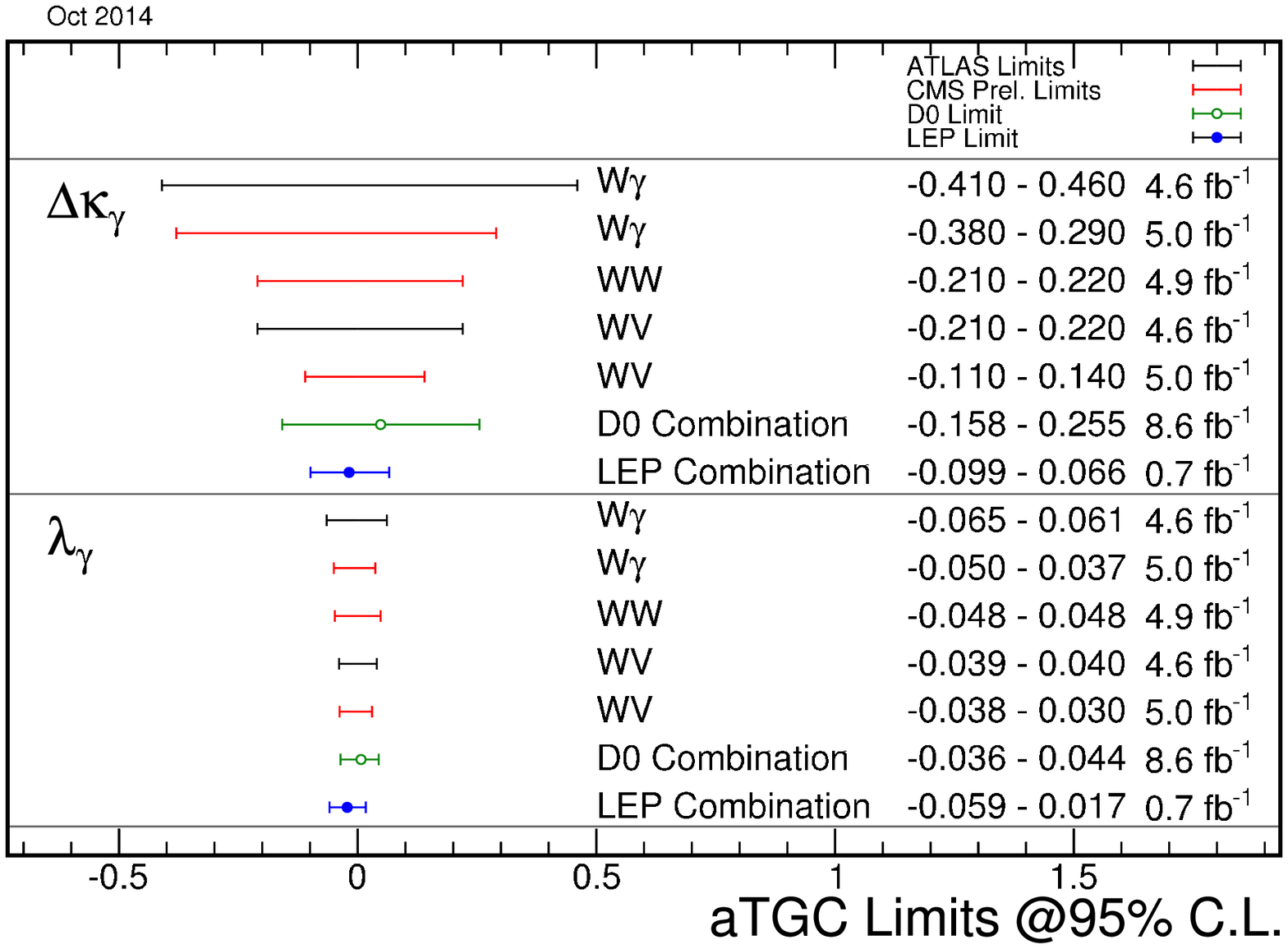,width=0.75\linewidth}
\end{center}
\vspace{-5mm}
\caption{Current limits on the anomalous couplings that contribute to the $WW\gamma$
vertex - compilation of results coming from LEP, TeVatron and LHC experiments.
Image reproduced from Ref.~\cite{atgc}.
}
\vspace{5mm}
\end{figure}

While all these parameters may be probed independently in experiment,
considerations of
gauge symmetry induce additional relations between them:

\begin{equation}
\lambda_Z = \lambda_\gamma,
\end{equation}

\begin{equation}
\Delta\kappa_Z = \Delta g_1^Z - \Delta\kappa_\gamma tan^2\theta_W.
\end{equation}

\noindent
From this it follows that, e.g., measurement of $WWZ$ couplings could be 
translated into $WW\gamma$ couplings on theoretical grounds.

\section{Other results of relevance for the study of $VV$ scattering}

For the correct assessment of reducible backgrounds, several other measurements
are as important.  Let us only mention the ones we will directly refer to in this
work.

Top production has been measured in several
final states, including different $W$ decay channels \cite{ttbar}.  The most accurate
inclusive $t\bar{t}$ production cross sections 
come from the dilepton final state.  CMS reports \cite{cmsttbar}

\vspace{3mm}

$\sigma (pp \rightarrow t\bar{t})|_{7TeV} = 162 \pm 2 (stat) \pm 5 (syst) \pm 4 (lumi)$
and

\vspace{2mm}

$\sigma (pp \rightarrow t\bar{t})|_{8TeV} = 239 \pm 2 (stat) \pm 11 (syst) \pm 6 (lumi)$
(from 5.3/fb of data).

\vspace{3mm}

\noindent
ATLAS measured \cite{atlasttbar}

\vspace{3mm}

$\sigma (pp \rightarrow t\bar{t})|_{7TeV} = 177 \pm 3 (stat) \pm^8_7 (syst) \pm 7 (lumi)$
and

\vspace{2mm}

$\sigma (pp \rightarrow t\bar{t})|_{8TeV} = 238 \pm 2 (stat) \pm 7 (syst) \pm 7 (lumi)$.

\vspace{3mm}

\noindent
The numbers are in good agreement with NNLO+NNLL calculations by 
Czakon et al.~\cite{czakon}, where
the quoted uncertainty of the latter is roughly the size of experimental errors.

A lot of other results indirectly relate to our subject.  
Important feedback is obtained in particular from forward physics where
jet multiplicity and kinematics obtained from common event generators can be
cross checked in detail against the data.
These things however play a rather
secondary role for us and we need not go through them here.

\section{The $VV$ interaction and why it is still interesting}

In the previous chapter
we have sketched the derivation of the Higgs boson using two independent
approaches: from the principle
of SU(2)$\times$U(1) gauge invariance and from the requirement of tree level unitarity
of all Standard Model processes.  
The paramount phenomenological manifestation of the underlying model is the existence
of a physical scalar particle which couples to all known particles of non-zero
mass in a completely determined way.  Such particle manifests itself in a twofold
way.  At energies available in the LHC to date it should be produced in 
proton-proton collisions via
different physical mechanisms, each of them yielding partly identifiable experimental
signatures, and decay into known Standard Model particles, with both production
cross sections and branching fractions completely determined by its mass.
A good candidate for such particle has
indeed been found.  The other manifestation will become available at
higher energy and help answer the main question: is this the same particle?

\subsection{Higgs mass and couplings in $VV$ scattering}

The second phenomenological manifestation of the Standard Model
Higgs boson resides in the high energy behavior of the $VV$ scattering amplitudes.
In the Standard Model the $HWW$ coupling is chosen such that it fully cancels
the quadratic divergencies that appear after combining photon and $Z$ exchange
graphs with the four-$W$ contact interaction.  Thus, it is precisely the same
particle which has been discovered at the LHC that
is supposed to provide these {\it exact} cancelations.
Violation of unitarity at some high enough energy will be the most
extreme (and unrealistic) manifestation of the still existing problem
should this cancelation not be the case.
But let us put questions of unitarity aside, as they in fact represent a technical issue.
The entire high energy behavior of the $VV$ scattering amplitude is a fully quantitative
question and depends on many inputs.
Total and differential cross sections for the scattering of longitudinally
polarized $W$ and $Z$
gauge bosons, at energies much larger than the masses of the latter, are a
major experimental field where consistency of the Standard Model has to be tested.
To this date we have practically no experimental data to confirm that the Higgs boson
indeed does its job, assigned to it by the Standard Model.

A simple tree level calculation of the process $W_L^+W_L^+ \rightarrow W_L^+W_L^+$
reveals two basic facts.  The total cross section as a function of the center of
mass energy behaves differently depending on both the Higgs mass and Higgs couplings.
As long as the $HWW$ coupling is exactly 1, expressed in units of the value predicted
by the Standard Model, the amplitude keeps rising up to the energy equivalent to the
Higgs mass (notice that for $M_H<2M_W$ it in practice never does so), then stays 
approximately flat.  Phase space causes the cross section fall for higher
energies.  In the absence of a Higgs boson the amplitude rises indefinitely and so
does the cross section.  It can be calculated that unitarity violation occurs
at about a 1.2 TeV energy and thus some new physics is bound to enter before this scale.
Put in a more physical language,
unless the scattering amplitude receives new contributions that reduce the
amplitude way before this point, the $WW$ interaction before the scale
of 1.2 TeV inevitably becomes strong.  The term ``strong" specifically means that
multiple rescattering is likely to occur.  This means a difference in the basic
dynamics of
electroweak symmetry breaking compared to the Standard Model case where it is
supposed to be ``soft" and a single Higgs boson exchange takes place instead.
We would talk then of a strongly interacting gauge sector.
An even more interesting scenario occurs if the $HWW$ coupling is different from 1.
As we already know, in such case the quadratic terms in the amplitude are not
completely canceled and this incomplete cancelation
must show up at a high enough energy.  In general,
the total cross section will rise up to the Higgs mass (not if $M_H<2M_W$),
fall past the Higgs mass
and rise up again at some energy.  The situation again calls for new physics
as the unitarity limit would be still inevitably hit at some energy whose precise
value depends on the value of the coupling.  
The unrealistic by now, extreme case of no Higgs boson at all is technically
equivalent to setting either an infinite Higgs mass or a zero coupling.
Relative to the SM prediction, the
cross section will be enhanced at all energies as long as the $HWW$ coupling is
lower than the SM one and will reveal an energy pattern consisting of a depletion
followed by a turning point and an enhancement if the $HWW$ coupling
is larger than the SM one.  This is because in the latter case there is an
overcancelation of the quadratic divergence by the Higgs graph which subtracts
from the constant term in the total amplitude.  At a certain
energy the quadratic term becomes dominant anyway and asymptotically the
cross sections for a given $g_{HWW}$ and for $1-g_{HWW}$ become the same.

\begin{figure}[htbp]
\vspace{-5mm}
\begin{center}
\epsfig{file=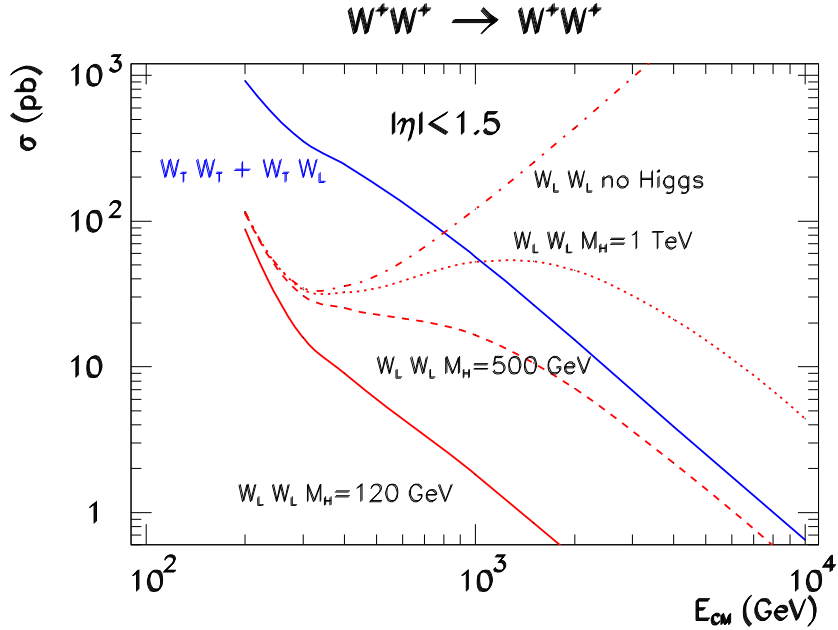,width=0.8\linewidth}
\end{center}
\vspace{-1cm}
\caption{The total $W^+W^+$ scattering cross sections as a function of the center of
mass energy for different final (and initial) state polarizations
and for different Higgs masses, including the limiting Higgsless case.
Assumed are two on-shell, unpolarized, colliding $W^+$ beams.  
A cut on the scattering angle that corresponds to pseudorapidity of $\pm 1.5$ with
respect to the incoming $W$ direction was applied.
The individual $W_TW_T + W_TW_L$ curves for each Higgs mass value coincide within
the width of the blue line.
Results of MadGraph \cite{madgraph} calculations.}
\end{figure}

Angular distributions of the scattered $W$'s are also sensitive to the mass
and couplings of the Higgs boson.  In the Standard Model with a light Higgs,
the scattering occurs predominantly at small angles.  A signature of any rise
of the total cross section at some high energy is visible as the appearence
of an additional component that tends to favor large scattering angles, with
a local maximum at 90$^o$.  Thus any deviation from the Standard Model in terms
of the Higgs couplings would be, quite similarly like different Higgs masses,
observable as a correspondent excess in the rate of $W_LW_L$ scattered at
large angles.  The excess is the more pronounced the higher the energy.
In the above demonstration of the principles, we have arbitrarily chosen same-sign
$WW$ scattering (in the next chapters we will see that this choice is in fact
well motivated), but the same basic qualitative features are expected of the other
scattering processes, involving $W^+W^-$, $WZ$ and $ZZ$
\footnote{$ZZ$ should be always understood as a sum of the amplitudes for the
$W^+W^- \to ZZ$ and $ZZ \to ZZ$ scattering processes.} pairs.

\begin{figure}[htbp]
\vspace{-5mm}
\begin{center}
\epsfig{file=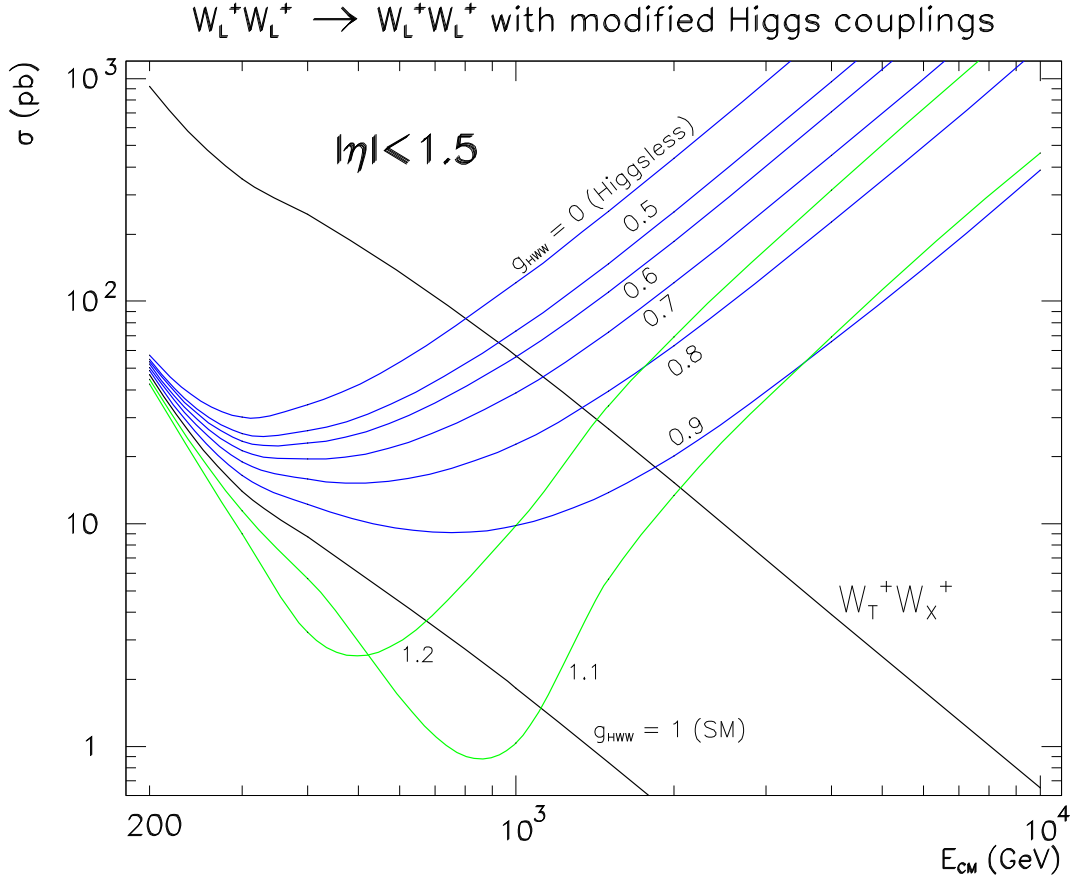,width=1.0\linewidth}
\end{center}
\vspace{-1cm}
\caption{The total $W_L^+W_L^+$ scattering cross sections as a function of the center of 
mass energy for different values of the $HWW$ coupling, $g_{HWW}$,
Assumed here are two colliding on-shell, unpolarized $W^+$ beams and a 120 GeV Higgs
boson.
Coupling $g_{HWW}$=1 (lower black curve) corresponds to the Standard Model. 
Blue curves represent $g_{HWW}<1$, the curve
for $g_{HWW}$=0 is equivalent to the Higgsless case.
Green curves represent $g_{HWW}>1$.
Also shown is the total cross section for $W_T^+W_X^+$ scattering
(upper black curve, subscript $X$ denotes any polarization, $T$ or $L$), 
its variations with the $HWW$ coupling are contained within the line width.
A cut on the scattering angle that corresponds to pseudorapidity of $\pm 1.5$ with
respect to the incoming $W$ direction was applied.  Results of MadGraph
calculations.}
\end{figure}

\begin{figure}[htbp]
\vspace{-5mm}
\begin{center}
\epsfig{file=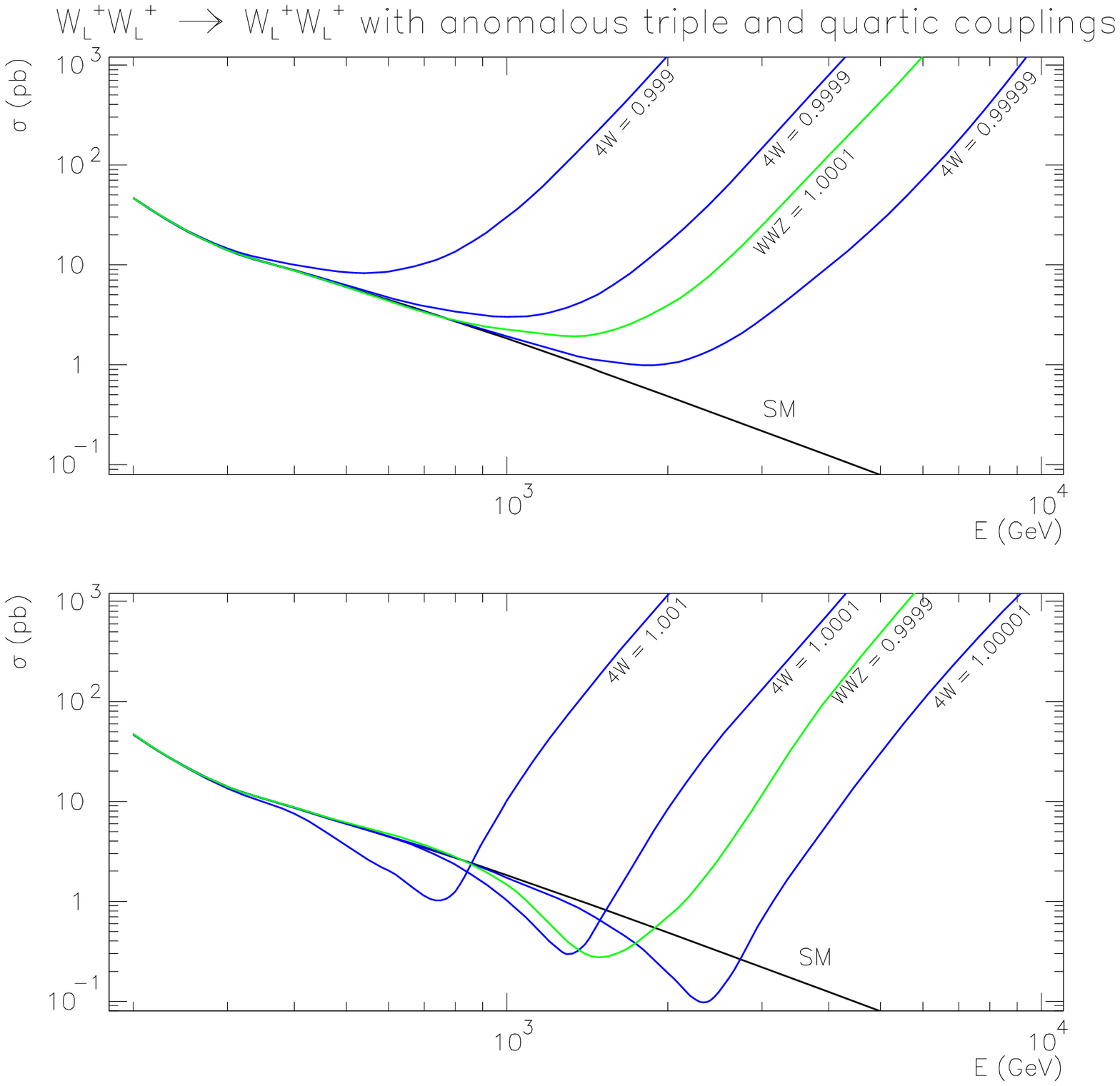,width=0.95\linewidth}
\end{center}
\vspace{-1cm}
\caption{Total $W_L^+W_L^+$ scattering cross section as a function of the center of
mass energy for different values of the $WWWW$ quartic coupling (labeled $4W$,
blue curves) and the $WWZ$ triple coupling (labeled $WWZ$, green curves).
The corresponding couplings are scaled by a constant factor relative to their
respective Standard Model values.
Assumed here are two colliding on-shell, unpolarized $W^+$ beams and a 120 GeV Higgs
boson.
A cut on the scattering angle that corresponds to pseudorapidity of $\pm 1.5$ with
respect to the incoming $W$ direction was applied.
Results of MadGraph calculations.}
\end{figure}

\begin{figure}[htbp]
\vspace{-5mm}
\begin{center}
\epsfig{file=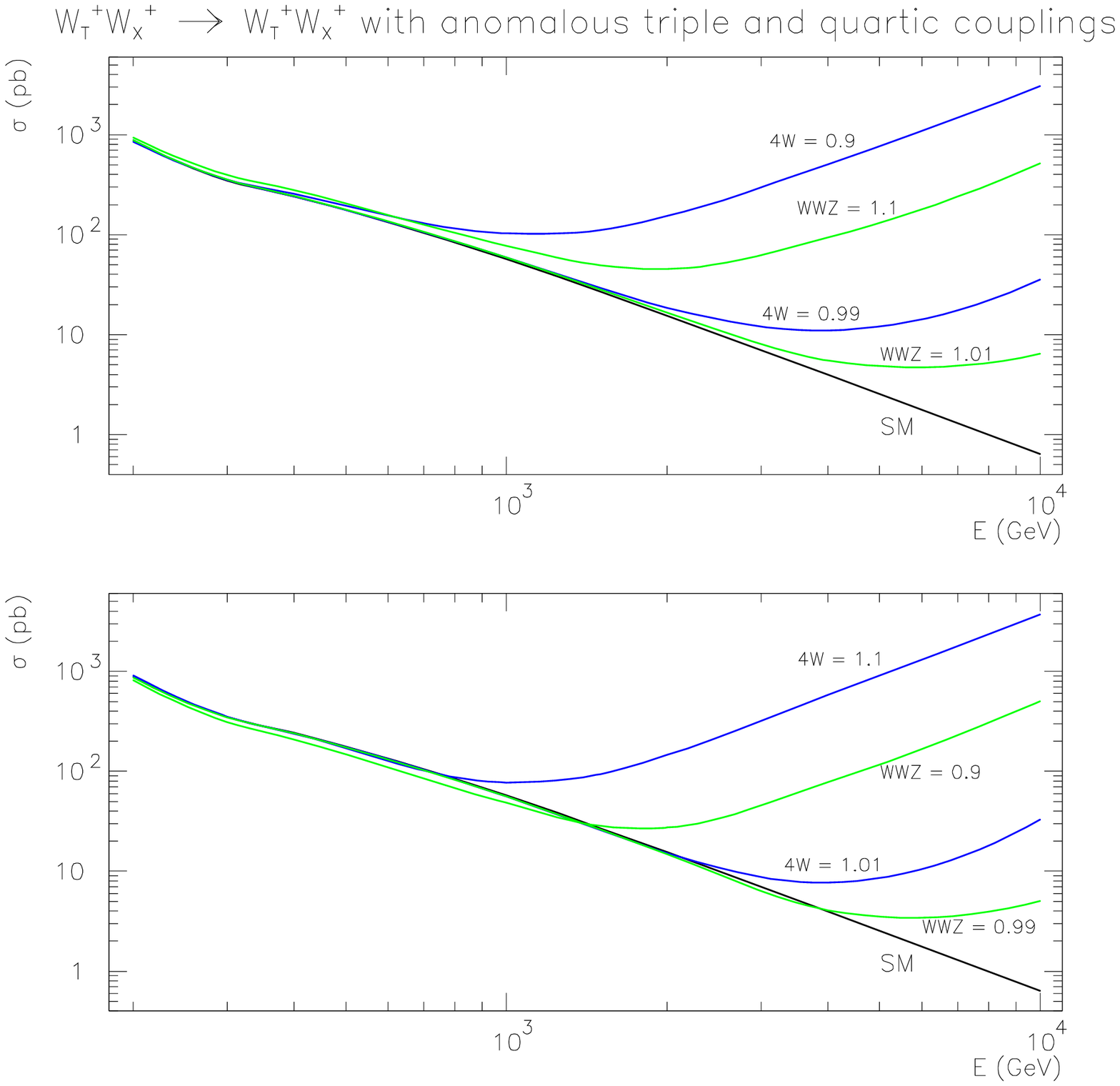,width=0.95\linewidth}
\end{center}
\vspace{-1cm}
\caption{The total $W_T^+W_X^+$ scattering cross section as a function of the center of
mass energy for different values of the $WWWW$ quartic coupling (labeled $4W$,
blue curves) and the $WWZ$ triple coupling (labeled $WWZ$, green curves).
The corresponding couplings are scaled by a constant factor relative to their 
respective Standard Model values.
Assumed here are two colliding on-shell, unpolarized $W^+$ beams and a 120 GeV Higgs
boson.
A cut on the scattering angle that corresponds to pseudorapidity of $\pm 1.5$ with
respect to the incoming $W$ direction was applied.  Results of MadGraph
calculations.}
\end{figure}

\begin{figure}[htbp]
\vspace{-5mm}
\epsfig{file=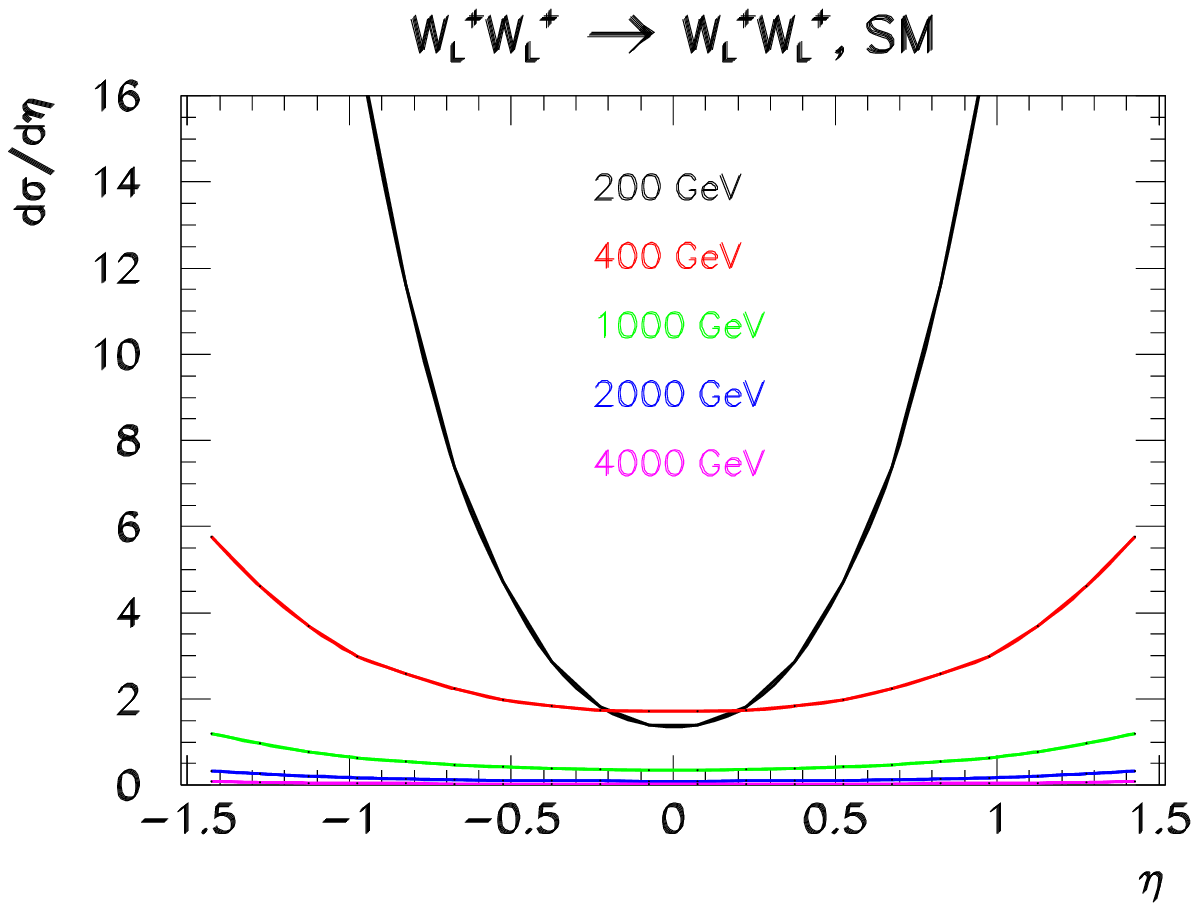,width=0.55\linewidth}
\hspace{-1cm}
\epsfig{file=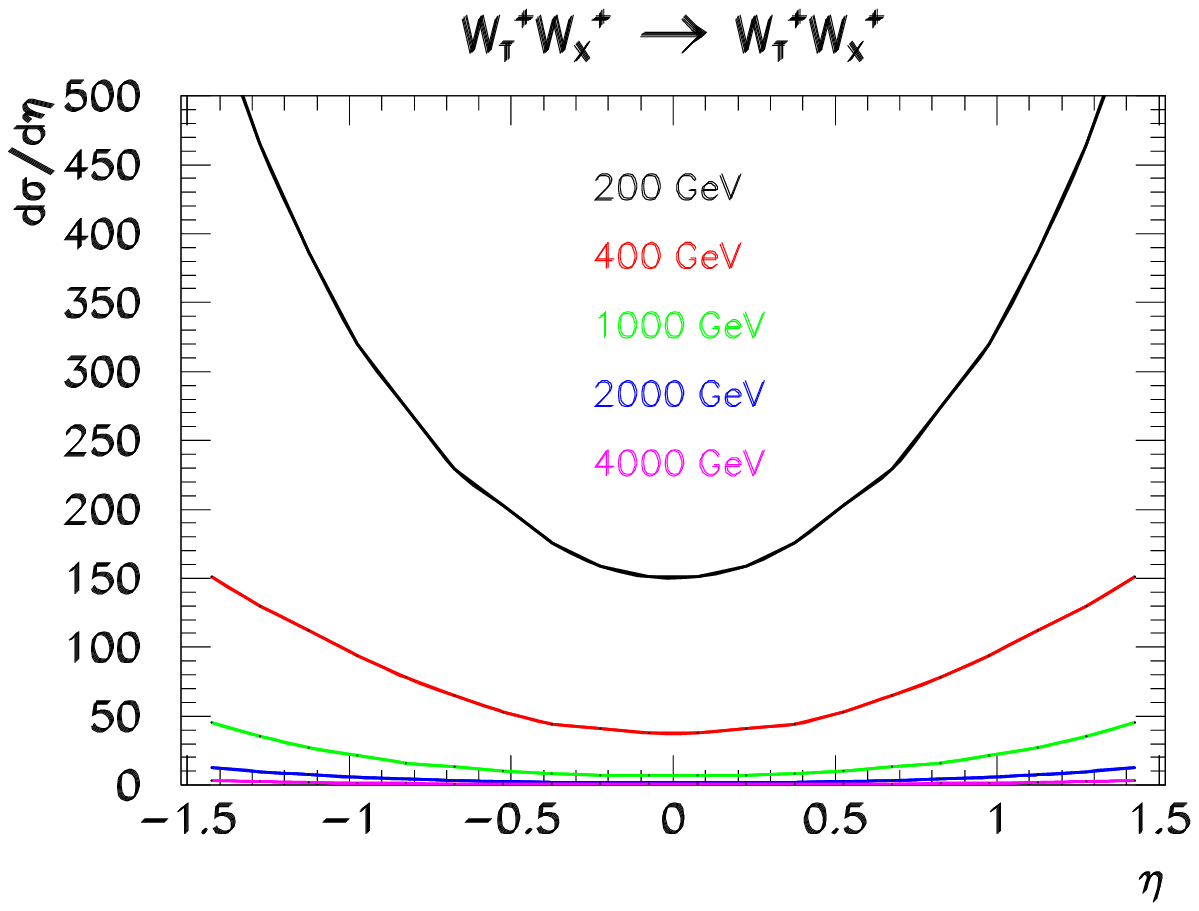,width=0.55\linewidth}
\vspace{-1cm}
\epsfig{file=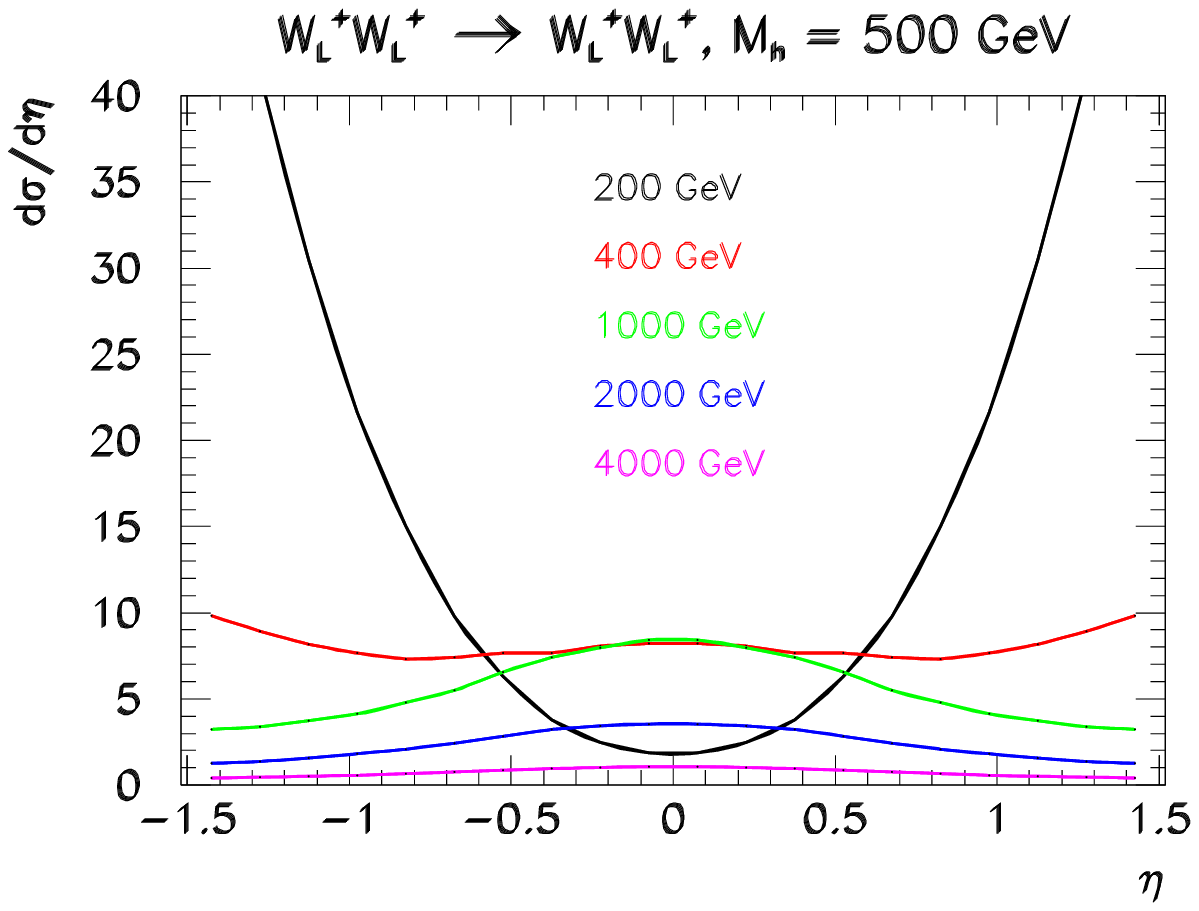,width=0.55\linewidth}
\hspace{-1cm}
\epsfig{file=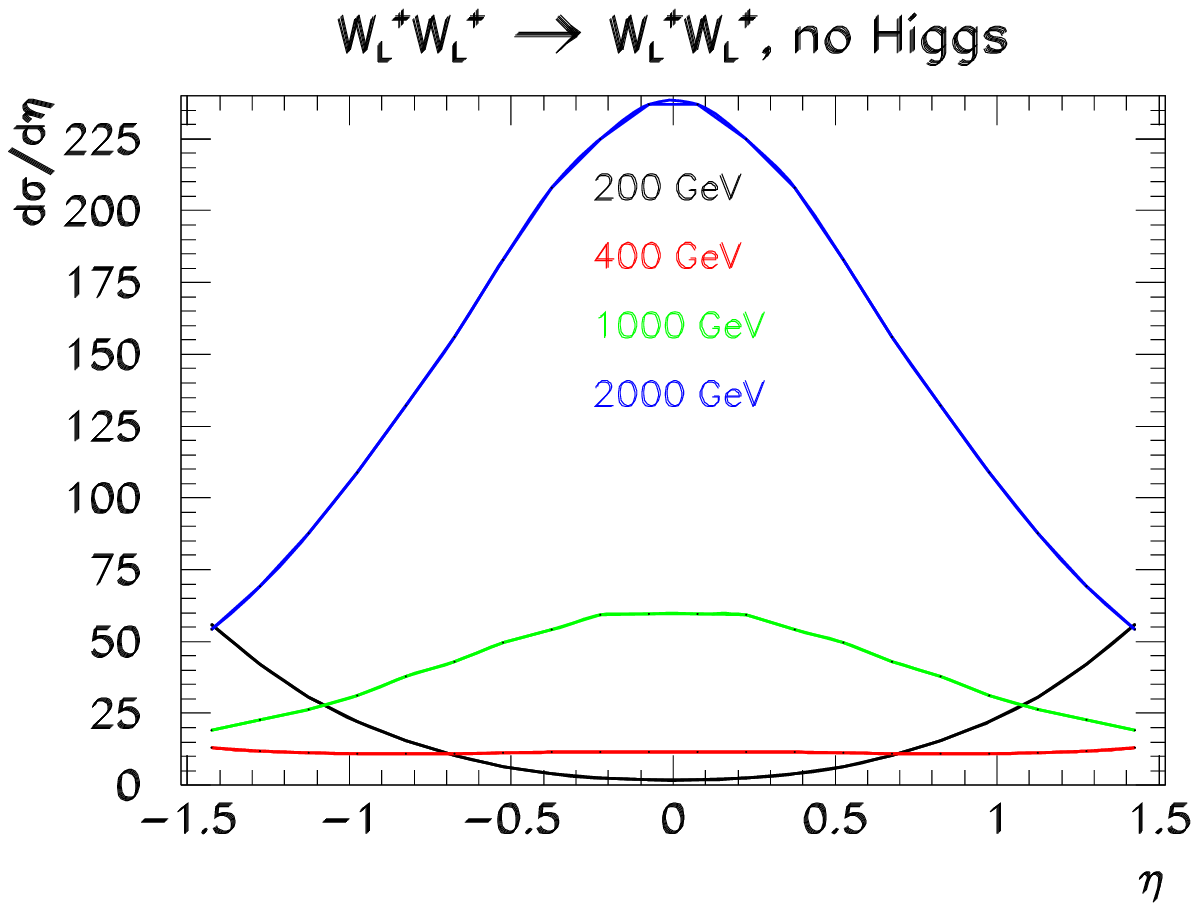,width=0.55\linewidth}
\vspace{1cm}
\caption{Examples of angular distributions of the scattered $W^+W^+$ pairs 
(pseudorapidities with respect
to the incoming $W^+W^+$ direction) at different center of mass energies,
depending on the value of the Higgs mass.
SM-like couplings were assumed in all the cases.
{\bf Top left:} $W_L^+W_L^+$ with a 120 GeV Higgs.
{\bf Top right:} $W_T^+W_X^+$ (here subscript $X$ denotes any polarization, $T$ or $L$).
{\bf Bottom left:} $W_L^+W_L^+$ with a 500 GeV Higgs.
{\bf Bottom right:} $W_L^+W_L^+$, Higgsless case.  The blue curve in the last plot
already involves
unitarity violation and therefore is unphysical.  Results of MadGraph calculations.}
\end{figure}

\begin{figure}[htbp]
\vspace{-5mm}
\epsfig{file=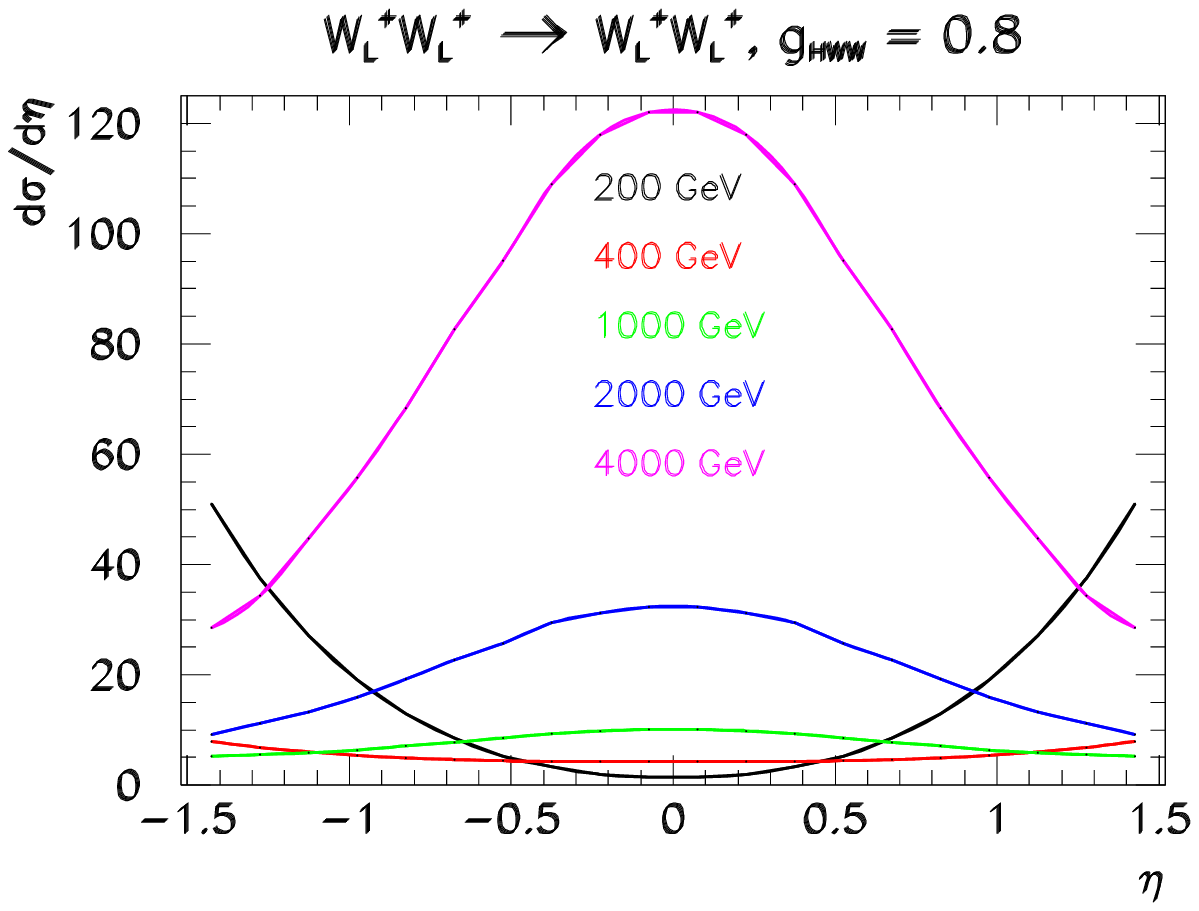,width=0.55\linewidth}
\hspace{-1cm}
\epsfig{file=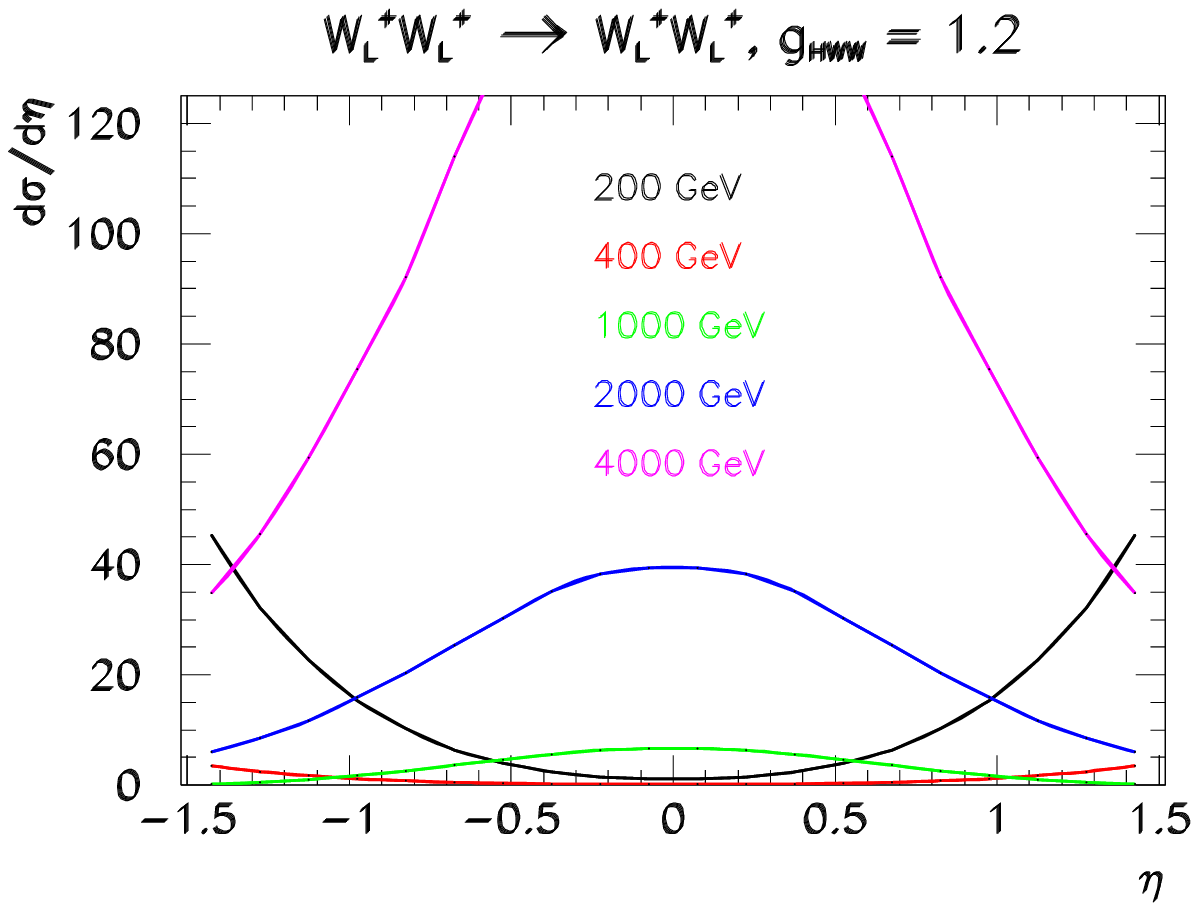,width=0.55\linewidth}
\vspace{-1cm}
\epsfig{file=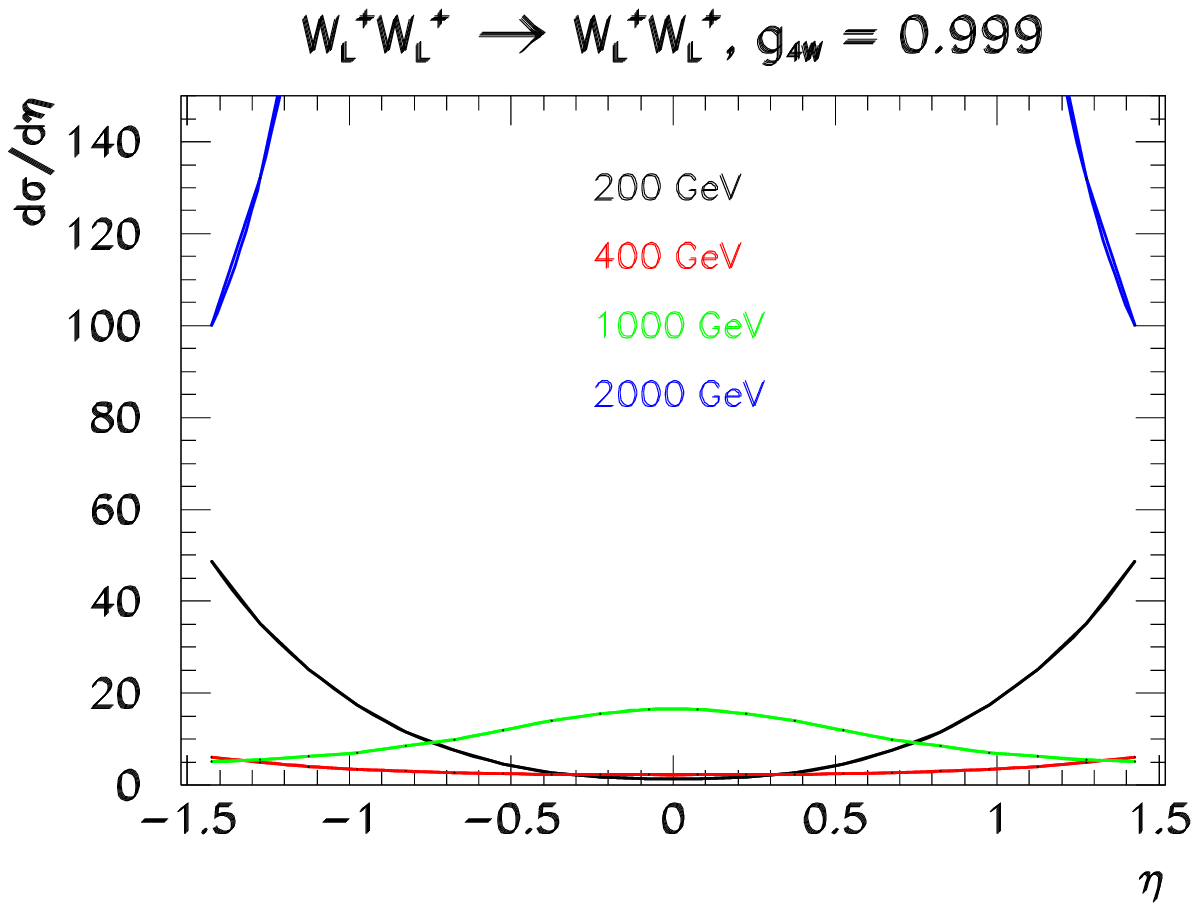,width=0.55\linewidth}  
\hspace{-1cm}
\epsfig{file=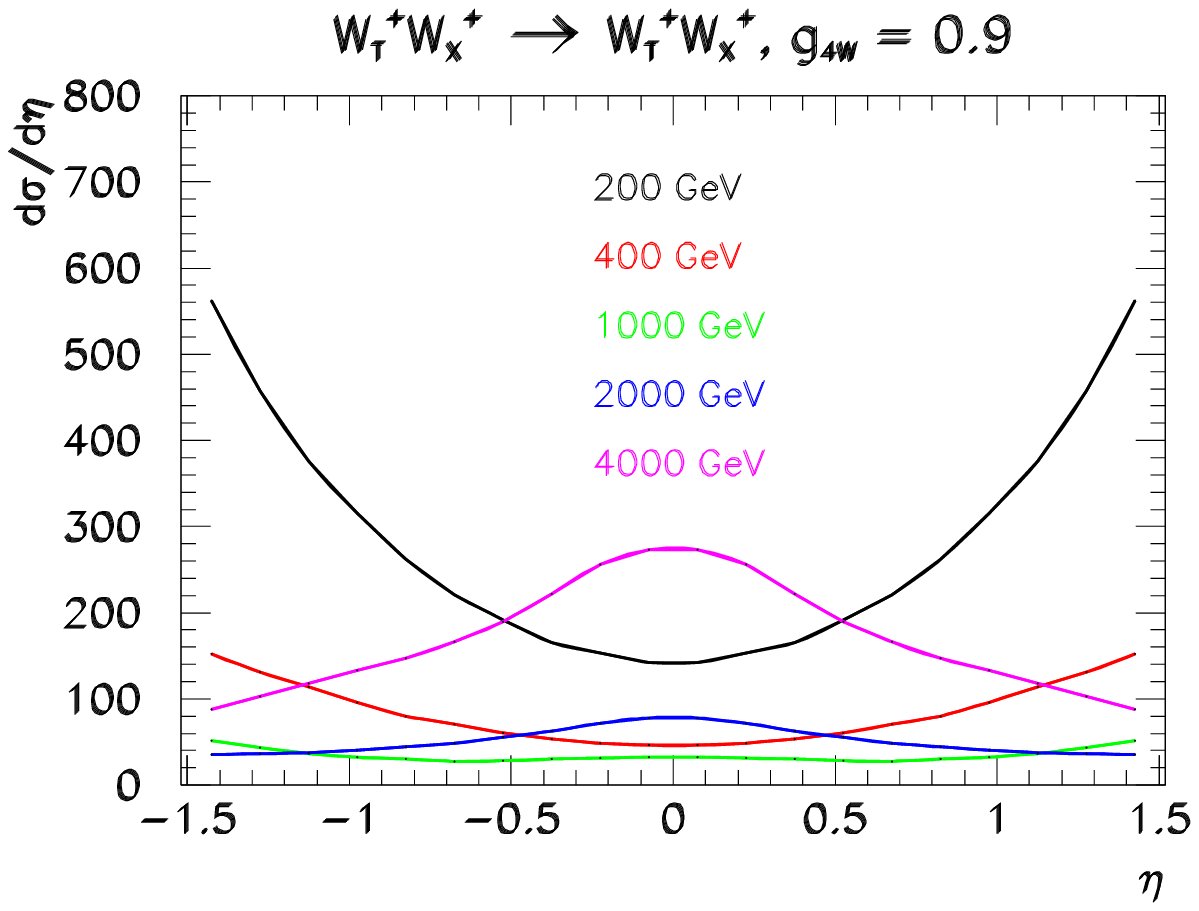,width=0.55\linewidth}  
\vspace{1cm}
\caption{Examples of angular distributions of the scattered of $W^+W^+$ pairs
(pseudorapidities with respect to the incoming $W^+W^+$ pair direction)
at different
center of mass energies, depending on the Higgs and gauge couplings.
A 120 GeV Higgs boson was assumed in all the cases.
{\bf Top left:} $W_L^+W_L^+$ with the $HWW$ coupling equal to 0.8 times its SM value.
{\bf Top right:} $W_L^+W_L^+$ with the $HWW$ coupling equal to 1.2 times its SM value.  
{\bf Bottom left:} $W_L^+W_L^+$ with the SM $WWWW$ coupling scaled by a
factor of 0.999 (the partially visible blue curve
involves unitarity violation and therefore is unphysical).  
{\bf Bottom right:} $W_T^+W_X^+$ with the SM $WWWW$ coupling scaled by a
factor of 0.9.  Results of MadGraph calculations.}
\end{figure}

\subsection{Gauge boson couplings in $VV$ scattering}

As mentioned in the previous chapter, the high energy behavior of vector boson
scattering amplitudes is sensitive not only to the Higgs couplings to vector bosons
(and Higgs mass), but also to the triple and quartic vector boson couplings.  As much as
the former are measured via Higgs partial width measurements, the latter can be
probed independently via measurements of diboson and triboson production.  Consistency
of the three types of measurements: Higgs couplings, multiboson production and vector
boson scattering at high energy is an important closure test for any consistent
physical theory and should be rigorously tested.

There is at least one fundamental difference between the phenomenology of
scaled Higgs couplings and that of non-SM gauge couplings.
The former manifests solely
in $W_LW_L$ pairs, ultimately as an enhancement with energy.
In the latter, there is always a combination of two effects.  One is still the
energy dependence of $W_LW_L$, which in this case is even steeper because the
leading divergence now goes like the fourth power of energy (to begin with, we are
assuming a simple scaling of the SM couplings by a constant factor), the other
is the overall energy-independent normalization constant which affects in
principle all helicity combinations in the same way.  This will be mainly
observable in $W_TW_T$ pairs, because they are the most abundant.  Because
however such normalization shifts will be much better measurable in the total
diboson production than in boson-boson scattering, this effect is of lesser 
interest for us.  Mixed $W_TW_L$ pairs will be
modified in both ways: in the overall normalization and as a rise at high energy
(remember that each $W_L$ intrinsically carries energy dependence!).  
Therefore, in the general case, both $W_TW_X$ as well as $W_LW_L$ may be of interest.
Moreover, angular distributions in vector boson scattering (VBS) processes
exhibit similar qualitative features for $W_TW_X$ and $W_LW_L$ pairs in the
scenario with a modified quartic coupling.
The leading divergence in a VBS process
is the same in case of an anomalous quartic coupling as
for an anomalous triple gauge coupling.  Put another way, for every anomalous
quartic coupling, there is an equivalent value of the triple couplings that
asymptotically produces the same effect.
As long as we restrict ourselves to pure VBS
processes and scaling individual SM couplings by constant
factors, energy dependence of $W_LW_L$ pairs still carry the most information.
This is because of their much steeper energy dependence which very quickly
dwarfs any effects in $W_TW_X$.  But, as we will see in the next chapter, a clean VBS
sample is impossible to isolate in a real experiment.  And new physics is likely
to modify different couplings in a correlated way.

\begin{figure}[htbp]
\vspace{-5mm}
\begin{center}
\epsfig{file=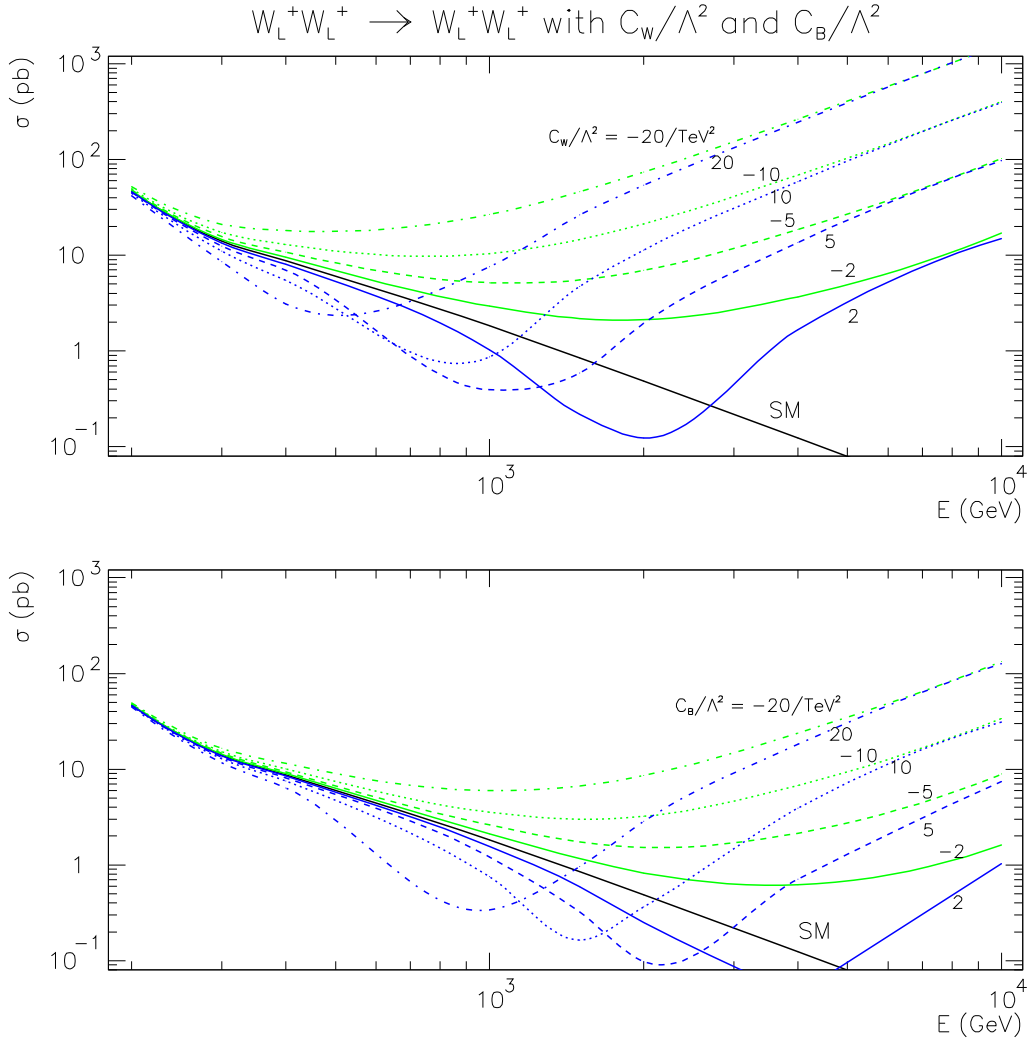,width=0.95\linewidth}
\end{center}
\vspace{-1cm}
\caption{The total $W_L^+W_L^+$ scattering cross section as a function of the center of
mass energy for different values of the relevant dimension-6 operators in the
$W$ Effective Field Theory approach.  Varied are: $C_W/\Lambda^2$ (upper plot)
and $C_B/\Lambda^2$ (lower plot).
Assumed here are two colliding on-shell, unpolarized $W^+$ beams and a 120 GeV Higgs
boson.
A cut on the scattering angle that corresponds to pseudorapidity of $\pm 1.5$ with
respect to the incoming $W$ direction was applied.  Results of MadGraph
calculations.}
\end{figure}

\begin{figure}[htbp]
\vspace{-5mm}
\begin{center}
\epsfig{file=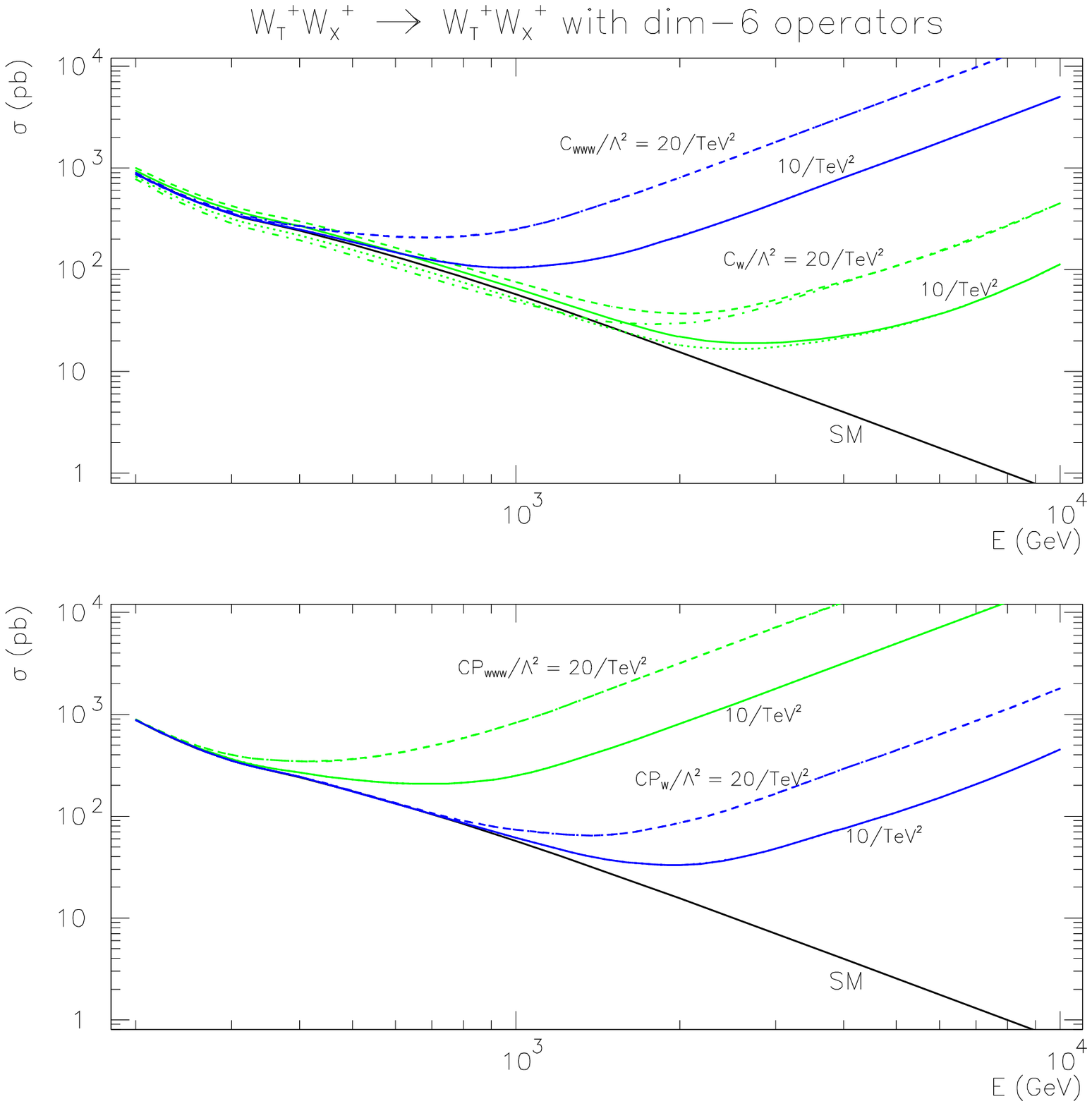,width=0.95\linewidth}
\end{center}
\vspace{-1cm}
\caption{The total $W_T^+W_X^+$ scattering cross section as a function of the center of
mass energy for different values of the relevant dimension-6 operators in the
$W$ Effective Field Theory approach.  
Varied are: $C_W/\Lambda^2$,
$C_{WWW}/\Lambda^2$ (upper plot), $C_{\tilde{W}}/\Lambda^2$ and
$C_{\tilde{W}WWW}/\Lambda^2$ (labeled $CP_W$ and $CP_{WWW}$, lower plot).
Assumed here are two colliding on-shell, unpolarized $W^+$ beams and a 120 GeV Higgs
boson.  The rises at high energy are due to the $W_T^+W_L^+$ combination, total
normalization effects are predominantly due to $W_T^+W_T^+$.
A cut on the scattering angle that corresponds to pseudorapidity of $\pm 1.5$ with
respect to the incoming $W$ direction was applied.  Results of MadGraph
calculations.}
\end{figure}

New physics may mainfest itself in new interactions between gauge bosons.
These interactions should show up indirectly as certain combinations of
modified effective gauge boson couplings and Higgs to gauge couplings.  We don't know
the underlying new physics, but we do have a theoretical machinery to parametrize
it in a model independent way.  This is where Effective Field Theory
comes back.  Once again, this general framework has enough flexibility to describe
the low energy phenomenology
of new physics regardless of what it really is.

A modern effective quantum field theory for physics beyond the Standard Model can be
written down in terms of an extended Lagrangian \cite{weft}

\begin{equation}
\mathscr{L} = \mathscr{L}_{SM} + \sum_i \frac{c_i}{\Lambda^2} \mathscr{O}_i +
\sum_j \frac{f_j}{\Lambda^4} \mathscr{O}_j + ...
\end{equation}

\noindent
where $\mathscr{O}_i$ are dimension-six operators,
$\mathscr{O}_j$ are dimension-eight operators,
the coefficients $c_i, f_j$ are dimensionless and $\Lambda$ is the energy scale of
new physics.  The Standard Model is recovered in the limit $\Lambda \to \infty$
and the entire model is bound to capture all the low-energy effects of physics
beyond the Standard Model.  By dimensional analysis one expects the lowest
dimensional, hence dimension-6, operators to be dominant, since all higher dimension
operators are suppressed by higher powers of $\Lambda$.  However, different gauge
boson interactions may or may not probe some or any of these operators, and so
going to dimension 8 is necessary for a more complete description (only
even-dimensional operators conserve lepton and baryon number).
Operators $\mathscr{O}_i$, $\mathscr{O}_j$ are constructed from known fields,
that is, particles
of the Standard Model.  Discovery of a new particle should result in revision
of the model and inclusion of additional operators.

All related BSM phenomenology is described in a way which depends only on
the ratios $c_i/\Lambda^2$ or $f_j/\Lambda^4$.
However, practical usefulness of an
effective quantum field theory is restricted up to energies of order $\Lambda$.
At energies higher than that, operators of arbitrary high dimension become important,
because they are no longer suppressed.  The quantity $\Lambda$ characterizes a
particular theory.
While we don't know the scale at which
new physics sets in, parameters of the form $c_i/\Lambda^2$ and
$f_j/\Lambda^4$ have calculable intrinsic validity bounds defined by the unitarity
condition.  These bounds fix the maximum allowed value of the scale $\Lambda$
that is relevant should new physics arise from any particular higher dimension
operator.  A prescription to apply $K$-matrix unitarization within the context
of the effective field theory has also been proposed \cite{kilian}.
In an alternative formulation, sometimes
called the Lagrangian approach \cite{hagiwara},
the anomalous couplings are taken to be constant Lagrangian parameters.
There is no explicit relation to the scale of new physics and this formalism is
applicable in the approximation in which these parameters do
not depend on energy.

The Effective Field Theory approach has been
gaining wide recognition in recent time, with more and more studies of sensitivity
to BSM being expressed in this language.
All possible independent dimension-6 operators constructed from the known fields
have been catalogued \cite{dim6}.
There are just three dimension-6 operators that conserve both $C$ and $P$
and affect the interactions of gauge bosons.  Following the notation used
elsewhere in literature, these can be written as:

\begin{equation}
\mathscr{O}_{WWW} = {\tt Tr}[W_{\mu\nu} W^{\nu\beta} W^\mu_\beta],
\end{equation}

\begin{equation}
\mathscr{O}_W = (D_\mu\Phi)^\dagger W^{\mu\nu}(D_\nu\Phi),
\end{equation}

\begin{equation}
\mathscr{O}_B = (D_\mu\Phi)^\dagger B^{\mu\nu}(D_\nu\Phi).
\end{equation}

\vspace{3mm}
\noindent
In the above, $\Phi$ is the Higgs doublet field and 

\begin{equation}
W_{\mu\nu} = \frac{ig}{2}\sigma^a (\partial_\mu W^a_\nu - \partial_\nu W^a_\mu +
g \epsilon_{abc} W^b_\mu W^c_\nu),
\end{equation}

\begin{equation}
B_{\mu\nu} = \frac{ig'}{2} (\partial_\mu B_\nu - \partial_\nu B_\mu);
\end{equation}

\noindent
$\sigma^a$ are Pauli matrices and $g$, $g'$ are the SU(2) and U(1) gauge couplings,
respectively.  Two additional operators
appear if we do not assume $C$ and $P$ conservation:

\begin{equation}
\mathscr{O}_{\tilde{W}WW} = {\tt Tr}[\tilde{W}_{\mu\nu} W^{\nu\beta} W^\mu_\beta],
\end{equation}

\begin{equation}
\mathscr{O}_{\tilde{W}} = (D_\mu\Phi)^\dagger \tilde{W}^{\mu\nu}(D_\nu\Phi).
\end{equation}

\vspace{3mm}
\noindent
Here the dual field strengths are defined as 
$\tilde{V}_{\mu\nu}=\frac{1}{2} \epsilon_{\mu\nu\rho\sigma} V^{\rho\sigma}$.

Typically, one such operator modifies more than one interaction vertex and
vice-versa, each interaction receives the contributions from more than one
higher-dimension operator.
There is some arbitariness in the way all these operators are defined.
Recently it was noted \cite{masso}
that the most useful formulation, at least from an experimental
point of view, could be one obtained by choosing a basis of higher-dimension operators
such that they match closely the measured processes, ideally in a one-to-one
correspondence.  Such approach would allow to study one
vertex at a time and spare from the additional work of combining data from different
processes in order to study the potential effects of a single operator.
This interesting approach is
still in the lounge, waiting for being implemented in commonly accessible event
generators and used in data analyses, and hence will not be applied in this work.

It is transparent that $VV$ scattering processes are not the best channels to
study operators involving modifications of triple gauge couplings.
Much better statistical significance can be obtained by measuring the total diboson
production, i.e., not necessarily in the VBS mode.  It is
actually non-VBS diboson production that produces the most stringent limits on
these parameters to present day, with currently existing data coming from LEP 
\cite{tgbc} and
TeVatron experiments,
as well as from Run 1 of the LHC.  However, it is vital to know how these
operators will affect the VBS measurements.

All of the above operators modify
triple vector boson couplings, in addition of some of them modifying the Higgs
couplings and/or the quartic vector boson couplings.  Namely, $\mathscr{O}_{WWW}$
modifies
also the quartic $WWWW$ coupling, while $\mathscr{O}_W$ modifies both the quartic
and the $HWW$ coupling.  By contrast, operator $\mathscr{O}_B$ affects neither, but it
may affect other couplings, like $HZZ$ or $HZ\gamma$.
The key point that we want to emphasize
and elaborate further on in this work is that each of these operators affects
$W_LW_L$ and $W_TW_X$ pairs differently.  This fact will have a paramount importance
in order to interpret correctly the results of future measurements, which will
- most probably - reveal a complicated combination of many effects (if anything!).

It is not difficult to tell which operators can affect $W_LW_L$, $W_TW_T$ or $W_TW_L$
vertices straight from their definition,
even without deep knowledge of quantum field theory.  A $W$ field can be
obtained either via a field strength $W_{\mu\nu}$ or via a Higgs field derivative.
Every appearance of the
field strength in the operator corresponds to transverse helicity.  Longitudinal
helicities
enter via covariant derivatives of the Higgs field ($D_\mu\Phi$).  
Consequently, $\mathscr{O}_B$ can affect only
the scattering of $W_LW_L$ pairs, while $\mathscr{O}_W$ affects all possible helicity
combinations.  Operator $\mathscr{O}_{WWW}$ has only field strengths in it, hence it
affects directly only vertices involving $W_TW_T$.  However, mixed pairs $W_TW_L$
get also affected
indirectly, via the $t$-channel scattering process with a $Z_T$
exchange, in which one vertex is bound to comprise only transverse helicity states.

Although a simple scaling of a triple or quartic gauge coupling by a constant
factor produces a
divergence that goes like $s^2$, gauge invariance enforces cancelation of the
$\sim s^2$ terms for all the dimension-6 operators \cite{corbett}.
Consequently, the leading divergences are always proportional to $s$.

In the language of higher dimensional operators, the Higgs to gauge couplings can
be furthermore modified via:

\begin{equation}
\mathscr{O}_{\Phi d} = \partial_\mu (\Phi^\dagger\Phi) \partial^\mu (\Phi^\dagger\Phi),
\end{equation}

\begin{equation}
\mathscr{O}_{\Phi W} = (\Phi^\dagger\Phi) {\tt Tr}[W^{\mu\nu}W_{\mu\nu}].
\end{equation}

\vspace{3mm}
\noindent
Both modify $HWW$ and $HZZ$ vertices, but not pure gauge couplings.  
The first of them affects only $HW_LW_L$ vertices and will be further considered by
means of a simple scaling of the $HWW$ coupling by a constant for better transparency.
The second one generates anomalous $HW_TW_X$ vertices and has no impact on
$W_LW_L$.  In addition, operator

\begin{equation}
\mathscr{O}_{\Phi B} = (\Phi^\dagger\Phi) B^{\mu\nu}B_{\mu\nu}
\end{equation}

\vspace{3mm}
\noindent
affects only $HZZ$, $HZ\gamma$ and $H\gamma\gamma$ of all the triple vertices
and so it can be probed via $ZZ$ scattering.

Here and in the remainder of this work we are
assuming that the Higgs boson is a pure scalar.  Possible admixtures from non-scalar
components can be parameterized by means of an additional set of higher dimension
operators.  They have been constrained by the LHC Higgs data at 7 and 8 TeV, using
a combination of
the most sensitive Higgs decay channels.  The limits were mainly driven by the
$ZZ$ and $\gamma\gamma$ channels, while standalone limits from $WW$ in the purely
leptonic decay mode are actually the weakest
because in this case crucial kinematic information escapes with the two undetected
neutrinos.
In fact, the entire visible $WW$ phenomenology very weakly depends on non-scalar
admixture effects within the limits driven by the other decay modes.
Although not necessarily negligible
on their own and although potentially important from the interpretative point of view, 
such effects cannot make any major impact on our considerations.

\vspace{5mm}

\begin{table}[htbp]
\begin{center}
\begin{tabular}{|c||c|c|c|c|c|c|}
\hline
Vertex, helicities & $\mathscr{O}_{WWW}$ & $\mathscr{O}_W$ & $\mathscr{O}_B$ &
$\mathscr{O}_{\Phi d}$ & $\mathscr{O}_{\Phi W}$ & $\mathscr{O}_{\Phi B}$ \\
\hline\hline
$HWW$, $W_LW_L$ & - & v & - & v & - & - \\
$HWW$, $W_TW_X$ & - & v & - & - & v & - \\
$WWZ$, $W_LW_L$ & - & v & v & - & - & - \\
$WWZ$, $W_TW_X$ & v & v & - & - & - & - \\
$WW\gamma$, $W_LW_L$ & - & v & v & - & - & - \\
$WW\gamma$, $W_TW_X$ & v & (v) & - & - & - & - \\
$WWWW$, $W_LW_L$ & - & v & - & - & - & - \\
$WWWW$, $W_TW_X$ & v & v & - & - & - & - \\
\hline
\end{tabular}
\end{center}
\caption{Sensitivity to dimension-6 operators of the individual gauge and Higgs to gauge
couplings that contribute to $WW$ scattering, decomposed into helicity combinations of
the interacting (initial and final) $WW$ pair.  Note that these are not necessarily
the helicities at a single vertex.  Helicity-flip contributions ($W_LW_L \to
W_TW_X$ and $W_TW_X \to W_LW_L$) have been ignored in this table.  For the $W^\pm W^\pm$
process these effects are only relevant at center of mass energies
near the $WW$ mass threshold and do not get enhanced by any of the dimension-6 operators.
The same is not necessarily true for the $W^+W^-$ process.
The entry marked as (v) stands for marginally sensitive, but not measurable.}
\label{tab:ewdim6}
\end{table}

It is also possible to reinterpret the anomalous couplings quoted in
section \ref{lhcatgc} in the language of the
coefficients of dimension-6 operators.  Based on Ref.~\cite{weft}, one gets
the following relations:

\begin{equation}
c_{WWW}/\Lambda^2 = \frac{2\lambda_\gamma}{3 g^2 m_W^2} = \frac{2\lambda_Z}{3 g^2 m_W^2},
\end{equation}

\begin{equation}
c_W/\Lambda^2 = 2 \frac{\Delta g_1^Z}{m_Z^2},
\end{equation}

\begin{equation}
c_B/\Lambda^2 =
2 \left[\frac{\Delta\kappa_\gamma}{m_W^2} - \frac{\Delta g_1^Z}{m_Z^2}\right] =
2 \frac{\Delta\kappa_\gamma-\Delta\kappa_Z}{m_Z^2}.
\end{equation}

\vspace{1cm}

\noindent
It should be stressed that the above relations hold so long as we expect
the dimension-6 operators be dominant.  Consideration of dimension-8 operators would
generally render them not valid anymore.

Precise determination of the corresponding limits on these coefficients from
the most up-to-date combination of all the existing LHC, TeVatron and LEP data is a
complicated task
that surpasses the scope of this work.  It is also inessential for us
in this moment.  In fact, most of these limits so far have not changed dramatically
since LEP times.  Their improvement will be possible with an order of magnitude
increase in integrated luminosity and doubled beam energy planned for LHC
Runs 2 and 3.
Without getting into too much detail and in accordance with the quoted relations,
we can safely assume the
allowed dimension-6 operator coefficients $c_{WWW}/\Lambda^2$, $c_W/\Lambda^2$ and
$c_B/\Lambda^2$ still be of order $\pm$1-10 TeV$^{-2}$.

On the other hand, a clean study of quartic gauge boson couplings can be
carried with interactions that do not have a triple vertex associated to it.  These are
described using dimension-8 effective operators.  Dimension-8 operators are not 
necessarily just
a higher order correction to dimension-6 operators.  Likely, they probe
different physics.  Anomalous triple couplings can result from averaging out
unknown heavy particles in loops.  Quartic couplings can be regarded as a window
to electroweak symmetry breaking.  They arise as a contact interaction
manifestation of heavy particle exchange.  It is quite possible that quartic
couplings deviate from the SM, but triple couplings do not.
The operators of direct relevance for us are:

\begin{equation}
\mathscr{O}_{S,0} = [(D_\mu\Phi)^\dagger D_\nu\Phi] \times [(D^\mu\Phi)^\dagger D^\nu\Phi]
\end{equation}

\noindent
and

\begin{equation}
\mathscr{O}_{S,1} =
[(D_\mu\Phi)^\dagger D^\mu\Phi] \times [(D_\nu\Phi)^\dagger D^\nu\Phi],
\end{equation}

\vspace{3mm}
\noindent
because they only modify the $WWWW$ and $WWZZ$ vertices.  A combination
$8\frac{c-1}{v^4} (\mathscr{O}_{S,0} - \mathscr{O}_{S,1})$,
where $v$ is the Higgs vacuum expectation
value and $c$ is a dimensionless number, corresponds to a simple rescaling of the
Standard Model quartic coupling by a factor $c$.

The above are the only two independent operators constructed solely from Higgs
field derivatives, and hence affecting only $W_LW_L$ pairs.  Additional dimension-8
operators can be constructed from field strength tensors and field derivatives
or from field strength tensors alone.  These are:

\begin{equation}
\mathscr{O}_{M,0} =
{\tt Tr} [W_{\mu\nu}W^{\mu\nu}] \times [(D_\beta\Phi)^\dagger D^\beta\Phi],
\end{equation}

\begin{equation}
\mathscr{O}_{M,1} =
{\tt Tr} [W_{\mu\nu}W^{\nu\beta}] \times [(D_\beta\Phi)^\dagger D^\nu\Phi],
\end{equation}

\begin{equation}
\mathscr{O}_{M,6} =
(D_\mu \Phi)^\dagger W_{\beta\nu}W^{\beta\nu}D^\mu \Phi,
\end{equation}

\begin{equation}
\mathscr{O}_{M,7} =
(D_\mu \Phi)^\dagger W_{\beta\nu}W^{\beta\mu}D^\nu \Phi,
\end{equation}

\begin{equation}
\mathscr{O}_{T,0} = {\tt Tr} [W_{\mu\nu}W^{\mu\nu}] \times
{\tt Tr} [W_{\alpha\beta}W^{\alpha\beta}],
\end{equation}

\begin{equation}
\mathscr{O}_{T,1} = {\tt Tr} [W_{\alpha\nu}W^{\mu\beta}] \times
{\tt Tr} [W_{\mu\beta}W^{\alpha\nu}],
\end{equation}

\begin{equation}
\mathscr{O}_{T,2} = {\tt Tr} [W_{\alpha\mu}W^{\mu\beta}] \times
{\tt Tr} [W_{\beta\nu}W^{\nu\alpha}].
\end{equation}

\vspace{3mm}
We have only listed here the operators that affect the same-sign $W^\pm W^\pm$
scattering process.  A full list of dimension-8 operators that can modify
quartic gauge couplings, including those which can produce anomalous quartic vertices
involving only $Z$'s and $\gamma$'s, that do not exist in the SM, can be found 
in Ref.~\cite{snowmassew}.
Numerical coefficients behind these operators (usually denoted as $f$ with the
appropriate subscripts) are largely unconstrained by
experiment.  The possibilities to study quartic couplings at LEP were very limited,
while the TeVatron did not offer enough energy and luminosity.
Vector Boson Scattering (VBS) at the LHC is the right place
to probe them.

\section{Beyond the Standard Model?}

Upon discovery of the Higgs boson, the Standard Model has been completed.
Is this really the end of the story?  
Volumins of theoretical papers have been written
to explain why the Standard Model cannot be the ultimate theory and we will not
repeat these arguments here.  And yet, for improbable
this may seem at first glance, the bare truth is that hardly any experimental result
in particle physics to the present date
can be said to support the idea that the Standard Model needs any
major change anywhere below the Planck scale!  Let us critically review what we
currently have.  The muon magnetic dipole moment
may be one such indication \cite{gminus2}.
The magnetic dipole moment is a measure of quantum effects that modify
the effective strength of a charged particle interaction with a photon.
These quantum corrections can be very precisely predicted in the framework of
Quantum Electrodynamics (QED).  Such calculations consistently reveal a lower
value than the experimental world average, the discrepancy is
currently at the level of 3.6$\sigma$.
This result, while very interesting, is still not significant enough, as well as
too isolated and indirect to be convincing on its own right.
The last 15-20 years brought an explosion of neutrino physics projects,
following the observation of neutrino oscillations by Super-Kamiokande.
But neutrino masses, regardless of what they ultimately are, including Dirac or
Majorana, can be in principle accomodated within the
Standard Model if only we relax the massless neutrino prejudice
which used to be sort of imposed by hand to the theory before 1996.  
In a minimalistic scenario it would only require giving neutrinos what
they always {\it could} have within the Standard Model framework and
not make any impact on the rest of the theory.
Finally, the much celebrated naturalness problem, i.e., keeping the
Standard Model Higgs boson light
despite its quadratically divergent
radiative corrections from fermionic loops, is possible simply by invoking
some kind of anthropic principle (technically by assuming
an enormous amount of fine tuning between the ``naked" mass and the radiative mass
shift).  Whether or not we find such solutions satisfactory from the
purely aesthetic point of view is, alas, a different question.
Yet other claims for the necessity of physics beyond the Standard Model
have been made on purely theoretical grounds, like within the frameworks of
Grand Unification Theories, Superstrings, etc., but they all
lack any experimental evidence.

More suggestive in this respect are in fact astrophysical observations.
Evidence of Dark Matter in the Universe is firmly established and does call for new
physics.  It could be argued, though, that in principle nothing forbids adding
extra particles to the Standard Model Lagrangian that
would completely decouple from the known particles except via gravitation, without
adding anything to
our undestanding of the known part of the world.  Overwhelming excess of matter
over anti-matter in
the Universe cannot be explained by Standard Model physics, either, at least
in its presently known form.  But it is still an open question whether this
asymmetry can be explained in terms of leptogenesis in the scenario of
a strongly $CP$-violating neutrino sector.
And that's really all we have.

Of the proposed extenstions of the Standard Model, Supersymmetry (SUSY) 
represents the best known class of models.  Originally proposed to tackle
the technical problem of loop corrections to the Higgs mass, over thirty years
later it still offers a wide
range of valid models which to this date are neither confirmed nor excluded
experimentally.  Results of the LHC Run 1 have rendered the simplest SUSY models,
such as the MSSM or the NMSSM, less popular, but more generalized models are
still in the mainstream of BSM searches.  SUSY has been said to be
the only known class of models that reduce {\it exactly} to the Standard
Model at low energy, so as to possibly reveal no hints of itself whatsoever at the
presently reachable energies.  Depending on one's point of view, this can be
found as much an advantage as a weakness.
In fact, if SUSY is true, there is not much to expect in the forseeable future
from $WW$ scattering, either,
in terms of deviations from the Standard Model.

A separate wide class of
alternative candidates for physics Beyond the Standard Model is
known as the Strongly Interacting Light Higgs (SILH) models \cite{partial}.
They are generally based on the assumption that electroweak symmetry breaking is
triggered by a light {\it composite} Higgs, which emerges from a new
strongly-interacting sector as a pseudo-Goldstone boson.  This implies the
existence, at some higher energy scale, of an additional particle spectrum,
characterized by a typical mass parameter $M >> M_H$ and a coupling constant
$g$, with $g_{SM} << g < 4\pi$.  The Higgs multiplet is assumed to belong to this
``strong" sector.
In the limit $g_{SM}=0$ the Higgs becomes an exact Goldstone boson.
Ordinary Standard Model particles couple weakly to the strong sector.
Models known as Little Higgs \cite{littlehiggs}, Littlest Higgs \cite{littlesthiggs}, 
Holographic Higgs \cite{holographic}, etc., are particular
variations of this general idea.

The effective Lagrangian corresponding to this class of models can again be written
down in a parametric form, where different physical scenarios correspond to
different values of the parameters in the Lagrangian.  In the low energy
approximation, corresponding to the energies accessible in the LHC, the
Lagrangian can be symbolically rewritten as a sum

\begin{equation}
\mathscr{L} = \mathscr{L}_{SM} + \mathscr{L}_H + \mathscr{L}_V,
\end{equation}

\noindent
where $\mathscr{L}_{SM}$ is our familiar Standard Model Lagrangian,
$\mathscr{L}_H$ describes
additional interactions involving the Higgs boson and $\mathscr{L}_V$ describes
additional interactions involving gauge bosons only.  These new interactions imply
modifications of the cross sections and branching fractions of the Higgs boson
relative to the predictions of the Standard Model.  In particular, Higgs couplings
to known fermions and bosons are somewhat different than in the Standard Model.
In an effective formulation, the whole Higgs-related phenomenology of SILH models can be
described via the choice of a few numbers that parameterize our ignorance of the
underlying physics.  It was shown that general rules of SILH select just three of
them as the most
important ones for LHC studies, which govern the leading effects expected in
Higgs physics.  In the following these are denoted as
$\xi = (vg/M)^2$ ($v$ = 246 GeV is the Higgs vacuum expectation value),
$c_y$ and $c_H$.  In terms of these parameters, the Higgs partial widths are
modified with respect to the Standard Model as follows:

\begin{equation}
\Gamma (h\rightarrow f\bar{f})_{SILH} = 
\Gamma (h\rightarrow f\bar{f})_{SM} [1 - \xi (2c_y + c_H)],
\end{equation}

\begin{equation}
\Gamma (h\rightarrow W^+W^-)_{SILH} =
\Gamma (h\rightarrow W^+W^-)_{SM} [1 - \xi (c_H-\mathscr{O}(g_{SM}^2/g^2))],
\end{equation}

\begin{equation}
\Gamma (h\rightarrow ZZ)_{SILH} =
\Gamma (h\rightarrow ZZ)_{SM} [1 - \xi (c_H-\mathscr{O}(g_{SM}^2/g^2))],
\end{equation}

\begin{equation}
\Gamma (h\rightarrow\gamma\gamma)_{SILH} =
\Gamma (h\rightarrow\gamma\gamma)_{SM} [1 - \xi Re(\frac{2c_y+c_H}{1+J/I}
+ \frac{c_H}{1+I/J} + \mathscr{O}(g_{SM}^2/g^2))].
\end{equation}

\noindent
Here $I$ and $J$ are loop functions describing Higgs radiative decays whose
numerical values depend mostly on the top quark mass.  Their
full definitions can be found in Ref.~\cite{partial}.  
The $\xi$ parameter naturally ranges between
0 and 1, the two limiting cases corresponding to the Standard Model and
technicolor theories, respectively.  Note that to the lowest order it is correct
to say that SILH phemomenology in comparison with the Standard Model
can be described as an overall modification of all the Higgs couplings to fermions
and another overall modification of all the Higgs couplings to gauge bosons.
For example, the ``fermiophobic Higgs" scenario is obtained by setting
$c_H$=0 and $\xi c_y$=1/2.
Extraction of $c_y$ and $c_H$ is a main task for precision measurements of the Higgs
production rate and branching fractions.  Given that $c_y$ and $c_H$ are
typically numbers of the order of unity, the size of possible
deviations from the Standard Model in terms of Higgs production cross sections
times branching fractions can amount even to $\sim$30\%.
Such effects are not excluded in the light of present data and their
existence will be subject to verification at the LHC with $\sqrt{s}$ = 13 TeV.

As we already know, an $HWW$ coupling different from the Standard Model one
reflects in the predicted high energy behavior of $WW$ scattering amplitudes.
Indeed, in SILH models the light Higgs unitarizes the amplitudes only partially
or, better saying, it only defers the unitarity crisis to higher energies.
We talk then of a partially strong $WW$ scattering.
Relevant cross sections still grow above the Higgs mass, albeit slowlier, with
an asymptotic behavior given in the lowest order of $g^2/M^2s$ by

\begin{equation}
A(W_L^\pm W_L^\pm \rightarrow W_L^\pm W_L^\pm) = -\frac{c_Hg^2}{M^2}s,
\end{equation}

\begin{equation}
A(W_L^+W_L^- \rightarrow W_L^+W_L^-) = \frac{c_Hg^2}{M^2}(s+t),
\end{equation}

\begin{equation}
A(W_L^\pm Z_L \rightarrow W_L^\pm Z_L) = \frac{c_Hg^2}{M^2}t,
\end{equation}

\begin{equation}
A(W_L^\pm W_L^\pm \rightarrow Z_LZ_L) = \frac{c_Hg^2}{M^2}s,
\end{equation}

\begin{equation}
A(Z_LZ_L \rightarrow Z_LZ_L) = 0,
\end{equation}

\noindent
and up to scale of $M$, where new physical states are bound to appear and do the
rest of the unitarization.  Notice that the above amplitudes are proportional
to the ones obtained within the framework of a Higgsless Standard Model and
in fact, up to the energy $M$:

\begin{equation}
\sigma(pp \rightarrow jjW_LW_L)_{SILH} =
(c_H\xi)^2 \sigma(pp \rightarrow jjW_LW_L)_{Higgsless}.
\end{equation}

The immediate conclusion that can be drawn at this point is that all the
previous studies of $WW$ scattering, which assumed a pure Higgsless Standard
Model as a phenomenological laboratory, are not entirely obsolete even
once the Higgs has been discovered.  Their
results remain completely valid as long as the predicted signal sizes are scaled
by an appropriate
factor dependent on the actual value of the $HWW$ coupling.

The following toy model well illustrates the phenomenological complementarity
between SILH and SUSY.
The most straightforward example of a general framework in which partially strong $WW$
scattering can take place is the two-doublet model (2HDM).  In this framework,
couplings of the
light and heavy Higgs scalars to the $W$ boson are given by $g_{SM} sin(\beta -\alpha)$
and $g_{SM} cos(\beta -\alpha)$, respectively, where $\alpha$ is the Higgs
mixing angle and $tan\beta$ is the usual ratio of vacuum expectation values.
If the factor $sin(\beta -\alpha)$ is sufficiently small and the heavy Higgs
sufficiently heavy, the relevant amplitudes can rise significantly in between
the energies corresponding to the masses of the light and heavy Higgses for
partially strong $WW$ scattering to take place.  The heavy Higgs will ultimately
unitarize this growth.
This, however, is generally not the case in models involving SUSY, e.g., in the
MSSM the heavier the heavy Higgs is the closer to unity the factor
$sin(\beta -\alpha)$ will be and vice-versa.  Thus no appreciable $WW$ scattering
can be expected in the MSSM.

Finally, it is always important to realize that different phenomenological features
may be directly linked to each other within certain classes of models, but need not
be so in general.  Searches for anomalous quartic couplings should be still carried.
Measurement of Higgs couplings that deviate from the Standard Model values does
not automatically imply the existence of partially strong $WW$ scattering.
The dynamics of electroweak symmetry breaking ultimately still will remain an
open question and can be possibly concluded
only via direct measurement of the $WW$ scattering cross section at high energy.

For the sake of completeness one should mention also another predicted
signature of SILH models, namely enhanced production of Higgs pairs
at high energy \cite{grojean}.  Measurement of double Higgs production can
have important implications for spotting out our position on the electroweak phase
diagram \cite{lambda}, but such effects can be hard to detect in practice 
and do not belong to our main topic.

The physics meaning of Higgs couplings larger than their Standard Model values
has been recently investigated as well.
The case of $HWW > 1$ would imply enhancement of the isospin-2 channel 
cross section, equivalent to doubly charged scalar exchange \cite{bellazzini}.

\chapter{$VV$ scattering at the LHC}

This chapter provides a comprehensive overview of the $VV$ scattering process
at the LHC from a phenomenological point of view.

\begin{figure}[htbp]
\vspace{-5mm}
\begin{center}
\epsfig{file=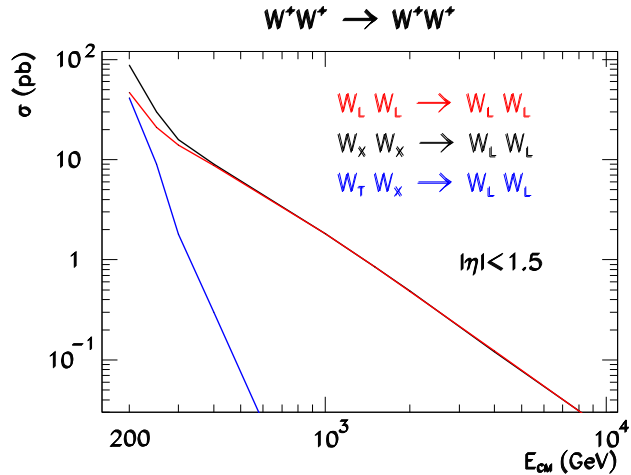,width=0.6\linewidth}
\end{center}
\vspace{-1cm}
\caption{Total $W^+W^+ \to W^+_LW^+_L$ scattering cross sections in the SM
as a function of the center of mass energy.
Shown are the individual contributions of different initial polarization states
to the final state consisting of purely longitudinal $W^+_LW^+_L$ pairs.
Subscript $X$ denotes any polarization ($T$ or $L$).
Assumed are two on-shell, unpolarized, colliding $W^+$ beams.
A cut on the scattering angle that corresponds to pseudorapidity of $\pm 1.5$ with
respect to the incoming $W$ direction was applied.  Results of MadGraph calculations.}
\label{wtwx2wlwl}
\end{figure}

Longitudinal $VV$ scattering carries the most direct, quantitative information about
the details of the actual mechanism of electroweak symmetry breaking.
The practical challenge lies in digging that information out.
As we have no $W$ beams, in any real experiment we have no control
over the polarizations of the interacting pair.  This means, assuming every helicity
state is taken with equal probability, that a $V_LV_L$ initial pair happens in
only 1/9 of all the $VV$ cases.  The majority of initial pairs are $V_TV_T$ and
$V_TV_L$ states, of little sensititvity to the Higgs parameters and indeed to
the Higgs existence at all.  Furthermore, because of the matrix element,
interacting $V_LV_L$
pairs make up no more than 5\% of the total interacting $VV$ pairs, assuming
the Standard Model is approximately correct and the Higgs boson light.
Variations of the $V_TV_T$ and $V_TV_L$ scattering cross sections as a function of 
the Higgs mass and $HVV$ couplings amount to some $\sim$3\% in total and moreover the
potential excesses or deficits have no clearly
defined preferred kinematic signature, making their isolation impracticable.
Thus the biggest part of all $VV$ interacting pairs is
merely a potential background in our search.
One way to proceed is to look for specific kinematic signatures associated to
a hard $2 \rightarrow 2$ scattering process and compare the yield of selected
events to the Standard Model expectation.  The underlying {\it assumption} is then
that any possible excess over the prediction is due to the additional $V_LV_L$
component.  This indeed was the approach taken in many early phenomenological papers
on the subject.  It is clear that such approach requires very good control
over the systematic errors related to the theoretical prediction.
Measurement of $V$ polarization based on the decay products is especially 
difficult for the $WW$ process
where crucial information escapes along with two neutrinos, although the
methodology is in principle known and applied in some analyses \cite{cmswhel}.
But in our kinematic regime of interest measurement of the final state polarizations
will be a challenge.  In what follows we will show that we can, nonetheless,
to some extent measure the polarizations of the initial state.
More often than not polarizations are actually conserved in the
scattering process, at least in what regards $W_LW_L$ versus all the rest.
And this conservation holds most strictly in the $W^\pm W^\pm$ process.
This means, in particular, that 
$W^\pm_LW^\pm_L$ pair in the final state can be produced
almost exclusively from an initial $W^\pm_LW^\pm_L$ pair (see Fig.~\ref{wtwx2wlwl}).  
The only exception to this rule lies in the region of center of mass energies
just above the double $W$ mass threshold, where helicity-flip effects are more likely
to occur, above all $W_TW_X \to W_LW_L$.  In the other direction, these effects are
negligible altogether because of the relative smallness of $W_LW_L$.
In any case, for center of mass energies above 400 GeV,
the admixture from helicity-flip effects is completely negligible.
This is the prime reason why it makes sense to separate $W_LW_L$ pairs in
the final state from the bulk of $WW$ interactions and study their distinctive
kinematic properties.

It is much more complicated to separate a clean
$W_LW_L$ scattering process in $W^+W^-$, for which still at an energy of a TeV
about 20\% of $W_LW_L$ pairs come from the process $W_TW_X \to W_LW_L$.

\section{Formal signal definition}

\begin{figure}[htbp]
\vspace{-2.5cm}
\begin{center}
\epsfig{file=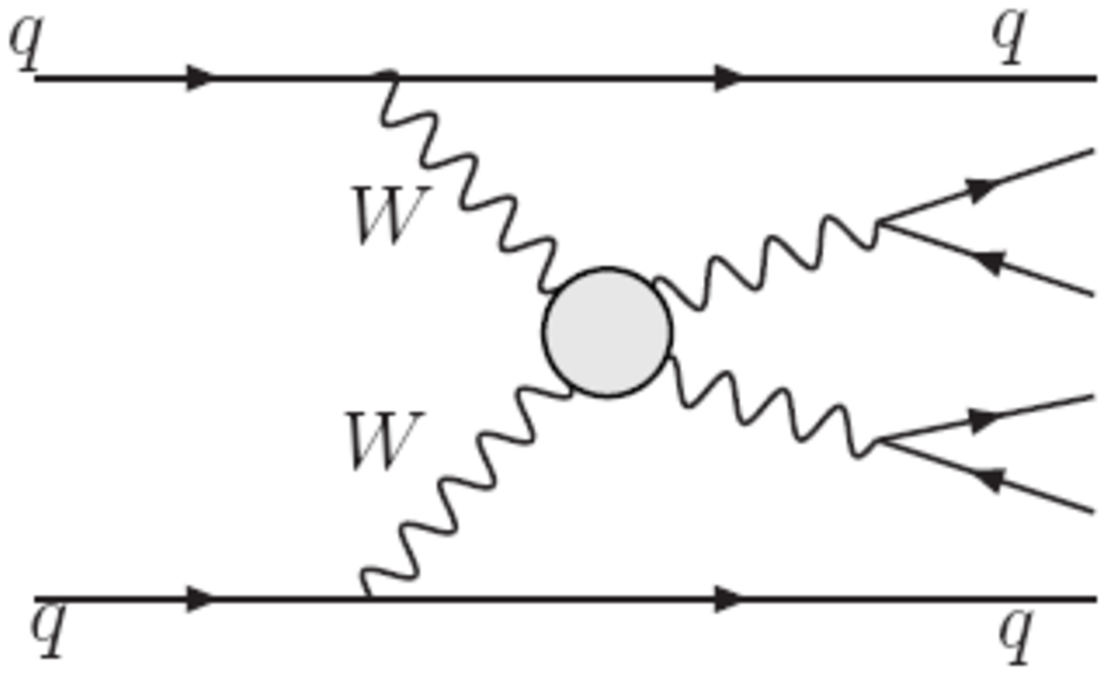,width=0.45\linewidth}\hspace{-3cm}
\epsfig{file=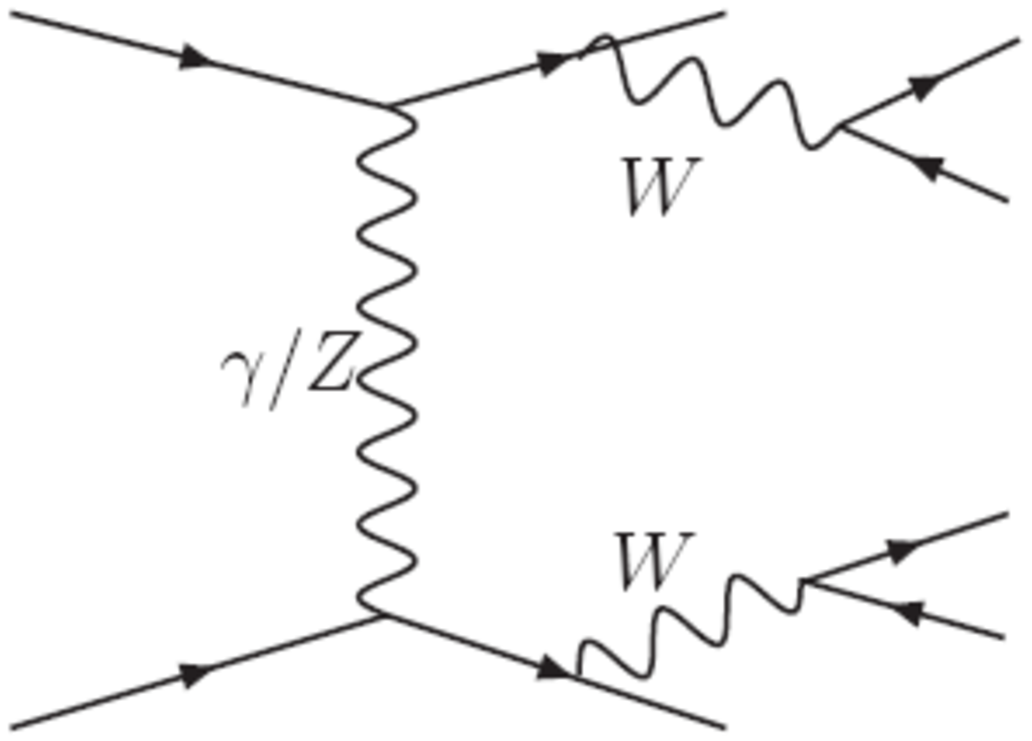,width=0.45\linewidth}\hspace{-3cm}
\epsfig{file=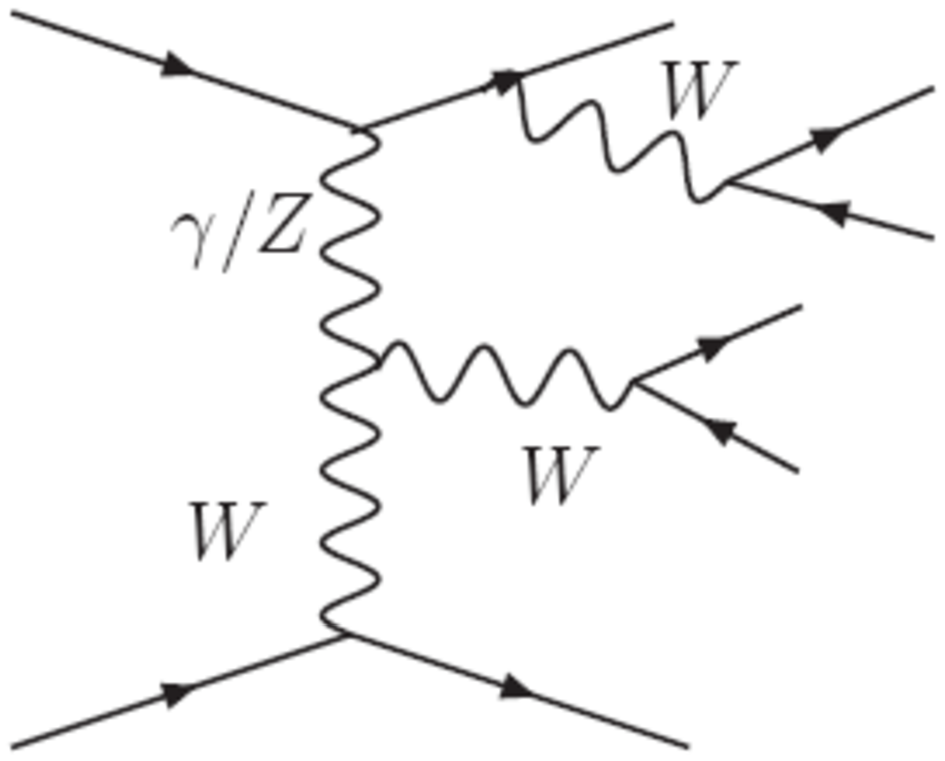,width=0.45\linewidth}
\end{center}
\vspace{-3.5cm}
\caption{Feynman diagrams of $WW$ scattering at the LHC (left) and 
examples of graphs contributing to the irreducible background (middle and right).
The scattering
graph contributes to the signal as long as $W = W_L$, otherwise it contributes
to the irreducible background as well.}
\label{wwatlhc}
\end{figure}

In a hadron collider, $WW$ scattering can occur via $W$ emission off two
colliding quarks.  A lowest order diagram of the process is shown in Fig.~\ref{wwatlhc}
(left).

The final state is characterized by the presence of two $W$ bosons
(more precisely: their respective decay products)
and two jets.  Regardless of how we technically define signal and background,
it is clear that in practice we have no control over whether a specific pair of
vector bosons has indeed interacted or not.  A whole other class of events in which 
two $W$ bosons are produced and do not interact will inevitably be present and
separable from the signal process on a statistical basis only, thus becoming the
bulk of irreducible background.
There are two different approaches often adopted in literature regarding how the
signal can be formally
defined.  The kinematic approach defines the signal in terms of the expected
kinematics of a hard $2 \rightarrow 2$ scattering process and considers
explicitly as signal events all those which fall into a predefined kinematical
(multidimensional) window.  As the kinematic limits are not sharp, the boundaries are
by necessity somewhat arbitrary and so no definition is really unique.
Indeed, many different signal definitions have been in use by experimentalists.
Moreover, by the same token, it is assumed that the part that does
not fall into the signal window and hence formally defines the irreducible background
does not depend on the Higgs sector parameters,
which in general need not be exactly true.  A feature of this approach is that the
Standard Model predicts some signal, too.  Deviations from the Stanard Model will
usually lead to different signal predictions; consistency of each prediction
with real data in the predefined kinematic window can be assessed.

A second, more generic approach defines the signal explicitly as the excess of $WW$
pairs over the prediction of the Standard Model, apparently regardless of the 
actual physical mechanism that leads to such excess.  The Standard Model in
itself, regardless of the actual physical process, is then the formal
definition of the total irreducible
background.  As we saw, this background will be composed mainly of $W_TW_T$
and $W_TW_L$ pairs.
To reemphasize this point, the Standard Model signal is equal to zero
{\it by construction}.  Signal is BSM.  
The first approach is of course closer to what
eventually will be done in a real experiment.  However, to study the problem
from a conceptual point of view, the second definition has at
least two important advantages.  First of all, it is unique as long as we
fix the Standard Model Higgs mass that we use to define irreducible background.
Second, it does not rely on any particular kind of interaction and there is
no signal region defined a priori.
The fact that we know what process is responsible for the possible appearence
of signal is a bonus we can make use of at a later stage, but not a prerequisite.
Note that {\it some} $W_LW_L$ scattering is naturally predicted
even in the Standard Model and so the signal graph in itself is not fully equivalent
to what we are for.  The correspondence between the two approaches is clear and the
translation of respective results into each other is conceptually
more or less obvious, although it has been sometimes the source of some confusion.

For a better understanding of the full process from a theoretical point of view,
one can decompose its complete parton level description into three
distinct, intrinsically connected parts involving $W$ emission, interaction
and decay.  However, one should always keep in mind that such
factorization is approximative, its practical applicability is a subject of
study and any potential conclusions we would like to draw
require independent confirmation in the 
exact evaluation of the full process.

\section{Computational issues and methods}

Factorization of the signal process into the subsequent steps of $W$ emission,
interaction and decay, as we would herewith like to do, is a useful means to study
certain features, but not always a satisfactory way of quantitative description.  
Quantum-mechanically, it is clear that all paths leading from the initial
$pp$ state (which can be decomposed into the many possible sub-states at the
quark and gluon level) to any specific final state, say, $jj\mu^+\mu^+\nu\nu$, must be
considered for the correct evaluation of the process.
Clearly, the
irreducible background must include not only graphs identical with the signal
graph, with a dominant contribution from transverse $W$'s, but likewise a large
number graphs
not involving any $WW$ interaction.  Either of the two categories of events is
not gauge-invariant on its own.  Strong interference effects may occur, depending
on the gauge, and so not only they cannot be treated separately,
but neither can the signal.
Correct calculation of signal and irreducible
background from first principles (i.e., Feynman rules) requires all these processes
added at the level of amplitudes.  Signal must be defined using the ``subtraction method"
which technically requires
the computation of two total cross sections for (e.g.)
$pp \rightarrow jj\mu^+\mu^+\nu\nu$:
one corresponding to the Standard Model, another one to the alternative scenario.
The signal ultimately comes from subtracting the former from the latter.  
Note that in general the signal can be positive inasmuch as it can be negative and
that both make physical sense.  And indeed, the signal {\it is} negative in certain
regions of phase space.
In addition to pure electroweak diagrams, $\sim\alpha^6$ in the lowest order
(up to the level of $W$ decay), background also includes
mixed, electroweak-QCD processes, $\sim\alpha^4\alpha_S^2$.  The minimal
collection of those correspond to gluon exchange graphs between the two interacting
quarks.  Depending on the chosen final state, the number of additional electroweak-QCD
diagrams can vary widely and so does therefore the total background cross section.
All in all, the lowest order calculation of the $pp \rightarrow jj\mu^+\mu^+\nu\nu$
process, which is
the simplest from the computational point of view, requires consideration of
5656 Feynman diagrams.

\begin{figure}[htbp]
\vspace{-1cm}
\begin{center}
\epsfig{file=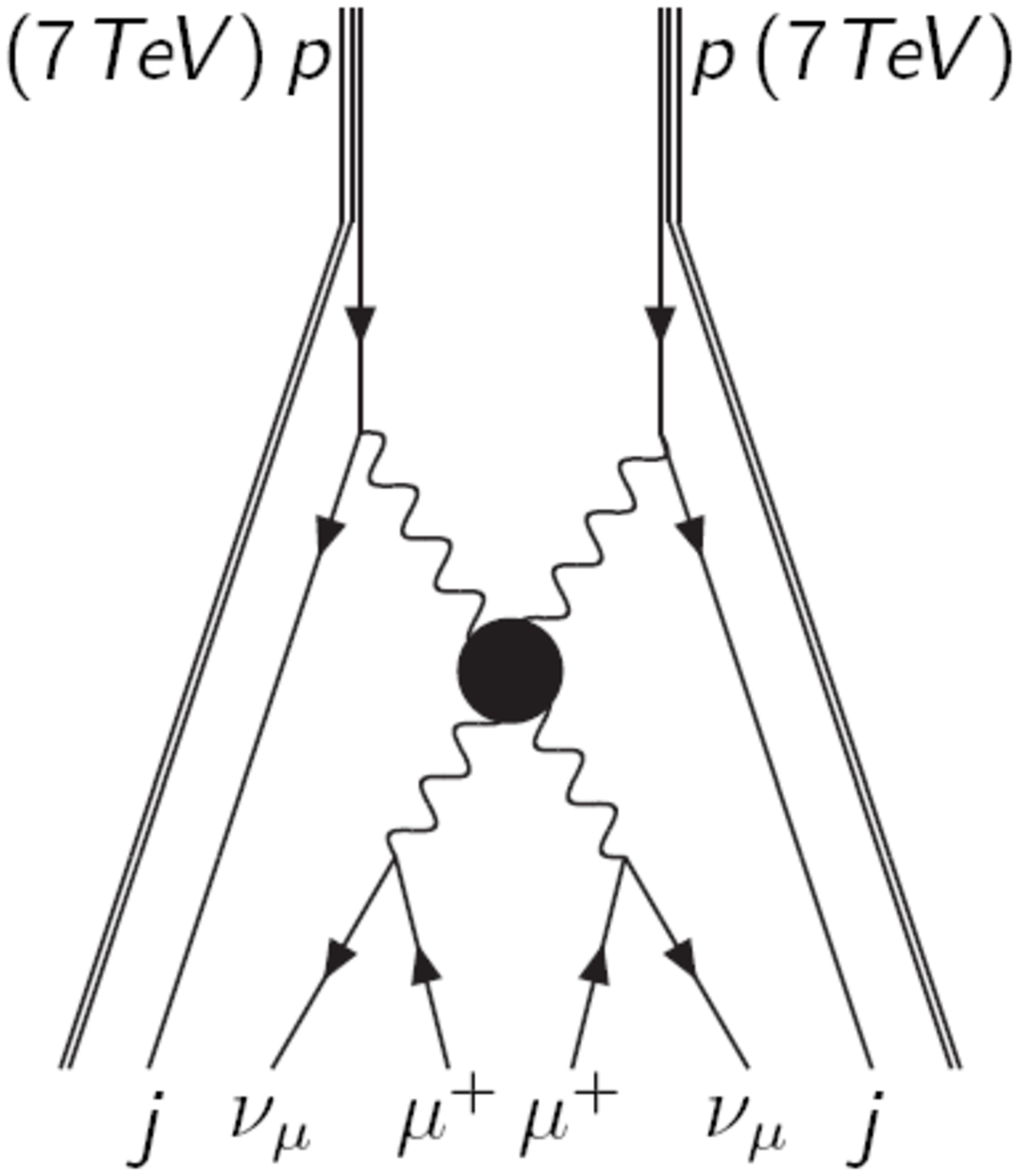,width=0.49\linewidth}
\epsfig{file=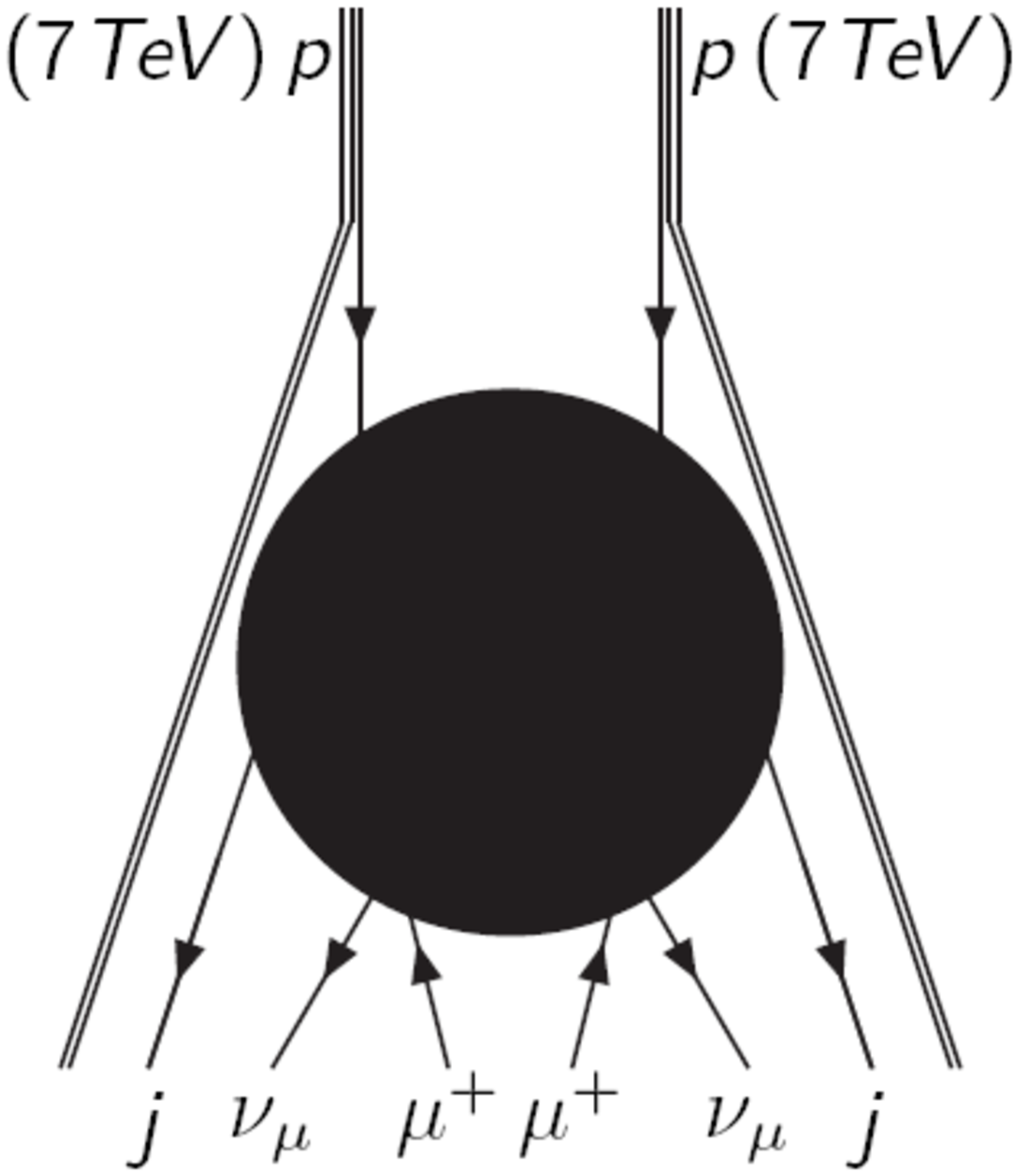,width=0.49\linewidth}
\end{center}
\vspace{-2.5cm}
\caption{Schematic representation of the signal process (left) and of the entire set
of processes which need be taken into account for the correct evaluation
of the signal (right).  Drawings by J.~Kuczmarski.}
\label{diag_qk13}
\end{figure}
\vspace{5mm}

The signal in the lowest order is a purely electroweak process.  However,
interference between scattering and non-scattering diagrams applies in principle also
for electroweak-QCD
ones.  The fact that signal (understood as BSM!)
can indeed be calculated ignoring any QCD contributions,
regardless of the relative amounts of the pure electroweak and electroweak-QCD
processes, is a present from nature rather than a rule of thumb.
Interference effects can be shown to cancel out to a good accuracy in the
difference through which we define the signal\footnote{Bear in mind of course that
what we are for in this chapter is an estimate of the magnitudes of signal and
background, not a precision measurement}.  This is because electroweak-QCD
events populate mostly a different region of kinematic phase space than the
purely electroweak signal events - the respective transverse momenta of scattered $W$'s
are nearly clean separated - and this conclusion holds even for $W^+W^-$
scattering where the total signal+background cross section is dominated by
QCD contributions by an order magnitude.  In our example process
$pp \rightarrow jj\mu^+\mu^+\nu\nu$, this allows to reduce the number of Feynman
diagrams necessary for the calculation of the signal to 5208 (however, another dedicated
calculation is then needed to determine the term to subtract, which is not equal to the
total irreducible background anymore).

\subsection{Effective $W$ Approximation and the Equivalence Theorem}

Older literature made extensive use of the so called Effective $W$ Approximation
\cite{ewa},
with its nice acronym E$W$A
(or more generally, Effective Vector Boson Approximation, EVBA).
Its main advantage is that it renders the lowest order signal graph gauge
invariant on its own, under some approximative assumptions.
The idea is similar to that of factorization for parton distribution functions.
The total cross section is described in terms of density functions for a polarized
$W$ being radiated by a fermion with a given momentum fraction, times the
scattering amplitude for two bosons carrying these momentum fractions.  The boson is
assumed to be radiated approximately collinearly at a high center of mass energy,
so it is close to the mass shell and we can neglect the fermion masses.  In
the amplitude of the scattering process it is then taken to be on shell.
This means that the treatment will only necessarily be valid when $\sqrt{s} >> M_W$
and so small virtualities of the gauge boson may be neglected.
The explicit expression for the density function can be derived from the
matrix element calculated for an on-shell boson being emitted off a fermion
as a function
of the fermion initial four-momentum and the momentum fraction carried by the
boson.  The total process cross section to the level of the scattered gauge bosons
is finally given by integrating the scattering
amplitude with the appropriate density function over the full range of
the momentum fraction.
The E$W$A provides an effective way to calculate the signal-like graphs standalone to a
typical accuracy of 20-30\%.  A substantial literature exists on the accuracy and
conditions of applicability of the E$W$A.
Although the validity of the approximations does not explicitly exclude any
helicity states, the E$W$A by construction disregards non-scattering contributions
and is therefore not able to predict the total irreducible background levels.
The latter still requires computation of the full set of diagrams.

The E$W$A technique is often coupled with the evaluation of gauge boson interaction
done using the Equivalence Theorem \cite{equiv}.
The Equivalence Theorem states that at an energy much larger than the vector boson mass,
the amplitude for a process involving interaction of longitudinally polarized vector
bosons on the mass
shell is given by the amplitude in which these vector bosons are replaced by the
corresponding unphysical Goldstone bosons.  Intuitively this is understandable
as a consequence of the Higgs mechanism.  The vector bosons get their
masses via absorbing the Goldstone bosons and so their longitudinal components
retain the properties of the scalar interactions.  Whether this naive intuition
is really correct or not,
the approximation is valid up to the leading energy term and it is applicable in
every order of perturbation theory.  The approximation is very useful because it is
technically much easier to calculate amplitudes involving massless scalars than those
involving massive vectors.  The ratio of the actual vector boson mass to the
center of mass energy of the interaction
defines its practical accuracy.

However, several authors emphasized the importance of using full matrix element
calculations in order to correctly reproduce the entire kinematics of the final state,
which lies at the basis of defining optimum selection criteria for the isolation of
longitudinal signal from transverse background.  The advantages of the E$W$A and
the Equivalence Theorem naturally waned once full matrix element generation tools
became available to the public and fast computer clusters alike.  Approximative
techniques to evaluate the $WW$ interactions are rarely used in modern studies.

\subsection{The ``production $\times$ decay" approximation}

In quantum physics,
full calculation of the process, say, $pp \rightarrow jj\mu^+\mu^+\nu\nu$ involves
summing over
all the possible paths leading to the final state.  Note that in such, formally
fully correct, treatment information on the individual
$W$ helicities is lost.  We don't even know whether there was a $W^+W^+$ intermediate
state at all at any time in the process.
As a matter of principle, helicity is well defined only for {\it on-shell} bosons.
To what extent the $W$'s after interaction are on-shell and hence to what extent
they can be sensibly assigned longitudinal or transverse polarizations at all, is
a crucial issue.  Experimentally, $W$ helicity manifests itself in angular
distributions of the decay products - for example the charged lepton from $W$ decay
with respect to the
mother $W$ direction.  It cannot be deduced on an event by event basis.

\begin{figure}[htbp]
\vspace{-1cm}
\begin{center}
\epsfig{file=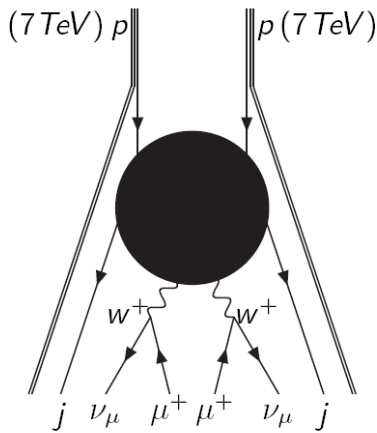,width=0.49\linewidth}
\end{center}
\vspace{-2.5cm}
\caption{Schematic representation of the full set
of processes which need be taken into account for the evaluation
of the signal in the ``production $\times$ decay" approximation;
cf.~Fig.~\ref{diag_qk13}.  Drawing by J.~Kuczmarski.}
\end{figure}
\vspace{5mm}

\begin{figure}[htbp]
\begin{tabular}{ll}
  \vspace{-7mm}
  \epsfig{file=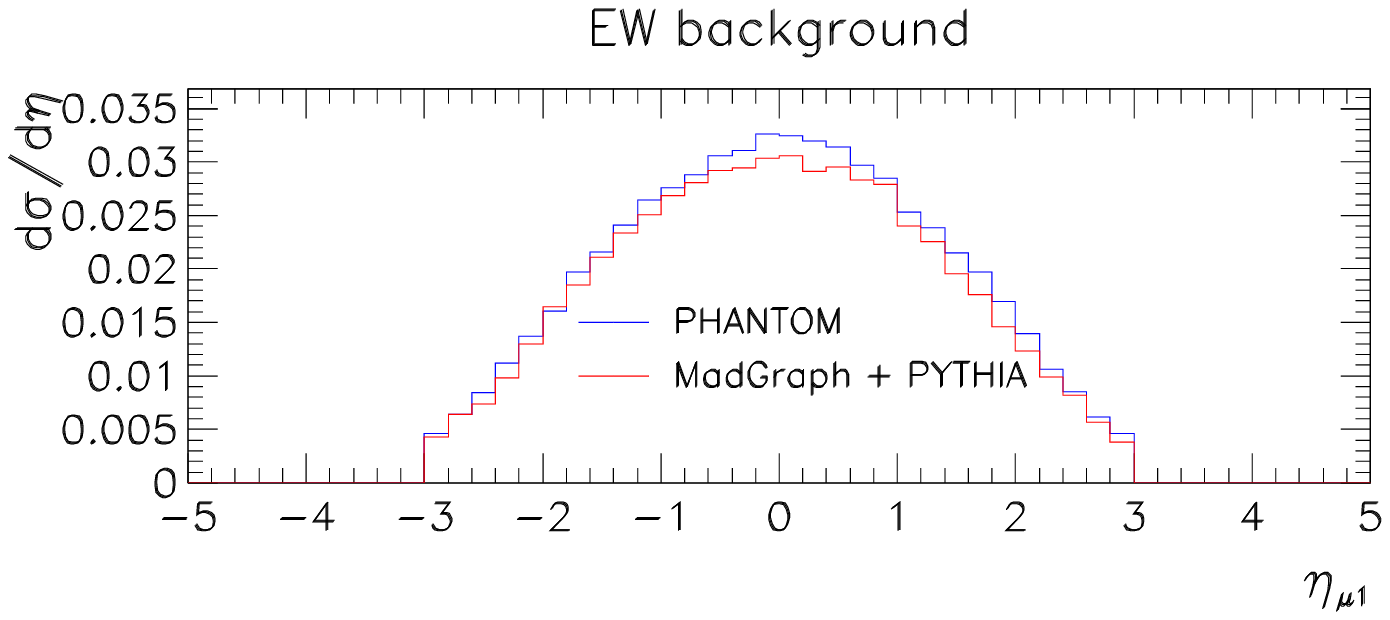,width=0.45\linewidth} &
  \epsfig{file=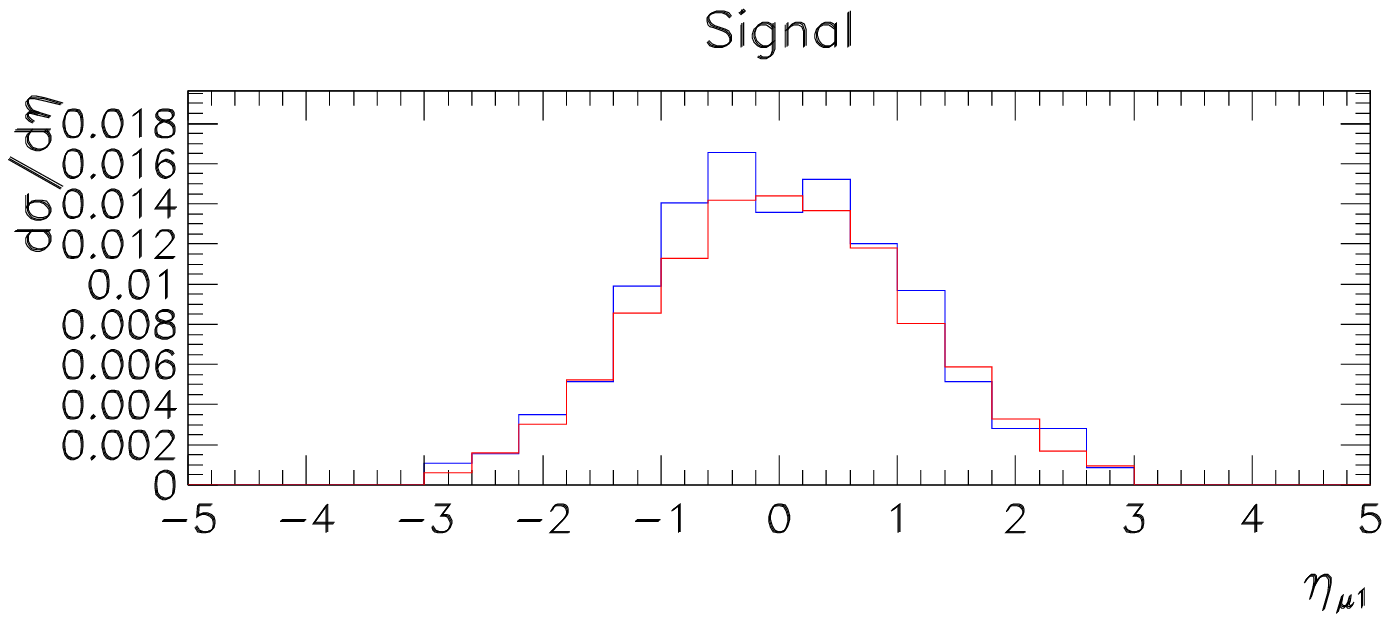,width=0.45\linewidth}
  \\
  a) & b) \\
  \vspace{-7mm}
  \epsfig{file=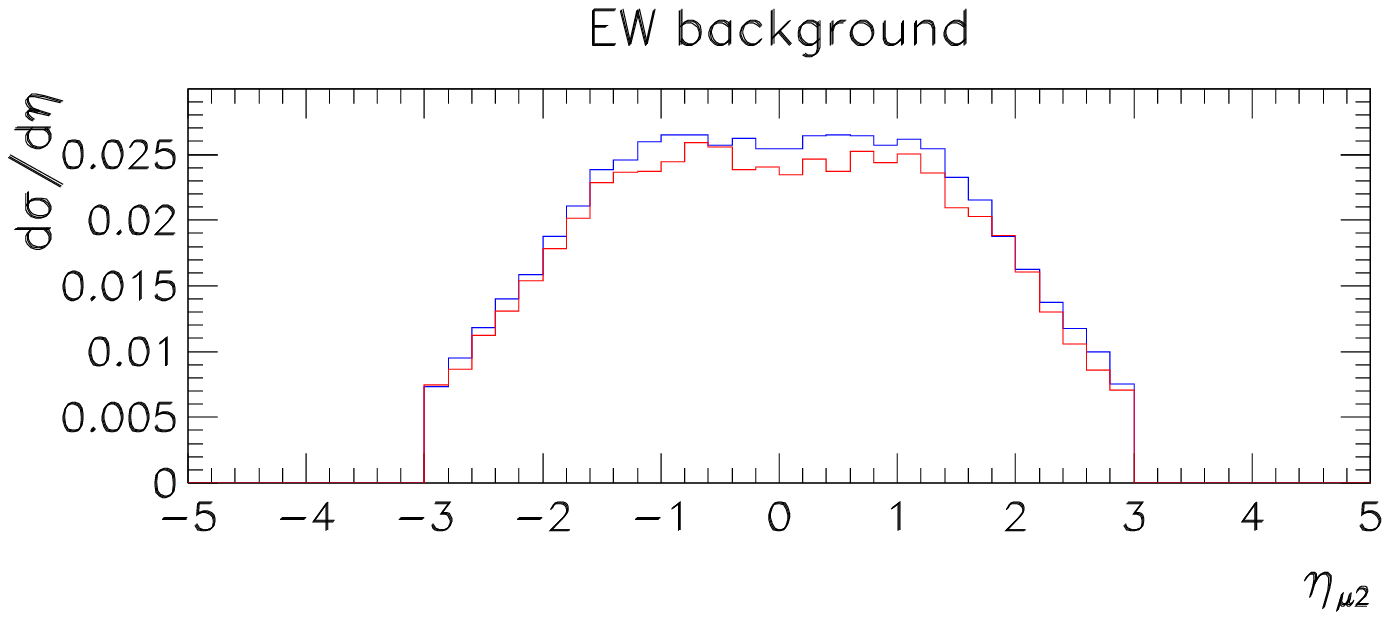,width=0.45\linewidth} &
  \epsfig{file=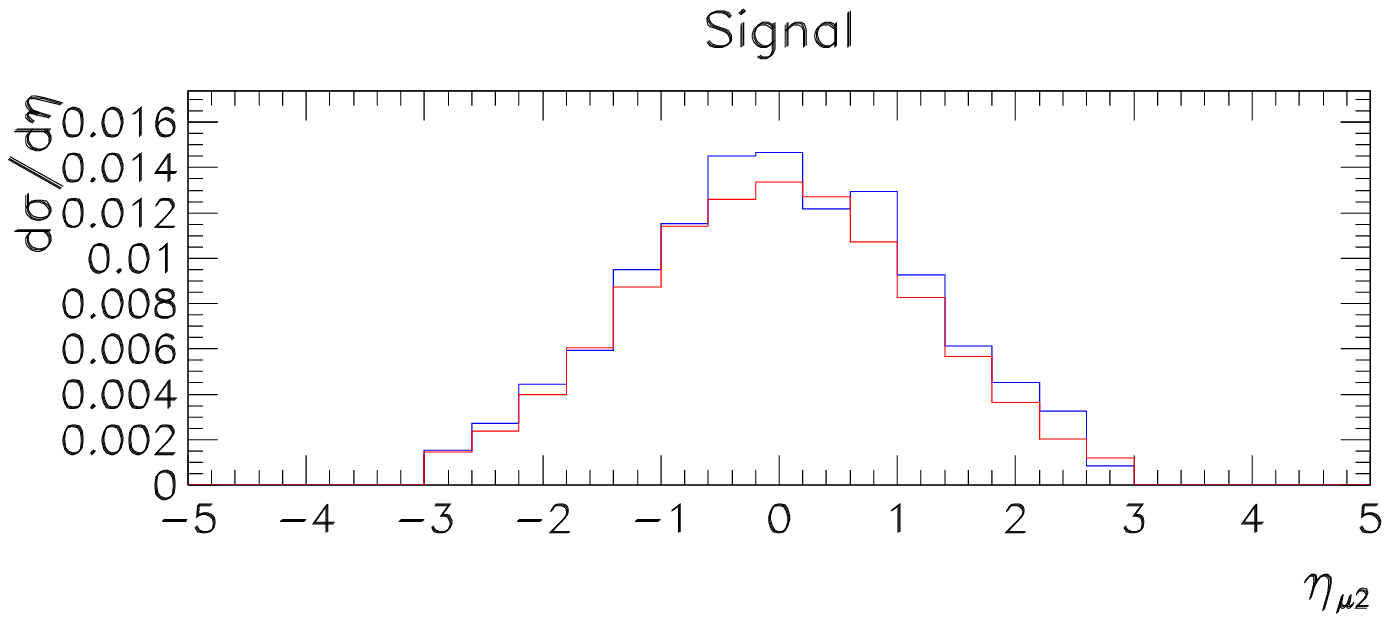,width=0.45\linewidth}
  \\
  c) & d) \\
  \vspace{-7mm}
  \epsfig{file=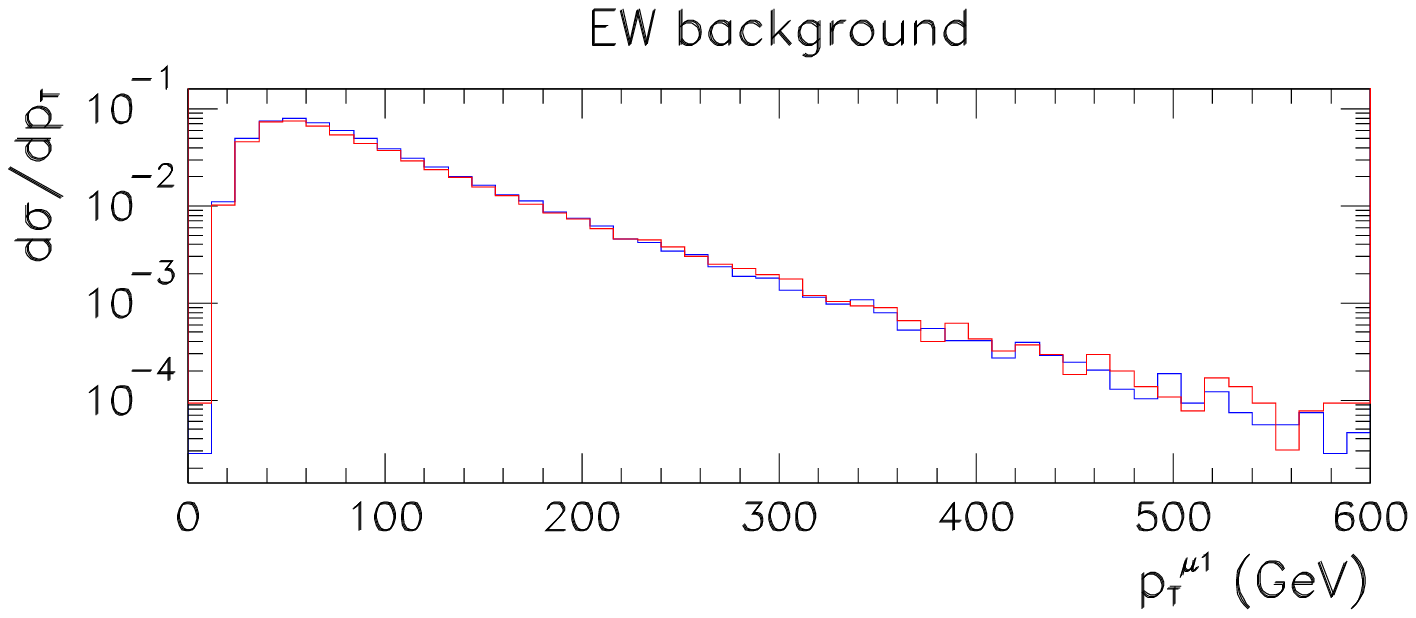,width=0.45\linewidth} &
  \epsfig{file=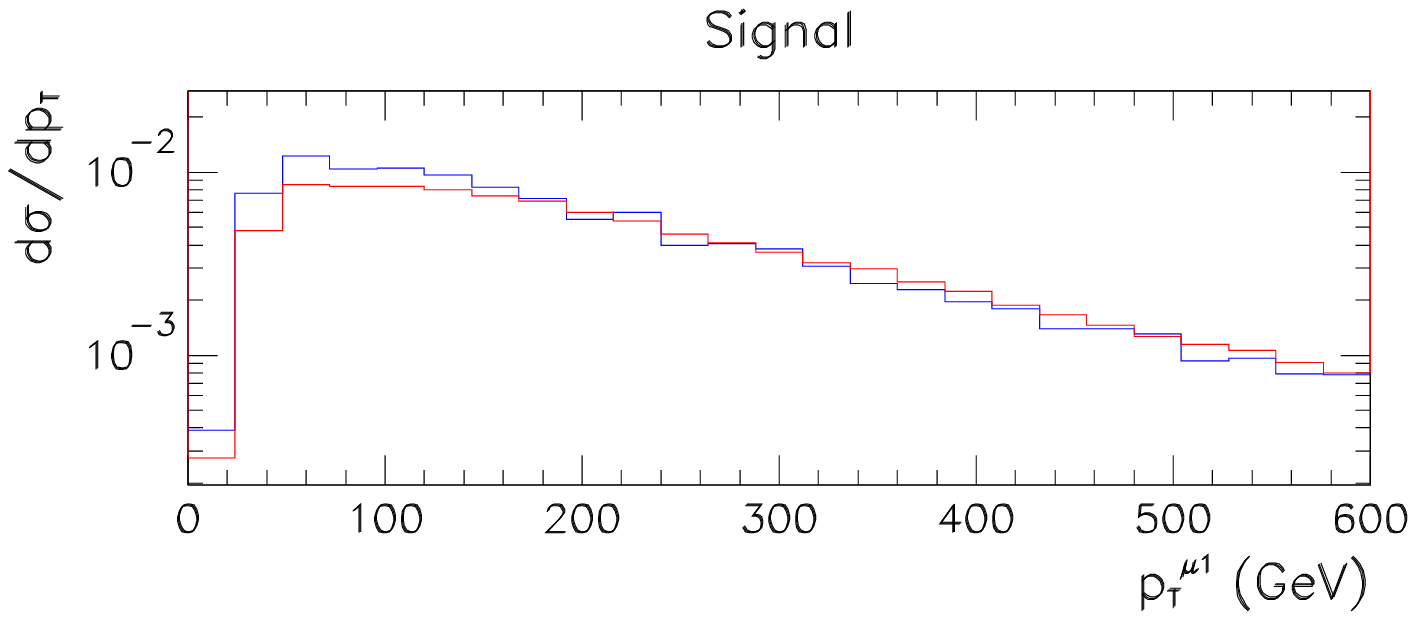,width=0.45\linewidth}
  \\
  e) & f) \\
  \vspace{-7mm}
  \epsfig{file=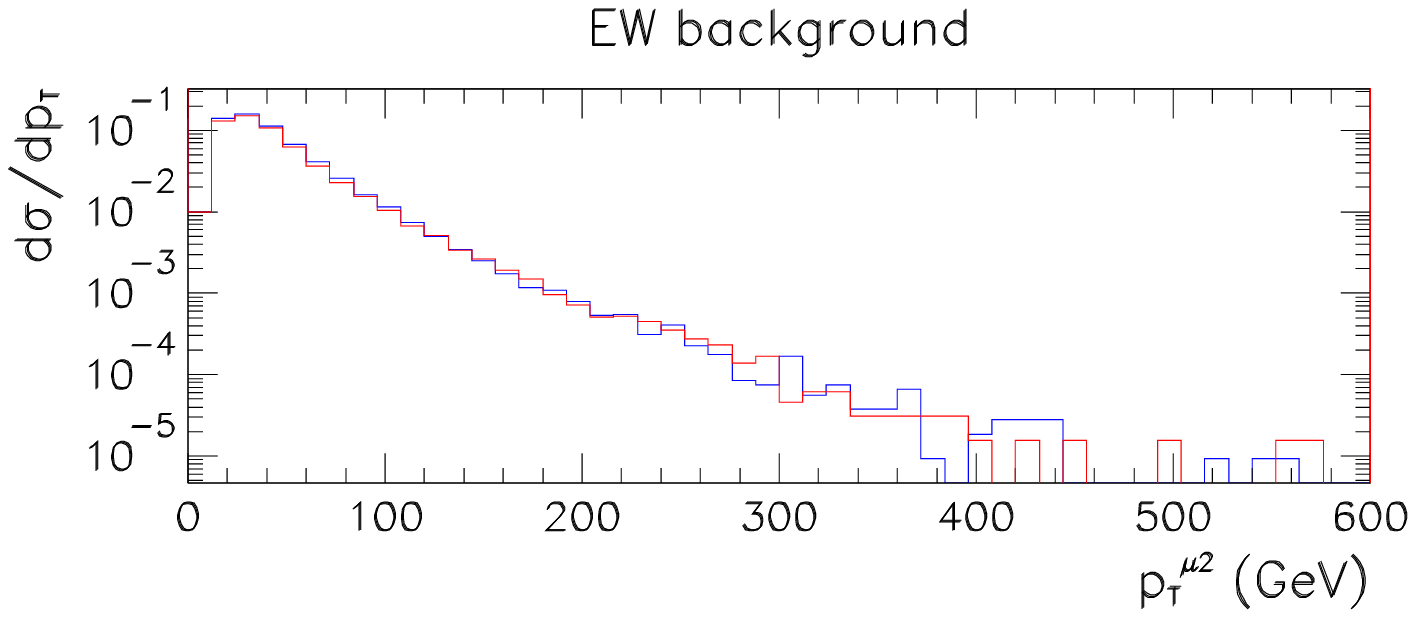,width=0.45\linewidth} &
  \epsfig{file=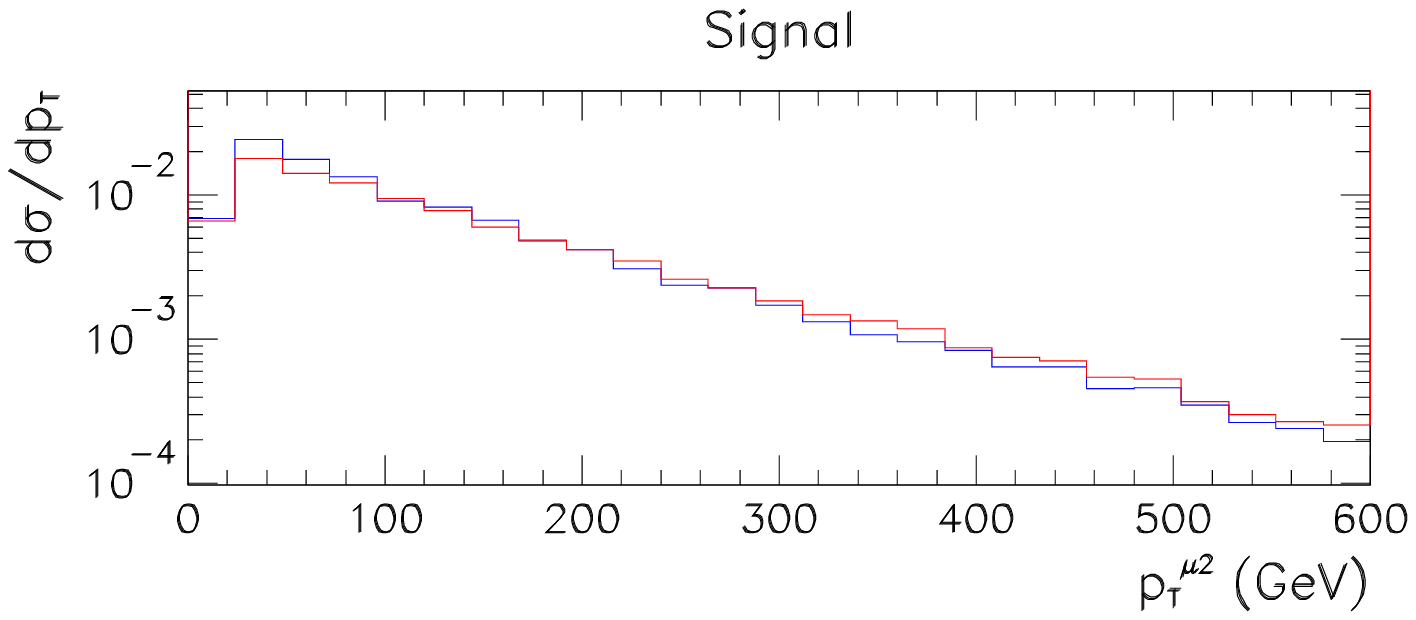,width=0.45\linewidth}
  \\
  g) & h) \\
  \vspace{-7mm}
  \epsfig{file=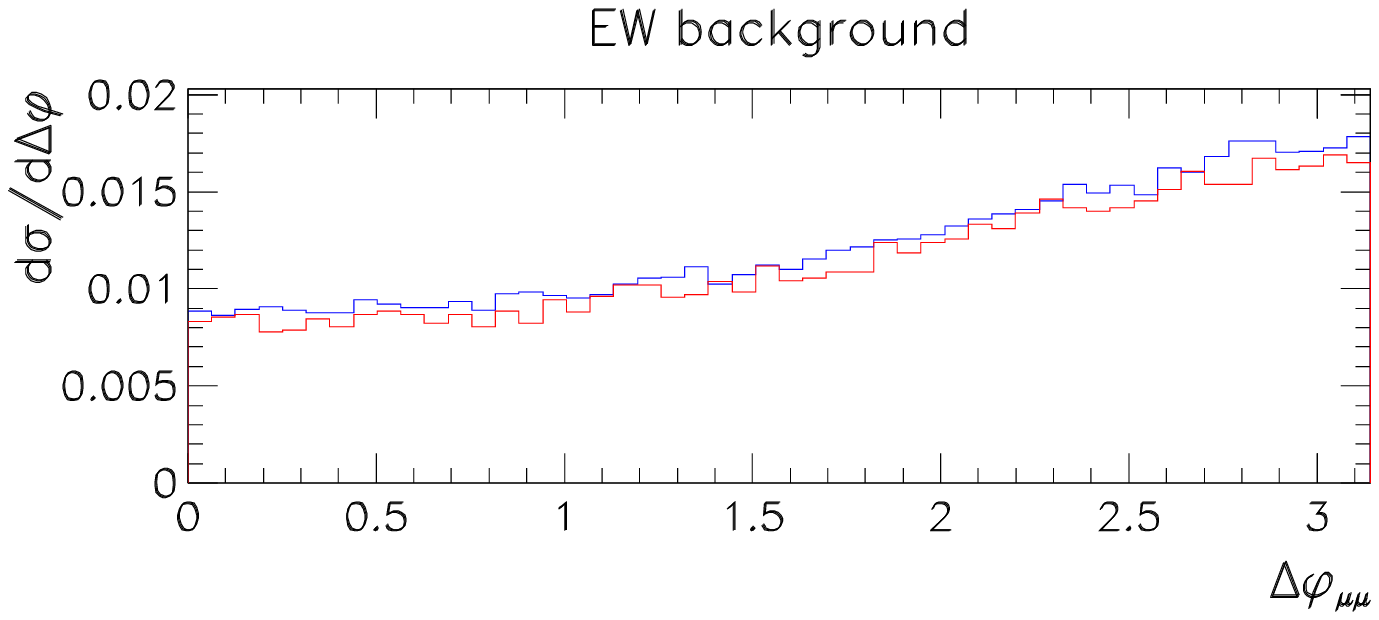,width=0.45\linewidth} &
  \epsfig{file=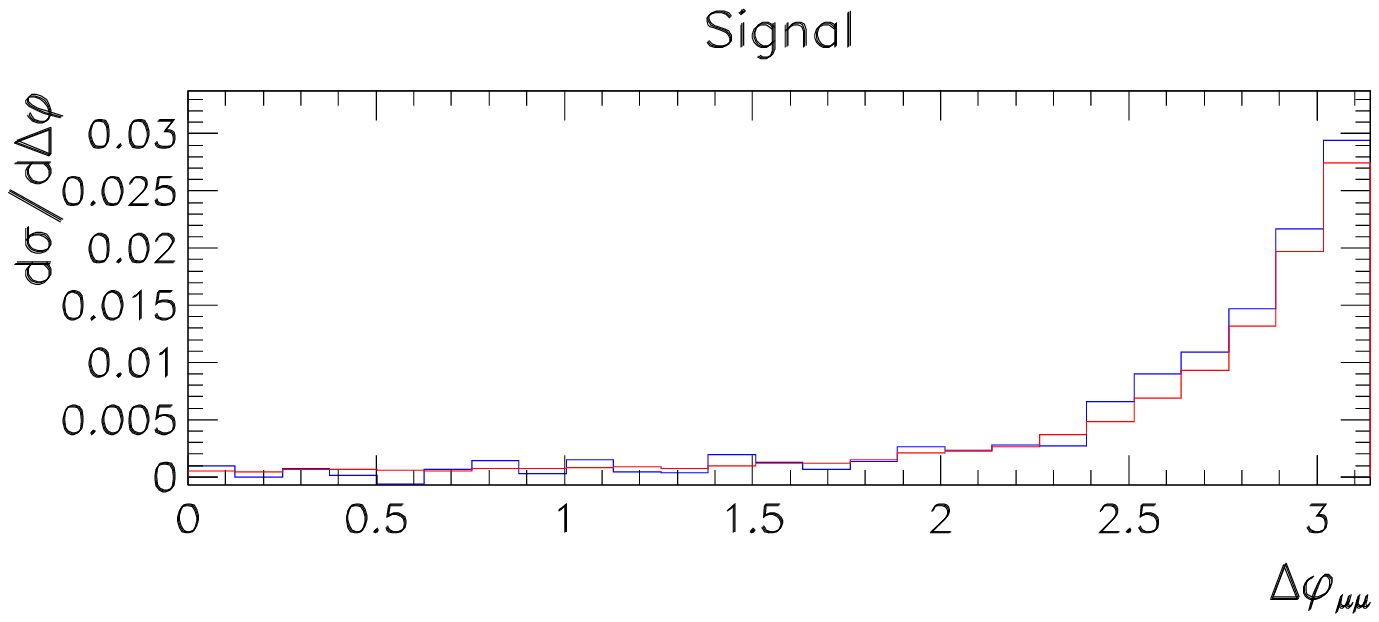,width=0.45\linewidth}
  \\
  i) & j) \\
  \vspace{-7mm}
  \epsfig{file=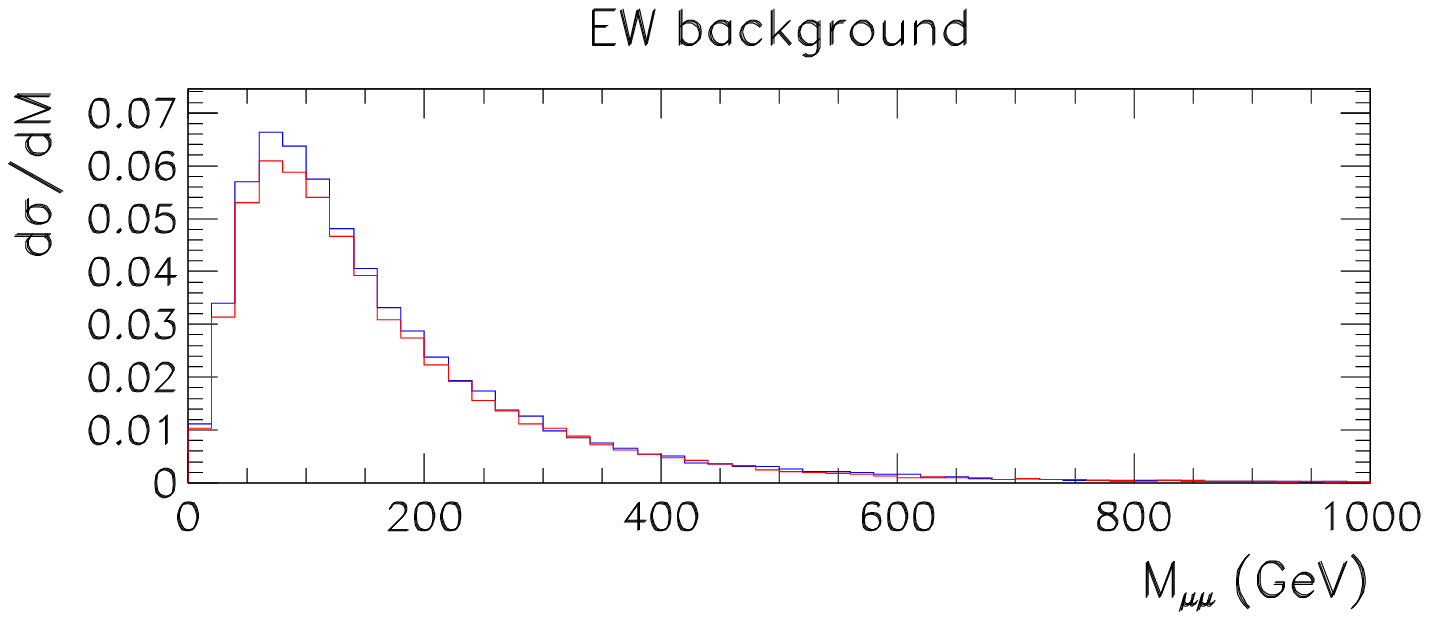,width=0.45\linewidth} &
  \epsfig{file=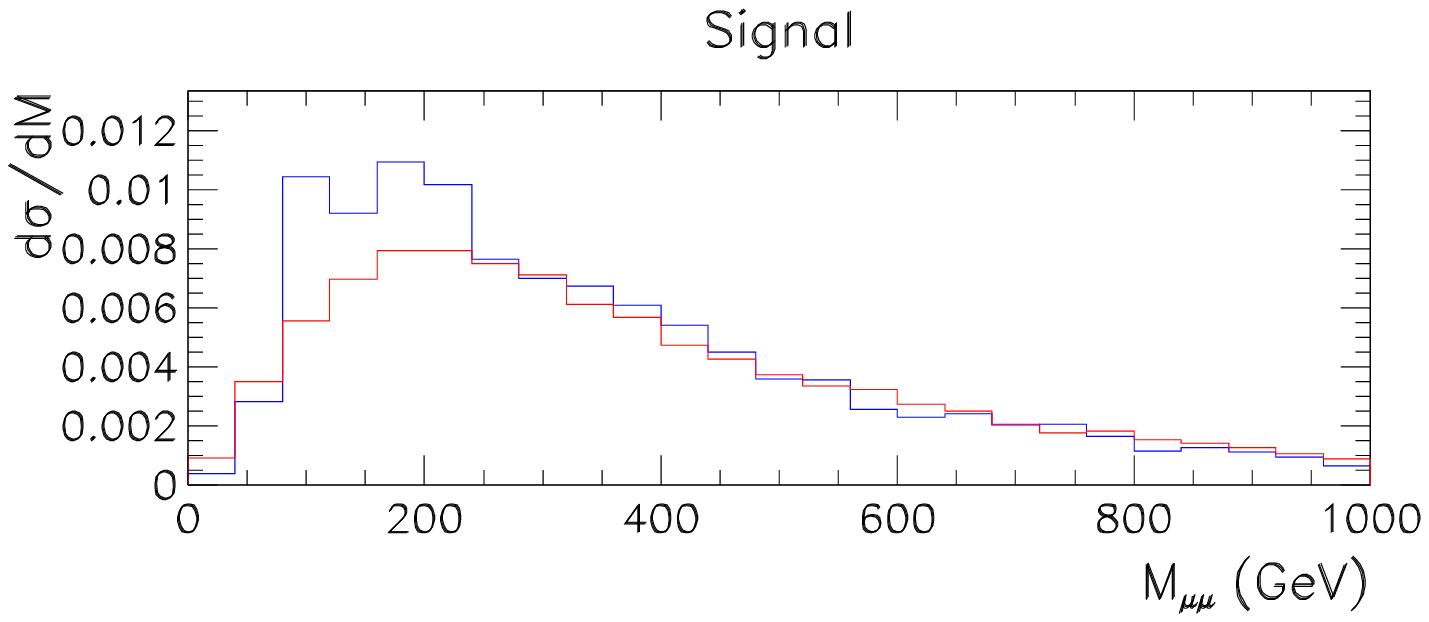,width=0.45\linewidth}
  \\
  k) & l)
\end{tabular}
\caption{Kinematic distributions of final state muons from the $pp
  \rightarrow jj\mu^+\mu^+\nu\nu$ process at 14 TeV, obtained using
  the $W$ on-shell approximation (labeled MadGraph+PYTHIA) and exact
  matrix element calculations (labeled PHANTOM).  Shown are:
  pseudorapidities of the two muons (a-d), their transverse momenta (e-h),
  distances in the azimuthal angle (i,j) and invariant masses (k,l).
  VBF topological cuts were applied, including $|\eta_\mu| < 3$ and
  $\Delta\eta_{jj} > 4$.
  In background calculations only electroweak processes were taken into
  account and the Higgs mass was set 200 GeV.}
\label{fig:prodxdecay}
\end{figure}

The full process $pp \rightarrow jj\mu^+\mu^+\nu\nu$ can be reasonably
expressed as a coherent sum of its $W_LW_L$ and $W_TW_T + W_TW_L$ contributions only
so long as we can
assign two $W$ helicities to each event, even if only on paper.  This is possible if
and only if the scattered $W$'s are produced near enough the mass shell, or equivalently,
off-shell effects, including graphs in which a $W$ boson is exchanged in the
$t$-channel, do not lead to significant changes of the measurable kinematics
of the final state.  Only under this assumpton can the process be approximately
factorized into steps consisting of $WW$ production and decay.
It was shown that in the kinematic region of interest for us, this
approximation indeed holds to better than 10\% both in shapes and normalizations,
which is quite enough for our purposes.  Because of this lucky fact,
the characteristic features of final states associated to $W_LW_L$
and $W_TW_T + W_TW_L$ can be studied separately of each other.  One can also hope
for a more detailed signal event
selection that will be based not solely on the scattering kinematics, but also
on the preferred $W$ helicities.
The on-shell approximation for the scattered bosons is sometimes referred to as
the ``production $\times$ decay"
approximation, as it technically allows to reduce computational work to the
reduced process
$pp \rightarrow jjW^+W^+$ (where the two bosons are {\it assumed} to be exactly
on-shell) in the first step and thus decrease the
number of Feynman diagrams to consider from 5656 to 1428.
Similar conclusions hold for $W^+W^-$ and $ZZ$ if only applied far enough from the
Higgs resonance region where the agreement expectedly breaks down.

Since in principle we can indeed sensibly assign specific polarizations to $WW$ final
states,
it is legitimate to restrict the formal signal definition explicitly to $W_LW_L$ pairs in
all the computations, at least in studies concerning the source of electroweak symmetry
breaking.  The practical advantage is one of avoiding large cancelations
in the signal definition coming from the dominant and largely
BSM-insensitive $W_TW_T + W_TW_L$ states.

In literature one also finds a similar approach under the name of Narrow Width 
Approximation.  Generally it has been
shown to work for Standard Model processes with a typical accuracy of $\Gamma/M$,
the ratio of the total width to the mass of the particle involved \cite{uhlemann}.
However, some implementations of the Narrow Width Approximation in event generators
consist merely of neglecting the non-resonant graphs, but with off-shell effects and
spin correlations kept, and thus are not fully equivalent to our approach.

\section{Emission of a gauge boson off a quark}
\label{wemission}

\begin{figure}[htbp]
\vspace{-3cm}
\begin{center}
\hspace{-1cm}
\epsfig{file=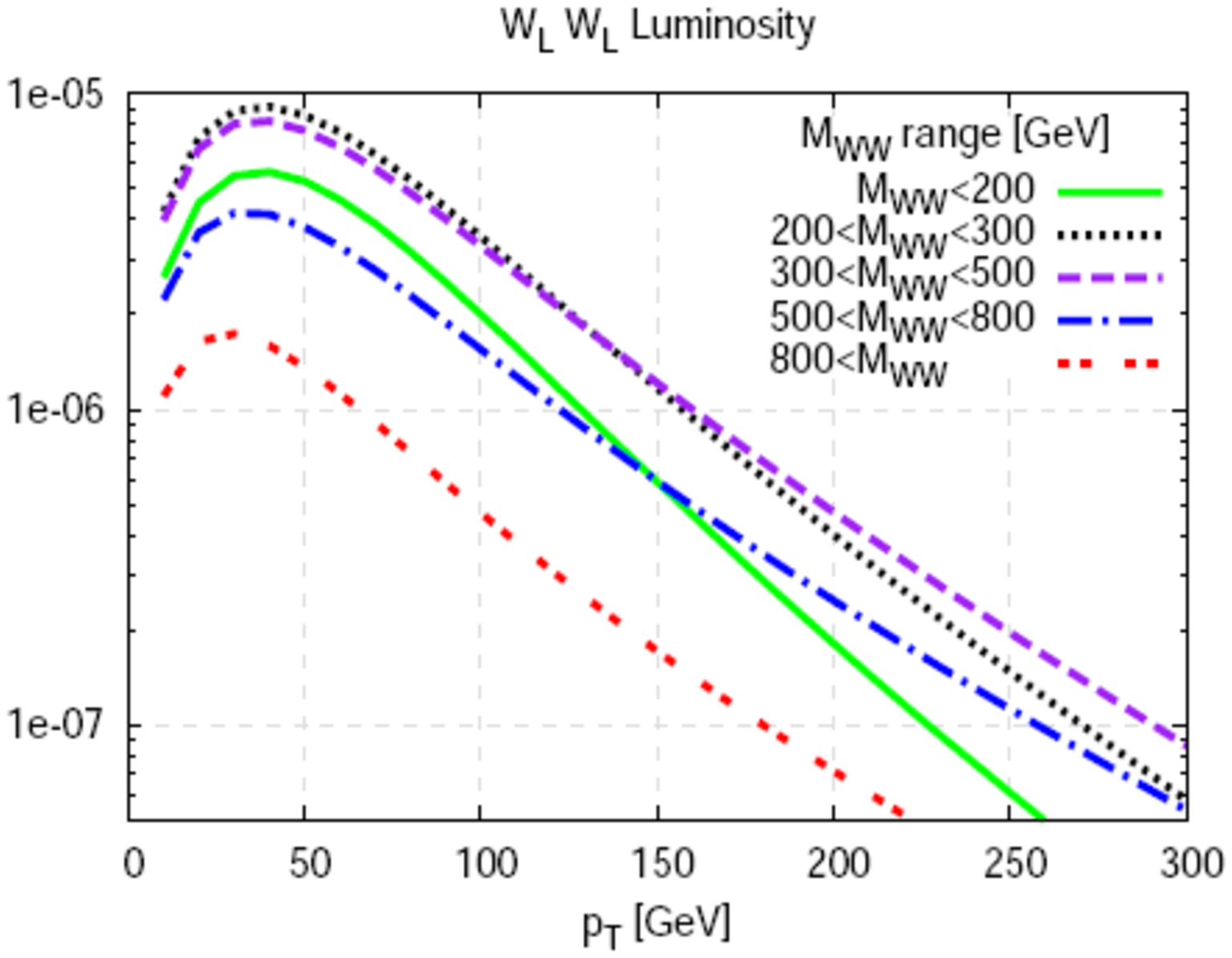,width=0.6\linewidth}\hspace{-2.5cm}
\epsfig{file=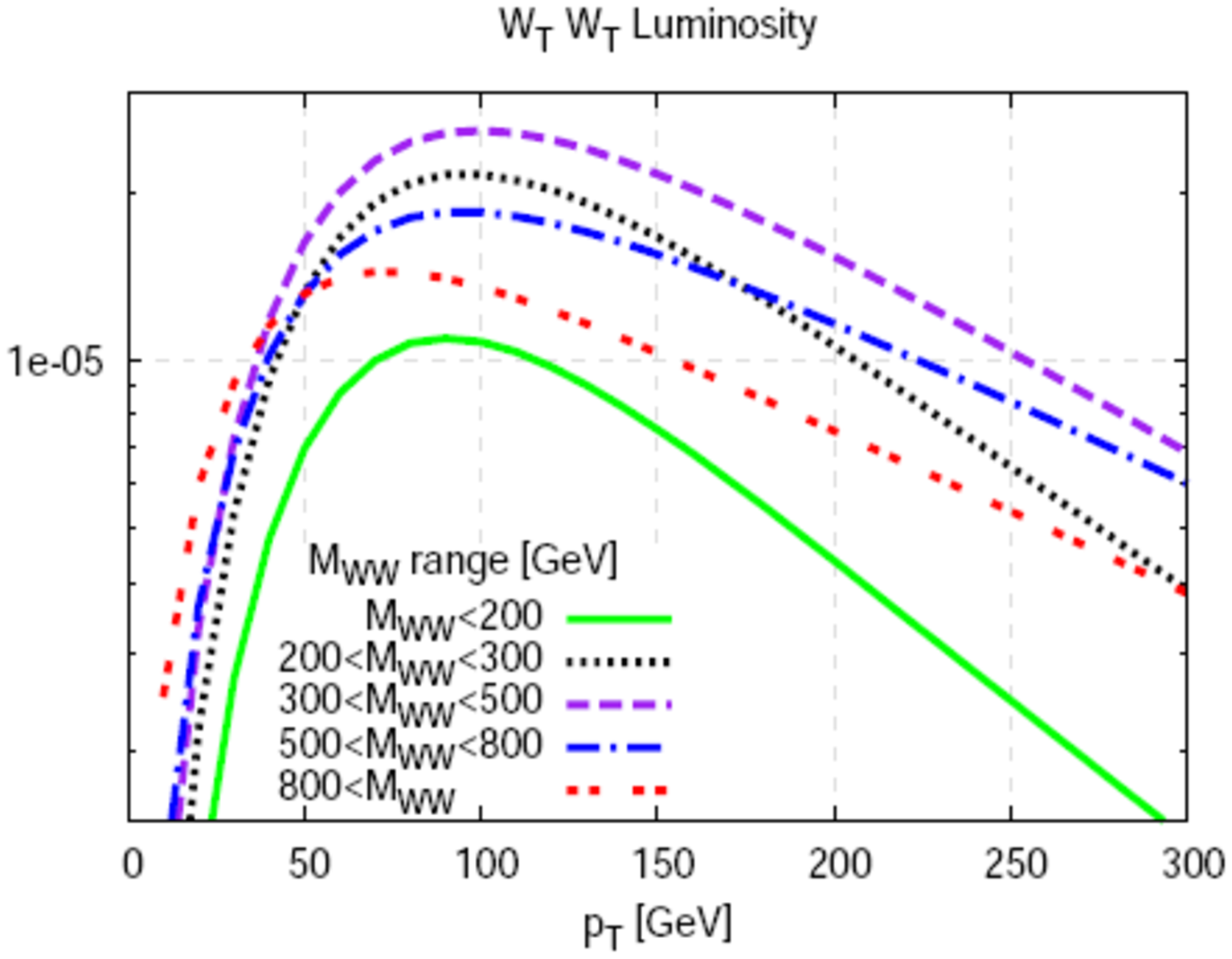,width=0.6\linewidth}
\end{center}
\vspace{-4cm}
\caption{Distributions of transverse momenta of the outgoing quarks
after the emission of a $W_L$ (left) and of a $W_T$ (right) in intervals
of the $WW$ invariant mass.  Assumed is a pair of colliding quarks,
each emitting a $W$ boson, no $WW$ interaction is taken into account.
Calculation done within the Effective $W$ Approximation.}
\label{ptjewa}
\end{figure}

The characteristic difference in the kinematics of the final states associated
to the emission of $W_L$ and $W_T$ off a quark are their different angular distributions.
The $W_LW_L$ and $W_TW_T$ luminosity spectra
calculated from pure emission probabilities from two quarks colliding head-on
at a fixed energy and
without any further interaction assumed, already reveal interesting differences
in their kinematic features - see Fig.~\ref{ptjewa}.  

\vspace{5mm}

The longitudinally polarized $W$ tends to be
emitted at a smaller angle (hence smaller transverse momentum) with respect
to the incoming quark direction than the transversely polarized $W$.
As a consequence, the final quark accompanying a longitudinal $W$ is more
forward than the one accompanying a transverse $W$.  This effect is more
pronounced the larger the invariant mass of the $WW$ pair, $M_{WW}$.  The transverse
momentum distributions of quarks associated with $W_L$ emission become narrower
as $M_{WW}$ increases and the peak of the distribution gradually moves to
lower values.  No such trend is associated with $W_T$ emission, except for
very large $M_{WW}$, where in case of a fixed incoming quark energy the effects
of overall energy and momentum conservation become a limiting factor.
These observations suggest that our potential to separate the $W_LW_L$ signal
from the $W_TW_T$ background increases with $M_{WW}$ already at the level
of emission.  Tagging two opposite forward jets in
a relatively narrow band of transverse momentum for a fixed value of $M_{WW}$
is the ideal technique to be used.  The practical problem in implementing
this conclusion in an experiment is that the absolute scale of transverse momentum
of the emitted $W$ is defined by the mother quark energy.  Events can be
efficiently discriminated based on the transverse momenta of the outgoing jets so
long as we have monochromatic quark beams\footnote{Obviously, we would be doing
much better in a lepton collider, if only it had a similar energy reach!}.

\section{Interaction of two gauge bosons}

Total cross sections and angular distributions in
the scattering process of two on-shell $W$ bosons, depending on their energies and
polarizations, were already discussed in the previous chapter.  
Here we have just learned that in
addition, since $W_L$'s tend to be emitted from a quark line in a more collinear way than
$W_T$'s, the $W_LW_L$
rest frame will be approximately equivalent to the lab frame as long as we
disregard highly asymmetric quark-quark collisions.  Excess over the predictions
of the Standard Model is therefore expected in the central region of the detector,
as far as the scattered $W$ directions are concerned.  Let us now build a naive toy model
of such process.  For a $WW$ pair coming from nearly collinear emissions and
scattered back-to-back
at a large angle, we can approximate

\begin{equation}
M_{WW} \approx 2 \sqrt{M_W^2 + p_T^{(1)} p_T^{(2)}},
\end{equation}

\noindent
where $p_T^{(1)}$ and $p_T^{(2)}$ are the transverse momenta of the scattered
$W$'s understood as unsigned scalar quantities.
The product $p_T^{(1)} p_T^{(2)}$ (or its square root, to be more precise)
is a measure of $M_{WW}^2$ and this
equivalence naturally works better for large $M_{WW}$.

\begin{figure}[htbp]
\vspace{1cm}
\begin{center}
\epsfig{file=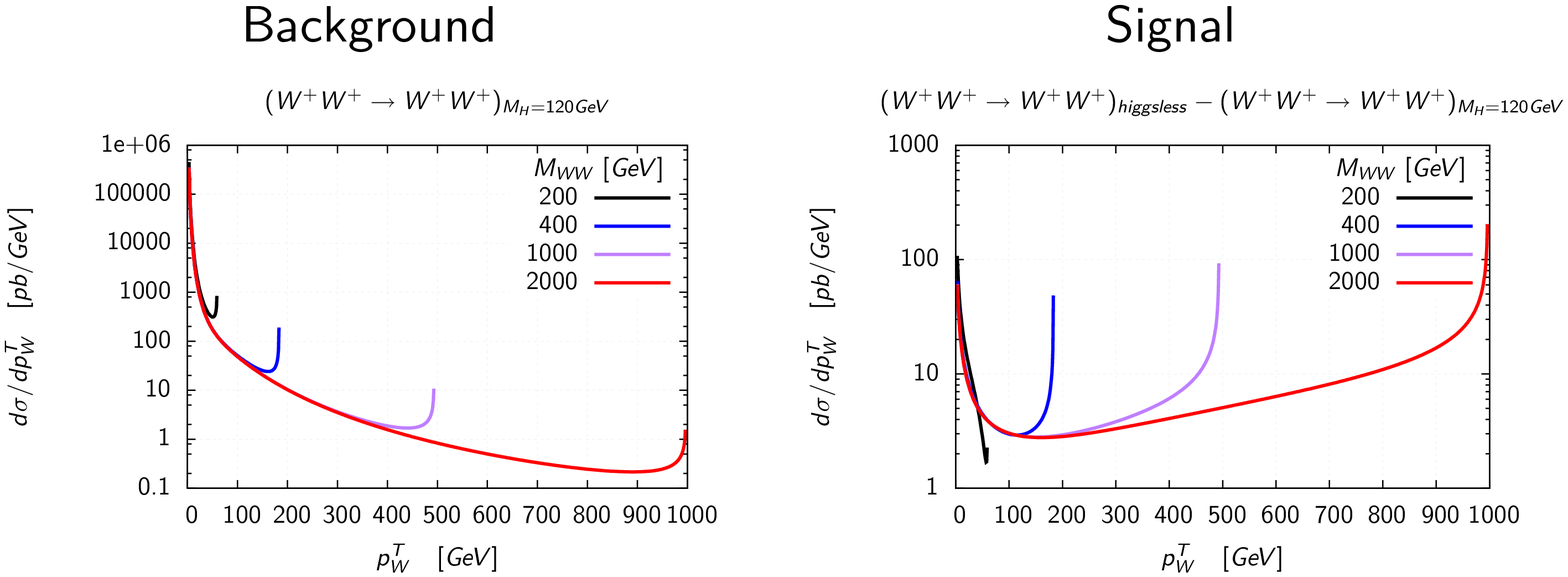,width=0.8\linewidth}
\end{center}
\caption{The $W^+W^+$ scattering cross sections for background and
signal as a function of the transverse momenta of the outgoing $W$,
for different center of mass energies ($M_{WW}$).
Calculation done within the Effective $W$ Approximation.}
\end{figure}
\vspace{5mm}

A large value of $p_T^{(1)} p_T^{(2)}$ is not only the kinematic region
where deviations from the Standard Model
are supposed to emerge (because of the $s$-divergence), but also independently
where $W_L$-associated jet kinematics is
more easily distinguishable from the $W_T$-associated jet kinematics.

\section{Gauge boson decay and possible final states}

The branching fraction of $W$ decay into any of the charged leptons with a
corresponding neutrino is $(10.80 \pm 0.09)$\%.  The $Z$ boson decays into
an oppositely charged lepton pair of a given flavor in $(3.366 \pm 0.002)$\% of the
cases and in $(20.00 \pm 0.06)$\% into neutrinos.  The rest are hadronic decays.
From an experimental point of view, this gives potentially many possible final
states of interest, depending on the particular scattering process and the
decay modes of the two bosons. 

\begin{itemize}

\item Purely leptonic

$pp \rightarrow jjW^+W^- \rightarrow jj l^+\nu l^-\nu$

$pp \rightarrow jjW^\pm W^\pm \rightarrow jj l^\pm\nu l^\pm\nu$

$pp \rightarrow jjW^\pm Z \rightarrow jj l^\pm\nu l^+l^-$

$pp \rightarrow jjZZ \rightarrow jj l^+l^-\nu\nu$

$pp \rightarrow jjZZ \rightarrow jj l^+l^-l^+l^-$

Leptonic $W$ and $Z$ decays are the preferred decay modes for a wide range of
measurements involving gauge bosons, and among other things, provide some of the
most sensitive means for Higgs studies.  Their main limiting factor is low
statistics induced by the small individual branching fractions.  Practical
viability
of these modes crucially depends on our background rejection capability.
This in general favors the non-zero total charge states $W^\pm W^\pm$, $W^\pm Z$,
and the four-lepton final state ($ZZ$), which is the only one where the full
kinematics of the process can be mesured.
On the other hand, both the $Z$ production cross section and its leptonic branching
fraction are lower than those of a $W$, hence production rates favor $W^+W^-$ followed
by $W^\pm W^\pm$.  The purely leptonic channels are often regarded as the
``gold-plated" modes in phenomenological literature because of their clean,
distinctive signatures and because the rough magnitudes of both the signal and the
main backgrounds
can usually be reasonably estimated without
involving a full detector simulation.

\item Semi-leptonic

$pp \rightarrow jjWW \rightarrow jj jjl\nu$

$pp \rightarrow jjZW \rightarrow jj jjl\nu$

$pp \rightarrow jjWZ \rightarrow jj jjl^+l^-$

$pp \rightarrow jjZZ \rightarrow jj jjl^+l^-$

These processes combine a hadronic decay of one gauge boson with a leptonic decay
of the other.  They are characterized by reasonable statistics and higher
reducible backgrounds.  Typically, control of the latter requires full detector
simulation to handle, e.g., the dominant backgrounds from processes involving production
of $W/Z$+jets with a jet misidentified as a lepton.  Additionally, at large $W/Z$ 
energies, the two jets originating from a hadronic decay tend to merge in the
detector which further reduces the signal isolation efficiency and adds extra
backgrounds to be considered.
Early studies usually revealed these channels be somewhat less promising,
overall, than purely leptonic.  However, improvements in event reconstruction in
LHC experiments and in particular
the use of novel techniques of ``jet pruning" \cite{pruning}
that allow to determine the
mass of the original object producing the jet and therefore distinguish QCD jets
from $W$ jets to a large accuracy, bring new interest to the semi-leptonic channels
again.  These techniques have been demonstrated to be applicable in the
$k_T$ and Cambridge-Aachen jet reconstruction algorithms \cite{jetalgos}, but
not in the default anti-$k_T$ algorithm used in CMS and ATLAS.  Since they have
been shown to offer great promise, reconsideration of the jet reconstruction
algorithm to be applied for VBS analyses is a potential possibility.
Clearly a lot of rework needs to be done
as dedicated reprocessing of all the past studies will be required, but in
the end the semi-leptonic channels may prove very useful to icrease the total
significance of the signal.

\item Purely hadronic

$pp \rightarrow jjWW \rightarrow jj jjjj$

$pp \rightarrow jjWZ \rightarrow jj jjjj$

$pp \rightarrow jjZZ \rightarrow jj jjjj$

Despite their large branching fractions, these processes are completely
overwhelmed by the multi-jet QCD background and in addition their study requires
full detector-dependent modeling of event reconstruction effects.  The purely hadronic
modes are therefore not considered for detailed
studies at this time.

\end{itemize}

In the above we assume $l=e,\mu$.  Decays into taus constitute yet a separate class of
specific final states and signatures, but due to the relative complexity and
lower identification efficiency they can be disregarded for the time
being.

In the search for the most promising channels to begin with, the crucial point
is the underlying physics and in particular existence or non-existence of heavy
Higgs-like resonances within the energy range of the LHC.  Their existence
of course favors the $W^+W^-$ and $ZZ$ channels and indeed these usually have been
given the most attention, also as a byproduct of Higgs physics.
Further on however we will mainly
focus on an alternative yet
plausible scenario that such resonances, if any, are too heavy for direct
detection at the LHC.  In this case, the non-resonant\footnote{Strictly
speaking, $W^\pm W^\pm$ is non-resonant as long as there are no bosonic
isospin triplets, and hence doubly charged bosons, in nature.} modes $W^\pm W^\pm$
(and $W^\pm Z$, in some sense) acquire not only equal importance, but as we will
further see, their relative exoticity can be well turned into an advantage.

\section{The uniqueness of $W^\pm W^\pm$}

It is now time to explain that the apparently arbitrary choice of the
$pp \rightarrow jj\mu^+\mu^+\nu\nu$ process as a particular example in many of our
earlier considerations was in fact well motivated.
The $W^\pm W^\pm$ final state, with its $\pm$2 total electric charge carries
unique features that make it of particular interest at the LHC.  
We have already seen that same-sign $WW$ scattering is the only process for
which the cross-talk amplitudes, $W_TW_X \to W_LW_L$ and $W_LW_L \to W_TW_X$,
are completely negligible,
mostly due to lack of any $s$-channel graphs that contribute
to the process.  The latter also has other consequences.
Contrary to other
diboson states with two accompanying jets, production of the $jjW^\pm W^\pm$ state in
the lowest order is dominated by only one physical mechanism at the quark level, namely
a quark-quark interaction associated with a $W^\pm$ emission from each colliding
quark.  Whether or not these two $W^\pm$ bosons do interact, information on their
polarizations stays encoded in the kinematics of the two outgoing quarks (recall
section \ref{wemission}), unless it is disturbed by a subsequent quark interaction.
If only we knew the energies of the colliding quarks,
appropriate cuts on the angles and
transverse momenta of the two tagging jets would increase the probability of choosing
a $W_LW_L$ state - regardless of their own final kinematics and the rest of the process.
By the same token, the only QCD contributions to the irreducible background
are graphs $\sim\alpha^2\alpha_S^2$ of the form of internal gluon exchange between
the two quarks.  Not only they are negligible in the calculation of the
BSM signal, as already shown, but their contribution to the background can be reduced
to below 10\% after basic topological VBF cuts.  Moreover, one can hope that
existence of a single physical mechanism will facilitate the search for more advanced
characteristic kinematic signatures of the signal.
Conversely, there are many more ways to produce a $jjW^+W^-$ final state at the LHC.

\begin{figure}[htbp]
\vspace{-3cm}
\begin{center}
\hspace{-1cm}
\epsfig{file=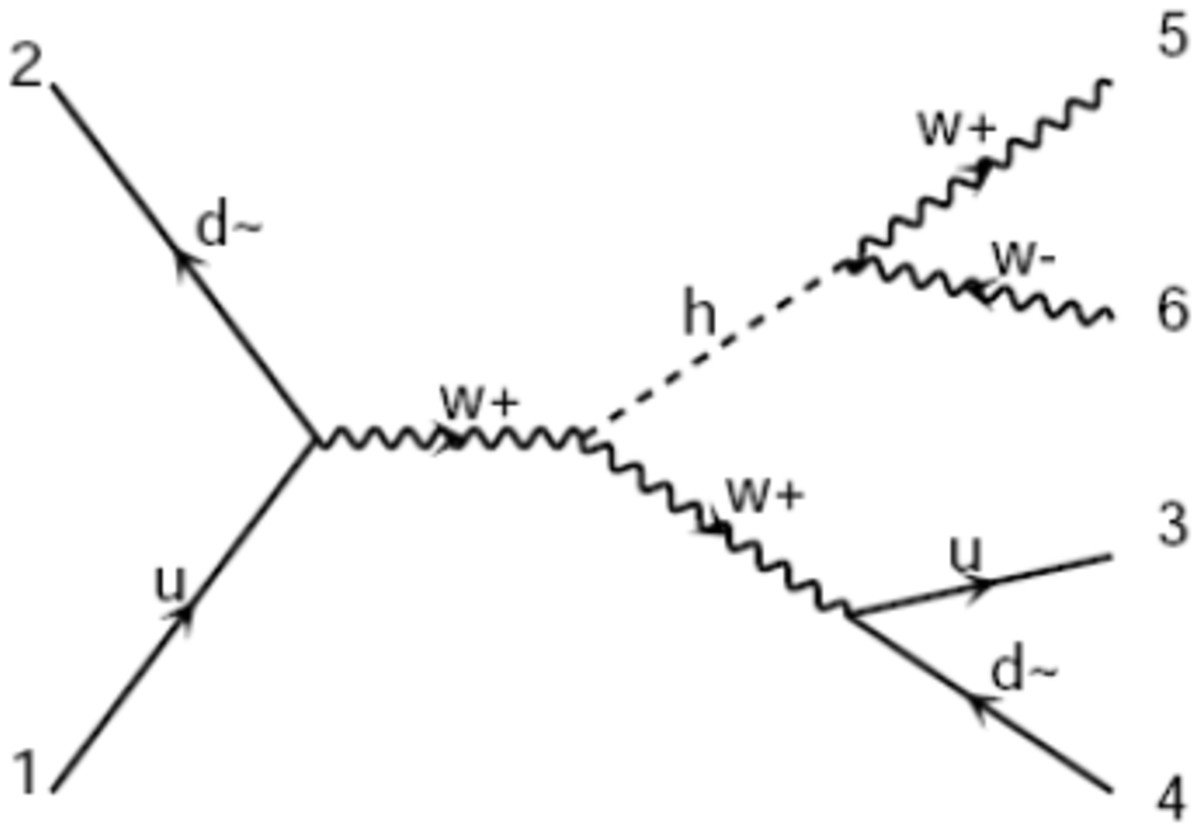,width=0.6\linewidth}\hspace{-2.5cm}
\epsfig{file=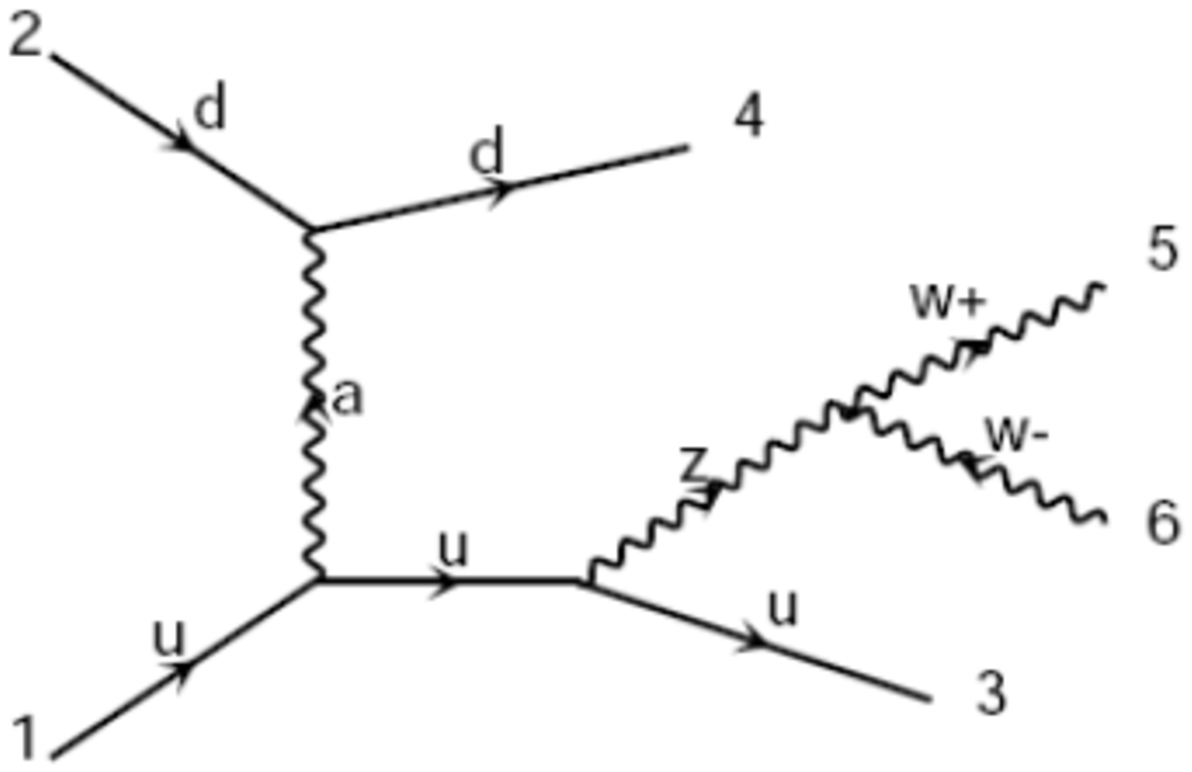,width=0.6\linewidth}
\end{center}
\vspace{-4cm}
\caption{Examples of Feynman diagrams of purely electroweak processes
that contribute to the process
$p p \rightarrow j j W^+W^-$, but have no equivalent for $jjW^+W^+$.
Events where both $W$'s originate from a decay of a neutral particle
contribute both to our definition of signal (left) and the irreducible
background (right), but kinematicwise do not allow the distinction
of $W$ polarizations.  The left graph is actually Higgs production via
Higgsstrahlung and is of little relevance once VBF selection criteria
are imposed.  However, huge additional contributions to the irreducible
background change significantly its overall kinematic distributions and mask
the part of the background which is related to single $W$ emissions from each
colliding quark.}
\vspace{5mm}
\end{figure}

A $W^+W^-$ pair can come from virtual $Z$ decay, as well as two
consecutive emissions off a single quark.  Even more importantly, the electroweak-QCD
background receives huge additional contributions from graphs involving
gluon-gluon and quark-gluon interactions.  In fact, processes $\sim\alpha^2\alpha_S^2$
dominate the total $jjW^+W^-$ production by an order of magnitude, prior to
kinematic cuts.
Usual non-VBF Higgs production graphs, e.g., those involving Higgsstrahlung followed by
Higgs decay into $W^+W^-$, also
contribute to the signal according to our working definition, but are of little
use kinematicwise when we go to higher energies.  All in all, signal in the $W^+W^-$
mode can be expected
less well kinematically separated from background, and background much larger.
Assuming the absence of new heavy Higgs-like resonances within the energy reach
of the LHC at 13/14 TeV, $W^+W^-$ is a more difficult choice.
For completeness one should notice that the choice of same-sign $W$ pair is also
a powerful shield against the overwhelming reducible background originating
from $t\bar{t}$ production.  Only second order effects, like leptonic $B$ decays
or lepton charge misidentification can lead to a non-zero $t\bar{t}$ background.
These aspects will be elaborated further on.

\begin{figure}[htbp]
\begin{center}
\hspace{-1.5cm}
\epsfig{file=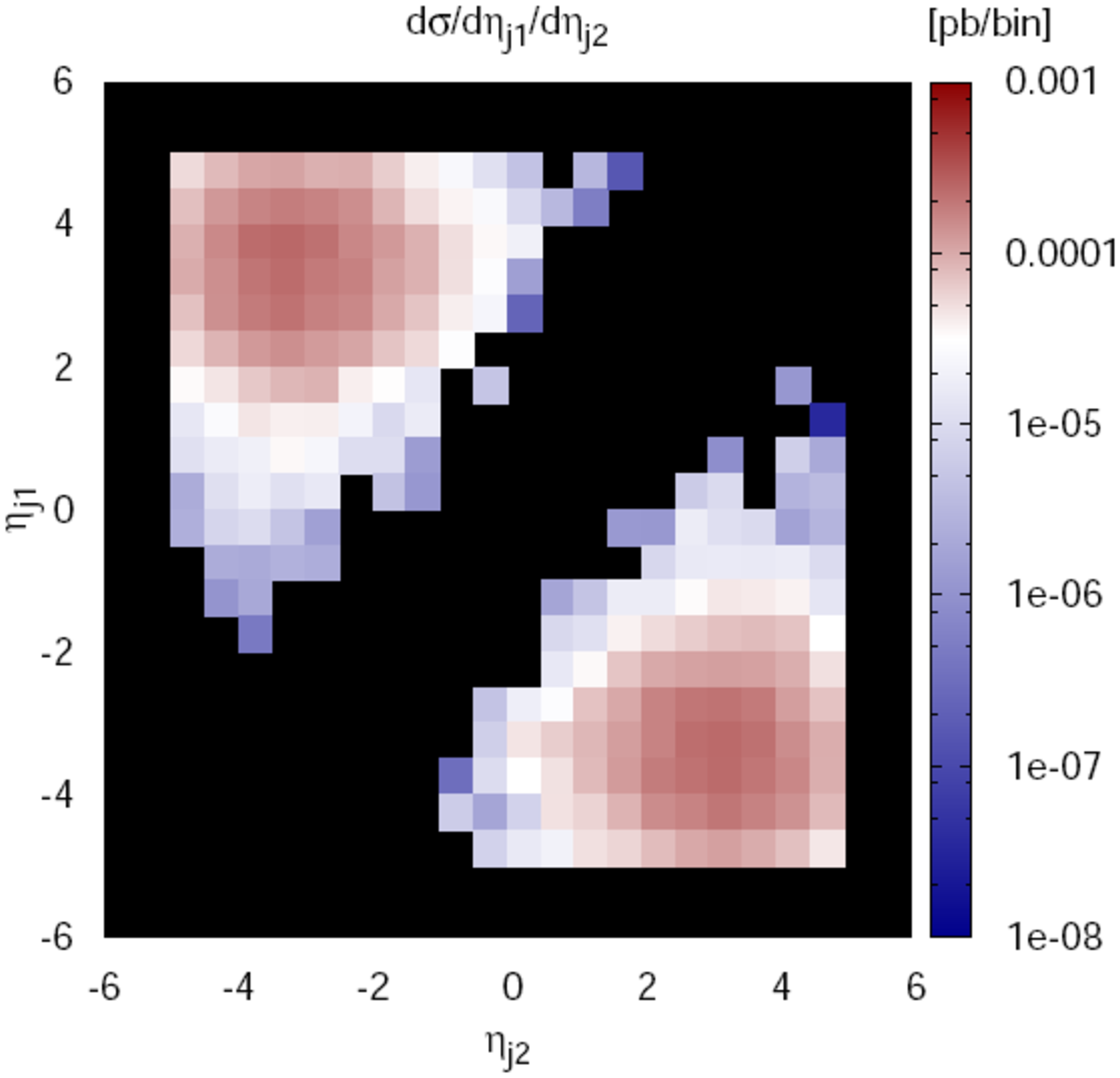,width=0.38\linewidth}\hspace{-6mm}
\epsfig{file=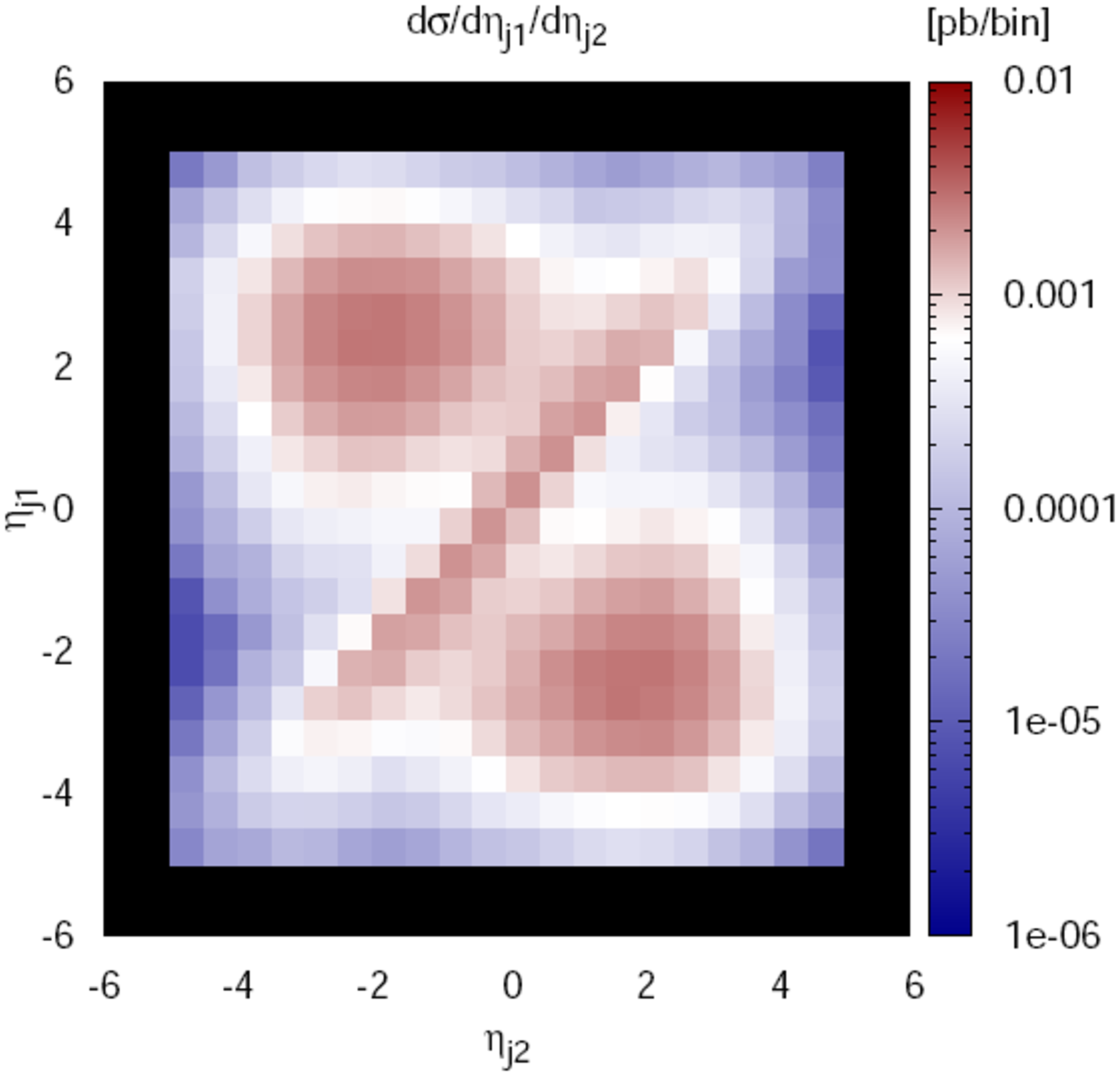,width=0.38\linewidth}\hspace{-6mm}
\epsfig{file=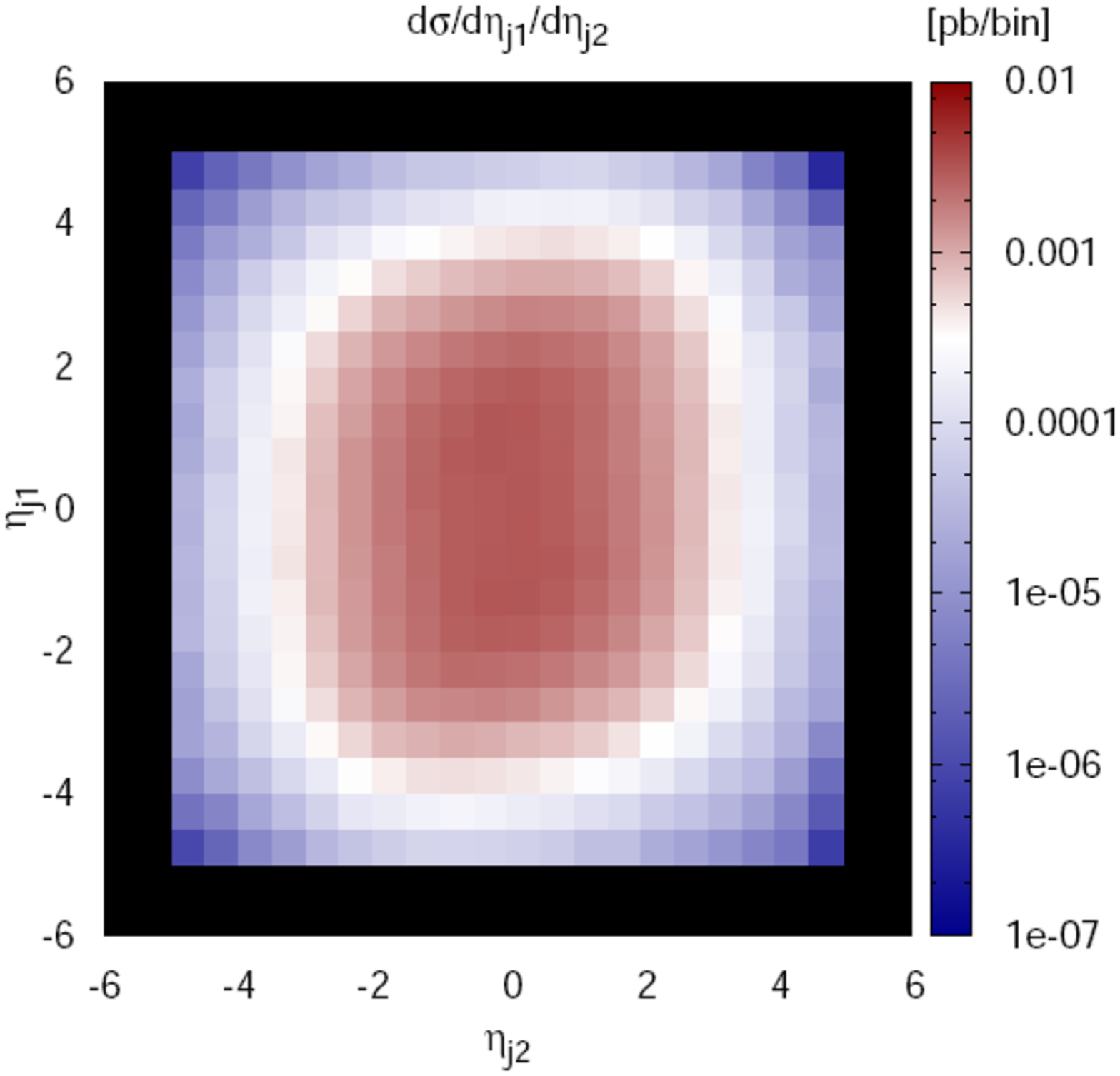,width=0.38\linewidth}
\end{center}
\vspace{-1.5cm}
\caption{Distributions of signal (left), electroweak background (middle)
and QCD background (right) for the process $pp \rightarrow jjW^+W^+$ at 14 TeV
as a function of jet pseudorapidities.  Results of parton-level MadGraph
calculations involving all processes $\sim\alpha^4$ (left and middle) and
all processes $\sim\alpha^2\alpha_S^2$ (right).  Interference between the
two classes was neglected for demonstration purposes.}
\label{etajetaj}
\vspace{5mm}
\end{figure}

\begin{figure}[htbp]
\vspace{-7cm}
\hspace{-2cm}
\epsfig{file=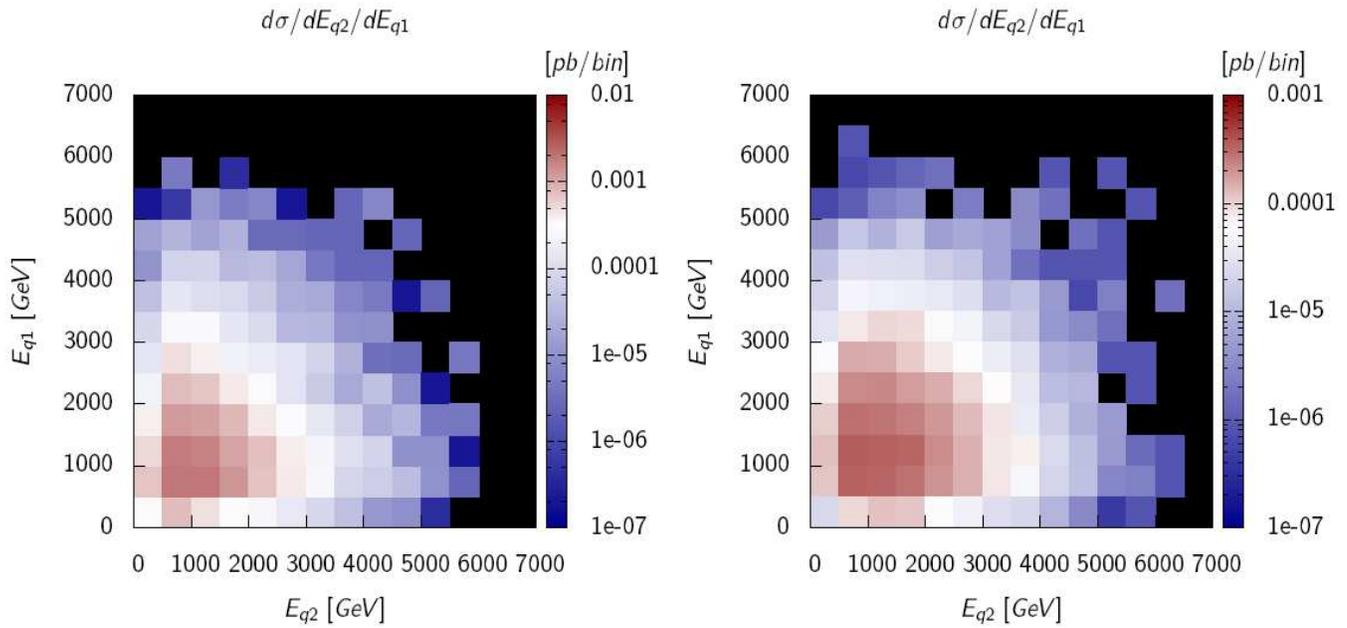,width=1.2\linewidth}
\vspace{-9cm}
\caption{Background (left) and signal (right) for the process
$pp \rightarrow jjW^+W^+$ at 14 TeV as a function of the energies
of the two incident quarks after imposing basic topological cuts discussed
in the text.  Results of a parton-level MadGraph calculation.}
\label{eq1eq2}
\vspace{1cm}
\end{figure}

Assuming the pure Higgsless Standard Model scenario as the theoretical basis
for the definition and numerical computation of the signal, the total signal
cross section for $pp \rightarrow jjW^+W^+$ at $\sqrt{s}$ = 13/14 TeV is
roughly an order of magnitude smaller than the irreducible background.
Basic kinematic signatures of the signal and irreducible backgrounds
in terms of
angular distributions of the two outgoing jets, confirm the usefulness of
the forward jet tagging technique (a basic VBF signature in the LHC) to isolate
signal from background.  The requirement of two opposite-sided, large pseudorapidity
jets, $2 < |\eta_j| <5$ and $\Delta\eta > 4$, suppresses the bulk of soft
parton-parton collisions.  It eliminates most of the electroweak background,
and even more efficiently the electroweak-QCD background - for an illustration
of the basic topologies of signal and backgrounds, see Fig.~\ref{etajetaj}.
This, together with another basic topological requirement of two $W$ bosons
within the acceptance of the detector (which can be approximated quantitatively
as $|\eta_W| < 2$) has an important effect at the quark level as it effectively
selects a very specific configuration of the colliding quarks.
Not only we have a single production mechanism - residual processes in which
both outgoing $W$'s originate from the same quark line, or not from a quark at all,
are now completely suppressed - but also a common production kinematics.
Energy distributions of the two quarks {\it before} interaction, usually
preferring the lowest energies, as dictated by proton PDF's, now begin
to peak quite strongly around
roughly $\sim$1/7 of the proton energy, as shown in Fig.~\ref{eq1eq2}.
At the quark level the whole process can now
be reasonably approximated by considering a symmetric collision of two nearly
monochromatic quark beams with $\sqrt{s}~\approx$ 2 TeV.  For kinematic calculations,
the complicated
proton-proton process with its 1428 tree level Feynman diagrams can be reduced
to the ``only" 102 diagrams corresponding to a pure quark level process
$uu \rightarrow ddW^+W^+$ at a fixed energy.

\begin{figure}[htbp]
\vspace{-3.0cm}
\begin{center}
\epsfig{file=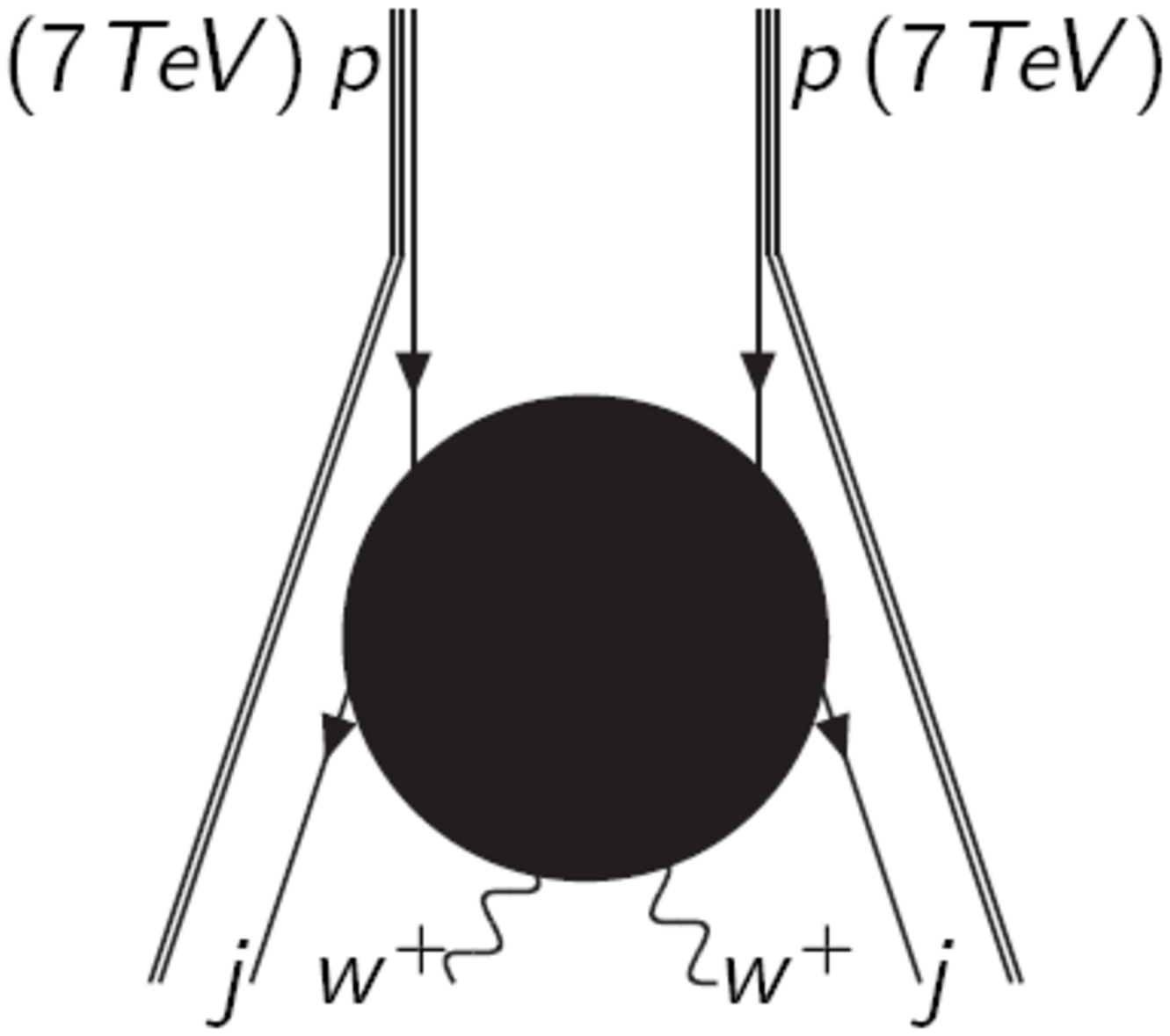,width=0.49\linewidth}
\epsfig{file=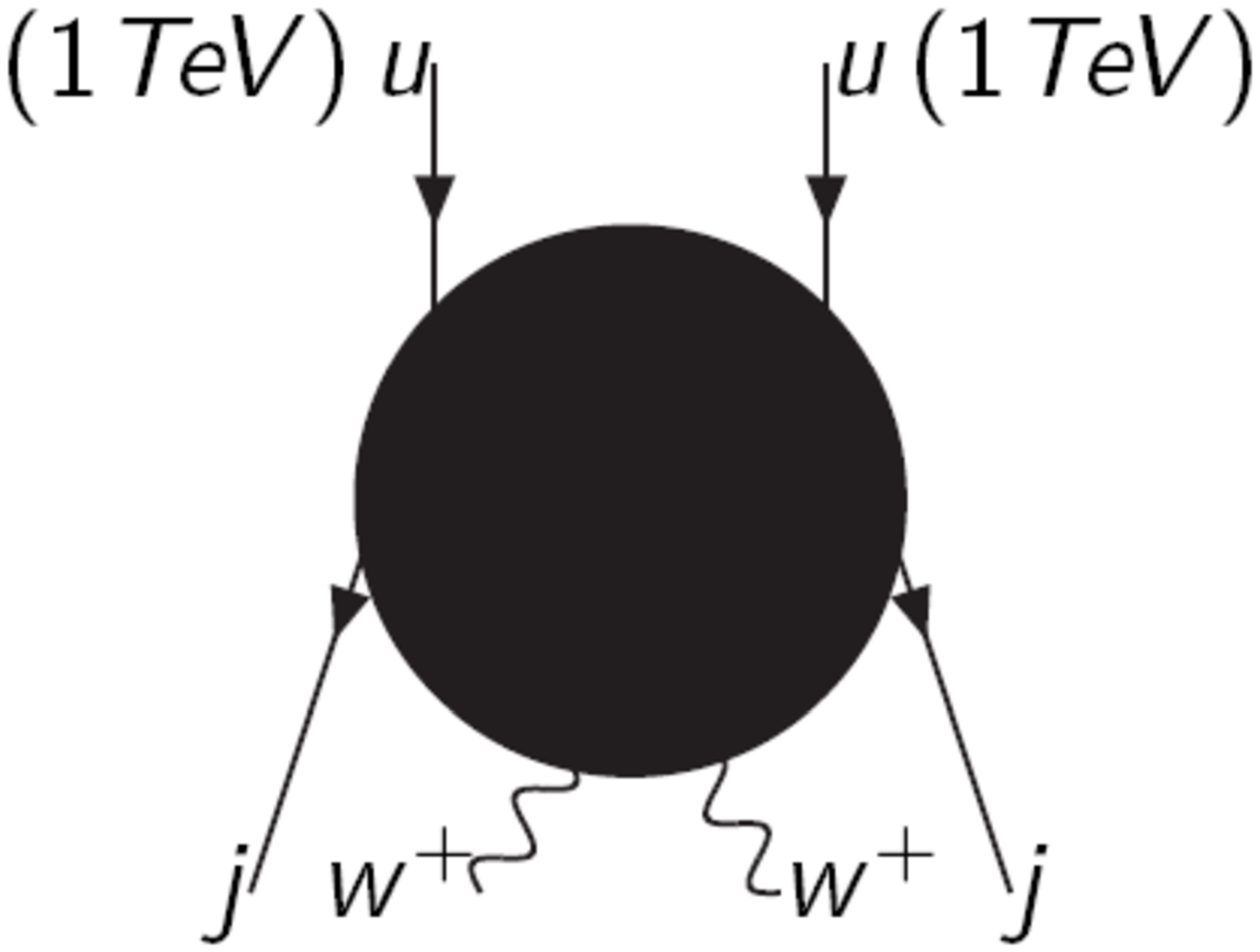,width=0.49\linewidth}
\end{center}
\vspace{-3.0cm}
\caption{Schematic representations of the full set of processes
which need be taken into account for the evaluation
of the signal in the ``production $\times$ decay" approximation (left)
and of the reduced set of processes which need be taken into account
to learn the basic kinematics of the signal process and of the irreducible
background (right).  Drawings by J.~Kuczmarski.}
\vspace{5mm}
\end{figure}

\begin{figure}[htbp]
\vspace{-3cm}
\hspace{-1cm}
\epsfig{file=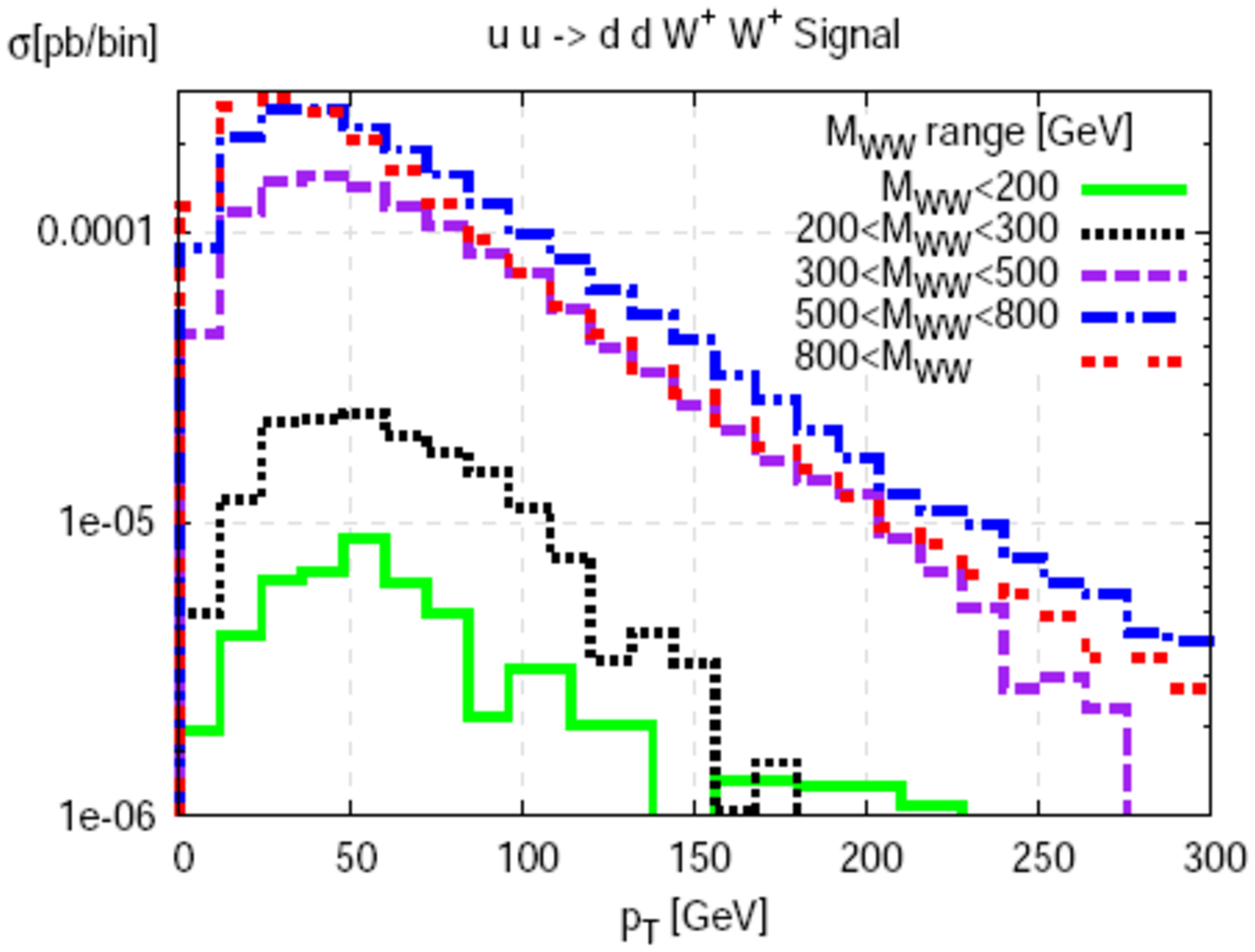,width=0.62\linewidth}\hspace{-2.5cm}
\epsfig{file=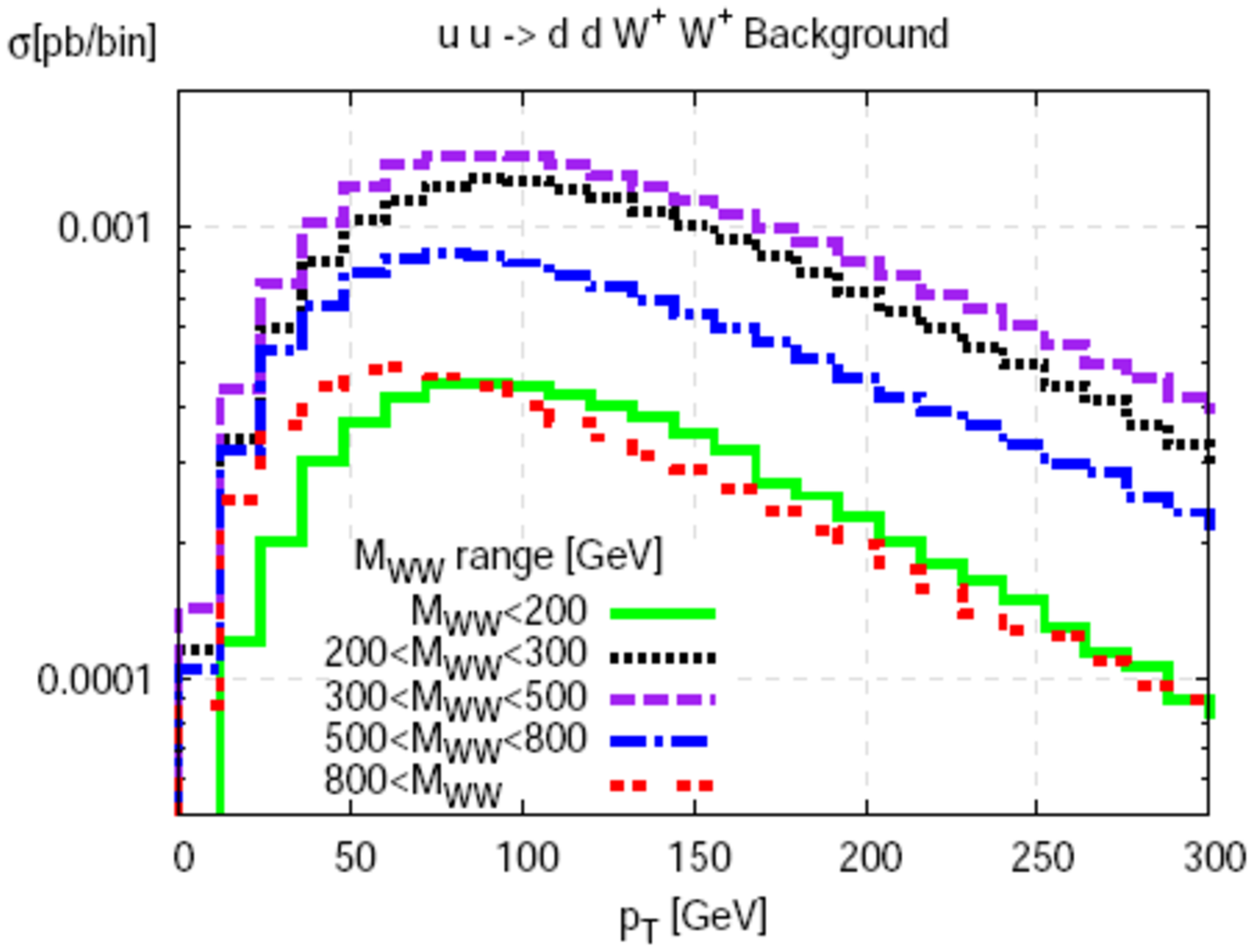,width=0.62\linewidth}
\vspace{-4cm}
\caption{Distributions of transverse momenta of the jets associated
to the emission of $W_LW_L$ signal (left) and $W_TW_T$ background (right)
in the quark level process $uu \rightarrow ddW^+W^+$ at an incident center of mass
energy of 2 TeV.
Results of MadGraph simulations.
Note similarity in gross features to the distributions resulting from
the pure emission process shown in Fig.~\ref{ptjewa}.}
\vspace{5mm}
\label{ptq}
\end{figure}

\begin{figure}[htbp]
\vspace{-4cm}
\hspace{-1cm}
\epsfig{file=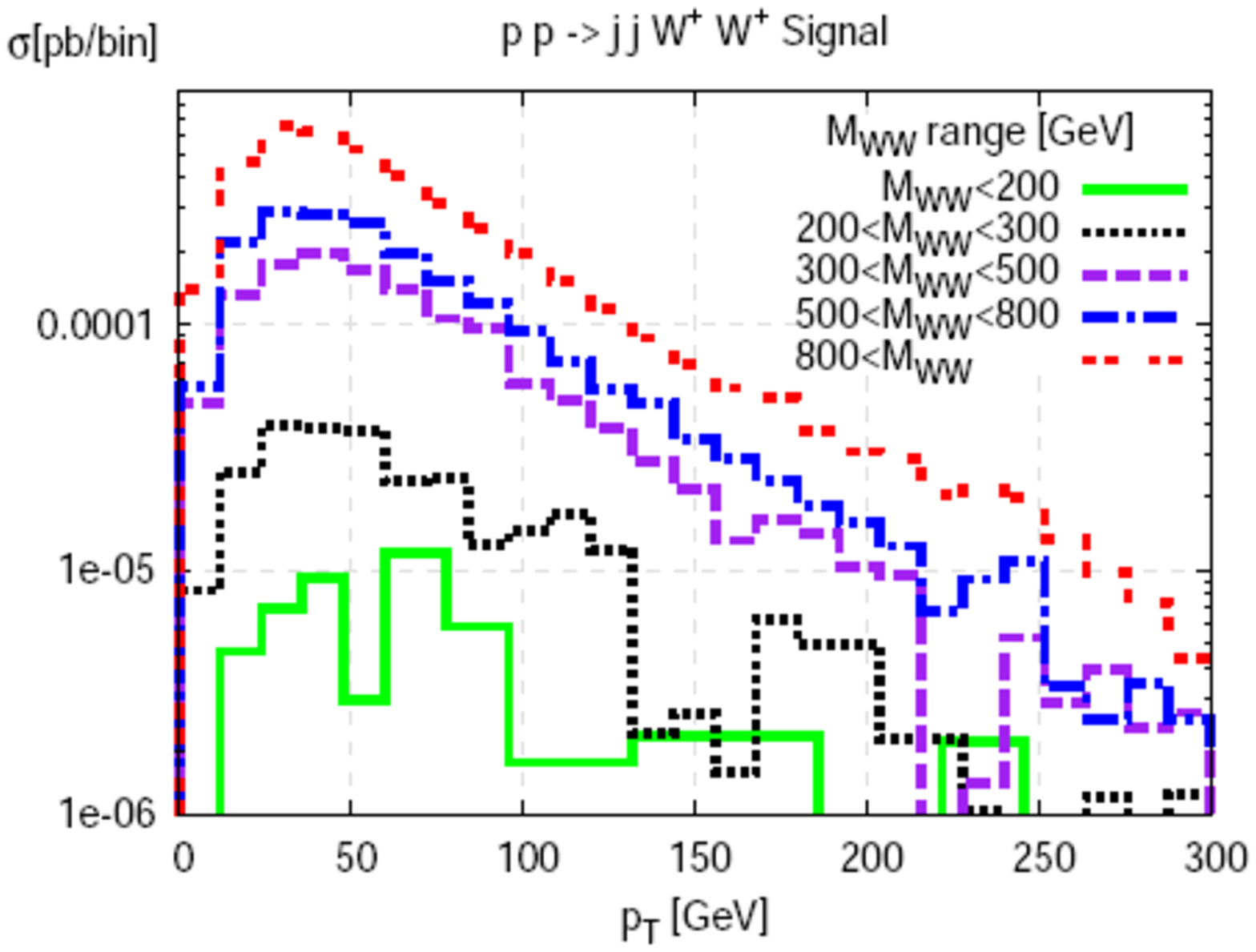,width=0.62\linewidth}\hspace{-2.5cm}
\epsfig{file=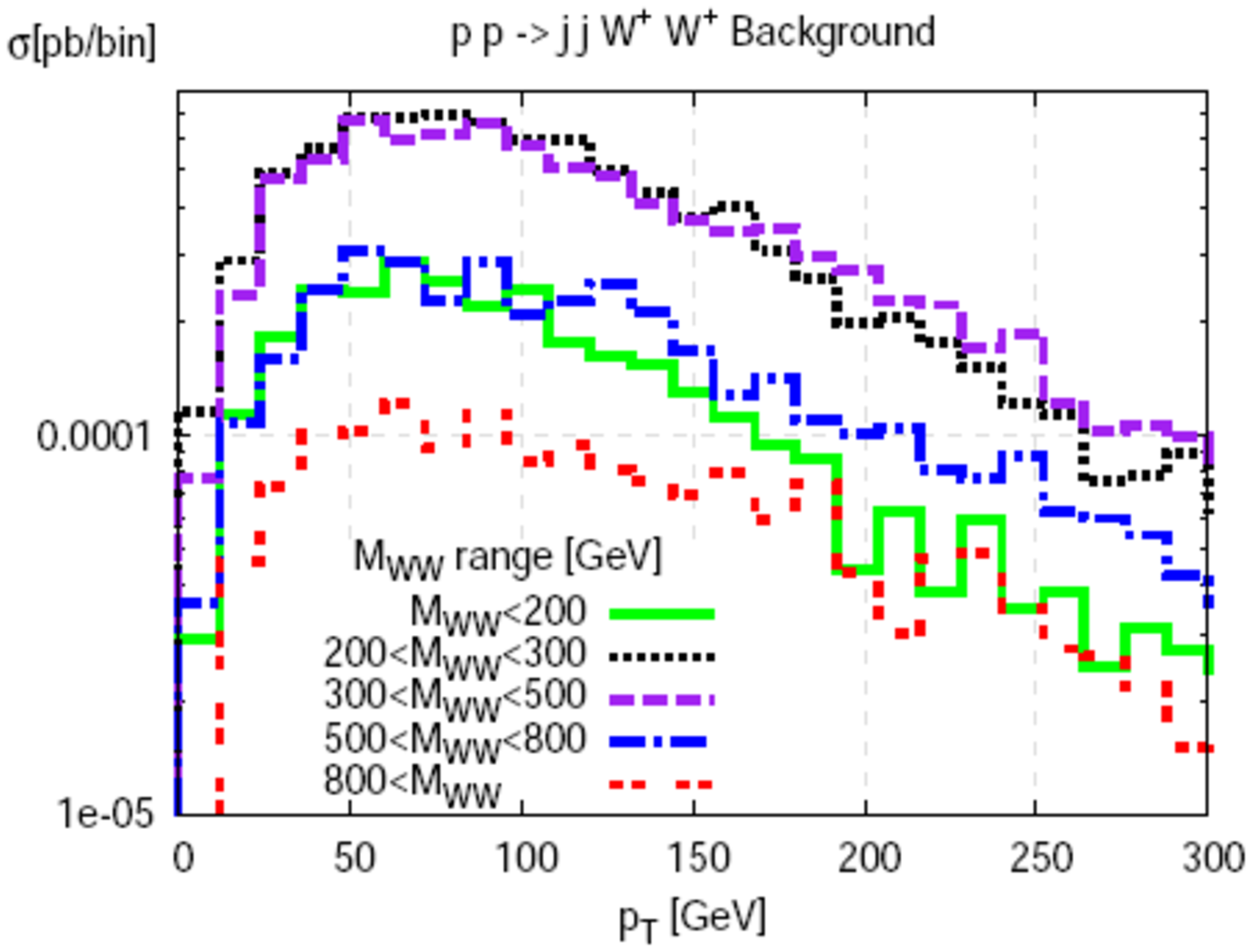,width=0.62\linewidth}
\vspace{-4cm}
\caption{Distributions of transverse momenta of the jets associated
to the emission of $W_LW_L$ signal (left) and $W_TW_T$ background (right)
in the full process
$p p \rightarrow j j W^+W^-$ at 14 TeV (bottom row).
Results of MadGraph simulations.
Note similarity in gross features to the distributions resulting from
the quark process shown in Fig.~\ref{ptq} and from the
pure emission process shown in Fig.~\ref{ptjewa}.}
\label{ptjptj}
\vspace{5mm}
\end{figure}

Comparison of the final state kinematics of the processes $pp \rightarrow jjW^+W^+$
at 14 TeV and $uu \rightarrow ddW^+W^+$ at 2 TeV, after no more than the basic VBF
topological cuts defined above, is very telling.  The kinematics of the
outgoing $W$ bosons are indeed very similar, with the respective pseudorapidity
and transverse momentum distributions merely being typically some 5-15\%
more smeared in the former.  Larger differences affect only the tails of
the transverse momenta of the jets and the two-jet invariant mass.

\begin{figure}[htbp]
\hspace{-5mm}
\epsfig{file=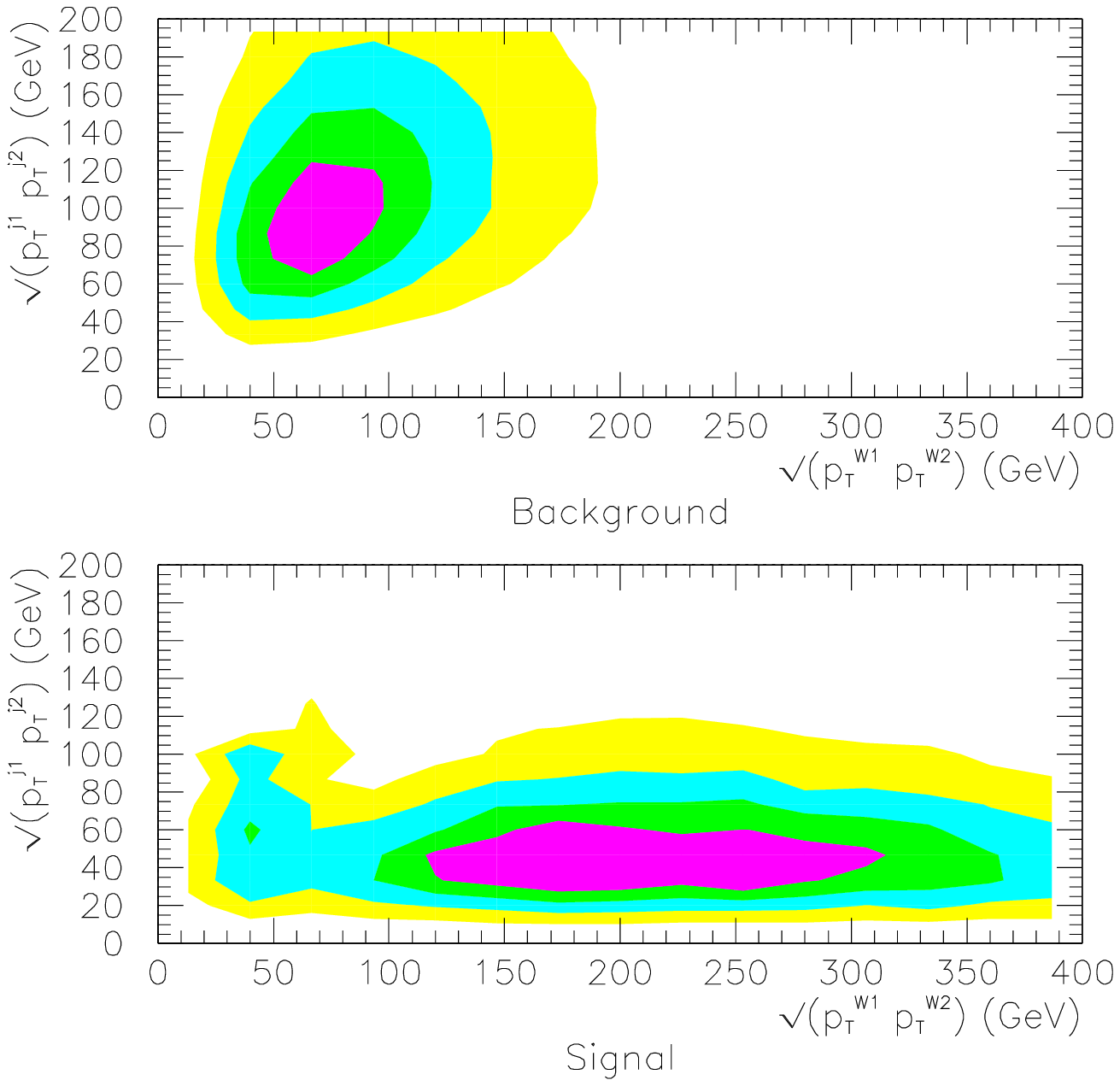,width=0.58\linewidth}
\hspace{-1cm}
\epsfig{file=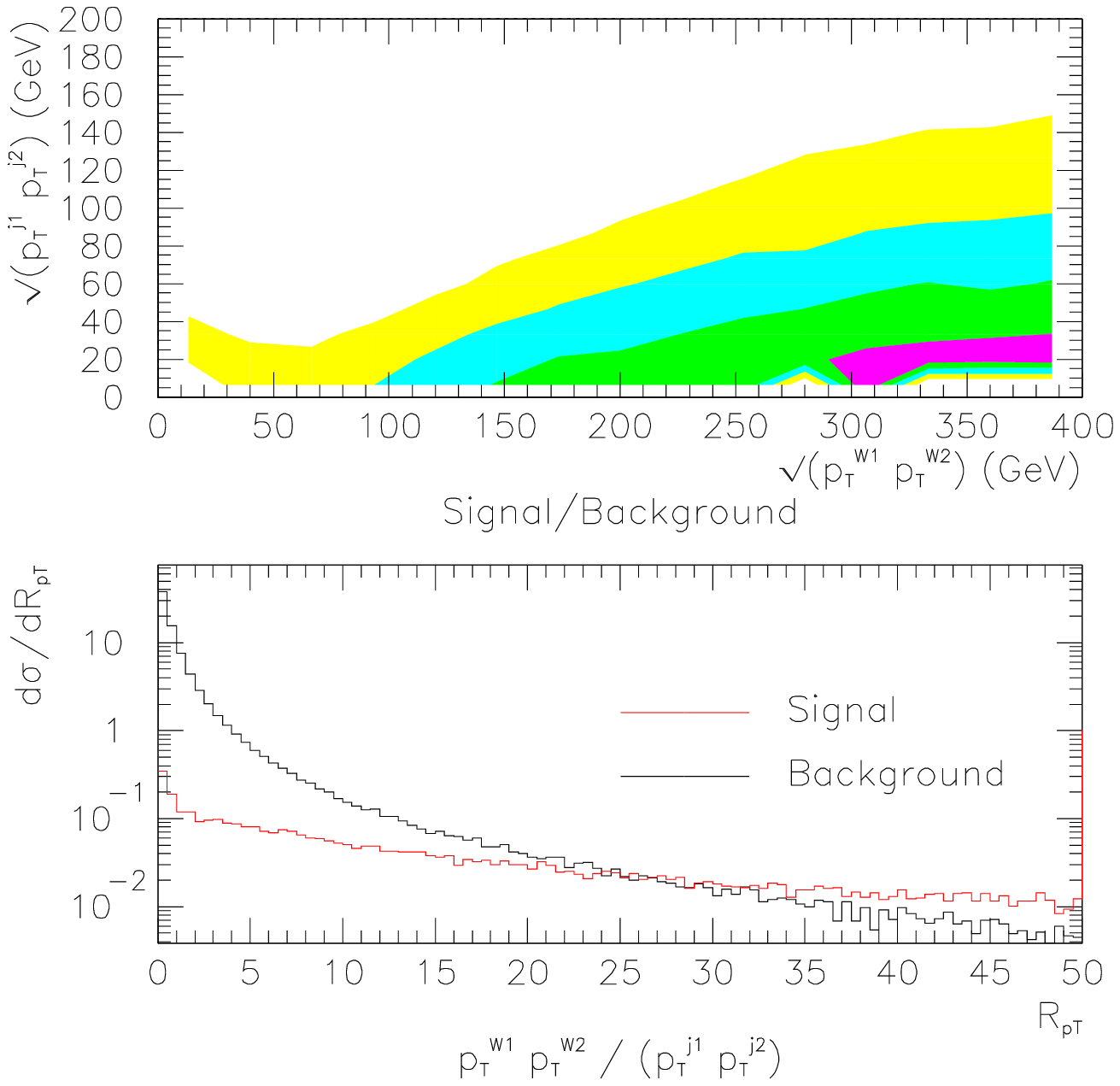,width=0.58\linewidth}
\vspace{5mm}
\caption{The kinematics of longitudinal signal and transverse background
in the pure electroweak quark-level process $uu \rightarrow ddW^+W^+$ at 2 TeV.
{\bf Left:} the differential cross sections of background and signal
in a two-dimensional space defined by the
square roots of the transverse momentum products of the two outgoing
$d$ quarks and the two outgoing $W$'s, $\sqrt{p_T^{d1}p_T^{d2}}$ versus
$\sqrt{p_T^{W1}p_T^{W2}}$; the color contours are equidistant
and the scale ranges from zero (white) to 0.004 fb/GeV$^2$ for
the background and to 0.0002 fb/GeV$^2$ for the signal (purple).
{\bf Upper right:} the signal to background ratio from dividing the two left
plots; the vertical scale is logarithmic for better visualization and
ranges from 0.03 (white) to 30 (purple).  {\bf Lower right:}
the distributions of the ratio $p_T^{W1}p_T^{W2}/(p_T^{d1}p_T^{d2})$
for signal and background.
No kinematic cuts were applied.
Results of a MadGraph calculation.  Signal was calculated by considering
longitudinal $W^+W^+$ pairs only and subtracting the SM-based distributions
from the Higgsless-based distributions.
}
\label{rptwpwp}
\vspace{5mm}
\end{figure}

The fact that most features of the final state kinematics can be approximated
with a picture of two colliding monochromatic quark ``beams"
has a very important phenomenological consequence.  Systematic differences
in the kinematics of the tagging jets associated to the emission of longitudinally
and transversely polarized gauge bosons can indeed be observed in an experiment.
Partonic structure functions inside a proton in the first step inevitably
smear out the measured transverse momentum distributions and hide the information
on $W$ polarity.  Remarkably, after basic VBF topological cuts this information
can be unveiled again.  Tree level calculations show that transverse momentum
distributions of jets in signal and background events indeed follow the same
qualitative trends as outlined before for the pure emission process of a
longitudinal and transverse gauge boson, as well as in the signal and
background in an ideal quark process, once VBF topological cuts are applied
on the former - see in particular Figs.~\ref{ptq} and \ref{ptjptj}.

\begin{figure}[htbp]
\hspace{-5mm}
\epsfig{file=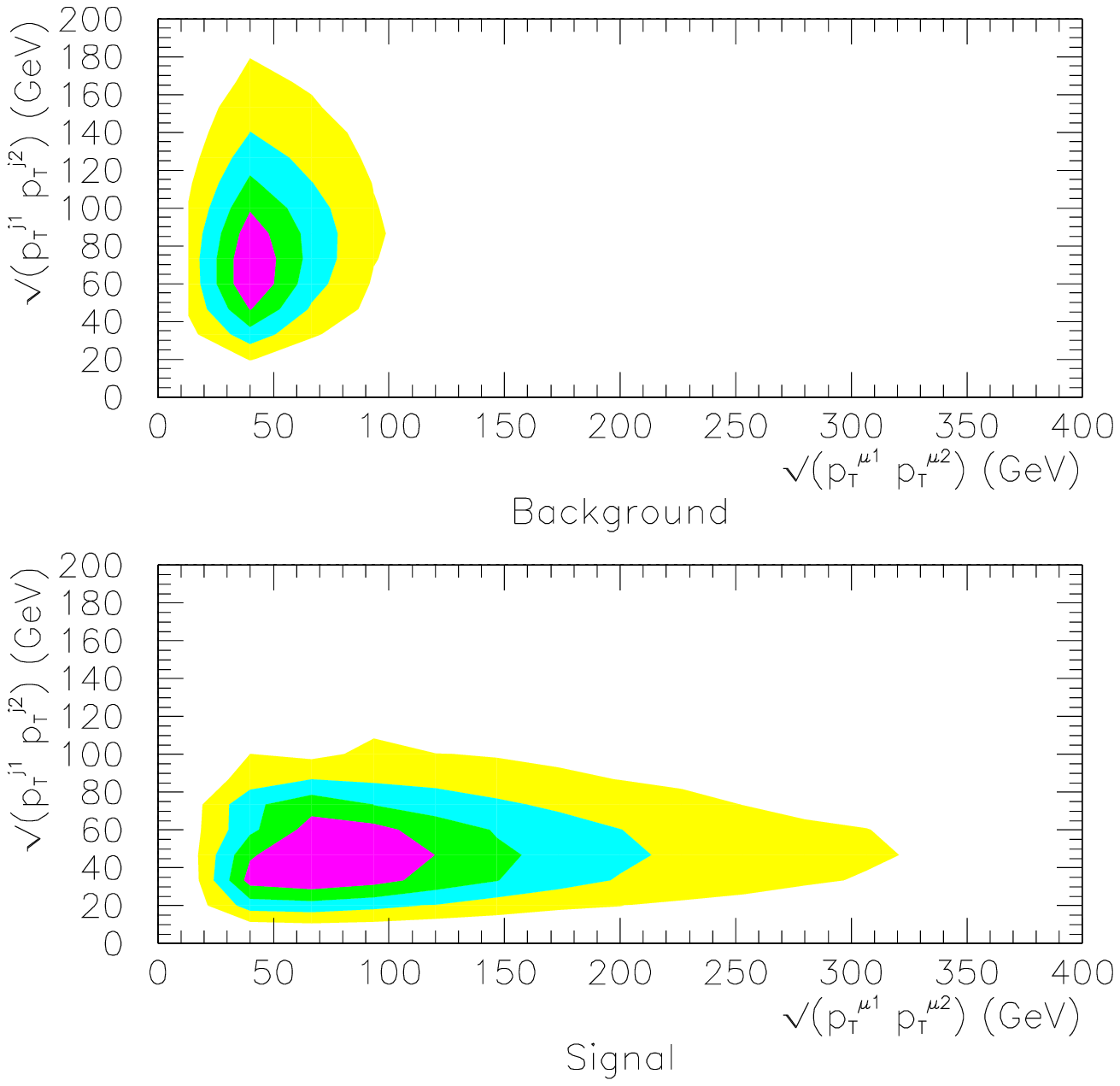,width=0.58\linewidth}
\hspace{-1cm}
\epsfig{file=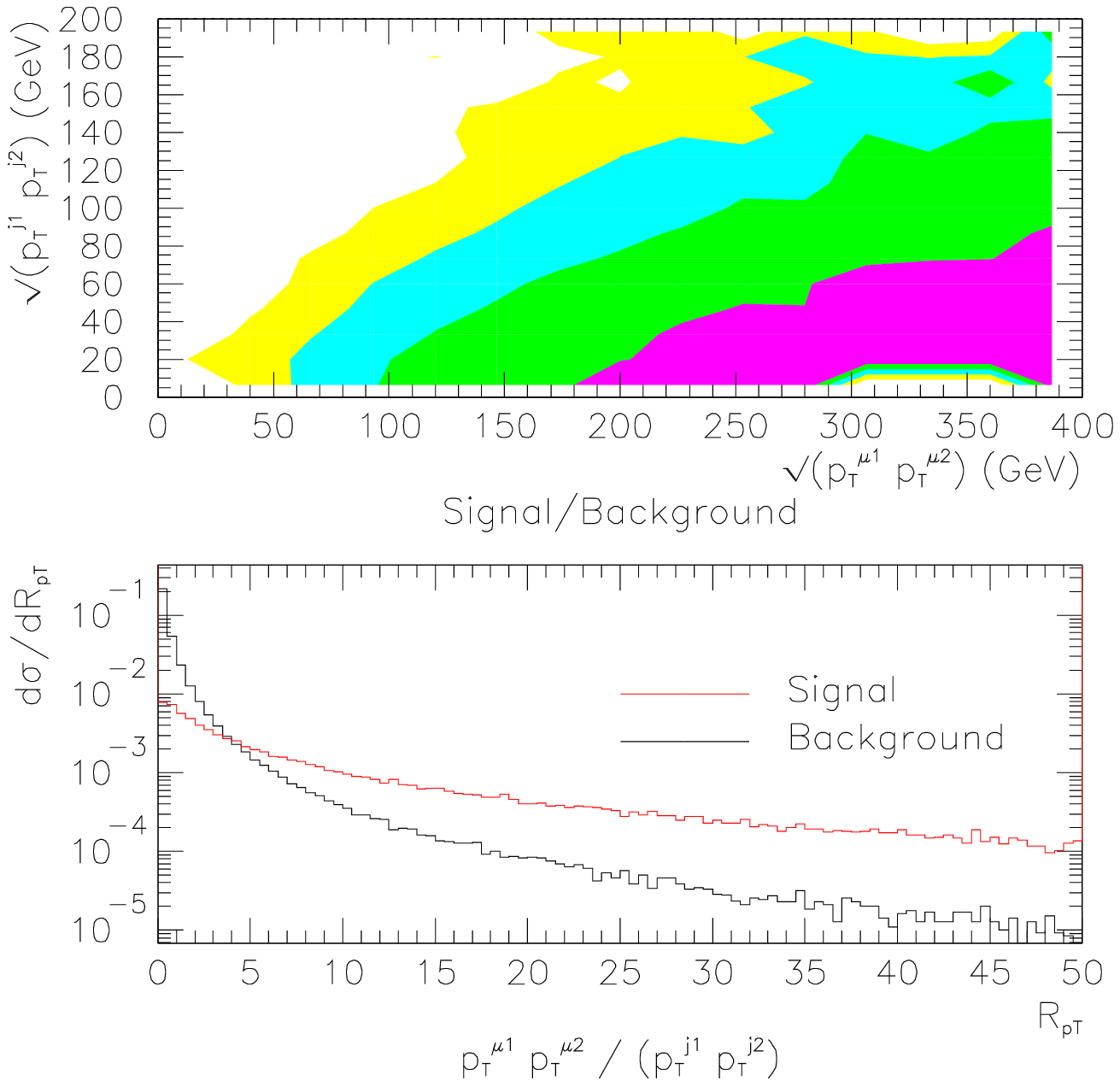,width=0.58\linewidth}
\vspace{5mm}
\caption{The kinematics of longitudinal signal and transverse background
in the full process $pp \rightarrow jj\mu^+\mu^+$ at 14 TeV after
applying basic topological VBF cuts only.
{\bf Left:} the differential cross sections of background and signal
in a two-dimensional space defined by the
square roots of the transverse momentum products of the two leading
jets and the two outgoing muons, $\sqrt{p_T^{j1}p_T^{j2}}$ versus
$\sqrt{p_T^{\mu 1}p_T^{\mu 2}}$; the color contours are equidistant
and the scale ranges from zero (white) to 0.48$\cdot 10^{-4}$ fb/GeV$^2$
for the background and to 0.7$\cdot 10^{-5}$ fb/GeV$^2$ for the signal (purple).
{\bf Upper right:} the signal to background ratio from dividing the two left
plots; the vertical scale is logarithmic for better visualization and 
ranges from 0.04 (white) to 40 (purple).  {\bf Lower right:}
the distributions of the ratio $p_T^{\mu 1}p_T^{\mu 2}/(p_T^{j1}p_T^{j2}$)
for signal and background.
Results of a MadGraph simulation, processed by PYTHIA 6 \cite{pythia} for the
effects of parton showering, hadronization and jet reconstruction,
and further processed by PGS 4 for the effects of finite resolution in
the measurement of jet and muon $p_T$ in a CMS-like detector.
Signal was calculated by subtracting the SM-based distributions
from the Higgsless-based distributions.
}
\vspace{5mm}
\end{figure}

\begin{figure}[htbp]
\hspace{-5mm}
\epsfig{file=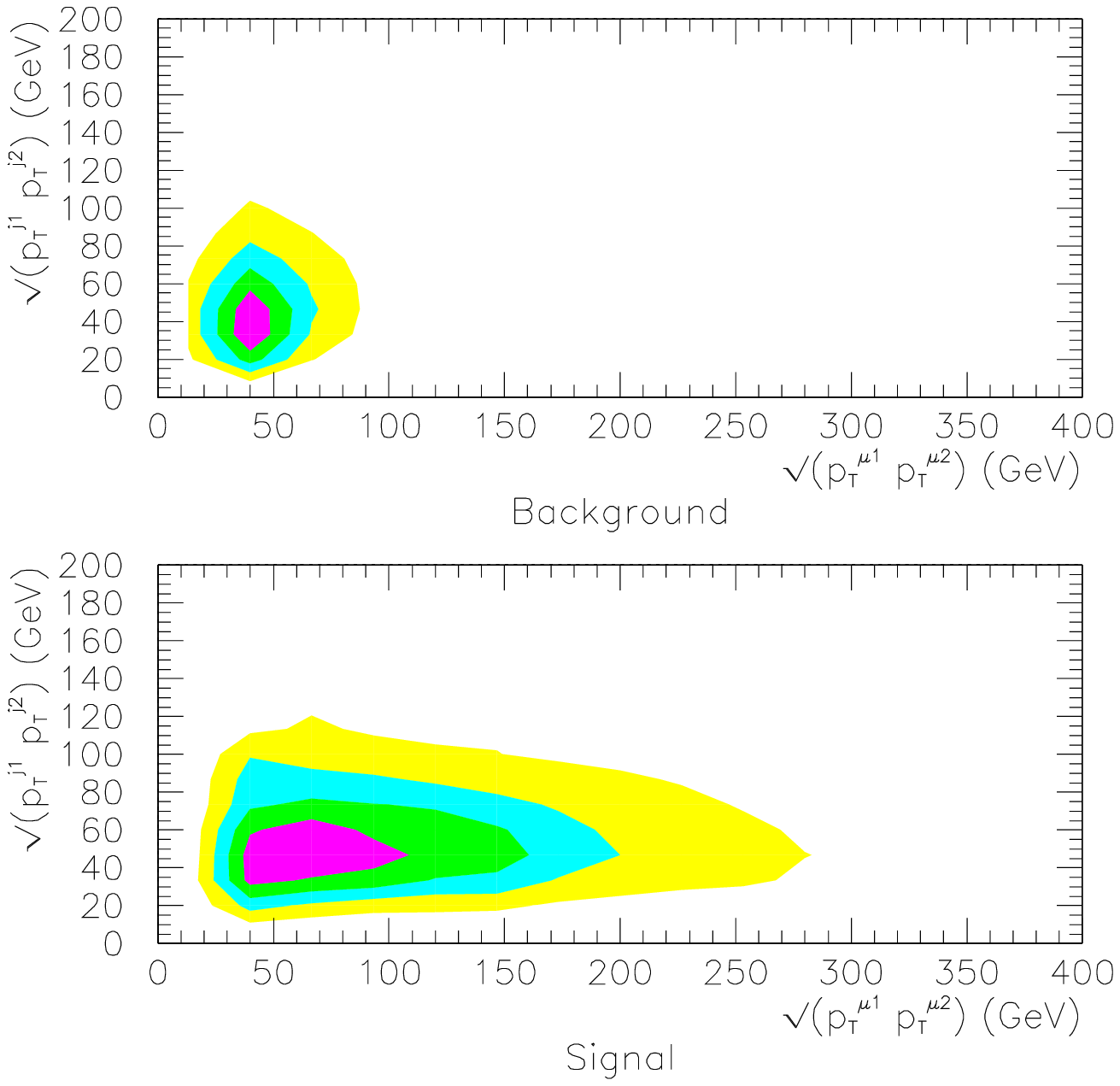,width=0.58\linewidth}
\hspace{-1cm}
\epsfig{file=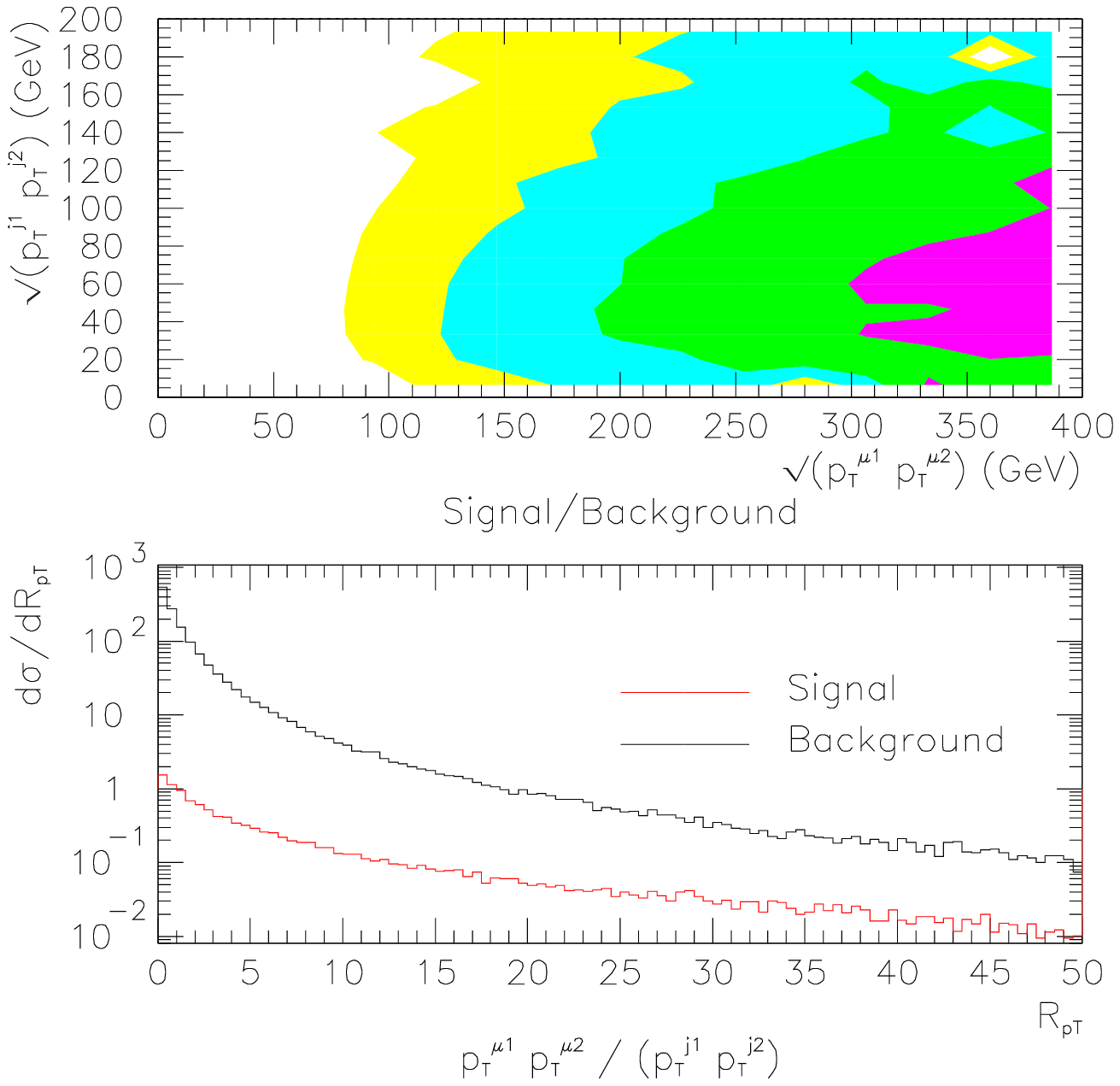,width=0.58\linewidth}
\vspace{5mm}
\caption{The kinematics of longitudinal signal and transverse background
in the full process $pp \rightarrow jj\mu^+\mu^-$ at 14 TeV after
applying basic topological VBF cuts only.
{\bf Left:} the differential cross sections of background and signal
in a two-dimensional space defined by the
square roots of the transverse momentum products of the two leading
jets and the two outgoing muons, $\sqrt{p_T^{j1}p_T^{j2}}$ versus
$\sqrt{p_T^{\mu^+}p_T^{\mu^-}}$; the color contours are equidistant
and the scale ranges from zero (white) to 0.0011 fb/GeV$^2$ 
for the background and to 0.12$\cdot 10^{-4}$ fb/GeV$^2$ for the signal (purple).
{\bf Upper right:} the signal to background ratio from dividing the two left
plots; the vertical scale is logarithmic for better visualization and
ranges from 0.0025 (white) to 2.5 (purple).  {\bf Lower right:}
the distributions of the ratio $p_T^{\mu^+}p_T^{\mu^-}/(p_T^{j1}p_T^{j2})$
for signal and background.
Results of a MadGraph simulation of $pp \rightarrow jjW^+W^-$, processed by
PYTHIA 6 for $W$ decay into muons, the
effects of parton showering, hadronization and jet reconstruction,
and further processed by PGS 4 for the effects of finite resolution in
the measurement of jet and muon $p_T$ in a CMS-like detector.
The original PYTHIA 6 source code was modified to account for the
correct, polarization-dependent, angular distributions for the decays
$W^\pm \rightarrow \mu^\pm \nu$.
Signal was calculated by considering longitudinal $W^+W^-$ pairs only
and subtracting the SM-based distributions
from the Higgsless-based distributions.
}
\vspace{5mm}
\end{figure}

\begin{figure}[htbp]
\hspace{-5mm}
\epsfig{file=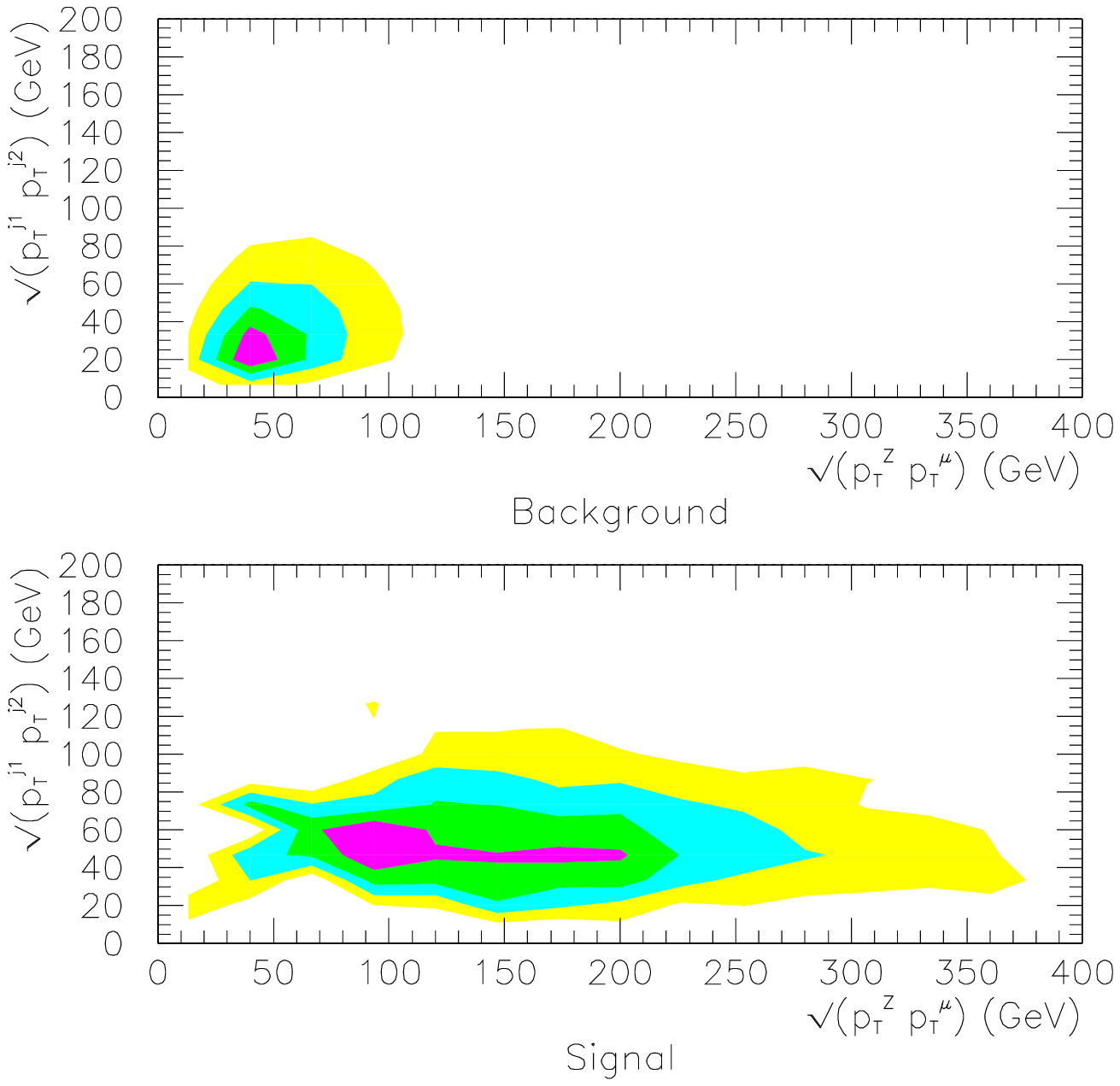,width=0.58\linewidth}
\hspace{-1cm}
\epsfig{file=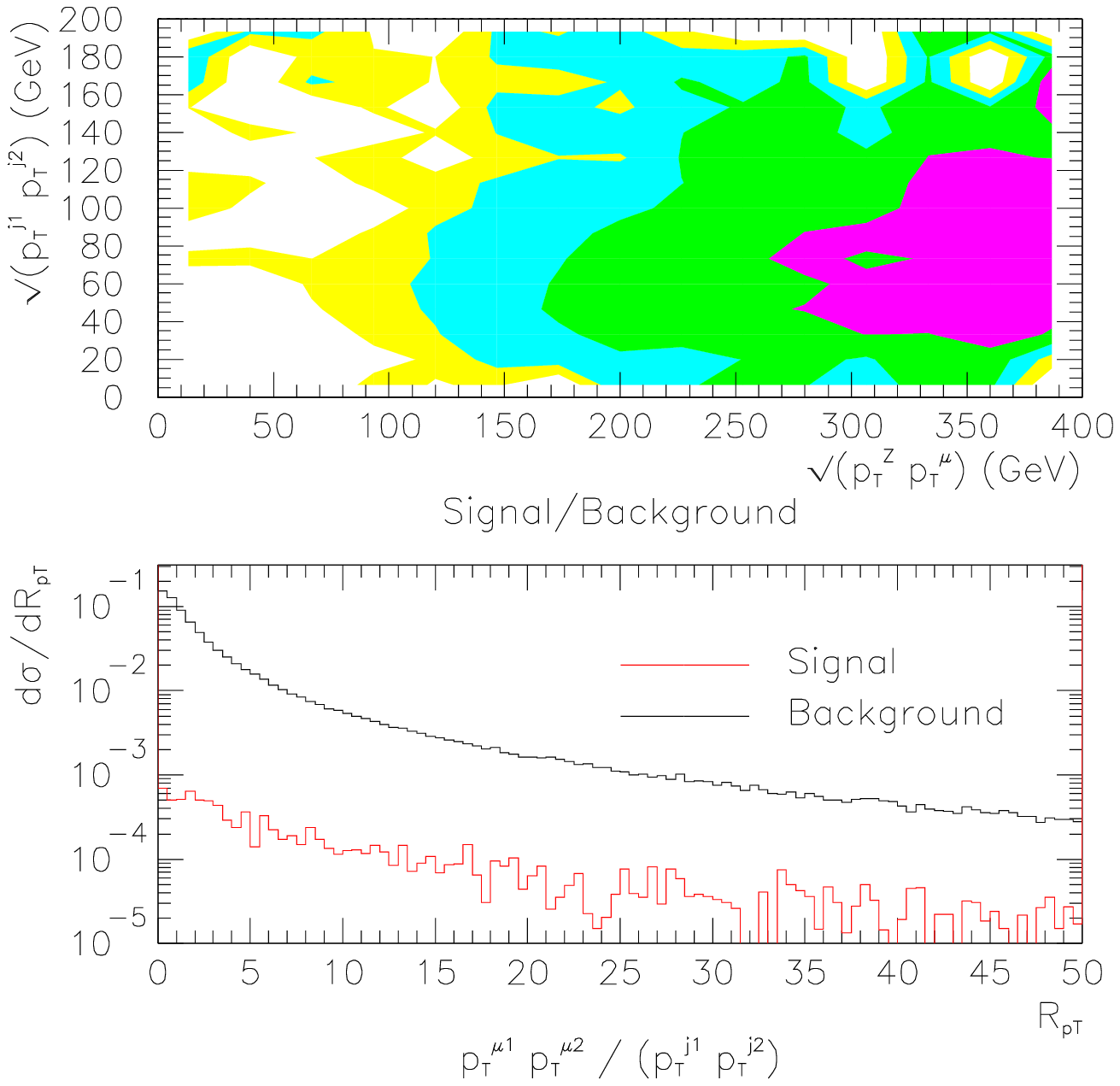,width=0.58\linewidth}
\vspace{5mm}
\caption{The kinematics of longitudinal signal and transverse background
in the full process $pp \rightarrow jjW^+Z \rightarrow jj\mu^+\mu^+\mu^-$ at 14 TeV
after applying basic topological VBF cuts only.
{\bf Left:} the differential cross sections of background and signal
in a two-dimensional space defined by the
square roots of the transverse momentum products,
$\sqrt{p_T^{j1}p_T^{j2}}$ versus
$\sqrt{p_T^{\mu^+}p_T^Z}$ where $p_T^Z$ stands for the total transverse
momentum of the two opposite-sign muons that reproduce the best $Z$ mass.
The color contours are equidistant
and the scale ranges from zero (white) to $2 \cdot 10^{-4}$ fb/GeV$^2$
for the background and to $3 \cdot 10^{-6}$ fb/GeV$^2$ for the signal (purple).
{\bf Upper right:} the signal to background ratio from dividing the two left
plots; the vertical scale is logarithmic for better visualization and
ranges from 0.004 (white) to 4 (purple).  {\bf Lower right:}
the distributions of the ratio $p_T^{\mu^+}p_T^Z/(p_T^{j1}p_T^{j2})$
for signal and background.
Results of a MadGraph simulation, processed by
PYTHIA 6 for 
parton showering, hadronization and jet reconstruction,
and further processed by PGS 4 for the effects of finite resolution in
the measurement of jet and muon $p_T$ in a CMS-like detector.
Signal was calculated by subtracting the SM-based distributions
from the Higgsless-based distributions.
}
\vspace{5mm}
\end{figure}

\begin{figure}[htbp]
\hspace{-5mm}
\epsfig{file=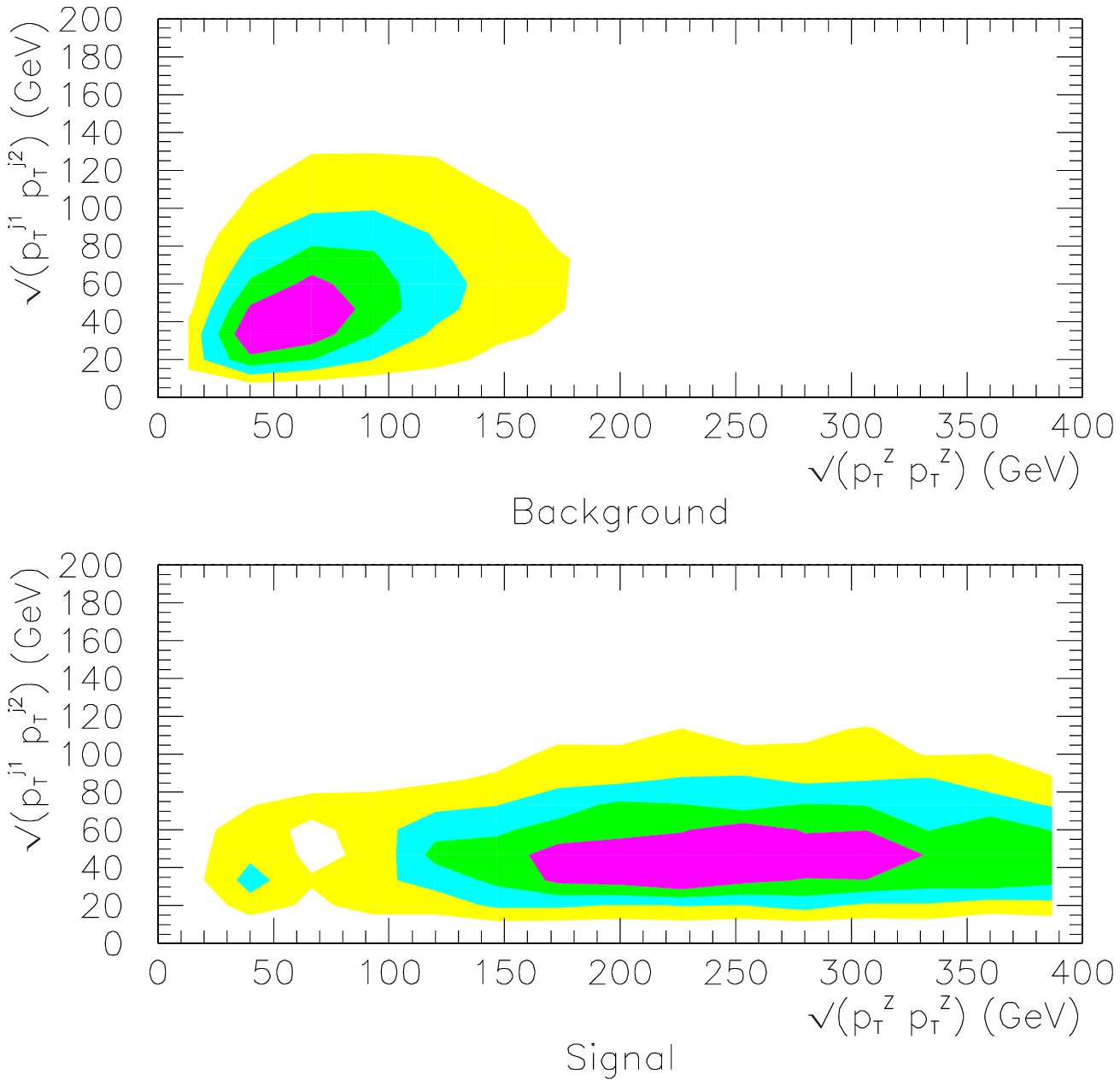,width=0.58\linewidth}
\hspace{-1cm}
\epsfig{file=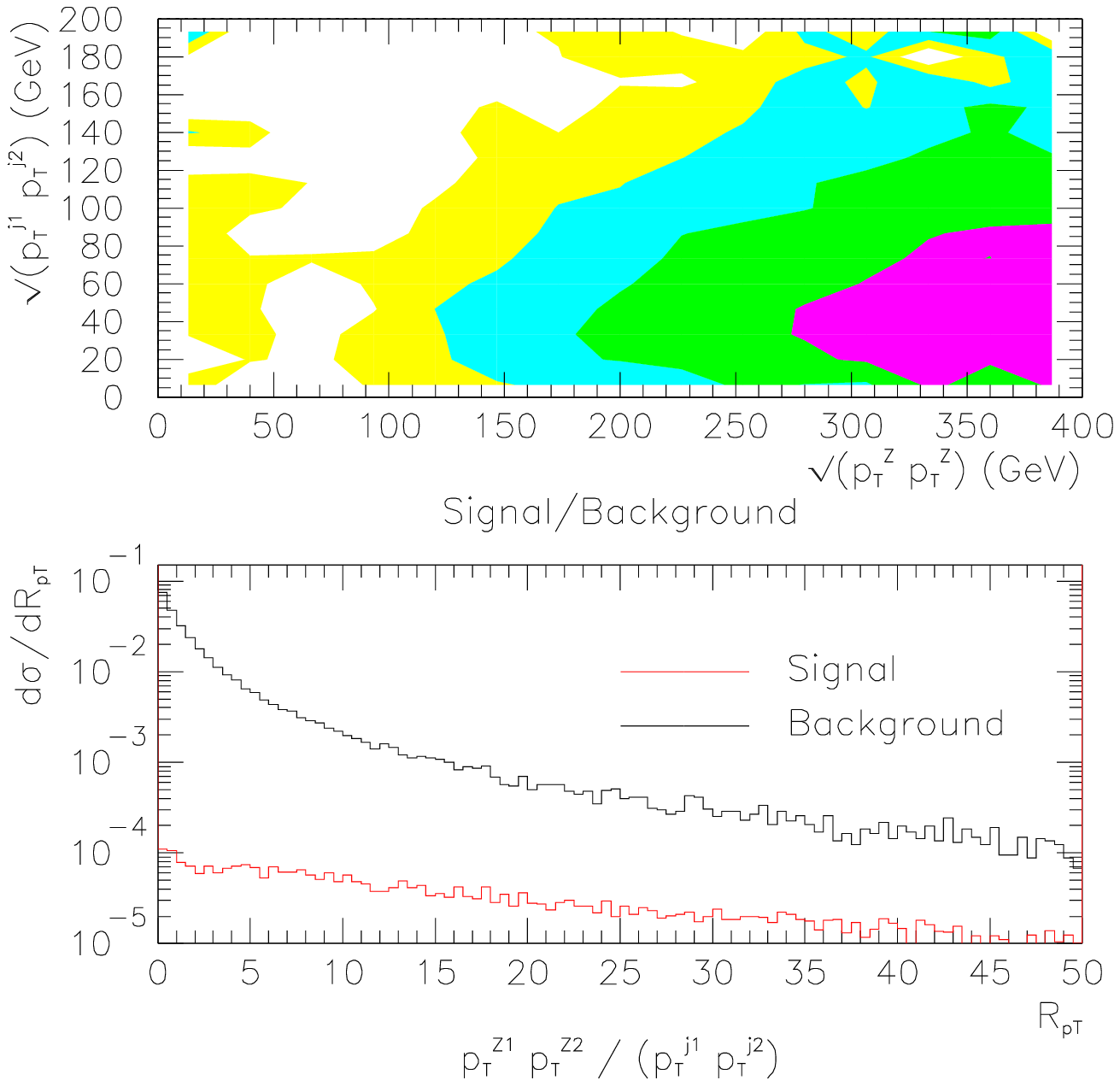,width=0.58\linewidth}
\vspace{5mm}
\caption{The kinematics of longitudinal signal and transverse background
in the full process $pp \rightarrow jjZZ \rightarrow jj\mu^+\mu^-\mu^+\mu^-$ 
at 14 TeV after applying basic topological VBF cuts only.
{\bf Left:} the differential cross sections of background and signal
in a two-dimensional space defined by the
square roots of the transverse momentum products of the two leading
jets and the two $Z$ bosons, $\sqrt{p_T^{j1}p_T^{j2}}$ versus
$\sqrt{p_T^{Z1}p_T^{Z2}}$ where the transverse momenta of the $Z$ bosons
were reconstructed from pairs of opposite-sign muons reproducing the best $Z$ masses.
The color contours are equidistant
and the scale ranges from zero (white) to $3 \cdot 10^{-5}$ fb/GeV$^2$
for the background and to $4 \cdot 10^{-7}$ fb/GeV$^2$ for the signal (purple).
{\bf Upper right:} the signal to background ratio from dividing the two left
plots; the vertical scale is logarithmic for better visualization and
ranges from 0.0008 (white) to 0.8 (purple).  {\bf Lower right:}
the distributions of the ratio $p_T^{Z1}p_T^{Z2}/(p_T^{j1}p_T^{j2})$
for signal and background.
Results of a MadGraph simulation of $pp \rightarrow jjZZ$, processed by
PYTHIA 6 for $Z$ decay into muons, the
effects of parton showering, hadronization and jet reconstruction,
and further processed by PGS 4 for the effects of finite resolution in
the measurement of jet and muon $p_T$ in a CMS-like detector.
Signal was calculated by considering longitudinal $ZZ$ pairs only.
This study did not include the correct, polarization-dependent, angular 
distributions for the decays $Z \rightarrow \mu^+\mu^-$.  Such effects cannot
nonetheless change any of our conclusions.
}
\label{rptzz}
\vspace{5mm}
\end{figure}

To summarize our findings, signal is characterized by emissions of longitudinal
$W$'s followed by their hard interaction.  Signatures of the first part
are two opposite tagging jets at large pseudorapidities and with relatively
{\it low} transverse momenta.  The second part induces a large $WW$ scattering
angle which translates into small $W$ pseudorapidities and large transverse
momenta.  Because their respective kinematics is severely constrained by
the sole physical mechanism and the energies of the colliding quarks,
signal and background events occupy rather restricted and largely separated regions
in the phase space of the four final state particles, and in particular
their transverse momenta.  
The signal cross section
is nearly flat over a large range of the product $\sqrt{p_T^{W_1} \cdot p_T^{W_2}}$
and much more rapidly falling with $p_T^{j_1} \cdot p_T^{j_2}$.
By contrast, the background cross section is much steeper in
$\sqrt{p_T^{W_1} \cdot p_T^{W_2}}$ than $p_T^{j_1} \cdot p_T^{j_2}$.
This is not unexpected, as we recall the expression $\sqrt{p_T^{W_1} \cdot p_T^{W_2}}$
directly correlates with the center of mass
$WW$ energy, $M_{WW}$.
The region of phase space where all four transverse
momenta are relatively low to moderate,
typically $p_T^W \sim $100-200 GeV and $p_T^j \sim $40-80 GeV,
is the region of largest kinematic overlap and thus is the most relevant
for a successful signal isolation.  
In this region, the condition

\begin{equation}
p_T^{W_1} \cdot p_T^{W_2} / (p_T^{j_1} \cdot p_T^{j_2}) = const
\end{equation}

\noindent
to a fair accuracy corresponds to a line of constant signal to background 
ratio (S/B).

As already mentioned when discussing hadronic decays, decays of energetic $W$'s
are highly boosted in the lab and decay products are emitted in a nearly collinear
way.  Our practical measure of $M_{WW}$ in an experiment is then the product of the
two transverse momenta of the visible charged leptons.  Asymptotically for high
energies, the latter is just a numerically scaled down (by a factor of 4)
version of the former.
We have hence arrived in a heuristic way
to the definition of an experimental dimensionless variable

\begin{equation}
R_{p_T} = \frac{p_T^{l_1} \cdot p_T^{l_2}}{p_T^{j_1} \cdot p_T^{j_2}}
\end{equation}

\noindent
whose fixed value indeed respresents a constant S/B to a good enough accuracy.

Correspondence between $R_{p_T}$ and the typical VBF signature is straightforward.
We recall that the latter includes two
central back-to-back leptons (in case of leptonic decays)
with high transverse momenta.  Here however the specific cut value for these
transverse momenta is now scaled with the values of transverse momenta of the
jets, in a way that is based on background rejection grounds.  Physically 
this can be said equivalent to adding the requirement of high $M_{WW}$ and longitudinal
polarizations.  
By contrast, the conventional VBF selection criteria are polarization-blind.
This unique combination of the four transverse momenta is in fact
more effective in separating signal from background than a combination of
selection criteria imposed separately on the individual transverse momenta, because
they scale with each other.

All the above considerations are equally true for $W^-W^-$ as for $W^+W^+$,
although quantitative details differ due to the presence of two valence $u$
quarks inside a proton.  The total production rate of $W^-W^-$ is approximately
one fourth of that of $W^+W^+$.  

By the same arguments it should be clear
that $R_{p_T}$ is a specific variable suited for the study of same-sign $WW$
scattering, but
not of other VBS processes, $W^+W^-$ in particular.  For a comparative study
of $R_{p_T}$ usefulness in different VBS processes, see Figs.~\ref{rptwpwp} 
thru \ref{rptzz}.  The meaning
of $R_{p_T}$ is not selection of a hard scattering process any more than conventional
VBF selections are.  Rather, it is rejection of background of
a specific type: the one related to
gauge boson emissions off two colliding quarks in which at least one of the
bosons is transversely polarized.  The uniqueness of same-sign $WW$ is that it is
the only process in which
this type of background can be made its main component.
This observation is very important.  For the purely leptonic decay modes, the whole
signal size, defined in terms of $W_LW_L$ pairs and the unitarity limit
is of order of 0.3 fb.  This means in any realistic scenario a low number of signal
events to begin with.  
Feasibility of signal detection, assuming luminosities measured in hundreds of inverse
femtobarns, is thus mainly determined by the background rejection potential.

\section{Reducible backgrounds and selected experimental issues}

By reducible background is meant all contributions that can mimick the signal in
a real experiment, but physically come from a different collection of particles in
the final state.  In other words, in an ideal detector the reducible background
could be zero, but is not because of finite detector performance and
event reconstruction capabilities.  Following the standard
background process classification used e.g.~in CMS, the most important
potentially dangerous reducible
background sources in the study of $WW$ scattering at high energies, for the
purely leptonic decay modes, are:
inclusive $t\bar{t}$ production, $W$+jets with a jet misidentified as a lepton and QCD
multijet events with two jets misidentified as leptons.  For the purely
leptonic decay modes, the key detector features
that determine the magnitude of the reducible background are the purities
of lepton reconstruction, including the charge, and to a lesser degree
the efficiency of $b$ quark tagging.

\begin{figure}[htbp]
\vspace{-4cm}
\hspace{-1.5cm}
\epsfig{file=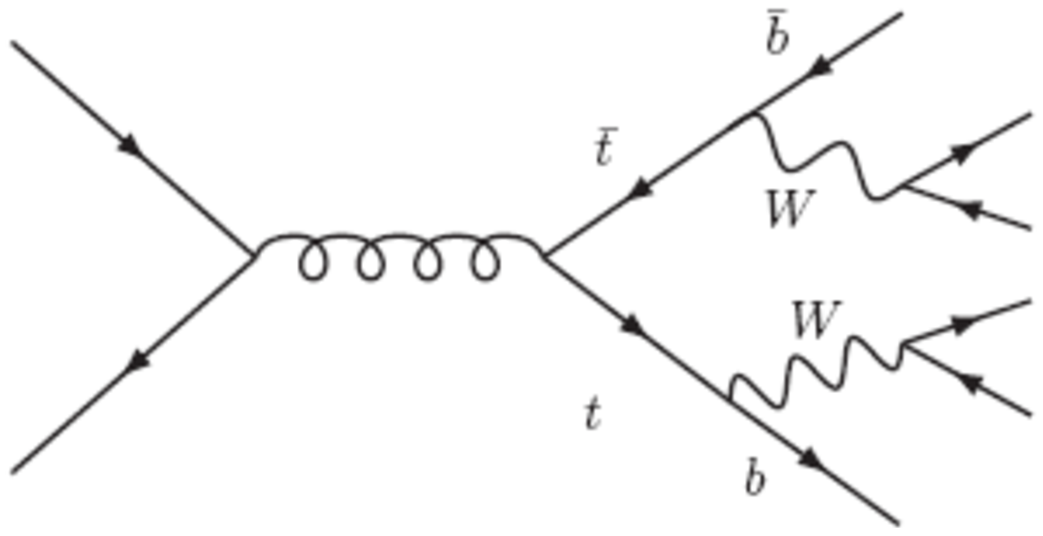,width=0.7\linewidth}\hspace{-4.5cm}
\epsfig{file=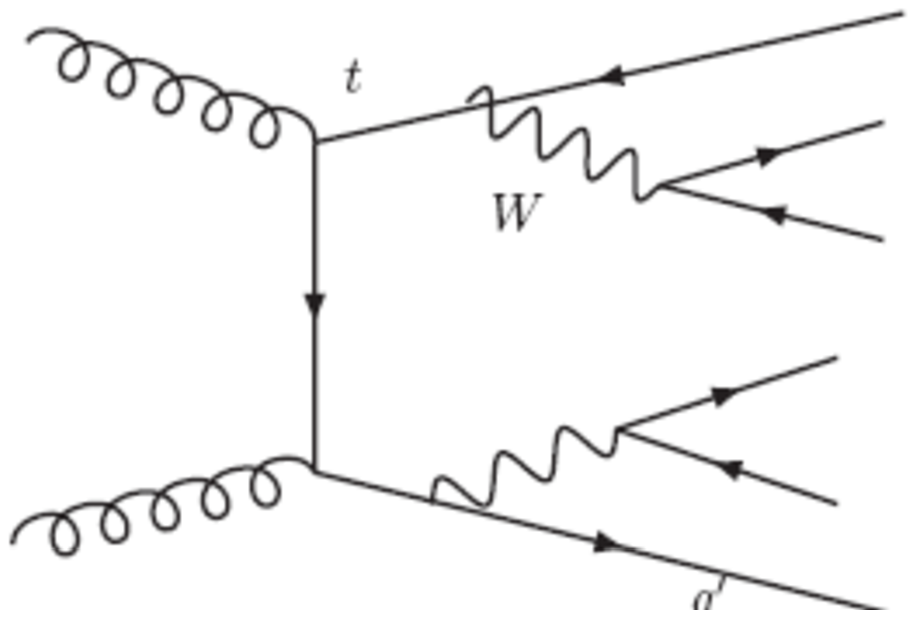,width=0.7\linewidth}
\vspace{-5.5cm}
\caption{Two lowest order graphs for the inclusive $t\bar{t}$ background.}
\end{figure}
\vspace{5mm}

It must be stressed immediately at this point that we are considering the
backgrounds that can be significant at 13/14 TeV after imposing all the 
discussed selecion criteria.  It is an obvious fact that use of looser
criteria, e.g., as a preselection for a multivariate type of analysis (MVA)
will translate into additional and differently composed background to consider.

Top pair production at the LHC overwhelms the $WW$ scattering signal by several
orders of magnitude.  
We have already noted that the same-sign $WW$ scattering mode is advantageous here
in that a $t\bar{t}$ pair produces in principle always an opposite-sign gauge boson pair,
along with two $b$-jets.
However, the initial $t\bar{t}$ production cross section is so much larger than our
potential signal that tiny effects associated to leptonic $b$ decays or a charge
mismeasurement of the lepton arising from $W$ decay, can lead to measurable
effects which cannot be disregarded.
As each top quark decays into a $W$ and a bottom quark,
it is clear that $b$-tagging efficiency plays an important role.  
A typical $b$-tagging algorithm in a collider experiment is based on the
most characteristic feature of $B$ mesons, namely their short lifetimes, identifiable
in the detector as subsequent decays occurring from a vertex which is displaced 
somewhat from the proton-proton interaction point.
Many $b$-tagging
algorithms have been developed and their performance studied in CMS, the most
commonly accepted being the Combined Secondary Vertex (CSV) algorithm \cite{cmsbtag}.
The algorithm relies on the reconstruction of secondary vertices together with
the track-based lifetime information in a jet.  For each track a 3-dimensional
impact parameter is computed from its minimum distance of approach to the vertex,
then tracks in a vertex are ranked based on a significance number equal to the
value of the impact parameter expressed in units of its uncertainty.
Likelihood discriminants to identify the jet as a $b$-quark are based on the
significance of usually the second-ranked (``High Efficiency") or sometimes
the third-ranked (``High Purity") tracks.  The threshold value is, as always,
arbitary and allows to choose an optimum working point for each analysis
based on the general performance
curve that correlates the tagging efficiency with tagging purity, the latter 
determined
in terms of the efficiency for tagging a $u-$, $d-$, $s-$, $c-$ or gluon jet.
Since tightening the tagging criterion quickly leads to an avalanche increase
of light quark mistagging, the final $b$-tagging efficiency is determined
mainly by the maximum acceptable tagging impurity.  From the CSV performance
curves we learn that signal losses can be kept up to or below 2\% overall, while 50\%
of genuine $b$ quarks get tagged.  For a $t\bar{t}$ event, with two $b$ quarks
in it, this means a reduction factor of 0.25.  Useful, but far insufficient
to keep the $t\bar{t}$ background to manageable levels.  Alternatively, a 0.10
reduction factor can be obtained by allowing of a 10\% loss of the signal.
The fact that $b$-tagging efficiencies decrease in the forward/regions regions
is not very disturbing because tag jets usually do not originate from $b$ quarks,
as we will see further on.  Because of steeply
increasing impurity rates, further adjustments of these numbers leave rather little
room for improvement.

\begin{figure}[htbp]
\begin{center}
\epsfig{file=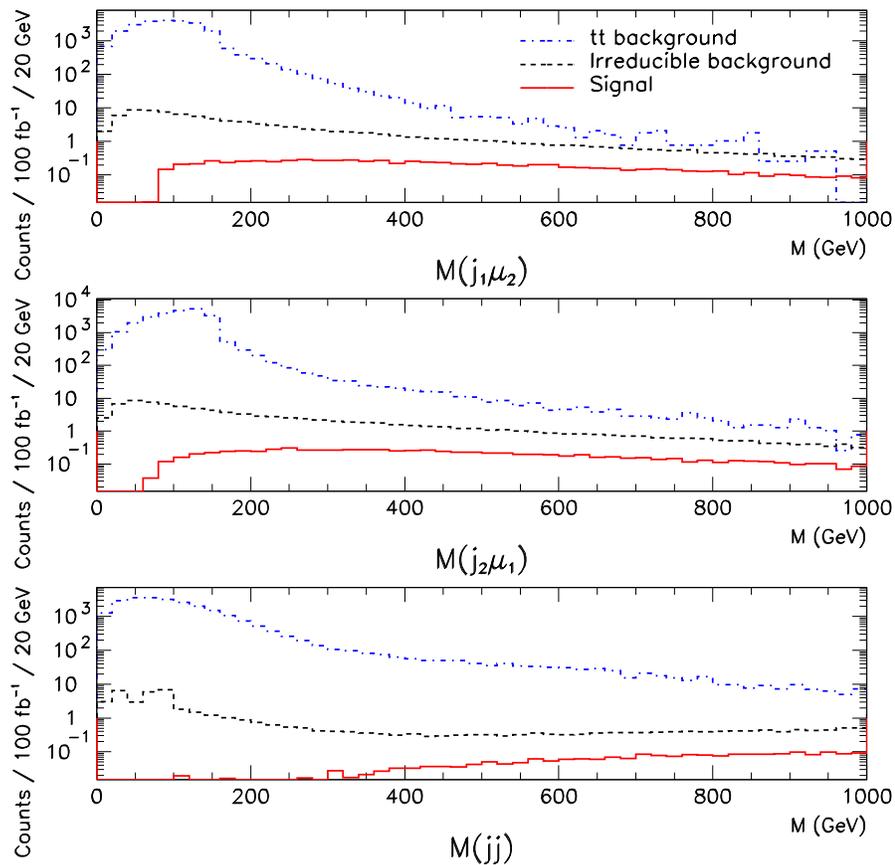,width=0.8\linewidth}
\end{center}
\caption{Distributions of the invariant mass of the two leading jets (top)
and of combinations of jets and leptons (middle, bottom) for the signal,
irreducible background and $t\bar{t}$ background in the $pp \rightarrow jjW^+W^+$
process at 14 TeV,
after applying basic topological VBF cuts.  Results of a MadGraph calculation,
processed by PYTHIA 6 for $W$ decay into muons, the
effects of parton showering, hadronization and jet reconstruction.
The original PYTHIA 6 source code was modified to account for the
correct, polarization-dependent, angular distributions for the decays
$W^\pm \rightarrow \mu^\pm \nu$.
}
\end{figure}
\vspace{5mm}

The bulk of the $t\bar{t}$ background must be in any case eliminated kinematicwise.
The top quark mass defines a natural upper bound for the invariant mass of
its visible decay products, in our case the jet and the lepton.  There is
of course an ambiguity here related to correlating the proper jet with the proper
lepton, overall kinematic constraints however favor a configuration in which
relative ranks (defined by the respective $p_T$ values) of the jets and the
lepton anticorrelate.  Which is to say, more often than not, the largest-$p_T$
lepton with the second-$p_T$ jet and the largest-$p_T$ jet with the second-$p_T$
lepton reproduce the top mass constraint.
Furthermore, as in a typical $t\bar{t}$ production event, the two $b$-jets
do not undergo any hard interactions, the two-jet invariant mass strongly prefers
much lower values than those typical of $WW$ scattering.  It was determined from 
simulation that the combination of simple cuts:

\vspace{3mm}

\hspace{5.5cm} $M_{l_1j_2} > 200$ GeV,

\hspace{5.5cm} $M_{l_2j_1} > 200$ GeV,

\hspace{5.5cm} $M_{jj} > 500$ GeV,

\vspace{3mm}

\noindent
together with $b$-tagging already reduces the $t\bar{t}$ background to manageable
levels as long as it is only driven via effects like charge mismeasurement or
leptonic $B$ decays.  Commonly used for reduction of the $t\bar{t}$ background
is the additional requirement of central jet veto.
Its usual form is removal of events with any additional jets of
$p_T$ larger than a predefined threshold and anywhere between the two tagging
jets in pseudorapidity.  
Because $WW$ scattering is a pure electroweak process,
little jet activity in between the two tagging jets is expected in signal.
Note however that a certain form of the central jet veto is already
applied via the requirement of two tagging jets.  By construction we require them
here to be the two leading jets in the event, hence ``central jet veto" proper
in practice means an additional cut only if at least one of the tagging jets has
$p_T$ below the chosen threshold.
Such cut provides another factor $\sim$0.25 in terms of
$t\bar{t}$ background rejection.

Detector efficiencies in terms of lepton charge determination, especially at
large transverse momenta ($p_T \sim$300 GeV), are relatively
poorly studied.  This is because beam energies of 7 or 8 TeV do not provide
much data in this region and most mainstream physics
analyses are very little sensitive to such effects anyway.  Finally, because
evaluation of potential backgrounds related to charge misidentification is
in practice done using various partly or wholy data-driven methods that do not
require explicit knowledge of the misidentification probability per se.
But the results from 7/8 TeV cannot be directly applied to 13/14 TeV because
the relevant $p_T$ and $\eta$ distributions differ.
Correct charge reconstruction is
generally easier in the central barrel region of the detector, which is
advantageous for us.
A simulation-based study done in CMS, in which the rate of muons
with the reconstructed charge not equal to the generated charge was measured,
revealed charge misreconstruction be at the level of $10^{-3}$ for muon $p_T$
up to 100 GeV and slowly rising above \cite{cmsmusign}.  
An earlier study of cosmic muons passing through
the whole detector in which muon charges were reconstructed separately in the top
and bottom halves and disagreed, revealed this disagreement be already at the 0.5\% level
for $p_T$ of the order of 300 GeV \cite{cmsmusign0}.  
The two results do not disagree badly and
taking into account the improvements in muon reconstruction between the times
of the two analyses, we can safely assume a 99.7\% muon sign matching probability as
fully realistic in our kinematic range of main interest.

The charge misreconstruction for electrons 
is known both from simulations and from data using a Tag-and-Probe technique
to be well below 1\% in the barrel region ($|\eta|<$1.5) for relatively low $p_T$
and gradually increasing with $p_T$ \cite{cmsereco} \cite{atlasereco}.  
The use of independent methods to estimate
the charge from a combination of various data from the central tracker detector and
the electromagnetic calorimeter was shown to significantly reduce the inefficiencies
of the standard Gaussian-Sum Filter (GSF) track curvature method.  The efficiency
still degrades somewhat with $p_T$, but 99\% gives the right order of magnitude
of what can be readily achieved as far as electron sign matching is concerned.
ATLAS also reports \cite{atlasereco}
sub-percent level inefficiencies of electron charge
reconstruction for the barrel once their ``tight identification" criteria are imposed,
while at the same time the efficiency of electron-ID (reconstruction+identification)
in the barrel region is larger than 85\% for $p_T >$50 GeV.

The jet $\rightarrow$ electron or photon $\rightarrow$ electron misidentification rates,
commonly called ``fake rates", are usually determined using data driven methods.
For an analysis of a given final state, a control sample is defined from data
which differs from the signal data in that the quality criteria used to formally define
an object of a certain kind, say, an electron or a muon, were loosened.  The control
sample is known to be composed mainly of ``fakes".  The fake rate in the signal
region is
calculated by scaling the measured background distributions with the measured
probabilities of each ``fake" to pass the nominal quality criteria, usually
as a function of its $p_T$ and $|\eta|$.  
Results of such methods are directly applicable only in the context of
their specific analyses.
On the other hand, generic but simulation-based studies exist in which misidentification
probabilities were measured relative
to any random jet of a given $p_T$ and $|\eta|$.  The results naturally depend
on the details of the electron selection criteria applied.  
In CMS, simulation work has shown \cite{cmsemisid}
that a combination of stringent electron identification criteria based on:

\begin{itemize}

\item track, electromagnetic and hadronic isolation, each defined as the $p_T$/$E_T$ sum
of all tracks/clusters lying within a $\Delta R$=0.5 cone around the reconstructed
electron, relative to the $E_T$ of the electron,

\item the geometrical matching of the track with the cluster in both pseudorapidity
and the azimuthal angle,

\item the ratio of the electromagnetic energy deposit to the electron momentum, $E/p$,
or alternatively, $|1/E - 1/p|$,

\item the electromagnetic cluster shape described in terms of the ratio of energy
deposits within a 3$\times$3 and 5$\times$5 cell collection centered around the
seed of the cluster,

\end{itemize}

\noindent
makes the probability that a jet gets reconstructed as an electron possible
to reduce to $(1.1 \pm 0.2) \cdot 10^{-4}$ on average and somewhat increasing
with $E_T$, as far as can be judged from a rather low statistics.
Moreover, due to the mechanisms of $W$+jets production at the LHC, in these events
only 27\% of those
``fake electrons" have the same sign as the $W$.  This translates into $W$+jets
background rates being nearly 3 times lower in $W^\pm W^\pm$ than in $W^+W^-$.
The same study suggests that the electromagnetic cluster shape
described in terms of the pseudorapidity spread of the shower
in 5$\times$5 cells around the seed has also a large discriminating power on top
of the other criteria and can reduce the fake rate by a further half.
The overall electron reconstruction efficiency using the identification criteria
that were ultimately applied in this study was 74\% overall, but larger than
80\% for $E_T >$ 50 GeV and larger than 90\% for $E_T >$ 100 GeV, which is of main
interest for us.
Given the large total $W$+jets and QCD multijet events cross sections,
keeping the fake rates low is imperative for the electron decay channels and so
even allowing a slight decrease in the reconstruction efficiency is the better choice.
Similarly, fake rates of photons misreconstructed as electrons were
determined from simulation to be (0.7 $\pm$ 0.1)\%, with an additional factor 2
reduction accounting for a particular choice of sign.
It is worth noting that all the abovementioned variables, and a few additional ones,
are used as discriminators in the standard electron-ID used in CMS analyses.
Based on early simulation studies,
ATLAS reported jet-to-electron fake rates of the order of $2 \cdot 10^{-4}$
for jet $p_T >$100 GeV, with the Boosted Decision Tree techniques used for electron
identification and isolation, while keeping high electron-ID
efficiency in this kinematic region \cite{atlasemisid1}.  
Since this study was based on a simulated dijet sample, it is not
possible to derive the charge correlation factor.  Another study 
\cite{atlasemisid2} reports
on the possibility of a further reduction down to the level of $\sim 10^{-5}$ 
at the expense of electron-ID
efficiency decreasing to 67\%.  Unfortunately both numbers are $p_T$-averaged.

In view of everything above, we can tentatively assume for further considerations
an average electron fake rate in our kinematic domain
of $\sim 10^{-4}$, times the appropriate sign factor, with a 90\% electron-ID
efficiency and a 99\% charge reconstruction efficiency.
However, one cannot completely trust Monte Carlo programs to study effects
related, e.g., to jet fragmentation at the LHC.  Only real data in the appropriate
kinematic domain will ultimately determine the impurities.
Further improvements in electron purity and sign matching, keeping a
reasonably high overall reconstruction and identification efficiency, must be
particularly encouraged and
followed with a special attention, since they can be the key for success in including
the electron-electron and mixed muon-electron decay channels to the $WW$ scattering
search at 13 TeV and can prove vital for the observation of signal.

Fake rates of hadrons misreconstructed as muons in principle include two distinct
effects.  The first of them are punch-thru pions which reach the
muon chambers.  They are usually associated with hadronic activity around the
fake muon track.  
The second class are real muons from pions or kaons decaying in the detector,
often referred to as ``non-prompt" muons.
These are recognizable by a characteristic kink in the track, visible at the point
of the pion or kaon decay.  
By choice of appropriate isolation criteria both effects can be suppressed to
a negligible level.  Measurements and simulations done within ATLAS \cite{atlasmumisid}
reveal a total fake rate for jets
of less than $\sim 10^{-5}$ and dropping with jet $p_T$, and about $10^{-3}$
for single tracks.  The latter however may be contaminated with real ``prompt" muons
from $W$ and $Z$ decays.  In our further considerations we will disregard these
backgrounds.

\begin{figure}[htbp]
\begin{center}
\epsfig{file=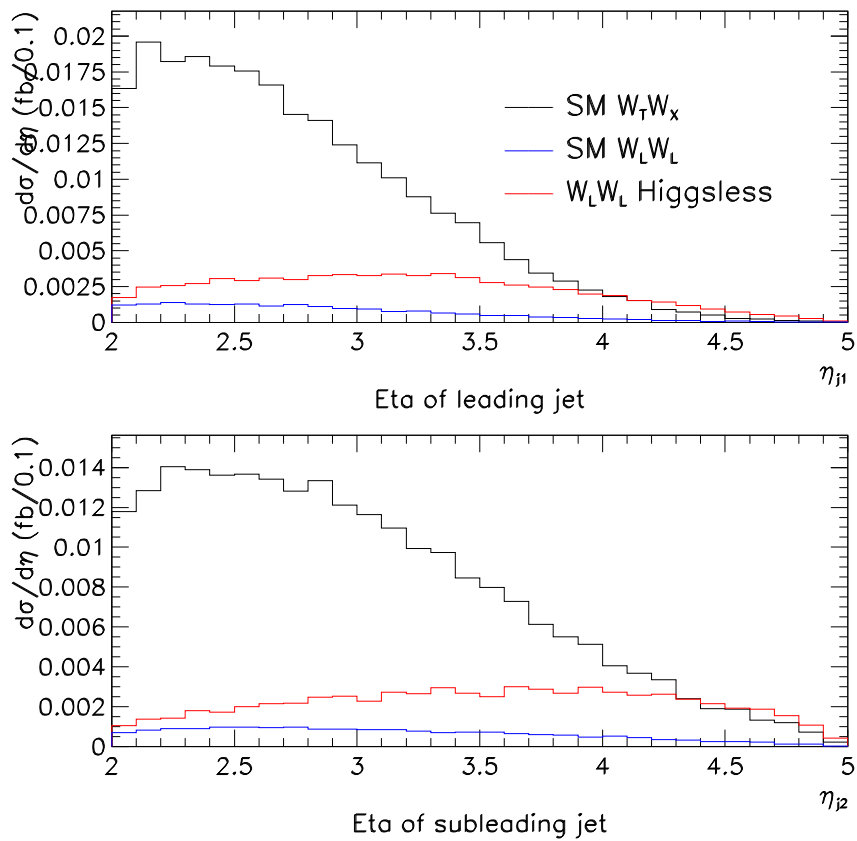,width=0.8\linewidth}
\end{center}
\caption{Pseudorapidity distributions for the leading (top)
and subleading (bottom) tagging jets in the
$pp \rightarrow jjW^+W^+$ process at 14 TeV, with leptonic $W^+$ decay,
after applying basic topological VBF cuts, namely
$\Delta\eta_{jj} > 4$ and $|\eta_l|<2.1$.  Shown are the SM spectra for
$W_LW_L$ and $W_TW_X$ pairs and the $W_LW_L$ spectra for the Higgsless signal.
Results of a MadGraph calculation,
processed by PYTHIA 6 for $W$ decay into muons, the
effects of parton showering, hadronization and jet reconstruction.
The original PYTHIA 6 source code was modified to account for the
correct, polarization-dependent, angular distributions for the decays
$W^\pm \rightarrow \mu^\pm \nu$.
}
\end{figure}

One last experimental issue that is definitely worth to mention at the present
moment, particularly in the context of the planned future upgrades of LHC
detectors, concerns jet reconstruction efficiency at large pseudorapidity.
In Standard Model VBF processes at 14 TeV, pseudorapidity distributions for tagging
jets peak between 2-3.  This holds approximately equally true for $V_TV_X$ as for 
$V_LV_L$ pairs.
This however does not imply that this region is of most interest for BSM
search.  Contrary, the best sensitivity to BSM effects is likely to be more forward.
E.g., non-SM Higgs couplings will reflect in a wide pseudorapidity range for the
tagging jets, going even all the way up to 5 for the subleading jet.
It is therefore important to have a good jet reconstruction in the entire 
pseudorapidity range
and keep high performance in the most forward region for the whole High Luminosity
LHC program.

\chapter{Simulation-based studies vs.~experimental results}

The processes of $VV$ scattering have been lying in the interest of physicists
for almost as long as the Standard Model itself.
Despite of there being many simulation-based analyses of $VV$ scattering at the LHC
with 14 TeV, both at a phenomenological level or involving elements of a full
experiment-specific
detector simulation, a vast majority of them need critical revisiting in accordance
to recent
developments in our experimental knowledge and in the available simulation tools.

To begin with, in most older studies it was Higgs boson existence that was 
considered the biggest unknown of the model.  Consequently, signal was calculated
either in terms of a pure Higgsless Standard Model, or a Higgsless Standard Model
where only the unitarity of scattering amplitudes was enforced by hand, or a
Standard Model with a very heavy Higgs,
or finally within the framework of a
particular alternative model of electroweak symmetry breaking.
Physicswise all these scenarios are now obsolete.
This however does not imply that older studies should be sent to oblivion.
It is rather straightforward to reinterpret the results of the former three
classes of works in terms of a 125 GeV Higgs with different $HWW$ or $HZZ$ couplings.
For that the relevant signal figures should to a good approximation
only be scaled down by calculable,
coupling-dependent factors.  In a similar manner it has been shown that in the regime
of the LHC at 14 TeV, the sole unitarity bound produces a 20-25\% reduction of
the signal figures compared to a pure Higgsless Standard Model.  Moreover,
in a non-resonant process like $W^\pm W^\pm$ this approach is in fact
approximately equivalent to assuming a heavy Higgs (with $M_H \approx$ 1.2 TeV),
the one notable difference being that the latter must then have a very large width,
which affects the scattered $W^\pm W^\pm$ mass spectrum in a non-trivial way.
More problematic is only reinterpretation of results that were obtained by assuming
particular
alternative models of electroweak symmetry breaking, but even there some of
the analysis methods that have been worked out may remain useful today.

Equally problematic are the old estimates of reducible backgrounds.  In phenomenological
studies such backgrounds are, more often than not, treated either qualitatively, e.g.,
by suggesting certain cuts to suppress them, or considered only partially.  
Optimistically, its supposedly largest
component was studied in more detail (typically, inclusive $t\bar{t}$ production).

Results based of full detector simulations for a specific experiment often differed
widely from results
of purely phenomenological analyses, especially in the semileptonic decay channels.
The former have been evaluated using detector-specific simulation
tools available at the time of their publication, which is, in the early stages of
software development for ATLAS or CMS, to focus on these two.
It is obvious that these tools have since improved paramountly.
But the improvements are usually difficult to quantify without redoing the whole
simulation.
Unfortunately this means that these older studies that involved full detector
simulation usually do not represent a valid reference
to assess the best current experimental sensitivity in the search for physics
beyond the Standard Model in $WW$ scattering.

First, however, let us briefly recall and review the leading past works in the subject,
focusing not that much on their numerical results, but rather
with a special emphasis on what things of all those older studies
remain completely valid today.

\section{Early calculations}

Physicists' interest in $VV$ scattering clearly predates the LHC.
Already in the early papers of Chanowitz et al.~\cite{chanowitz1}
it was noticed that scattering of
same-sign longitudinally polarized $W$'s is the most sensitive probe of
effects related to the mechanism of electroweak symmetry breaking.
Several other authors, including Barger et al.~\cite{bargerhan}
and independently Dicus et al.~\cite{dicus} studied
in detail the process of $W^+W^+$ scattering in the context of the planned 
Superconducting Super Collider (SSC).
Their numerical results do not
have a direct importance for us, but some of their qualitative observations are
strikingly up to date.  Among other things, they proposed kinematic cuts to keep the
$t\bar{t}$ background under
control and stressed the importance of jet transverse momenta in the
separation of the longitudinal $WW$ signal from the transverse $WW$ background.
In particular, to tackle the latter, cuts on the {\it maximum} allowed jet $p_T$
were discussed.
Early studies exist also for electron-positron colliders \cite{herrero}.

Systematic studies of the $WW$ scattering phenomenology in the particular
context of the LHC started later in
the 1990's.  Their main physical focus was observation of signal related
to different scenarios of electroweak symmetry breaking on the assumed absence of a
Higgs boson.  Despite their main underlying physics assumptions are now implausible,
a large amount of knowledge is still contained in these studies and a lot of
this knowledge remains valid in the context, e.g., of searches for
new heavy resonances
or other experimental signatures in the absence of such resonances.
To a large extent we can still follow the general guidelines presented in
those papers.

The early paper of Barger et al.~\cite{bargerhan}
provided the justification on theoretical grounds
of some basic signal selection criteria
in the context of heavy Higgs searches at the LHC.  It is here that introduced and
justified on theoretical grounds was the idea of a central jet veto
as a primary criterion to distinguish QCD-related
backgrounds from the purely electroweak signal.

Especially enlightening from the phenomenological point of view, although again
focused on various Higgsless scenarios, are the works
of Bagger et al.~\cite{bagger}.  They developed the general methodology, introduced the
``subtraction method" for the mathematical definition of the signal and recapitulated
on the basic experimental
signatures.  They proposed original sets of kinematic cuts optimized for all the 
individual VBS processes
separately and finally, they showed a comparative study of possible signals and
backgrounds 
(not only irreducible) after each step of the full event selection.
They also noticed the experimental advantages of
purely leptonic $W$ and $Z$ decay modes in their analyses and termed them
as ``gold-plated".
Let us recall some of their main conclusions that are still valid today.  
From the different analyses by Bagger et al.~it follows that depending on the
actual physics scenario, any of the different scattering processes: $W^+W^-$,
$W^\pm W^\pm$, $W^\pm Z$ or $ZZ$, may turn out to be the most promising one,
or even a combination of all of them could be required.  Models which
predict heavy scalar resonances were found most easy to study in the $ZZ$ and $W^+W^-$
processes (in agreement with everything we have said so far),
heavy vector resonances should show up more efficiently in the $WZ$ process, while
very heavy resonances or scenarios with no such resonances at all would manifest only as
an increase of the total event yield at large invariant mass and this increase is the
most pronounced in no else than $W^\pm W^\pm$.  The significance of the $ZZ$ channel
is driven mainly
by the $l^+l^-\nu\nu$ final state rather than the cleaner, but lower rate $4l$
final state.
The former is nonetheless contaminated by a detector dependent background coming from
$Z$ + QCD jets events, which have not been explicitly taken into account in this
analysis, except from assuming it be suppressable by applying a cut on missing
transverse energy (MET).
The significance of the $W^+W^-$ process in its turn crucially depends on the
efficiency of suppressing the overwhelming inclusive $t\bar{t}$ background using
such techniques as $b$-tagging and central jet vetoing.  
In these works, the respective signals were calculated using the Effective $W$
Approximation and the Equivalence Theorem, as well as assuming
particular scenarios of electroweak symmetry breaking, alternative to the Higgs model.  
All the analyses were carried at a purely partonic level.
Quantitative estimates
of the required luminosity to observe a non-SM signal in the different channels
vary from below 100 to 250/fb
for the LHC running at 14 TeV, but because of the approximative character
of the relevant calculations they should be taken with care.
All the signal scenarios involve strong $WW$ scattering, in which they resemble
a Higgsless Standard Model with reinforced unitarity.
Even though the authors perform an essentially counting experiment, with an analysis
which is not optimized for resonance search (in most cases the resonance is
very broad anyway),
differences of more than a factor 3 in the required luminosities give a rough
idea of the degree of model dependence of the signal significances and hence
of all their quoted results.  The process
$ZZ \rightarrow 4l$ is perhaps the most interesting both because of its low
background and because it offers the best event reconstruction and hence full
determination of the nature of the heavy resonance, but it also requires the largest
luminosity for observation.  It was
estimated to be around 300/fb of LHC running at 14 TeV to
observe a 99\% CL signal, which corresponds to roughly 4$\sigma$.
The sensitivity of the $W^\pm Z$ process to non-SM physics was shown rather marginal.
The authors suggest in fact focusing on the Drell-Yan process to enhance the
significance of $W^\pm Z$ in the search for heavy vector resonances.
Meanwhile, the $W^\pm W^\pm$ process fares poorly in scenarios with heavy scalars, but
somewhat surprisingly turns to be the most efficient in scenarios with heavy vectors,
in addition to non-resonant ones.
Typical luminosities required to observe signal at a 99\% CL
oscillate roughly around 200/fb.
Of other interesting observations that are worth recalling, the authors stress
that signal is contained mainly at relatively low $p_T$ of the tagging jets,
typically $p_T \sim M_W/2$.  Large jet $p_T$ thresholds usually applied in
various analyses of LHC data because of
pile-up related background would therefore translate
into low signal detection efficiency.  As a possible alternative,
single jet-tagging was proposed, which of course would come at the expense of
background rejection efficiency.  However, good signal efficiency could be achieved
with double-jet tagging if only the jet $p_T$ threshold could be lowered to 15 GeV.

Unfortunately, the studies by Bagger et al.~came too much ahead of their time
and many of their important conclusions, perhaps because of the obsolete by now
computational techniques they applied, got largely forgotten before the LHC
started operation.  It is time now to rediscover the findings of this work
and reevaluate them with modern and fully up to date simulation tools.

Other classic works include the ones by
Dobado et al.~\cite{dobado}, focused specifically on $ZZ$ and $WZ$ production.  Their
studies were carried within the Electroweak Chiral Lagrangian approach.
They mainly elaborated on a unitarization technique
based on the Inverse Amplitude Method, in which new dynamic resonances appear
in $VV$ scattering and enforce unitarization.
They also applied the Equivalence Theorem and used the Effective $W$ Approximation 
in their calculations of the relevant
VBS processes.  A tentative analysis was
presented at the level of undecayed gauge
bosons, by further assuming a 100\% efficiency in their reconstruction, and moreover
only irreducible backgrounds were taken into account.  Since both their signal
and background treatments are highly approximative, their numerical results cannot be
considered but purely qualitative.  In what's important for us, however, they do 
confirm the importance
of Drell-Yan production rather than VBS for the $WZ$ process.

Chanowitz et al.~\cite{chanowitz}
in a series of follow-up papers focused on $W^+W^+$ and $WZ$ processes.
The authors observed a complementarity of the $W^+W^+$ and $WZ$ processes
as a function of the mass scale of the hypothetical new, heavy vector resonances.
The combination of the two was shown to guarantee the ``no-lose theorem", meaning
that signal would be always observable one way or another, i.e., in at least
one of the two processes.  
For the calculation of the $WZ$ signals they considered Drell-Yan as well as VBS.
Here too calculations were based on the Chiral Lagrangian Model and,
as in the previous analyses, the Equivalence Theorem and the Effective $W$
Approximation were used to evaluate VBS processes.  Only irreducible backgrounds
were explicitly considered.  However, their work was
the first to mention the potential
importance of detector dependent backgrounds related to lepton sign mismeasurement.
They also followed up on the issue of separating the final state polarizations, but
focused on purely leptonic cuts for this purpose.  The reason was simple: specific
of their analysis was the treatment of VBS and Drell-Yan together.  They omit
some typical VBS cuts, like forward jet tagging, which would kill their
Drell-Yan signal.  Under these conditions, they finally found
an LHC luminosity of 140/fb guarantee the ``no-lose" condition
with a significance of at least 3$\sigma$, which perphas does not have a
direct meaning for us.

In a ground breaking paper, Butterworth et al.~\cite{butterworth}
found that semi-leptonic decay
modes could be as promising as purely leptonic.  They considered only the $W^+W^-$
scattering process (note however than in semi-leptonic decays one of the $W$'s has no
measured charge, so in reality a sum of $W^+W^-$ and $W^+W^+$ is automatically
implied) and calculated the signal in several models within the Electroweak
Chiral Lagrangian approach, that corresponded to the existence of heavy scalar or
heavy vector resonances, as well as no resonances at all.  The calculation was done
using a modified version of the PYTHIA generator which indirectly involves the
Effective $W$ Approximation.
Background evaluation included $t\bar{t}$ production and radiative $W$+jets events,
calculated likewise within PYTHIA.  The authors concluded that signal could be
measured after 100/fb of LHC data, but stressed that the final word would only
come from real measurements of the reducible backgrounds.  Indeed, it was later
shown that uncertainties related to background modelling
in the semi-leptonic decay modes were tantalizing.
Although this was probably the first simulation-based phenomenological analysis which
involved a toy jet reconstruction procedure and envisaged the use of jet
substructure to tackle the problem of jet merging from highly boosted $W$ bosons
decaying hadronically, the real performance of this procedure may depend
on additional effects, e.g., detector resolution, not studied in this analysis.
Using a newer PYTHIA version with
improved parton showering and a dedicated event reconstruction software used
by CMS at the time, it was found \cite{zych} that these predictions were
way too optimistic.  It also indicated
that a lot of work was still required on the detector and reconstruction side.

Some of the many other studies of the phenomenology of $VV$ scattering
before Higgs discovery are listed under Ref.~\cite{others}.

A lot of early simulation work, that in addition included simulated detector response
and event reconstruction, was done within the ATLAS collaboration \cite{atlas}.
In these studies, signals were calculated using PYTHIA
and the various backgrounds using such generators as MadGraph and MC@NLO.
The main focus was $WZ$ in different semi-leptonic and purely leptonic decay
modes and $WW$ in the semi-leptonic decay mode in which case the signal and
backgrounds were
evaluated together with $WZ$, where $W \rightarrow l\nu$ and $Z \rightarrow jj$
(in a real experiment, processes tend to be naturally grouped by final state
as seen in the detector).
They also produced a result for $ZZ \rightarrow l^+l^-\nu\nu$, but only in the
scenario of a Higgs-like resonance with a mass of 500 GeV.
Unfortunately, no results from purely leptonic decays of $WW$ have been shown and
no $W^\pm W^\pm$ in particular.  Their analyses included standard VBS selection
criteria, not specifically optimized for gauge boson polarization.  However,
their $p_T$ threshold for the identification of tag jets vary between 10 and 20 GeV,
ensuring reasonably high acceptance for longitudinal bosons.
The study focused mainly on Higgsless, resonant scenarios, many of the
considered resonances were relatively light and so the results must be regarded
as out of date today.
Interestingly, the only non-resonant scenario considered in this study (there is 
however no information about the exact parameter values used within PYTHIA to
simulate this kind of signal)
did not lead to promising results and was not even included in the table of results
that concluded the study.

\section{Recent works and post-Higgs discovery developments}

A new generation of $VV$ scattering studies commenced with the introduction of
universal, commonly accessible physics calculation tools, like MadGraph, CompHEP, ALPGEN,
PHASE/PHANTOM or VBFNLO, which calculate the full matrix elements for a given
process.  They replaced PYTHIA-based and other signal calculations done only in the
Effective $W$ Approximation.  At the same time, since those generators often did not
have any alternative models of electroweak symmetry breaking explicitly
implemented, signal calculations necessarily required a more generic, model-indpendent
approach.  Such approach was in pratice provided by considering a pure Higgsless
Standard Model, or a Higgsless Standard Model with effective unitarization.  This
could be implemented either as a sharp cutoff or else assuming that
the relevant scattering amplitudes saturate just before reaching the unitarity limit.
Incidentally, the latter is phenomenologically similar to non-resonant Higgsless
models.  Therefore many of these newer studies are directly relevant for the case of
a light Higgs boson with modidied couplings and
no new resonances within the mass range of the LHC.  They require in principle
only a scale factor for an effective translation into a physically valid and up
to date scenario.

One of the earliest $WW$ scattering analyses that did not make use of approximative
computation
techniques was the work by Eboli et al.~\cite{eboli}, 
in which purely leptonic decays were studied
in the $WW$ process and in all charge combinations.  They find a full calculation
of the scattering amplitudes necessary not only for a correct cross section
evaluation, but also to describe accurately all correlations between final state
particles.  A notable feauture of the presented analysis was the most complete
available treatment of inclusive $t\bar{t}$ background, it included contributions
from processes with up to two associated QCD jets computed at the matrix element
level.  The inadequacy of considering only pure $t\bar{t}$ production in the
lowest order was shown.  Conceptually the work was focused on a study
of anomalous quartic vector boson couplings in which two exclusive working hypotheses
were considered in what regards Higgs existence.  For a discussion of the main
concept and of the obtained results, we will still come back to this work
in the next section.

Of the newer analyses at the phenomenological level, the works by Ballestrero 
et al.~\cite{ballestrero}
clearly stand out and they also effectively triggered a lot of further,
detector level work within the CMS collaboration.  In 
a series of papers they studied both the semi-leptonic and purely leptonic
decay channels.  Calculations were done with the newly created PHANTOM program 
\cite{phantom} which
computes complete tree level matrix element amplitudes for $2 \rightarrow 6$ fermionic
processes to
the orders $\mathscr{O}(\alpha^6)$, $(\alpha^4\alpha_S^2)$ and $(\alpha^2\alpha_S^4)$,
wherever appropriate.  All the analyses were carried in a manner which closely
resembles realistic experimental analyses.  Scattering processes were grouped
by final state.
Signal was defined in terms of a VBF-like kinematics in the purely electroweak
process where the final event yields were compared in the Higgsless and
light Higgs cases.  The fact that they typically assume $M_H = 200 GeV$ 
for the Standard Model case is a rather minor issue.
Additionally, processes $\mathscr{O}(\alpha^4\alpha_S^2)$ and 
$(\alpha^2\alpha_S^4)$ accounted for all the extra, non-scattering, background.
In principle, the scope of background processes that can be taken into account in
this way includes the most basic $t\bar{t}$ production process without additional
quarks or gluons.  But as we already mentioned as will further see in the next
chapter, such treatment is
insufficient for an accurate account of inclusive $t\bar{t}$ background for VBF
processes, since the bulk of events that can survive VBF cuts comes in fact
from higher order diagrams.  The applied
selection criteria selected general VBF events and were not optimized for
longitudinal $W/Z$ polarization.  Moreover, a high $p_T$ threshold for tag
jets was used, as had become already routine e.g.~in Higgs searches.
All the analyses were carried at the partonic level and background treatment,
as mentioned above, in practice included only irreducible backgrounds.
The final sensitivity was evaluated from event counting or from a shape analysis of
the invariant mass
spectrum of the visibile gauge boson decay products.
Not surprisingly, the most interesting results came from the purely leptonic states.
Again here the same-sign dilepton channel ($W^\pm W^\pm$) was shown to provide the best
discrimination between different scenarios, closely followed by
$ZZ \rightarrow l^+l^-\nu\nu$ and $W^+W^-$.  Obviously, quantitative comparisons
are subject to further change once all reducible backgrounds are properly included.
The analysis of semi-leptonic decays can only be treated as a demonstration
because the jet merging issue which affects the decays of highly boosted gauge bosons
in a real detector was
not addressed in this study and because final state radiation, leading to
additional jet combinatorics, was not simulated.
Perhaps the most interesting result of Ballestrero et al.~from our (biased)
perspective resides in that their works were among the first
ones to explicitly consider the Strongly Interacting Light Higgs models as
an alternative to either Higgsless or the Standard Model.
Their results suggested a decrease of the signal size (redefined here for our purposes
as the enhancement with respect to the Standard Model) by a factor 3-4
when compared to a pure Higgsless scenario.  Since effectively the only relevant feature
of the considered SILH scenario was a modification of the $HWW$ and $HZZ$
couplings by a factor numerically close to 0.7, compared to the SM,
their result presented in this way can be treated as model independent.
It is in fact a particular example of a scale factor that is necessary to apply
to all the former Higgsless-based studies to render them fully realistic.

The CMS collaboration produced a full set of results, corresponding to the many
different final states, obtained using the general
prescription of Ballestrero et al., with the addition of the dedicated CMS event
reconstruction software \cite{torino}.  
The final states that were considered corresponded
both to semi-leptonic and purely leptonic decay modes of $W^+W^-$, $W^\pm Z$, $ZZ$ and
$W^\pm W^\pm$.  The results were admittedly not very encouraging.  However, e.g.,
the analysis of the same-sign channel was clearly suboptimal.
Moreover, as already stressed, the work was done at the time of rapidly changing
CMS reconstruction and analysis tools and cannot be taken as the final word.
Very derisable would be to have these data reanalyzed with
the most recent versions of the CMS software and using the most efficient
selection criteria for a complete and up to date evaluation.

In another analysis done at the phenomenological level,
Zeppenfeld et al.~\cite{zeppenfeld}
studied leptonic decays of the $W^+W^-$, $ZZ$ and $W^\pm Z$ 
scattering pairs.  They calculated signal and backgrounds using the VBFNLO
generator program, where signal was defined in terms of a 1 TeV Standard Model
Higgs or alternatively via a Warped Higgsless model with heavy vector resonances.
Their most important conceptual innovation from today's perspective was that their
background treament included realistic modelling of
$t\bar{t}$+jets in addition to irreducible backgrounds.  For the former
they developed an original simulation-based approach which is
suitable for VBF analyses.  Initial and final state radiation processes were
simulated and double counting was avoided by defining mutually exclusive
topological requirements for processes with 0, 1 and 2 QCD jets generated at the
matrix element level.
Consequently, they found $t\bar{t}$+jets the most
important remaining background in the $W^+W^-$ channel, in contrast to what
was assumed in many other studies.  
We will review their method in further detail in the next chapter.
A toy jet reconstruction by recombination of the final state partons was also applied.
Quite consistently with most
previous studies, they found $W^\pm Z$ the preferred channel for the vector
resonance scenario and $W^+W^-$ closely followed by $ZZ \rightarrow l^+l^-\nu\nu$
for the heavy Higgs scenario.  They finally found a very high signal significance
after collecting 300/fb of data at 14 TeV within the considered scenarios.
This analysis also lacks separate consideration of the $W^\pm W^\pm$ process.

The importance of the $W^\pm W^\pm$ process as the one which guarantees the best
realistic signal to background ratios, and the possibility to further improve
signal selection criteria by careful study of specific signatures of $W_LW_L$
and $W_TW_X$ separately, including lower thresholds on the $p_T$
of tag jets, was rediscovered in the paper by Doroba et al.~\cite{doroba}.
Many of the old observations of Bagger et al.~were reconfirmed using a full tree level
matrix element calculation of signal and backgrounds, including $t\bar{t}$+jets,
and including rough estimates of some additional detector effects related to jet
reconstruction and lepton charge misidentification.  Signal in this work
was defined in terms
of a Higgsless Standard Model with the unitarity condition implemented
by applying appropriate weight factors to generated events with a $WW$ mass
larger than 1.2 TeV.  Translation of all the results into the case of a light
Higgs boson with modified couplings is straightforward.

From 2012 onwards it has become clear that all we can realistically hope for 
in $VV$ scattering are the effects of non-SM Higgs (and gauge) couplings.
Some general guidelines for a rough recomputation of all the predicted signal sizes
were presented by Cheung et al.~\cite{cheung}.  
They considered all the different $VV$ scattering processes and
calculated scale factors to be applied to results of former Higgsless studies
as a function of the actual $HWW$ and $HZZ$ couplings.  Only purely leptonic
$W$ and $Z$ decays were taken into account.  Computation of signal sizes was
carried at the parton level and
using the framework of the two Higgs doublet model (2HDM).

Most recently, various studies have been concentrated on possible methods to
enhance signal significance via improvement of
data analysis techniques as well as event reconstruction tools.
Improvements in the analysis can be expected by means of
applying novel techniques to explore the full shapes of
signals and backgrounds in the multidimensional phase space spanned by the
entire kinematics of visible particles in the final state.  Measured multidimensional
distributions can be compared to predictions arising from particular
theoretical models calculated from the matrix elements.  A likelihood function
can then be defined to quantify the consistency of data with a predefined model.
Such approach was
used in a study by Freitas and Gainer \cite{freitas}.  The analysis they propose
falls into the category
of Multivariate Analyses (MVA), which have become the standard in contemporary
experiments like ATLAS or CMS, suplementing or in many cases completely
superseding the respective
cut-based analyses.  In fact, most Higgs related results published by CMS or ATLAS
to date have versions of MVA's at their bases.
The potential of discerning various theoretical models
is quantified by a $\Delta\chi^2$ calculated for any two hypotheses.
Focusing on $W^+W^+$ scattering at $\sqrt{s}$ = 14 TeV, the authors found a
significant improvement in the LHC potential to discern SILH models from the
Standard Model by using their own version of the Matrix Element Method (MEM).
The reference in this study was a one-dimensional analysis of the
lepton-lepton invariant mass spectrum,
as is routinely practiced in data analyses in HEP experiments.
Expressed directly as a function of the SILH parameter
$\xi c_H$ which governs the modification of Higgs couplings to gauge boson in
the lowest order, the expected $\Delta\chi^2$ rises approximately linearly from 0
to 10 as the value of $\xi c_H$ increases from 0 to 1.  Note that effectively $\xi c_H$=0
is equivalent to the Standard Model, while $\xi c_H$=1 is equivalent to no Higgs.
Meanwhile, a similar $\Delta\chi^2$ obtained
by considering solely the two-lepton invariant mass spectrum was found larger than 1
only for unrealistically large deviations from the Standard Model, beyond
$\xi c_H >$0.7.  A purely counting experiment offers of course yet lower
sensitivity.  However, one should remember that $\Delta\chi^2$ in a counting
experiment is bound to depend on the selection criteria and $\Delta\chi^2$
in any analysis that does not
exploit the full final state kinematics is bound to depend on the event
preselection used to measure the analyzed spectrum.  For further discussion on 
the correspondence of counting experiments with MVA's in the analysis of
$VV$ scattering, see next chapter.

Also in the context of MVA's, the subject of semi-leptonic decay modes was
recently revitalized by Cui and Han \cite{cuihan}.  
Their main theoretical focus was also SILH models versus the Standard Model.
They considered $WW$ scattering, with
all charge combinations included, and explored the jet substructure to
separate the signal from various background processes, including reducible
backgrounds such as $t\bar{t}$+jets and $W$+jets.  A detailed study of jet
substructure provides an effective means not only to distinguish boosted $W$
jets from QCD jets to a large accuracy, but also to account for the different $W$
polarizations between signal and irreducible background.
Wherever signal consists of longitudinally polarized gauge bosons, the two
partons from hadronic $W$ decay tend to be emitted more perpendicularly with
respect to the $W$ direction than in background events.  Put another way, the $p_T$
ratio of the
two partons from signal events tends to be larger than the corresponding
ratio from background events.  For the purely electroweak processes of the
Standard Model, the $p_T$ share of the two partons is usually highly asymmetric,
while it is typically more balanced in the signal.  In order to distinguish
boosted $W$ jets from QCD jets, the authors use the current state-of-the-art
methods.  They are based on the fact that a boosted $W$ has two hard subjets
(i.e., geometrical regions where hadronic energy is concentrated), while a QCD
jet has a single hard subjet.  Subjets can be identified using dedicated techniques
known as filtering, pruning or trimming.  Additionally, $W$ decay products have no color
altogether.  On the other hand, QCD jets carry color charge and are color-connected
to other partons in the event.  This reflects in different transverse jet profiles -
QCD jets are typically much more diffuse.  The most powerful way of separating
signal from background is to combine different variables describing jet
substructure: the masses and transverse momenta after jet pruning, planar flows,
jet cone size dependencies,
etc., to form an effective discriminator for the Boosted Decision
Tree method.  Overall, they found signal for the simplest Higgsless case possible
to observe at more than 5$\sigma$ (from $S/\sqrt{S+B}$) after 100/fb of data at 14 TeV,
which is better than reported for
the purely leptonic decay modes.  The corresponding result
for SILH models scales like $(c_H \xi)^2$.  There is however considerable
uncertainty related to their quantitative background evaluation.
Although they considered both $t\bar{t}$+jets and $W$+jets, as well as the
irreducible $jjWW$ background from processes $\sim\alpha^4\alpha_S^2$, to demonstrate
the principles of operation of the $W$-jet tagging procedure, their final
background numbers are likely to be underestimated.  This is because these
backgrounds were obtained solely via parton showering from the basic
$t\bar{t}$, $W$ + 1 jet and $WW$ processes, respectively, generated with PYTHIA,
when this approach is known to be insufficient.
The efficiency of the $W$-jet tagging algorithm may also depend on detector
resolution and pile-up.  Nevertheless, this work clearly indicates the direction.
It proves that with
the current reconstruction and analysis tools, and envisaging possible further
refinements in the coming years, semi-leptonic decay channels will
offer additional discovery potential and should not be neglected.

\section{The quartic coupling perspective}
\label{tqcp}

Three leading order graphs contribute to $W^\pm W^\pm$ scattering in the
Standard Model (five to $W^+W^-$), including Z/$\gamma$ exchange, the $WWWW$
(quartic) contact interaction and Higgs exchange.  Before
Higgs discovery, and even later until Higgs couplings were known to enough precision,
it was the Higgs exchange graph that represented the major puzzle in the entire
picture and so a measurement
of $WW$ scattering could be practically considered equivalent to Higgs probing.
As the by now discovered Higgs boson continues to fit Standard Model
predictions with better and better precision, a shift of viewpoint is gradually
taking place in what regards the physical motivation of studying $VV$ scattering
processes at the LHC.
By assuming Higgs couplings known, e.g., exactly equal to their values predicted
in the SM or to the values measured in the LHC, and by moreover assuming that
no new physics be directly observed
within the energy range at consideration, we can
revert this reasoning and reformulate the problem in terms of
the quartic $WWWW$ coupling.  Indirect signs of new physics may include a
modification of the effective four-$W$ interaction term, leading to an anomalous
coupling value.  Such deviation would violate the cancelation of the leading
$\sim s^2$ terms in the $WW$ scattering amplitudes between the contact
interaction graph and the Z/$\gamma$ exchange graph, producing a divergence
proportional to the fourth power of energy.
The observed energy dependence would therefore in principle be different than 
in the case of modified Higgs couplings.  However, deviations from the SM are
likely to show up in more complicated forms than as a simple scaling factor
applied on the SM value.  Generally, anomalous quartic couplings may be
generated as a contact interaction approximation of heavy particle exchange.
The specific form of the operator that effectively contributes to
the quartic vertex, plus the energy scale at which new physics sets in and
places a natural cutoff for the relevant calculations, is a key question in order
to assess the expected energy dependence.
The actual value of the
$WWWW$ quartic coupling is currently very poorly constrained by experiment.
As we noticed before, $VV$ scattering, along with triboson production, provide
the most direct probes of the quartic couplings.

Interestingly, prior to Higgs discovery some authors studied Higgsless models in the
language of effective anomalous quartic couplings.  The correspondence is
straightforward.  Deviations from a pure Higgsless SM, possibly arising from
heavy particle exchange or some kind of strong dynamics, were effectively
parameterized as an anomaly in the quartic gauge vertex.
As an example, based on such approach Godfrey \cite{quartic} noticed early on that
for the lowest dimension operators that do not include photons the LHC will provide
the most constraining measurements compared to e$^+$e$^-$, e$^-$e$^-$, $\gamma\gamma$
or e$\gamma$ colliders.  Furthermore, he found same-sign $W^\pm W^\pm$ scattering
be the best process to study the quartic couplings.
A similar approach was taken by Belyaev et al.~\cite{belyaev}, who basically redid the
work of Bagger et al.~in the language of anomalous quartic couplings.
Today these studies are however of historical interest only.

By contrast, the work of Eboli et al.~\cite{eboli}
retains its actuality because they have
considered the case of a 120 GeV Higgs as one of their two reference scenarios
in the study of quartic couplings.  It was also one of the earliest papers
where the full analysis was carried within the language of the Effective Field
Theory with higher dimension (dimension-8 in this case) operators.
As already mentioned, these authors
considered the $W^\pm W^\pm$ and $W^+W^-$ processes and purely leptonic decays.
The analogy to former Higgsless studies is evident.
They propose a full collection of selection
criteria, which is in fact a variation of the familiar Higgsless selection criteria.
They included: $|\eta_j|<$4.9 and $\Delta\eta_{jj}<$3.8, $\eta_j^{min} < \eta_l <
\eta_j^{max}$, MET $>$ 30 GeV and $p_T^l >$ 30 (100) GeV for same-sign (opposite-sign)
leptons.  Additional cuts: $M_{jj}~>$ 1000 GeV, a central jet veto and
$\phi_{e\mu}~>$ 2.25 rad, were applied for $W^+W^-$ only.
$W$ helicities were not distinguished, but a moderate jet $p_T$ threshold for
the tag jets (20 GeV) ensured good acceptance for $W_LW_L$ pairs.
Signals were calculated using the MadGraph generator with self-added modifications
to include the anomalous terms.  It must be stressed that in this analysis
background evaluation included only irreducible
background and inclusive $t\bar{t}$ production for $W^+W^-$
and irreducible background only for $W^\pm W^\pm$.  Moreover, all the analysis
was done essentially at the parton level, with some experimental resolutions
and estimates of reconstruction efficiencies simulated on top.  The results
are therefore likely to be too optimistic as far as overall background rejection
is concerned.  Unitarity constraints were satisfied by imposing a sharp cutoff
at $M_{WW}$ = 1.25 TeV.
Final results were extracted by merely counting the
total event yield in the signal window.
From a combination of both processes and assuming an integrated LHC luminosity of
100 fb$^{-1}$, they came to predict the following 99\% CL limits:

\vspace{3mm}

\hspace{5cm} $-22~<~\frac{f_{S,0}}{\Lambda^4}$ TeV$^{-4}~<~24$,

\vspace{2mm}

\hspace{5cm} $-25~<~\frac{f_{S,1}}{\Lambda^4}$ TeV$^{-4}~<~25$,

\vspace{3mm}

\noindent
under the assumption of only one non-vanishing coefficient at a time.  In practice,
combination of $W^\pm W^\pm$ and $W^+W^-$ is crucial, because the two
coefficients studied from each process separately show strong
anticorrelation, especially in the same-sign process.
This was the first such detailed study that explicitly focused on
the LHC sensitivity to anomalous
quartic couplings and used the language of higher-dimension operators.  Although
it may need minor updates in several places, it still remains the most complete
phenomenological analysis of its kind.

\section{$VV$ scattering in LHC measurements at 8 TeV}

Practically no simulation-based studies of $VV$ scattering on a detector-independent
level exist
for $pp$ collisions at 7 or 8 TeV.  This is not just because these beam energies were
not really considered at the early stages of LHC planning, but rather decided
later on as a compromise between current technical possibilities and physics needs.
The main reason is that it was known from rough order of magnitude estimates that 7/8 TeV 
would in fact not suffice to carry a truly conclusive measurement in terms of possible
physics beyond the SM.  After Higgs discovery, this became
even more clear.  This is why existing VBS-like measurements done at 8 TeV are still
largely disconnected conceptually from all the simulation work presented above and
it is not always a trivial task to realize how they in fact relate.
Here we will examine first results from ATLAS and CMS concerning $VV$ scattering
that were obtained from an analysis of the 8 TeV data.  In doing this our main
point of interest will not be what the results tell us about physics, but rather
what we can learn for future analyses at a higher beam energy.

Preliminary results on vector boson scattering at 8 TeV have been produced both
by ATLAS \cite{atlasww8} and CMS \cite{cmsww8}.  
As the main ideas behind these two studies are closely related
and both analyses came out at a similar time (as usual, though, the ATLAS paper
came first), we will discuss both of
them simultaneously, making appropriate distinctions only when relevant.
In both cases
searches were carried for a loosely defined VBS-like signature in the same-sign $WW$
scattering process and the purely leptonic decay channel.  To partially tackle
the inescapable problem of low statistics, applied VBS-like selection was in both
cases minimal.  The signature consisted of two reconstructed same-sign leptons
(each of which
could be either an electron or a muon) passing all the respective ``high-quality"
criteria and at least two jets within detector
acceptance.  The definitions of detector acceptance were marginally different
for ATLAS and CMS, but could not play any major role in the final result.
Minimum $p_T$ of 20 (25) GeV was required in ATLAS (CMS) for the
leptons and of 30 GeV for the jets.  In addition, a minimum missing transverse energy
of 40 GeV was required to account for the two neutrinos.  The only additional
selection criteria applied on the data in order to separate pure electroweak
$jjW^\pm W^\pm$
production from processes involving gluon exchange
was a cut on the jet-jet
invariant mass, $M_{jj} >$ 500 GeV, and rapidity separation $|\Delta y_{jj}| >$ 2.4
(ATLAS) or 2.5 (CMS).  Subscript {jj} refers always to the two leading (highest-$p_T$)
jets in the event.  These criteria are also instrumental in suppressing
various sources of reducible background, chiefly inclusive $t\bar{t}$ production.
At an average lepton $p_T$ that corresponds to the beam energy
of 8 TeV, the charge of the muon is measured accurately to negligible levels.
However, inefficiency of electron charge determination can produce non-negligible
detector background coming from $e^+e^-$ pairs copiously produced at the $Z$ boson peak.
For this reason, the $ee$ invariant mass was required to lie outside a band
of 10 (ATLAS) or 15 (CMS) GeV around the $Z$ mass.  This cut affected only
the electron-electron decay channel.  
Moreover, for any lepton pair its mass had to be larger than 20 GeV in ATLAS
and 50 GeV in CMS.  This is perhaps the most significant difference between the
two analyses and was dictated by the respective detector capabilities in what
concerns in particular the contamination from jets misreconstructed as leptons.
Remaining inclusive $t\bar{t}$ background, entering via both lepton sign-flip effects
and leptonic $b$ quark decays, was
effectively eliminated by standard $b$ quark vetoing.
Details of the $b$-tagging techniques were developed independently by the two 
collaborations, but both are based on combining the information from impact
parameter significance of the individual tracks with explicit secondary vertex
reconstruction.
The bulk of background coming from $WZ$ or $ZZ$+jets production was reduced
by a veto on a third lepton.  Here the cut depends on the efficiency and purity
of lepton reconstruction and must be optimized in a detector-dependent
manner.  Consequently, it was somewhat stricter in ATLAS than in CMS: it involved any
additional reconstructed
leptons with $p_T > 6/7$ GeV or 10 GeV, respectively.
In addition to the above, the ATLAS analysis used minimum angular separation criteria
between the two leptons and between leptons and jets, the meaning of which
is rather marginal.

Signal is defined as electroweak $jjW^\pm W^\pm$ production within a kinematic
region consistent
with vector boson scattering.  There is no explicit
distinction between $WW$ scattering and non-scattering $WW$ production at
any stage of the two analyses.  In fact, the ATLAS paper is conservatively entitled
{\it ``Evidence of Electroweak Production..."} and makes no mention of VBS
anywhere in the paper abstract.  Naturally, signal definition includes SM
contributions and so it contains what in all our
previous considerations has been called irreducible background.
There is a subtle way in which signal is not exactly the same thing in the
two analyses.
In the CMS analysis, there is no distinction between pure electroweak
$jjW^\pm W^\pm$ production and QCD mediated (gluon exchange) processes within
the selected kinematic window.  Both are treated as an integral part of the
signal.
The ATLAS analysis explicitly separates QCD production of $jjW^\pm W^\pm$
as another class of background, as opposed to a purely electroweak process.
How much of each we have in the sample is deduced from simulation.  The question is
not completely straightforward because the two types of processes interfere 
constructively,
i.e., the total cross section is larger than the sum of the individual pure
cross sections.  This interference is of about 15\% of the electroweak signal
and depends on the choice of scale.
ATLAS calculates the contribution from the interference terms
by subtracting the coherent sum of pure electroweak and QCD process from the
complete calculation including the interference.  It is then added
to the pure electroweak process as part of the signal prediction.
They define a wider ``inclusive" kinematic region in order to verify that
QCD contributions are indeed suppressed in the proper signal region.
In the signal region, the QCD contribution amounts to roughly 10\% of the
total $jjW^\pm W^\pm$ production.  Such differences are in fact within the
statistical errors of the signal sample collected in these studies.
The future practical solution to the above problems would be in applying selection
criteria tight enough so that any QCD contribution, including the interference,
would become negligible altogether.  This is, however, not a viable option for
the present energy and the accumulated statistics.

For the background estimates both collaborations developed original methods which
differ rather widely.
Whenever simulations are used, typically, leading order
generators were applied (MadGraph, POWHEG, SHERPA or ALPGEN), and the results
were normalized in terms of a constant factor to the next-to-leading order in QCD
cross sections obtained, e.g., with VBFNLO.  
Uncertainties of the order of 10\%
were found within the signal kinematic window for the main backgrounds.
Differences in the respective LO generators used by ATLAS and CMS are unlikely to play
a major role, but the respective PDF and QCD scale choices
are in fact one of the main components of the systematic errors.
CMS background predictions are for the most part data-driven.  The so-called
``non-prompt"
lepton backgrounds, originating from leptonic decays of heavy quarks, hadrons
misidentified as leptons ($W+$jets), and electrons from photon conversions in the
detector were deduced from a control sample defined by one lepton which passes
the full lepton selection criteria and another lepton which fails these criteria,
but passes a ``loose lepton" selection.  Fake rates for such loose leptons to
pass the nominal lepton criteria were then calculated and applied to the signal
region.  Similarly, the $WZ$ background with two accompanying jets is
predicted from a data control region requiring an additional lepton with
$p_T >$ 10 GeV.  Other background sources included triboson production,
sign-flip effects and double parton scattering.  They amounted to less than
10\% of the total background and were estimated from
simulation.
ATLAS background predictions for $WZ$ and $ZZ$+jets (labeled ``prompt"), as well
as photon conversion background, were driven from full detector
simulations and
cross checked with the data in several same-sign dilepton control regions.
Sign-flip backgrounds (part of which they include in the ``conversion" category)
and backgrounds involving leptons reconstructed from jets (collectively
denoted as ``other non-prompt") were estimated directly from data.

Interesting is the significant difference in the final background composition
between ATLAS and CMS.  ``Prompt" backgrounds, composed in 90\% of $WZ$ production
with a lost third lepton (either not reconstructed or falling outside detector
acceptance), amount to as much as two thirds of the total background at ATLAS.
By contrast, it is less than 20\% in CMS.  In absolute numbers, the remaining
$WZ$ background is over 7 times higher in ATLAS than it is in CMS (7.5 $\pm$ 1.1 
vs.~1.0 $\pm$ 0.1 (stat.) events, respectively).  
Moreover, the effect is present consistently in all the lepton flavor combinations:
$e^\pm e^\pm$, $\mu^\pm \mu^\pm$ and $e^\pm \mu^\pm$.  Such difference cannot
be explained solely by physics, i.e., details of the applied selection criteria,
although stronger cuts on
$|\Delta y_{jj}|$ and $M_{ll}$ adopted by CMS contribute in the right direction.
We recall that
the third lepton veto used to suppress $WZ$ was stricter in ATLAS than in CMS
in terms of the $p_T$ threshold.  However, different efficiencies of lepton 
reconstruction and third lepton veto,
connected to the respective ``tight" and ``loose" lepton identification
criteria
applied by the two experiments, must be doubtlessly causing the resulting
discrepancy.
By contrast, the amounts of ``non-prompt" (including conversion) background
predicted in the two experiments are roughly similar: 6.3 $\pm$ 1.1 events in ATLAS,
4.2 $\pm$ 0.8 (stat.)
events in CMS.  We recall here that CMS used a stricter cut on the minimum
lepton-lepton mass to protect from this kind of backgrounds.  Different fake
rates of hadrons misidentified as leptons may explain the observed differences.
One thing is evident from this comparative study.  Background levels and
compositions crucially depend on tiny detector-specific effects related to the
reconstruction and identification algorithms of different physics objects.  They
are difficult to predict from pure physics principles, unless in restricted
kinematic regions chosen so to render most of these effects negligible.
We will recall this when trying to draw updated predictions for 14 TeV.

Predicted signal levels are calculated from full detector simulations
that use the MadGraph generator for CMS and
POWHEG-BOX for ATLAS.
Signal predictions in the SM amount to
8.8 $\pm$ 0.2 (stat.) events for CMS and 15.2 $\pm$ 0.8 for ATLAS, where
we have redefined the ATLAS signal to include QCD for better consistency with CMS.
As before, the difference may be at least partly due to signal selection criteria, in
particular different cuts on $M_{ll}$ and $|\Delta y_{jj}|$; to a
lesser degree
$|M_{ee}-M_Z|$ and $|\eta_e|$ which affect only the electron decay channels.
A 5\% difference exists in the total integrated luminosity recorded by the two
detectors (20.3 fb$^{-1}$ ATLAS vs.~19.4 fb$^{-1}$ CMS).
The rest of the difference is nailed down to be due to theoretical uncertainties,
reconstruction efficiencies and resolutions.  In particular,
the most important systematic uncertainties in the signal predictions are those
related to the choice of PDF's (7.7\%) and the QCD scale (5\%), jet energy scale
and resolution (5\%) and lepton efficiency (3\%).

To quantify the statistical significance of the signal, events yields were
examined in eight separate intervals formed by 4 bins of $M_{jj}$ times two
lepton charges.
The observed (expected) signal significance in CMS is 2.0 $\sigma$ (3.1 $\sigma$).
In ATLAS, the corresponding numbers are 3.6 $\sigma$ (2.8 $\sigma$).
The apparently large discrepancy between the actually observed numbers of events
in the signal region: 12 events in CMS vs.~34 events in ATLAS is
consistent within the errors with all the earlier predictions.

It is trivial to convince oneself that ATLAS and CMS 8 TeV results neither confirm nor
disconfirm Higgs existence, let alone give any clue of the relevant Higgs
coupling.  Likewise, they are hardly sensitive to anomalous triple gauge couplings
within their present experimental bounds.  Instead, they can be interpreted in terms
of the first experimental bounds on the quartic $WWWW$ coupling.
The fact that both analyses used a relatively high jet $p_T$ threshold to protect
from pile-up jets means that $W_LW_L$ pairs were disfavored.  This additionally
reduces the sensivity to the $HWW$ coupling and also to those higher dimension
operators which modify only longitudinal gauge boson interactions.
The CMS collaboration obtained 95\% CL limits on all nine dimension-8 operators
that lead to anomalous contributions to the $WWWW$ coupling.  Respective signal 
predictions
were derived from MadGraph-generated samples in which one anomalous parameter
was varied at a time.
Limits were based
on the measured lepton-lepton invariant mass spectrum of events that pass the
full selection.  Other considered spectra included the jet-jet invariant mass
the four-body mass and the leading lepton $p_T$, but did not reveal improvements in
sensitivity to the parameters
in question.  Here there was no specific optimization with the respect to the
individual anomalous parameters, e.g., in what concerns the $WW$ helicity 
combinations they directly affect.
The effect of these parameters on the background is marginal and was neglected.
This concerns also the $WZ$ background, since in our signal window it is dominated
by QCD contributions rather than $WZ$ scattering.
In addition to the uncorrelated limits on individual parameters, limits were
derived in the two-dimension space of $f_{S,0}/\Lambda^4$ vs.~$f_{S,1}/\Lambda^4$.
Obtained contours show a strong anticorrelation between these parameters.
Because of this the limits on individual parameters obtained by one-dimensional
projections of the correlated limits are weaker than their
uncorrelated limits by at least a factor of $\sim$5.  The reason is straightforward.
Each scattering process probes in fact specific combinations of anomalous parameters.
And the other way around, improvement on the correlated limits can be only
achieved by combining data from different scattering processes: $W^+W^-$, $WZ$
and $ZZ$, which probe different combinations of the same parameters.  Data at 8 TeV
are however of not enough statistical power to study the other processes.

\begin{figure}[htbp]
\begin{center}
\vspace{5mm}
\hspace{-1cm}
\epsfig{file=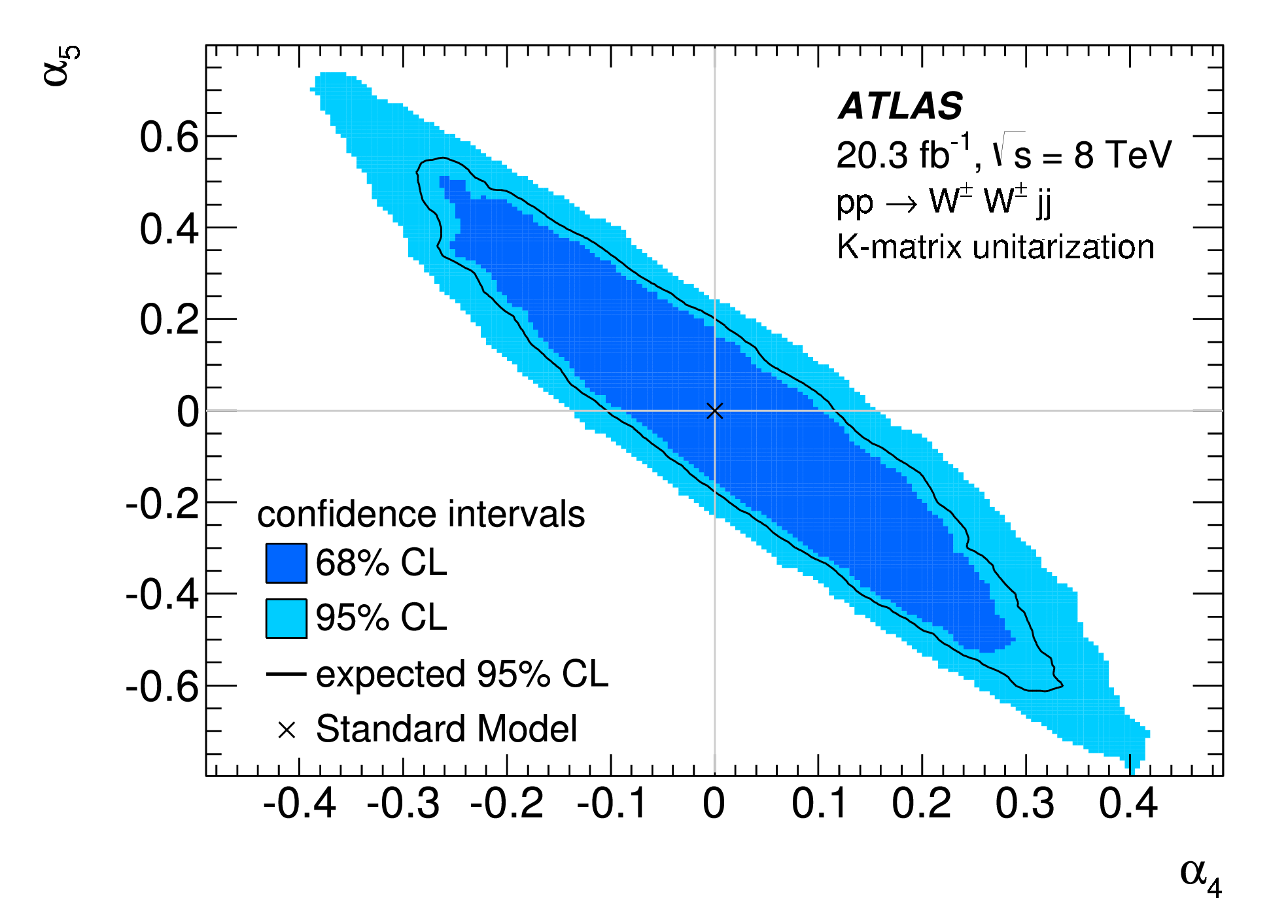,width=0.52\linewidth}
\epsfig{file=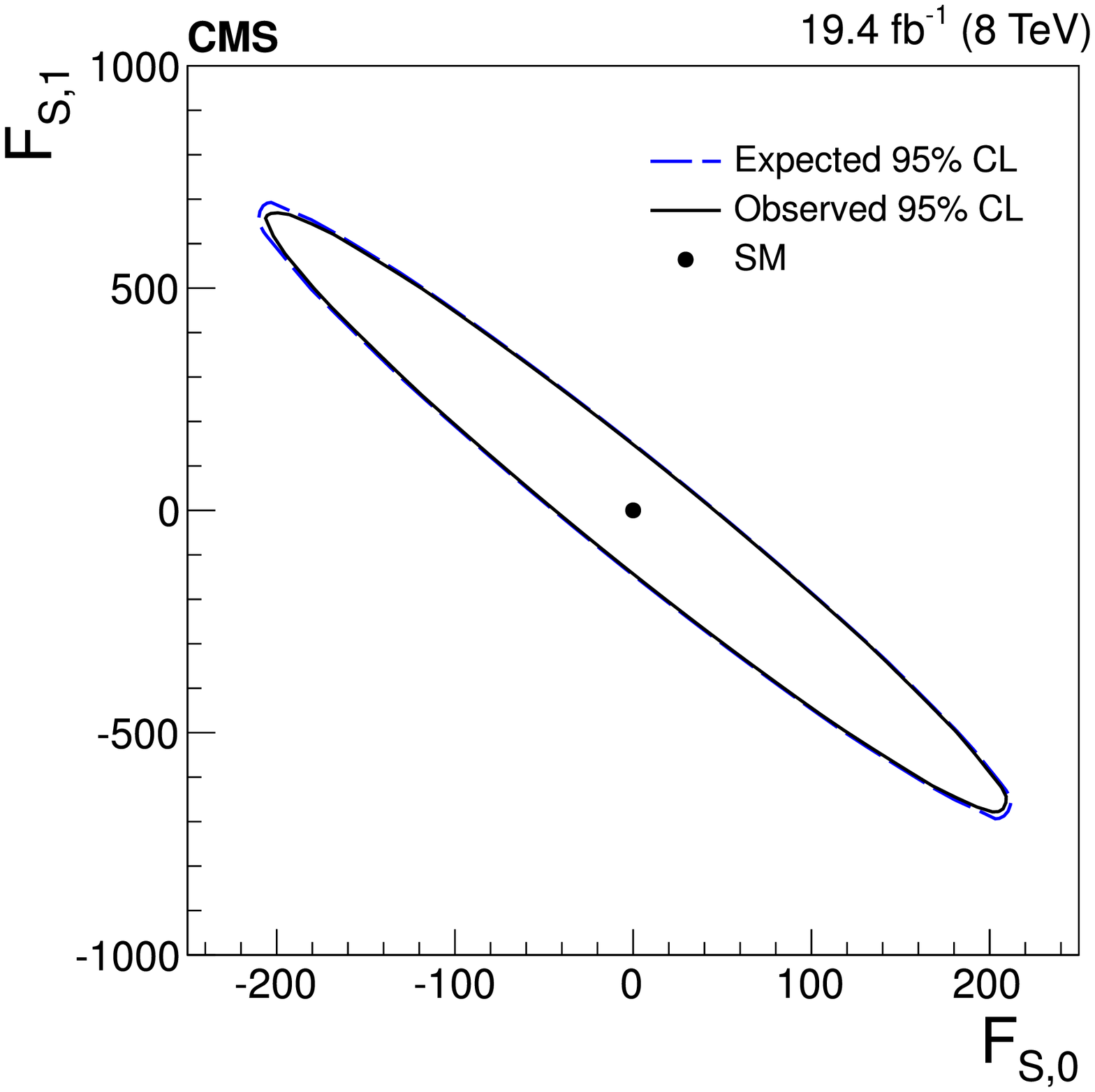,width=0.45\linewidth}
\end{center}
\vspace{-5mm}
\caption{Left: 
observed and expected two-dimensional limits on the operator coefficients
$a_4$ and $a_5$ from an analysis of the process $pp \to jjW^\pm W^\pm$ at 8 TeV 
done by the ATLAS collaboration - image reproduced from Ref.~\cite{atlasww8}.
Right: 
observed and expected two-dimensional limits on the operator
coefficients $f_{S,0}$ and $f_{S,1}$ from an analysis of the process
$pp \to jjW^\pm W^\pm$
at 8 TeV done by the CMS collaboration - image reproduced from Ref.~\cite{cmsww8}.
}
\vspace{3mm}
\end{figure}

The ATLAS collaboration derived limits on the $WWWW$ coupling expressed in
terms of the $a_4$ and $a_5$ parameters of the Electroweak Chiral Lagrangian.
The underlying model, as currently implemented in the WHIZARD generator, includes
a 125 GeV Higgs boson in addition to the traditional dimension-4 operators
foreseen within the framework of the EWChL, and so non-zero values of $a_4$
and $a_5$ can be in principle reinterpreted as equivalent to anomalous quartic couplings.
While formally data may still be interpreted in this language, such treatment
has not gained wide recognition in the physicists' community.
A more practical problem is that non-zero $a_4$ and $a_5$ still induce unitarity
violation and provide no built-in mechanism to restore unitarity.  Results
do depend on the arbitrarily chosen unitarization scheme, which is an
intrinsic uncertainty of the model.  ATLAS used the K-matrix unitarization procedure
and did not quantify the theoretical uncertainties related to this particular choice.
Effective equivalence of the phenomenological impact of non-zero
($a_4$, $a_5$) with
that of dimension-8 operators from the Effective Field Theory approach was
demonstrated.  The relationships for the $jjW^\pm W^\pm$ process
are supposedly the following \cite{snowmassew}:

\begin{equation}
a_4 = \frac{v^4 f_{S,0}}{8\Lambda^4},
\end{equation}

\begin{equation}
a_5 = \frac{v^4 (f_{S,1}-f_{S,0})}{16\Lambda^4},
\end{equation}

\noindent
where $v$ is the usual Higgs vacuum expectation value.
With the above one can verify that the far endpoints of the 95\% CL contours from ATLAS,
approximately $a_4=\pm$0.4 and $a_5=\mp$0.7, correspond to
$f_{S,0}/\Lambda^4 \approx \pm$ 870 TeV$^{-4}$ and $f_{S,1}/\Lambda^4 \approx \mp$ 2180
TeV$^{-4}$,
several times weaker bounds than from the CMS analysis.  Why such discrepancy?
Partly because ATLAS does see an excess of events with respect to SM predictions,
while CMS sees a deficit.
Another reason is that the CMS methodology does not assume
any physical cutoff $\Lambda$ for the evaluation of the anomalous signals.
The underlying assumption that new physics does not directly show up to the
presently available energy is natural in the light of no new physics having
been actually observed in the LHC so far.
However, $f_{S,0}/\Lambda^4$ and $f_{S,1}/\Lambda^4$ may lead to unitarity violation
within the quoted limits.  Put another way, even a BSM signal equivalent to
hitting the unitarity limit at the highest available energy
could not be observed with the present data.  By saying earlier on that
the results neither confirm nor disconfirm Higgs existence, we effectively meant
exactly the same thing.
Data at 8 TeV do not provide enough sensitivity to establish really
physically meaningful limits on the studied parameters.  Formally calculable limits
reflect the applied unitarization procedure or lack of it and one should be
extremely careful in drawing physics conclusions.  The exercises done by ATLAS and CMS
serve as a demonstration of principles and technical preparation for future
measurements at higher energies.  They set up and test the methodology.  They reveal
the main experimental and theoretical issues to
be addressed in such measurements.  But their physics meaning on its own is
for the time being quite limited.

The CMS collaboration derived also limits on the production cross section times
branching fraction of a doubly charged Higgs decaying into $W^\pm W^\pm$.
Doubly charged Higgs bosons are expected in models that contain a Higgs triplet
field.  In such models, the $W^\pm W^\pm$ scattering process would be a
resonant one, in contrast to the SM and its most popular proposed extensions.

\chapter{What can the LHC measure}

After delivering 5 fb$^{-1}$ of proton-proton data at 7 GeV and 20 fb$^{-1}$
at 8 TeV, the LHC entered the first long shutdown (LS1) phase from 2013 till
the end of 2014 and has been due to upgrades.  LS1 included a large number of
simultaneous activities concerning both the injectors and LHC itself, aimed
to ensure reliable operation at nominal parameteres from 2015.  Most importantly
the center of mass energy will now be nearly doubled and become 13 TeV.  Early
plans assumed a center of mass energy of 14 TeV and a lot of earlier simulation work
was in fact done
under this assumption.  As physics is unlikely to change significantly between
13 and 14 TeV, these studies are mostly still valid and in this work we will discuss
13/14 TeV simulation results in a complementary way, without making clear distinctions.
The main priorities of LS1 are to repair and consolidate the interconnects, bring
all necessary equipment up to the level needed for 6.5 TeV per beam, repair leaks
and other maintenance work required after 3 years of operation.
Upgrade and maintenance activities in the machine are accompanied by
concurrent upgrade and maintenance activities on part of the individual detectors.

The LS1 will be the first long shutdown of the LHC, part of a long term draft plan
which foresees operation until 2035, with several subsequent operation periods
and long shutdowns.  Run 2 of the LHC is due to start early in 2015 and last for
the next 3 years with an intermediate luminosity of 10$^{34}$ cm$^{-2}$s$^{-1}$.
Long shutdown 2 (LS2) is planned from mid-2018 until the end of 2019.
After that, Run 3 of the LHC will proceed with nominal energy and nominal
luminosity of 2 $\times$ 10$^{34}$ cm$^{-2}$s$^{-1}$.  The amount of proton-proton
data collected in Runs 2 and 3 is conservatively expected to be 300 fb$^{-1}$.
Given the experience from Run 1 and the excellent machine operation which
surpassed conservative expectations already in the second year of running, it may
possibly turn out even larger.  After 2022, the machine will be due for another
major upgrade for an order of magnitude increase of luminosity, while keeping
the same beam energy.  
This future phase is refered to as the High Luminosity LHC (HL-LHC).
Long shutdown 3 (LS3) is planned to last from 2023
until late 2025 for the LHC and from 2024 until mid-2025 for the injectors.
The following three machine operation periods, interspaced with long shutdowns 4 and
5, aim at delivering 3000 fb$^{-1}$ of
proton-proton collisions until 2035.  This is the ultimate aim of the LHC.

In this chapter we will try to answer the question of what can the LHC, operating at
13/14 TeV, measure in the various $VV$ scattering processes, having in mind
everything we have learned so far both on the theory side and from existing
measurements.  Because the physics motivation to study VBS processes has
significantly changed only in the last couple of years and is still in the process
of reformulation, up to date analyses
are not so abundant and a comprehensive review of fully valid predictions for the LHC
is rather hard to find.  In an attempt to fill the hole,
we will herewith sketch some analyses
which are supposed to be completely consistent with all our present knowledge,
yet not involving anything
more than common simulation tools.  For full transparency and in order to avoid
usage of any experiment specific software, the analysis will
be kept as simple as possible from the point of view of the applied analysis tools.
It will be a cut-based analysis.  While we do not assume that the
future final analyses by the ATLAS
and CMS collaborations will indeed be done in this way, such simple analysis is 
accurate enough
for our purpose, which is to evaluate the order of magnitude of possible signals
from different BSM sources, shed light on the LHC potential to identify a physics
scenario from the sole study of VBS processes, as well as to identify some of the main 
challenges and limiting factors
in what concerns background rejection.  Whenever possible and applicable, we will
follow the ideas of earlier works by many authors, but all the results will be 
independently recalculated with modern simulation tools.  This will include
some detector resolution effects, so long as the latter do not involve full detector
simulation.

Our main focus here will be on the purely leptonic decays.  This choice is
mainly motivated by pragmatism - these channels do not suffer from complicated,
QCD-related systematic uncertainties in event reconstruction and, as we have seen,
the final signal to background ratio,
with all major detector dependent effects included, is driven by 
merely a few experiment-specific factors: the purity of electron
reconstruction, charge measurement efficiency for electrons and muons at high
$p_T$, and the efficiency of $b$-tagging.
All these effects can be to a rough
accuracy described in terms of simple numbers, without necessarily applying
the entire methodology of event reconstruction used in a real experiment.
All other systematics can be either assumed known or play a lesser role.

Our baseline to define the BSM signal and tune the necessary selection criteria
will be the Higgsless Standard Model, as
we inherit from most of the classic studies.  However, the ultimate goal is to find how
this translates into realistic scenarios with modified Higgs couplings and anomalous 
triple and quartic gauge boson couplings, with all the relevant similarities
and differences being taken into account.

\section{Modeling of the signal and irreducible background}

In previous chapters we have discussed in detail the formal definitions, the
computational methods and issues for the complete calculation of the signal and
irreducible background.  A full set of signal selection criteria that are applicable
to future data at $\sqrt{s}$ = 14 TeV follows directly from
our previous considerations:

\begin{itemize}

\item at least two jets with $2 < |\eta_j| < 5$ and $\eta_{j_1}\eta_{j_2} < 0$,

\item exactly two isolated same-sign/opposite-sign leptons with $\Delta\varphi > 2.5$,

\item $M_{jj} > 500$ GeV,

\item $M_{l_1j_2}, M_{l_2j_1} > 200$ GeV,

\item $b$ quark veto,

\item optional: central jet veto,

\item $R_{p_T} > 3.5$ or a suitable combination of cuts that selects large $p_T^l$,
small $|\eta_l|$ and large $M_{ll}$.

\end{itemize}

\noindent
We recall that the first two criteria are basic topological VBF cuts,
the next four are dedicated $t\bar{t}$ suppression cuts (not exactly - requirement
of large
$M_{jj}$ also suppresses the irreducible background) and only the {\it last} item
represents cuts, only one in the same-sign case, that
separate the longitudinal $W$ signal from the transverse $W$ background.

\begin{figure}[htbp]
\begin{center}
\epsfig{file=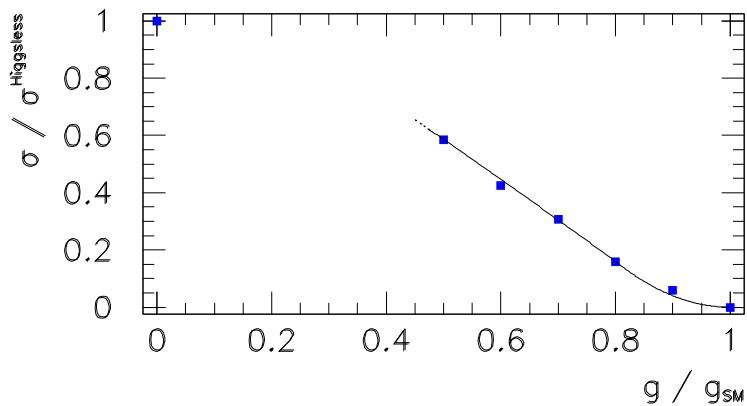,width=0.75\linewidth}
\end{center}
\vspace{-1cm}
\caption{Actual signal cross section relative to the Higgsless cross section
as a function of the actual $HWW$ coupling relative to its Standard Model
value.  Signal sizes were determined after applying all the signal selection
criteria discussed in the text.
The points provide a good first approximation of how to scale the results
of all the former $WW$ scattering studies which used the Higgsless hypothesis
to evaluate signals in order to reinterpret them in terms of
a Higgs with modified couplings.  Result of parton-level MadGraph
calculations for the process $pp \rightarrow jjW^+W^+$ with $W^+ \rightarrow \mu^+\nu$.
}
\label{sscafa}
\end{figure}
\vspace{1cm}

From analyses available to date it can be inferred that
the order of magnitude $W^+W^+$ signal cross section, where signal is
defined in terms of a pure Higgsless Standard Model, and by further assuming purely
leptonic decays ($l=e,\mu$), is
close to 0.12 fb.  This number has been recalculated using MadGraph 5.
It includes effects associated to
hadronization, final state radiation and jet reconstruction using an imitative
simple jet cone algorithm.  Sheer signal size is similar for $W^+W^+$ and for $W^+W^-$.
Physically realistic scale factors range from 0.8 for a pure
Higgsless with unitarity or a heavy Higgs, to 0.31, 0.16 and 0.06 for a light Higgs
boson that couples to the $W$ with a strength equal to, respectively, 0.7, 0.8
and 0.9 of what is predicted in the Standard Model \cite{kusmierczyk}
(see Fig.~\ref{sscafa}).
Because of the amplitude interference patterns, signals for $HWW$ couplings
larger than unity (in SM units) are generally lower than for their mirror values.
Finite detector resolutions in the measurement of
$p_T$ or $\eta$ of the leptons and jets effectively play the role of a
further $\sim$10\% reduction.  
Signal in the anomalous gauge coupling scenarios must be calculated
independently and will be shown later on.
Irreducible background levels for $W^+W^+$ are of the order of 0.05 fb in a
conventional cut analysis and can be shrinked at least to 0.02 fb by applying
an $R_{p_T}$ cut instead or even more sophisticated correlated variable techniques.
The latter will be much closer to what eventually can be achieved using a
Multivariate Analysis in which the entire final state kinematics is exploited.
For $W^+W^-$ the irreducible background amounts to about 0.11 fb and it is unlikely
to improve in a significant way.
Signal for $W^-W^-$ is about a factor 4 lower than for
$W^+W^+$, but backgrounds are at similar levels.

\begin{figure}[htbp]
\begin{center}
\epsfig{file=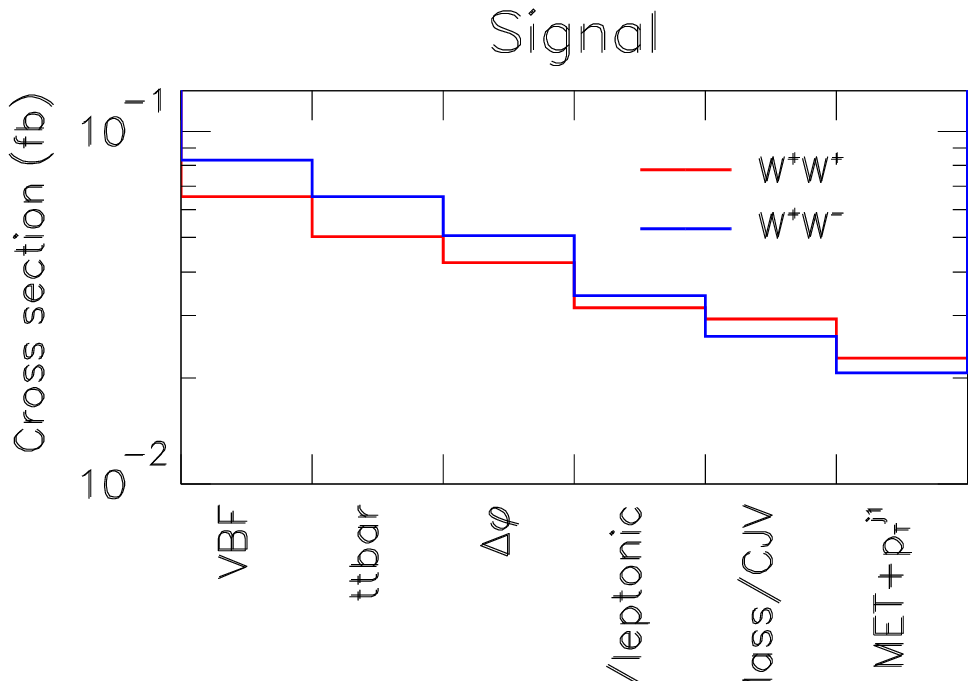,width=0.75\linewidth}
\epsfig{file=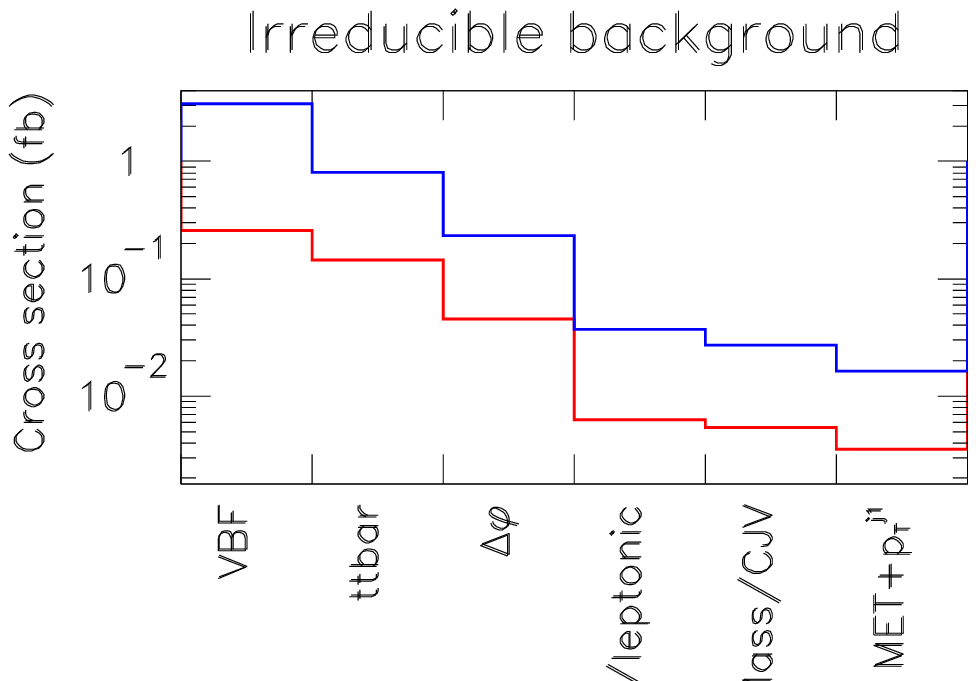,width=0.75\linewidth}
\end{center}
\caption{{\small Total cross sections for the signal and irreducible
background after each subsequent class of cuts proposed in the analysis.
Red histograms are for
$pp \rightarrow jjW^+W^+$, blue histograms for $pp \rightarrow jjW^+W^-$;
in both cases $W$ decay into muons is assumed.  
The signal is calculated under the Higgsless hypothesis.
Results reflect pure
event kinematics, all detector efficiencies and inefficiencies (where
appropriate) are assumed 100\%.  The meaning of the cut labels is the following:
$\bullet$ {\bf VBF:} $2<|\eta_j|<5$, $\eta_{j1}\eta_{j2}<0$ and $|\eta_\mu|<2.1$,
$\bullet${\bf ttbar:} $M_{jj} > 500$ GeV and $M_{j1\mu 2}$, $M_{j1\mu 2} >$ 200 GeV,
$\bullet${\bf $\Delta\varphi$:} $\Delta\varphi > 2.5$,
$\bullet$ {\bf $R_{p_T}$/leptonic:} $R_{p_T} > 3.5$ for $W^+W^+$ or
$p_T^{\mu 1}+p_T^{\mu 2} > 300$ GeV and $M_{\mu\mu} > 300$ GeV for $W^+W^-$,
$\bullet$ {\bf Mass/CJV:} $M_{\mu\mu} > 250$ GeV for $W^+W^+$ or central jet veto
with $p_T > 25$ GeV for $W^+W^-$,
$\bullet$ {\bf MET+$p_T^{j1}$:} missing $E_T > 60$ GeV and $p_T^{j1} > 30$ GeV.
Results of a MadGraph calculation,
processed by PYTHIA 6 for $W$ decay into muons ($jjW^+W^-$ only), the
effects of parton showering, hadronization and jet reconstruction
and further processed by PGS 4 for the effects of finite resolution in
the measurement of jet and muon $p_T$ in a CMS-like detector.
The original PYTHIA 6 source code was modified to account for the
correct, polarization-dependent, angular distributions for the decays
$W^\pm \rightarrow \mu^\pm \nu$.  The corresponding results for the decays
of $W$ into electrons are typically consistent within a few per cent
and/or statistical fluctuations and are not shown here.
}}
\label{cuts1}
\end{figure}

Total cross sections for the signal (calculated within the Higgsless scenario)
and irreducible background in the
$W^+W^+$ and $W^+W^-$ processes, after each subsequent class of selection
criteria discussed in the text, are shown in Fig.~\ref{cuts1}.

\section{Modeling of the inclusive $t\bar{t}$ production background}
\label{modttbar}

Reducible backgrounds in real experiments are typically determined using partly
or wholy data-driven methods.  This and the following sections discuss pure
simulation-based results and are not intended
as a model for a future analysis of experimental data.  Their purpose is
merely to establish a suitable methodology to estimate all these backgrounds
in simulation-based studies, before they can be cross checked against the data.
The methods described here should be accurate enough to assess
the orders of magnitude of the relevant backgrounds
and to study the main challenges related to background reduction.

Calculations of the $t\bar{t}$ background are affected by large QCD-related
uncertainties.
The inclusive $t\bar{t}$ production cross sections in proton-proton collisions
at $\sqrt{s}$ = 7 and 8 TeV has been measured by both ATLAS and CMS.  These numbers
provide the only currently available direct experimental bond to reduce
the theory-based systematic uncertainties for the predicted $t\bar{t}$
cross sections at 14 TeV.  As the number of relevant Feynman diagrams grows
rapidly with the
order in $\alpha_S$, the whole process cannot be accurately modelled in the
lowest order plus allowing initial and final state radiation.
The leading parton level subprocesses that are complete missed
in such approximative treatment are graphs leading to an additional
quark-jet in the final state, $pp \rightarrow t\bar{t}q$
\footnote{Following the common convention,
by $pp$ we always mean the sum of all the
corresponding interactions at the parton level, i.e., quark-quark, quark-gluon or
gluon-gluon, while the remnants of the protons are ignored.  Hence, e.g.,
$pp \rightarrow t\bar{t}$ does not mean baryon number violation.}.
These two classes
of events do not involve any double-counting.  
Their coherent sum reasonably reproduces the total cross sections at 7 or 8 TeV,
as measured in the LHC.
A more satisfactory description, developed especially for the study of inclusive
$t\bar{t}$ production as a background to VBF processes, is based on explicitly
considering three processes at the
tree level: $pp \rightarrow t\bar{t}$,
$pp \rightarrow t\bar{t}j$ and $pp \rightarrow t\bar{t}jj$ ($j$ denoting quarks and
gluons alike) plus initial and final
state radiation.  These processes represent the leading order contributing diagrams
of inclusive $t\bar{t}$ production for the cases where 2, 1 or 0 tagging jets arise from
$b$ quarks, respectively.  The three different topological configurations select
three mutually exclusive subsamples and thus double-counting is automatically avoided.
It is actually the latter two classes that define the amount of $t\bar{t}$
background.  The contribution from $pp \rightarrow t\bar{t}$ with both $b$ quarks
becoming tagging jets is minimal.  That the two abovementioned methods produce
consistent results for 14 TeV has been verified.

From completed simulation-based studies, that include also CMS-like detector
resolution effects, it can be inferred that the total
top production background falling within the kinematic phase space defined by
all the abovementioned signal selection criteria: basic VBF cuts, $t\bar{t}$
suppression cuts and the $R_{p_T}$ cut,
can be roughly parameterized as

\begin{equation}
B_{t\bar{t}} = 12~fb \cdot (1-\epsilon_{b-tag})^2 \cdot (1-\epsilon_{sign})
\cdot \epsilon_{CJV}.
\end{equation}

\noindent
Here the normalization factor includes the branching fractions of $W$ decaying
into electrons or muons and the proper selection efficiency, $\epsilon_{b-tag}$ is the
average efficiency of $b$-tagging,
$\epsilon_{sign}$ is the average efficiency of lepton charge reconstruction and
$\epsilon_{CJV}$ is the central jet veto factor, if applied.
For example, setting $\epsilon_{b-tag}$=0.5, $\epsilon_{sign}$=0.995 and
$\epsilon_{CJV}$=1,
as expected for the same-sign mode, one gets $B_{t\bar{t}} \sim$ 0.015 fb.
It should be noted that with the above numbers the predominant contribution indeed
comes from
charge misreconstruction.  Leptonic $B$ decays are suppressed by a combination
of kinematics and isolation criteria to much below this level.
Another subclass of the inclusive $t\bar{t}$ production background that affects the
same-sign mode is $W^+t\bar{t}$ production, where one lepton comes from $W$ decay,
another from top decay.  This background was shown negligible after applying
standard signal selection cuts.
For $\epsilon_{sign}\sim$0 and $\epsilon_{CJV}\sim$0.25, as is in the opposite-charge
mode, $B_{t\bar{t}} \sim$ 0.75 fb, and more stringent cuts on the lepton
transverse momenta can reduce this number by perhaps an additional factor 2
while leaving a major part of the signal intact.  It is clear at this point that,
unless heavy resonances are present within the reach of $\sqrt{s}$ = 14 TeV,
only $W^+W^+$ carries the potential of signal levels above background fluctuations
assuming luminosities measured in hundreds of inverse femtobarns.

The total $t\bar{t}$ cross section after each of the analysis cuts discussed
in the text is shown in Fig.~\ref{cuts2} (top plot).

\section{Modeling of the $W$+jets backgrounds}
\label{modwjets}

Jets misreconstructed as electrons are the primary source of these backgrounds.
The lowest order process of this kind that can mimick the signal is $W$ + 3 jets,
where in principle any of the three jets can be the fake electron.  The
kinematic
regime we are probing by applying the signal selection cuts strongly favors large-$p_T$
leptons.  As a direct consequence, events in which the leading jet (where, as usual, we
rank objects in a given class according to their $p_T$) is the one that gets
misreconstructed
make up over 90\% of all the cases of $W$+jets events falling kinematically within
the signal phase space.  The subleading jet as the fake electron accounts for just
about the rest of it.  For the same reasons, it is inessential
to consider additional samples with more than three jets at the generation level.
Given the large total cross section for $W$+jets at the LHC, the purity
of electron reconstruction is a crucial number.
The final amount of $W^+$+jets events mimicking the signal can be predicted
as being roughly

\begin{equation}
B_{W^++jets} = 5~pb \cdot \epsilon_{j-fake} \cdot f_{+/-},
\end{equation}

\noindent
in total,
where $\epsilon_{fake}$ is the overall probability of a jet being reconstructed as an
electron satisfying all the quality selection criteria and $f_{+/-}$ is
the sign matching factor.
For $\epsilon_{fake} \sim 1.1 \cdot 10^{-4}$ and $f_{+/-}$=0.27 ($W^+W^+$), this
gives 0.08 fb.  For $f_{+/-}$=0.73 (opposite-sign), it is
about 0.2 fb.  It is not a negligible number and, not so unexpectedly, it is
a more important background source than top production for the same-sign mode.
However, this background is bound to affect different final states differently.
Half of the total $B_{W+jets}$
is due in the $jjee$ final state, the other half in the $jje\mu$ final state
(where signal is twice the size of the $jjee$ signal) and no contribution is
possible to the $jj\mu\mu$ final state.

The $W^-$+jets background is typically a factor 2 lower due to the charge
asymmetry in $W$ production at the LHC.  We assume then additional contributions
of 0.1 fb for the opposite-sign and 0.04 fb for $W^-W^-$.

Another class of background is related to a fake electron being reconstructed
from a photon with an associated track.  The leading order process than can
generate such events is $Wjj\gamma$.  Its total cross sections is much lower 
than for $W$ + 3 jets and additional kinematic and combinatorial factors make
it in fact negligible.  This background is of the order of

\begin{equation}
B_{Wjj\gamma} = 0.18~fb \cdot \epsilon_{\gamma -fake}.
\end{equation}

\noindent
where $\epsilon_{\gamma -fake}$ is the overall probability of a photon misreconstructed
as an electron satisfying all the quality selection criteria and of the required charge.
For $\epsilon_{\gamma -fake} \sim$ 0.0035 we get $B_{Wjj\gamma} <$ 0.001 fb.

The total $W$+jets cross section after each of the analysis cuts discussed
in the text is shown in Fig.~\ref{cuts2} (second plot).

\section{Modeling of the QCD multijet background}
\label{modqcd}

The leading order background process of this kind is $jjjj$ with two of the four
jets misreconstructed as electrons.  Huge cross sections for QCD processes at
the LHC
compensate the low probability of having two simultaneous fakes and so this background
can prove overwhelming.  This result may seem surprising at first glance, but in fact
it is dictated by the very specific kinematic correlations we are looking for,
significantly different from the ones typically observed in regular gauge boson 
physics analyses done on the 7 and 8 TeV data.  
Again here, the kinematic regime we probe favors large $p_T$
and therefore fakes generated by the two leading jets account for 80-90\% of all
events satisfying the complete selection criteria, while the rest to a
sub-percent level comes from the combination of the first with the third jets
being reconstructed as fake electrons.  For the same reasons it is also here
inessential to consider higher order processes.
In order to render the QCD
multijet background manageable, we further assume the following combination
of cuts to be applied in the $jjee$ final state only:

\vspace{3mm}

\hspace{5cm} $MET >$ 60 GeV,

\hspace{5cm} $M_{ee} >$ 250 GeV,

\hspace{5cm} $p_T^{j_1} >$ 30 GeV.

\vspace{3mm}

\noindent
The meaning of the first two cuts is straightforward.  For the third cut,
note that here in most cases $j_1$ denotes really the {\it third} jet.
Extra cuts bring a substantial reduction of the background while keeping 70\%
of the $jjee$ signal, or equivalently well over 90\% of the total signal.
In total,

\begin{equation}
B_{jjjj} = 6.5~nb \cdot \epsilon_{j-fake}^2 \cdot f_{+/-},
\end{equation}

\noindent
where $\epsilon_{j-fake}$ is the probability of a jet being reconstructed as an
electron satisfying all the quality selection criteria
and $f_{+/-}$ is a combinatorial factor equal to 0.25
for each same-sign mode and 0.5 for opposite-sign.  By assuming 
$\epsilon_{j-fake} \sim 10^{-4}$
we end up at $B_{jjjj} \sim$ 0.016 fb for same-sign and 0.032 fb for opposite-sign, 
still not negligible numbers.  However, knowing that this background only concerns
the $jjee$ final state, as does the $W$+jets background concern the
$jjee$ and $jje\mu$ final states in fixed proportions, comparison
of the selected event yields will be an additional tool to disentangle the
various background sources and isolate the signal (provided enough statistical
power).
Yet another piece of valuable information will be provided by the study of the
$W^-W^-$ mode.

The total QCD multijet cross section after each of the analysis cuts discussed
in the text is shown in Fig.~\ref{cuts2} (third plot).

\section{$WZ$ and $ZZ$ as backgrounds to $WW$}

Several previous analyses, in particular the ones by Chanowitz et al.~\cite{chanowitz}, 
hinted on
the possibility that continuum $WZ$ production with one lepton which escaped detection,
could be as well an additional significant background to $W^\pm W^\pm$.  
The subject was brought up again in the recent analyses by ATLAS \cite{atlasww8}
and CMS \cite{cmsww8}.
The validity of this assertion strongly depends on the applied selection criteria.
In our case, the amount of remaining $W^+Z$ background with at least
two associated jets after cuts gets reduced to
about 0.04 fb altogether, i.e., regardless of whether the negatively charged lepton
from $Z$ decay gets reconstructed or not.  The geometrical condition of this third
lepton falling outside of the accepted pseudorapidity range of $|\eta|<$2.1,
translates into a further suppression by more than an order of magnitude,
to about 0.003 fb, and this number gives the final
estimate of the $W^+Z$ background in the analysis of $W^+W^+$.  For $W^-W^-$, the
relative contamination from $W^-Z$ is about a factor 2 larger from pure
combinatorics.  Similar is the $WZ$ contamination to $W^+W^-$, here
however both $W^+Z$ and $W^-Z$ can contribute.
Signal from $W^\pm Z$, if any, eventually adding up to signal from $WW$ is
of course a bonus rather than a problem.

\begin{figure}[htbp]
\begin{center}
\epsfig{file=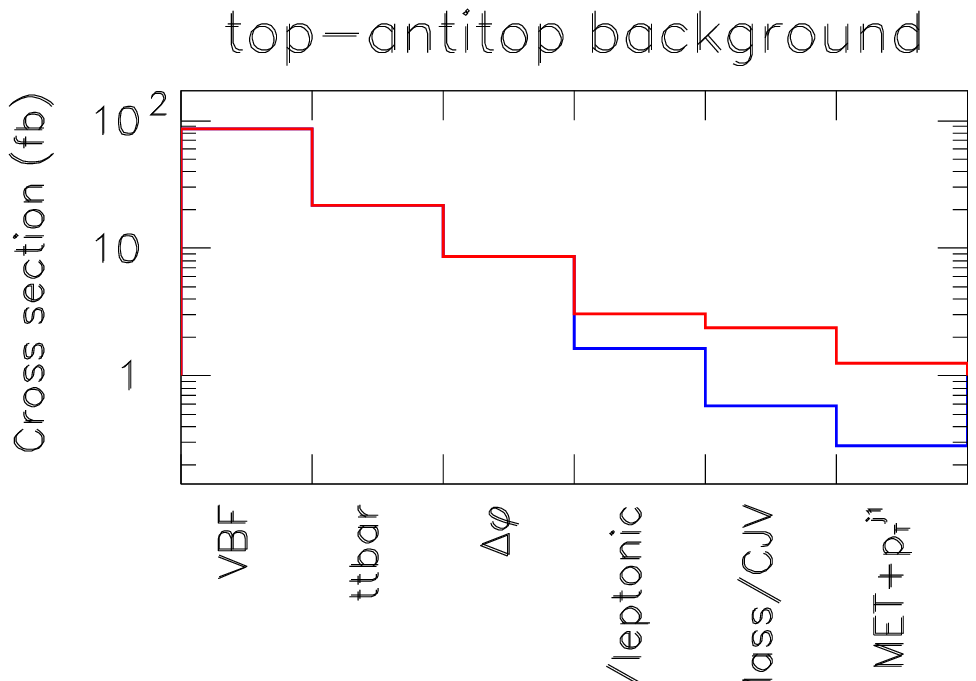,width=0.5\linewidth}
\epsfig{file=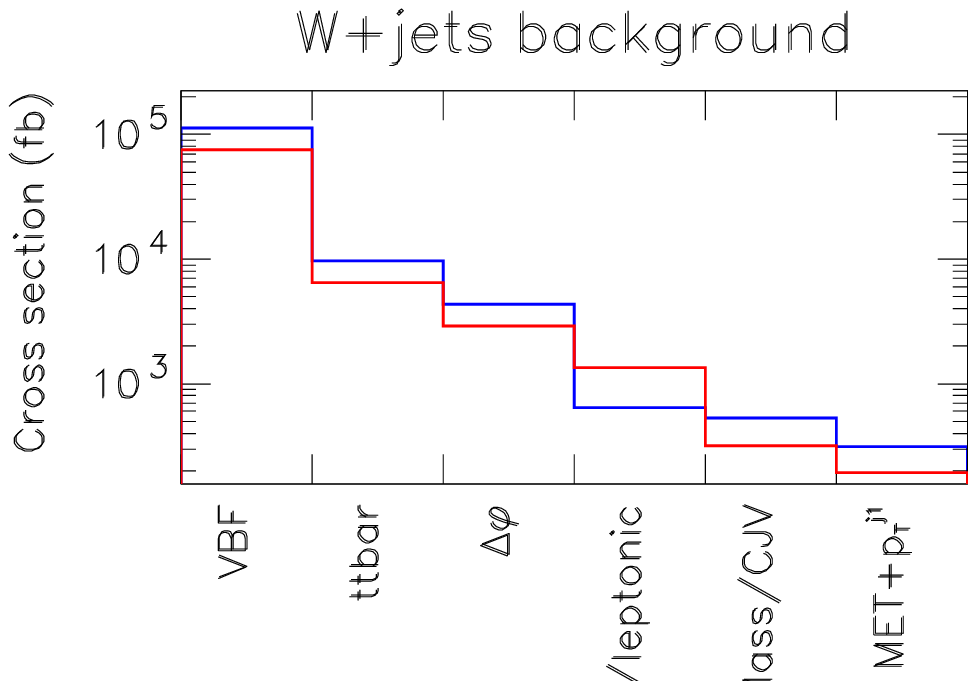,width=0.5\linewidth}
\epsfig{file=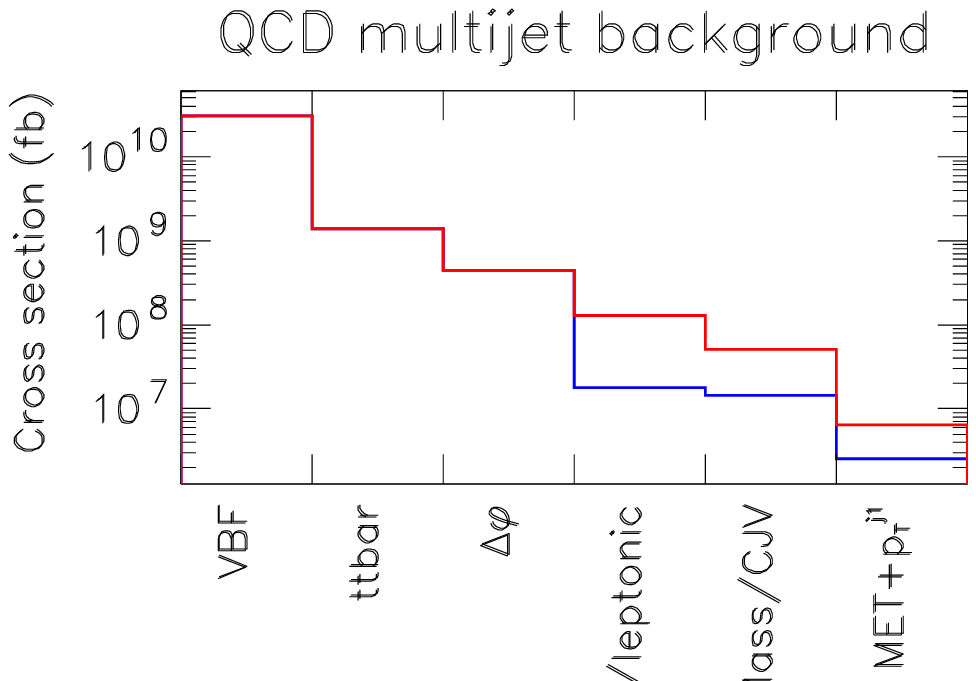,width=0.5\linewidth}
\epsfig{file=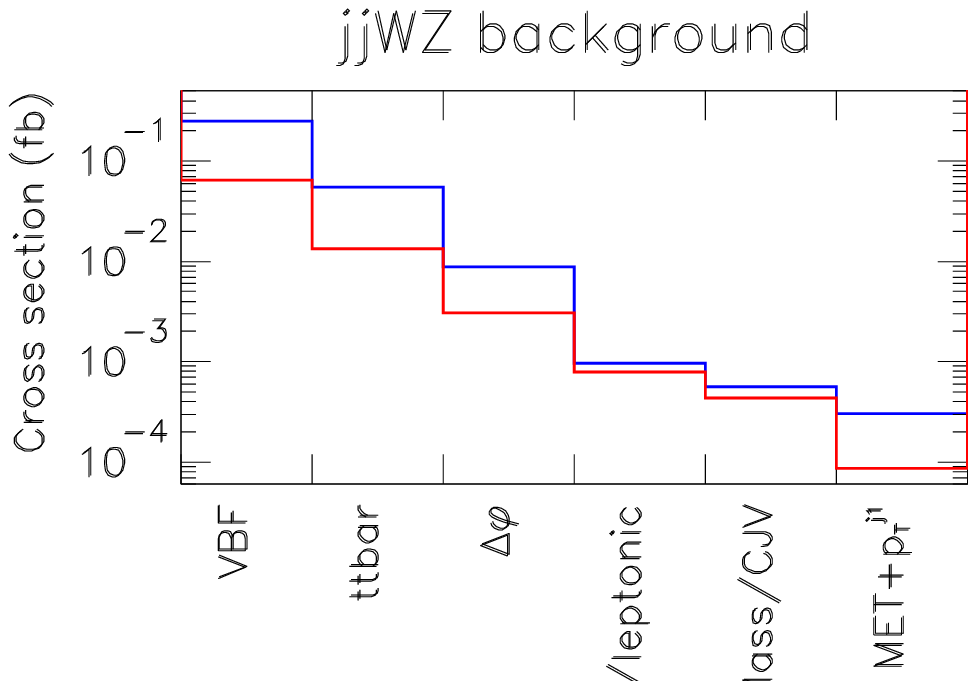,width=0.5\linewidth}
\end{center}
\vspace{-2mm}
\caption{{\small Total cross sections for the different main kinds of reducible
background: inclusive $t\bar{t}$, $W$+jets,
QCD multijet and $WZ$,
after each subsequent class of cuts proposed in the analysis.  
Red histograms are for
$pp \rightarrow jjW^+W^+$, blue histograms for $pp \rightarrow jjW^+W^-$;
in both cases $W$ decay into muons is assumed.  Results reflect pure
event kinematics, all detector efficiencies and inefficiencies (where
appropriate) are assumed 100\%.  For details regarding this calculation
and the precise meaning of cut labels see caption of Fig.~\ref{cuts1}.
}}
\label{cuts2}
\end{figure}

The total $WZ$ cross section after each of the analysis cuts discussed
in the text is shown in Fig.~\ref{cuts2} (bottom plot).
Contaminations from $ZZ$ are still smaller.

\section{$WZ$ and $ZZ$ as signals}
\label{wzzzsig}

To estimate the amount of BSM signal for the $WZ$ and $ZZ$ processes, we
follow existing literature on the subject,
and the work of Bagger et al.~\cite{bagger} in particular.  We can recall and confirm
here some of their most elaborated and relevant signal selection criteria that were
shown to exploit specific kinematic features of each of these processes in order to
enhance S/B.  In addition to requiring standard VBF topology and applying cuts
against inclusive $t\bar{t}$ background (for $ZZ$ only a cut on $M_{jj} > 500 GeV$
applies), the process specific cuts are the following.  For $WZ$:

\begin{itemize}

\item $M_Z - 10$ GeV $< M_{l^+l^-} < M_Z + 10$ GeV,

\item $M_T(WZ) > 500$ GeV,

\item $p_T^Z > \frac{1}{4} M_T(WZ)$,

\item $MET > 50$ GeV,

\item $p_T^l > 40$ GeV.

\end{itemize}

\noindent
For $ZZ \to 4l$:

\begin{itemize}

\item $M_Z - 10$ GeV $< M_{l^+l^-} < M_Z + 10$ GeV for both lepton pairs,

\item $M_{4l} > 500$ GeV,

\item $p_T^Z > \frac{1}{4} \sqrt{M_{4l}^2 - 4M_Z^2}$ for each $Z$,

\item $p_T^l > 40$ GeV.

\end{itemize}

\noindent
For $ZZ \to l^+l^-\nu\nu$:

\begin{itemize}

\item $M_Z - 10$ GeV $< M_{l^+l^-} < M_Z + 10$ GeV,

\item $M_T(ZZ) > 500$ GeV,

\item $p_T(ll) > \frac{1}{4} M_T(ZZ)$,

\item  $MET > 250$ GeV,

\item $p_T^l > 40$ GeV.

\end{itemize}

\noindent
In the above, $M_Z$ is the PDG $Z$ mass, while all other symbols refer to
reconstructed quantities.  The transverse masses are defined as follows:

\begin{equation}
M_T^2(WZ) = [\sqrt{M^2(lll)+p_T^2(lll)}+MET]^2-[\vec{p_T}(lll)+\vec{MET}]^2,
\end{equation}

\begin{equation}
M_T^2(ZZ) = [\sqrt{M_Z^2+p_T^2(ll)}+\sqrt{M_Z^2+MET^2}]^2-[\vec{p_T}(ll)+\vec{MET}]^2.
\end{equation}

\noindent
Background is expected to be dominated by irreducible SM background for $ZZ$
and additionally $Zt\bar{t}$+jets production for $WZ$.
Under these assumptions and applying the cuts described above, background levels
amount approximately to 0.027 fb for $W^\pm Z$, 0.003 fb for
$ZZ \to 4l$ and 0.009 fb for $ZZ \to 2l2\nu$.  Higgsless signals would be of
the order of 0.009 fb, 0.005 fb and 0.012 fb, respectively.
For $WZ$, background is relatively large and its kinematic separation from the
signal, if by the latter we understand non-SM Higgs couplings, is marginal.  
This forces to use
strict selection criteria which in turn would require very high luminosity to
be successfully applied.  The $ZZ$ modes are relatively clean, especially
the $4l$, but clearly suffer of low statistics.  

\begin{figure}[htbp]
\begin{center}
\epsfig{file=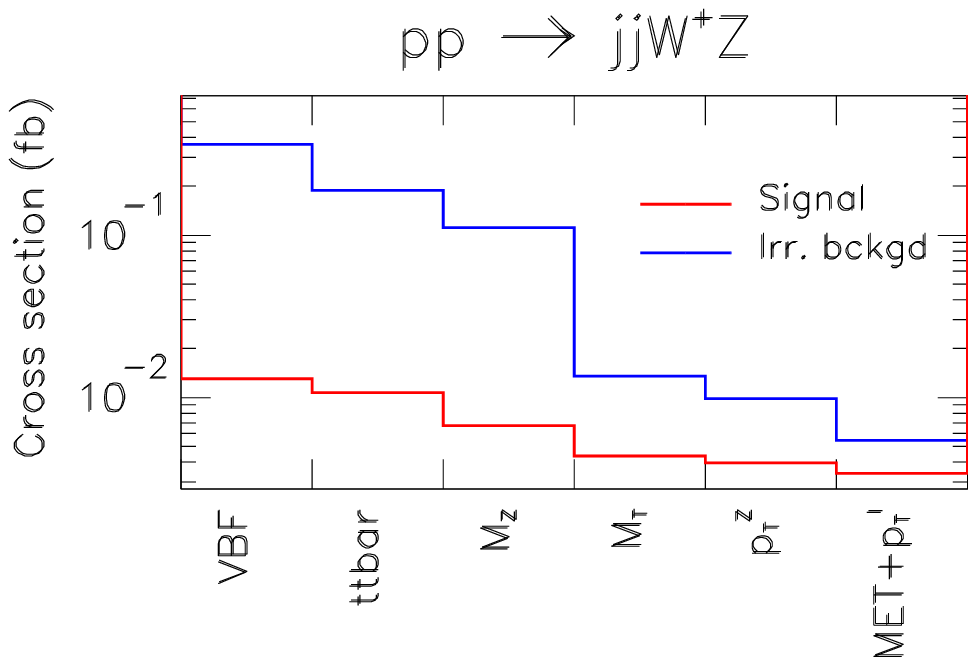,width=0.6\linewidth}\vspace{-8mm}
\epsfig{file=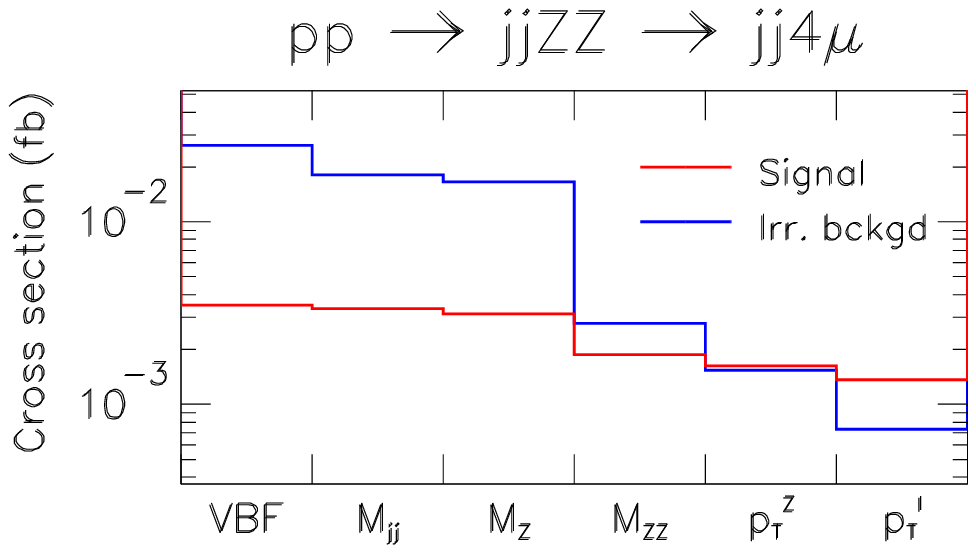,width=0.6\linewidth}\vspace{-8mm}
\epsfig{file=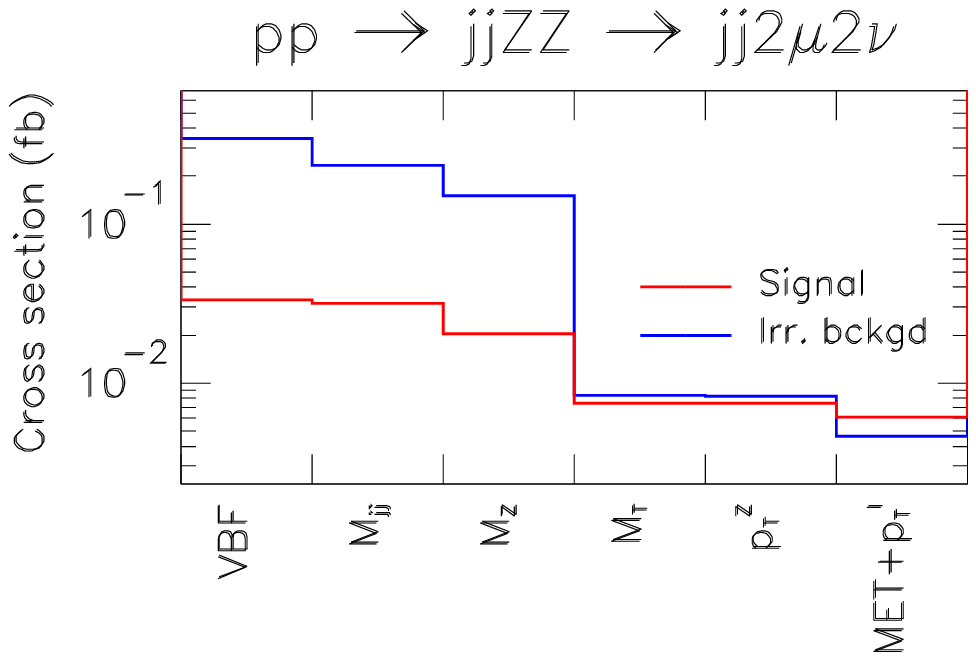,width=0.6\linewidth}\vspace{-6mm}
\end{center}
\vspace{-5mm}
\caption{Total cross sections for the signal (red histograms) and irreducible
background (blue histograms) after each subsequent class of cuts proposed in the
analysis for the processes $pp \rightarrow jjW^+Z \rightarrow jj\mu^+\mu^+\mu^-$ (top),
$pp \rightarrow jjZZ \rightarrow jj\mu^+\mu^-\mu^+\mu^-$ (middle) and
$pp \rightarrow jjZZ \rightarrow jj\mu^+\mu^-\nu\nu$ (bottom) at 14 TeV.
The signal is calculated under the Higgsless hypothesis.
The meaning of the cut labels is the following: 
$\bullet$ {\bf VBF:} $2<|\eta_j|<5$, $\eta_{j1}\eta_{j2}<0$ and $|\eta_\mu|<2.1$,
$\bullet$ {\bf ttbar:} $M_{jj} > 500$ GeV and $M_{j1\mu 2}$, $M_{j1\mu 2} >$ 200 GeV,
$\bullet$ {\bf $M_{jj}$:} $M_{jj} > 500$ GeV,
$\bullet$ {\bf $M_Z$:} reconstructed $Z$ mass(es) within 10 GeV,
$\bullet$ {\bf $M_T$:} transverse mass (defined in detail in section \ref{wzzzsig}) 
$> 500$ GeV,
$\bullet$ {\bf $M_{ZZ}$:} $M_{4\mu} >$ 500 GeV,
$\bullet$ {\bf $p_T^Z$:} $p_T^Z > \frac{1}{4} M_T$ for $jjW^+Z$ and $jjZZ \to jj2l2\nu$
or $> \frac{1}{4}
\sqrt{M^2_{4\mu} - 4 M^2_Z}$ for $jjZZ \to jj4l$,
$\bullet$ {\bf MET:} missing $E_T > 50$ GeV for $jjW^+Z$ or 250 GeV for $jjZZ$,
$\bullet$ {\bf $p_T^l$:} $p_T^l > 40$ GeV.
Results of a MadGraph calculation,
processed by PYTHIA 6 for $W$ decay into muons ($jjZZ$ samples only), the
effects of parton showering, hadronization and jet reconstruction
and further processed by PGS 4 for the effects of finite resolution in
the measurement of jet and muon $p_T$ in a CMS-like detector.
}
\label{cuts3}
\end{figure}

The total cross sections of the signal and irreducible background for
$WZ$ and $ZZ$  after each of the analysis cuts discussed
in the text are shown in Fig.~\ref{cuts3}.
In the event of absence of new
heavy resonances within reach, these processes are unlikely to improve our
knowledge of the Higgs sector.

\section{Key uncertainties}

A phenomenological analysis based on
signal and background calculations done by matrix element generators at the tree level
is affected by specific uncertainties.  These come partly from theory itself and partly
from imperfect knowledge of detector related effects.  A detailed analysis of
all the systematic erros is rather inessential at this point, but we can outline
the most important limitations to the accuracy of our predictions.  Not
accidentally, some of them will translate into the limiting factors at the time of
carrying the real measurement.

Total cross sections
for proton-proton processes calculated in a given order in perturbative expansion
are sensitive to the choice of
such things as the set of parton distribution functions (PDF's) and the QCD factorization
and renormalization scales.  Typically, the choice of PDF's by itself does not change
numerical results by more than 5\%.  The factorization scale corresponds
to the resolution at which the proton is being probed.  When calculated to all
orders in perturbative QCD, the hadronic cross section is independent
of the scale.  But at any finite order it must depend logarithmically on it
\cite{factscale}.
Moreover, the dependence is usually significant at low orders in perturbation
theory.  The way to obtain a reliable prediction is to calculate
higher-order corrections until the factorization scale dependence is reduced.
It was shown that calculations of diboson production in the vector boson scattering
configuration carried at the next-to-leading order (NLO) in QCD are very weakly
dependent on the scale.  The residual uncertainty is of 2.5\% in a typical VBF
kinematics (for $W^+W^+$).  Meanwhile, results of leading order (LO) calculations
can be made coincide with the former by a choice of the factorization scale equal to
the momentum transfer of the $t$-channel electroweak boson \cite{jager}.  This solution
has recently been implemented as an option in MadGraph 5.
Deviations induced by setting the scale to a fixed value, e.g., the $Z$ mass are
of order of 10\%.  Even more sophisticated recipes are currently devised \cite{newzepp}.
These will allow further reduction of scale related uncertainties for future
studies and data analyses.

Furthermore,
a key problem in making precise perturbative QCD predictions is to set the proper
renormalization scale of the running coupling.  A poor choice of the renormalization
scale can manifest itself as a strong dependence on the ratio of the
NLO cross section to the LO cross section (the so called K-factor) \cite{renoscale}.
As our process of interest is of purely electroweak nature, the signal predictions
are affected only via uncertainties in the modeling of parton hadronization
and final state radiation.  In studies that do not involve detailed detector
simulation, these are anyway dwarfed by imperfect modeling of
jet reconstruction procedures, calorimeter efficiencies and resolutions.
In the prediction of QCD-related background, variations depending on the scale
can easily amount to 30\% without dedicated hard work.  In our case we take clear
advantage of the fact that these backgrounds are expected to be small after all
cuts.  For inclusive $t\bar{t}$ production we have also an experimental ansatz
since the total cross sections have been measured at 7 and 8 TeV by both ATLAS
and CMS.  It is in any case the inclusion of $t\bar{t}j$ and $t\bar{t}jj$ processes
in the first place that ensures the total cross sections are consistent with the
measurements within the errors of the latter.  These errors are of the order
of 5-10\%.

The overall smallness of background in the same-sign process is an advantage
at the time of the measurement, but a relative disadvantage for phenomenological studies.
The background is difficult to predict because it depends
primarily on a combination of tiny detector effects
rather than physics calculable from first principles.  In case of $t\bar{t}$
this is not only knowledge of exact $b$-tagging efficiencies as a function of
jet $p_T$ and $\eta$, but as we saw in the same-sign $WW$ channel, the result
is mainly driven by the efficiency of lepton charge identification.  At the
present moment this is only taken into account as an order of magnitude estimate.
Surely, a small change in efficiency can produce a significant
effect on the S/B ratio.  Likewise, $W$+jets and QCD mulitjet backgrounds enter via
the tiny effects of jets misidentified as leptons.  These are strongly
detector- and software-dependent and only an order of magnitude
estimate can again be made at this point.  And in any case they must be considered
in simultaneous relation with the efficiency of lepton-ID for genuine leptons.
It is of little use to evaluate such effects in more detail on purely phenomenological
grounds.
However, we can at least define kinematic conditions under which the abovementioned
backgrounds are small enough that their precise values can be measured using
data-driven methods at the proper time, but will not jeopardize the entire analysis.
Ultimately,
exact background levels and compositions will
differ from experiment to experiment (in our case from ATLAS to CMS).

In all the numerical predictions that are presented in this section,
things like detector efficiencies, if only different
from unity by more than a few per cent, were taken into account by simply scaling
the final cross sections.
Basic detector resolutions \cite{cmsmureco} \cite{cmsereco} \cite{cmsjetreco}
can be simulated with dedicated simulation tools,
namely the PGS program \cite{pgs}.  
In general, PGS-level results are over 10\% lower than
PYTHIA-level (generator+hadronization) results.  Meanwhile, results in the electron decay
channels are consitent with those in muon decay channels, from which they differ
only in the assumed resolution, to a few per cent and hence this number can be used
as a rather conservative upper limit of the resolution-related uncertainty.
Differences between jet reconstruction algorithms and the effects of choosing
a cone/cluster size parameters $R$ are negligibly small.
We recall that although the current standard jet definition
used, e.g., in CMS is set by the anti-$k_T$ alghorithm \cite{antikt},
which is not an available
option in PYTHIA 6, but has been implemented in PGS 4, it is likely to be changed
in the future for specific analyses in order to improve $W$ tagging in hadronic decays.
For the purpose of the studies presented here, either the $k_T$ or a simple cone
algorithm \cite{jetalgos} with $R=0.5$ was used.
High pile-up conditions may additionally
degrade the efficiency of tag-jet selection.  The issue of lowering the tag-jet
$p_T$ threshold, important from the point of view of $W_L$ selection, requires
a dedicated study within a fully realistic pile-up simulation.  Such study is
currently under preparation.  Triggering
efficiencies are not taken into account anywhere in this study, but purely
leptonic decay modes are obviously advantageous in this respect.  They do not
require any dedicated VBF trigger based on hadronic signals.  Instead, triggering
on a single lepton should be quite enough and the fact that any of the two
required leptons may fire the trigger makes trigger efficiency a minor issue.
Since we want high-$p_T$ leptons, trigger thresholds should not be a problem,
either.

\section{Higgs couplings in $VV$ scattering}

\begin{figure}[htbp]
\begin{center}
\epsfig{file=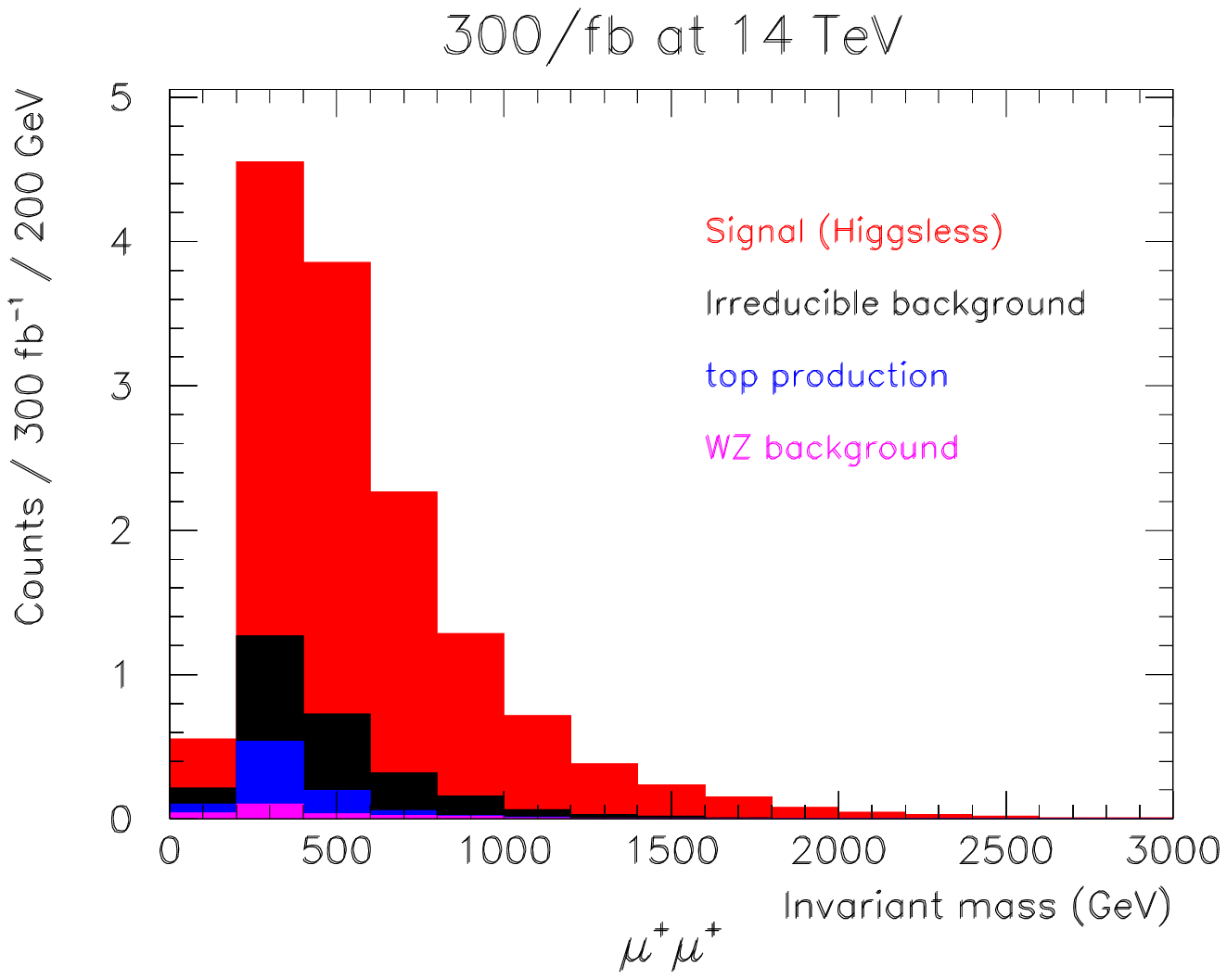,width=0.52\linewidth}\hspace{-1cm}
\epsfig{file=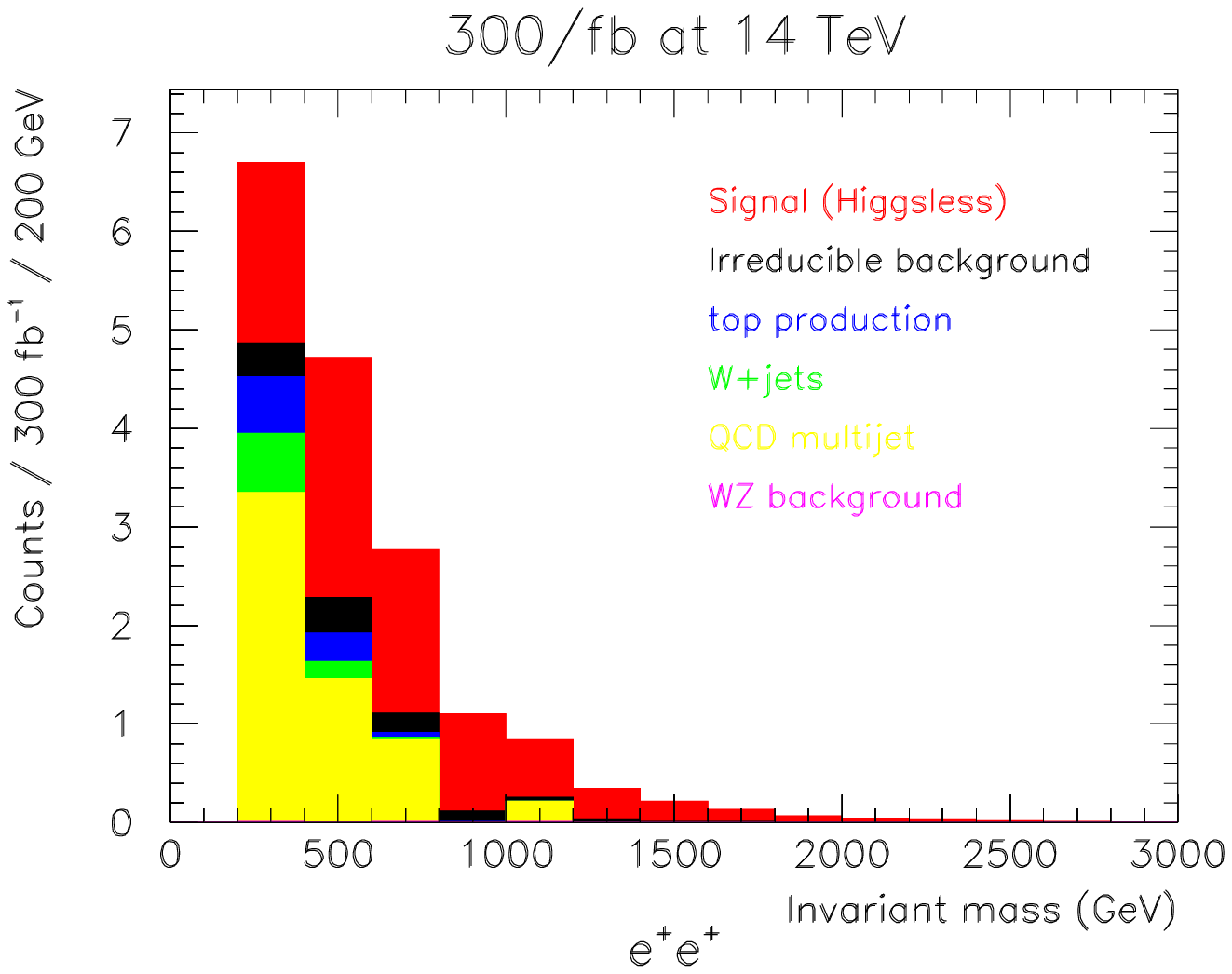,width=0.52\linewidth}
\end{center}
\caption{Invariant mass distributions of the two leptons resulting from
the process $pp \rightarrow jjW^+W^+$ at 14 TeV, with $W^+ \rightarrow \mu^+\nu$
(left) and with with $W^+ \rightarrow e^+\nu$ (right).
Shown are the signal calculated under the Higgsless hypothesis and various
contributions to the background, normalized to 300/fb of data.
Applied were respectively all the signal selection criteria foreseen for the
same-sign muon
channel (cuts 1-4 from Fig.~\ref{cuts1}, up to and including $R_{p_T}$)
and for the same-sign electron channel (all cuts 1-6 listed in Fig.~\ref{cuts1}).
Signal was calculated by subtracting 
the SM $jjW^+W^+$ sample (by definition also identical with irreducible background)
from the Higgsless $jjW^+W^+$ sample.  The top production background
was simulated as described in section~\ref{modttbar}.  The $WZ$ background was
obtained from a dedicated $jjW^+Z$ sample with subsequent $W$ and $Z$
decays into muons.
The $b$-tagging efficiency was assumed 50\% for a single $b$ quark,
for the muon charge mis-ID probability a constant value of 0.3\% was taken,
all other efficiencies and purities were assumed 100\%.
The $W$+jets background was deduced from dedicated $jjjW^+$ and $jjW^+\gamma$ samples,
where either any of the jets or the photon was assumed to be misidentified as an
electron.
The QCD multijet background was deduced from a dedicated $jjjj$ sample
where any pair of jets was assumed to be simultaneously misidentified as electrons.
The probability of a jet faking an electron was assumed $1.1\cdot 10^{-4}$
with a further 27\% probability of sign matching and that for a photon 0.7\%
with a 50\% probability of sign matching.
For a more detailed explanation of the procedure see sections~\ref{modwjets} and
\ref{modqcd}.
The electron charge mis-ID probability
for the evaluation of the top background was assumed to be 1\%.
Results of MadGraph simulations, processed by PYTHIA 6 for
$W$ decay into leptons (top production only), the effects of
parton showering, hadronization and jet reconstruction
and further processed by PGS 4 for the effects of finite resolution in
the measurement of jet and lepton $p_T$ in a CMS-like detector.
The original PYTHIA 6 source code was modified to account for the
correct, polarization-dependent, angular distributions for the leptonic $W$ decays.
}
\label{jjw+w+1}
\end{figure}

\begin{figure}[htbp]
\begin{center}
\epsfig{file=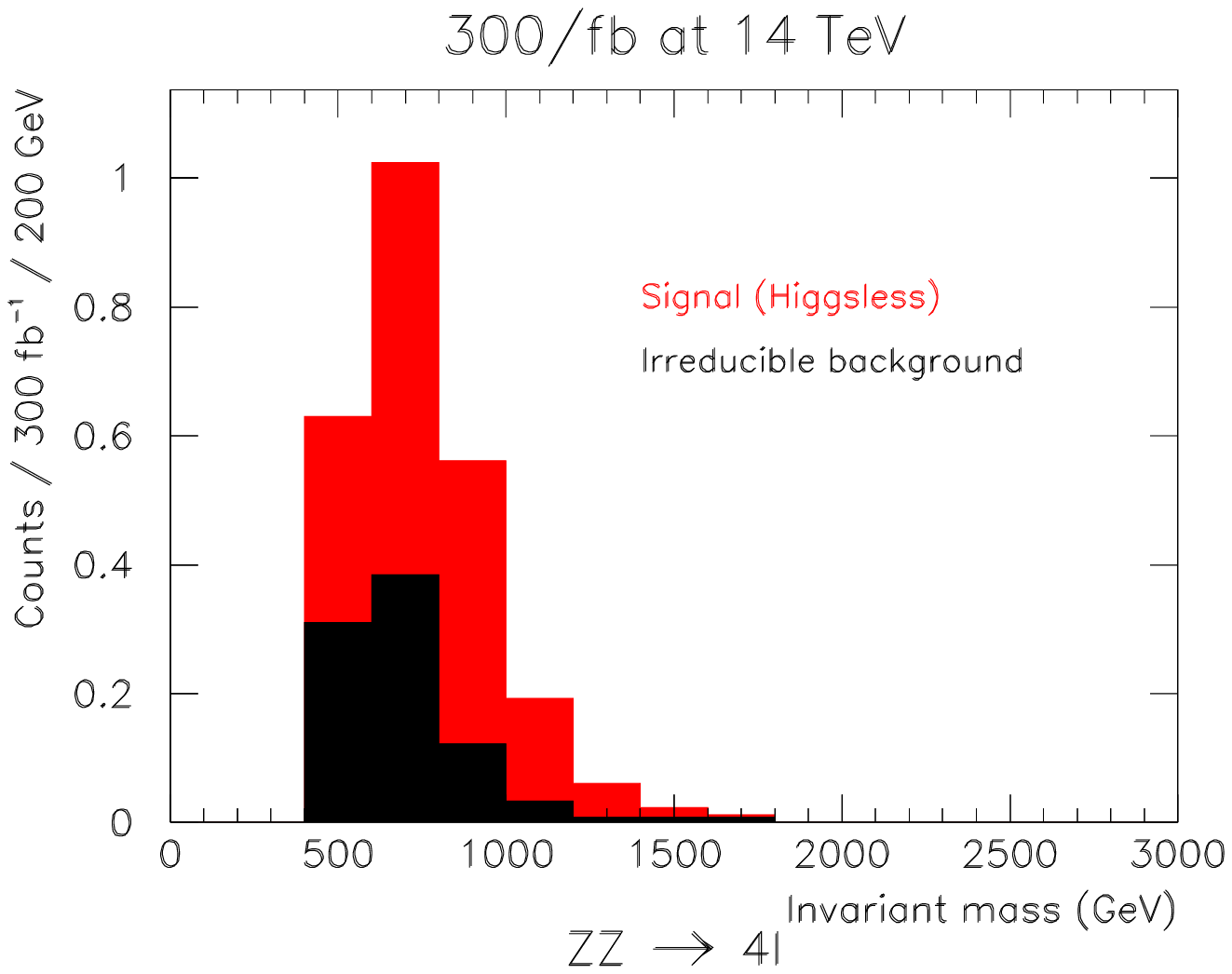,width=0.55\linewidth}
\epsfig{file=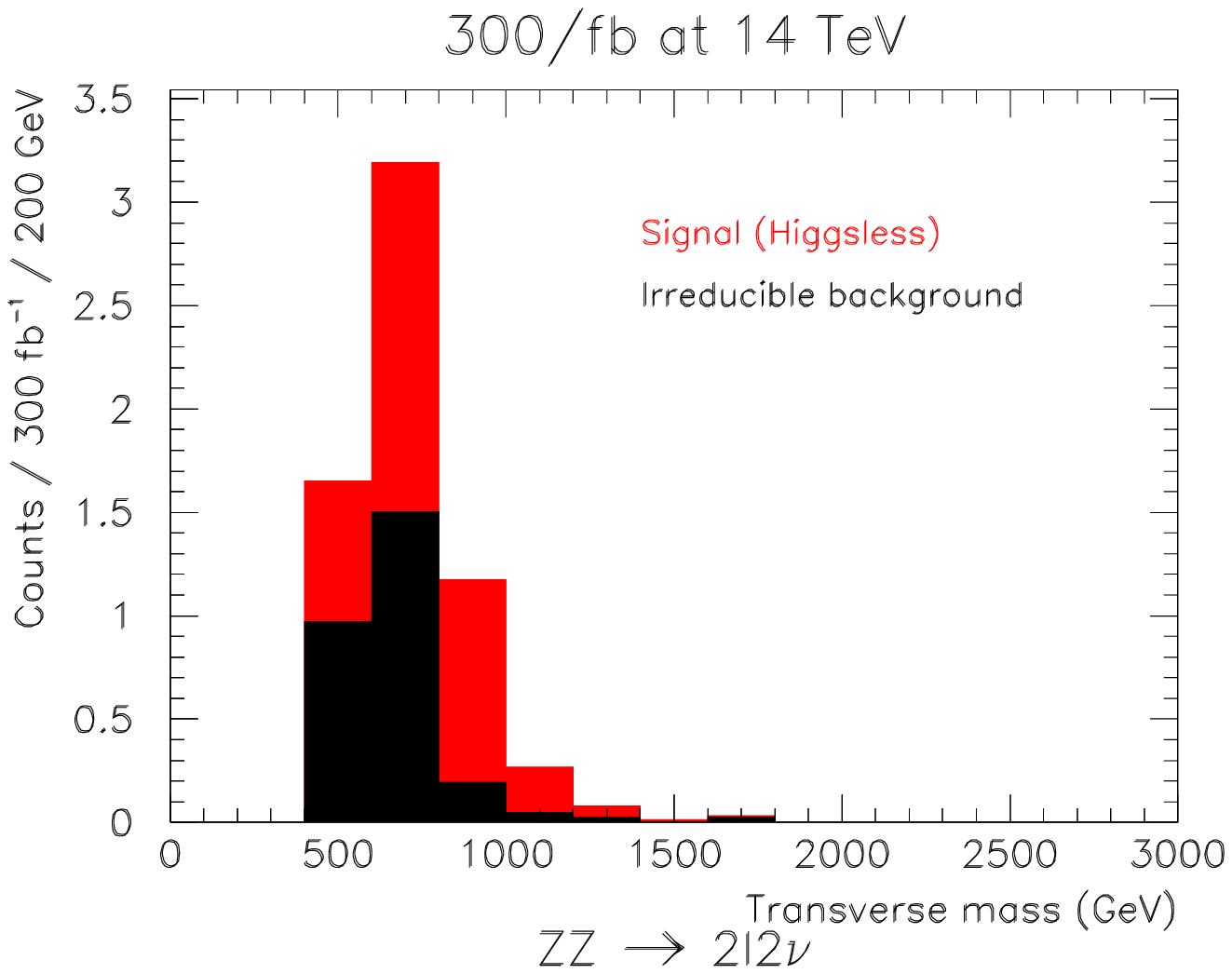,width=0.55\linewidth}
\end{center}
\caption{Invariant mass distrubution of the four leptons resulting from
the process $pp \rightarrow jjZZ \rightarrow jjl^+l^-l^+l^-$ (upper plot)
and transverse mass distribution (defined in in section \ref{wzzzsig}) from the process
$pp \rightarrow jjZZ \rightarrow jjl^+l^-\nu\nu$ (lower plot) at 14 TeV.
Shown are the signal calculated under the Higgsless hypothesis and the
irreducible background, normalized to 300/fb of data.
Applied were all the signal selection criteria foreseen for the $ZZ$ four-lepton
channel and for the $ZZ
\rightarrow l^+l^-\nu\nu$
channel, respectively; no distinction was made between lepton flavors (e or $\mu$).
Signal was calculated by subtracting
the SM $jjZ_LZ_L$ sample from the Higgsless $jjZ_LZ_L$ sample.
All the relevant efficiencies and purities were assumed 100\%.
Results of MadGraph simulations, processed by PYTHIA 6 for
$Z$ decay into leptons, the effects of
parton showering, hadronization and jet reconstruction
and further processed by PGS 4 for the effects of finite resolution in
the measurement of jet and lepton $p_T$ in a CMS-like detector.
This study did not include the correct, polarization-dependent, angular 
distributions for the leptonic $Z$ decays.  Such effects are nonetheless unlikely
to change any of our conclusions.
}
\label{jjzz}
\end{figure}

The lepton-lepton invariant mass or transverse mass spectra (where applicable)
of the signal and backgrounds,
after aplying all the selection criteria, are shown for all the $VV$ scattering
processes in Figs.~\ref{jjw+w+1}
thru \ref{jjzz}.  Respective signals were calculated within the Higgsless SM
scenario.
Assuming optimistically the $HWW$ coupling be 0.7, 0.8 or 0.9 of its SM value,
which is still not ruled out by experiment, for $W^+W^+$ in the purely leptonic decay
we can hope for a signal size of the order
of 0.040, 0.020 or 0.008 fb, respectively, after all selection criteria;
similar for $W^+W^-$, and about a quarter of that for $W^-W^-$.  Total background levels
may amount to 0.1 fb, 1.1 fb and 0.07 fb, respectively.
Sticking to $W^+W^+$ alone, this means roughly 12, 6 or 2 signal events after
collecting 300 fb$^{-1}$ of data over 30 Standard Model events.
In terms of anomalous Higgs-to-gauge couplings and having in mind the present
experimental bounds derived from Higgs measurements from LHC Run 1, it already
looks unlikely that $WW$ scattering could provide a quantitative measurement on its
own right.  
In order to observe frail hints of BSM anytime before the LHC
enters in its High Luminosity regime (2025), it will be necessary to combine different
processes and different decay channels.  
Nonetheless, consistency cross checks to $\sim$20\% with precision
measurements of
Higgs production rates and decays will certainly be attainable and they should still
be considered an important part of the physics program for LHC Runs 2 and 3.
Variations of the $HWW$ coupling of less
than 20\% will only be accessible with 3000 fb$^{-1}$ of data.
And of course, the more the Higgs boson appears SM-like, the more confined gets
$WW$ scattering to the role of a consistency cross check with limited precision,
as opposed to a true BSM search.  

In the $ZZ$ channel, the primary focus will be direct search for new resonances.
On the absence of such, BSM signal arising from a scaled $HZZ$ coupling
may consist of a handful of events even after 3000 fb$^{-1}$.  
Not unexpectedly, the $l^+l^-\nu\nu$ final state offers in principle more statistics than
$4l$, but is harder to analyze.  Here, however, special effort is required
to include the semileptonic decays into the game.  Under strict requirements of two
tagging jets in the endcaps, two additional jets in the barrel that reproduce the
$Z$ mass and no additional jet activity, QCD background levels may turn out controlable.
Dedicated simulations are missing at the present moment.

The $WZ$ channel probes in principle both $HWW$ and $HZZ$ couplings in a combined
way, but its sensitivity is marginal.  BSM signal levels are insufficient to be
measurable even with 3000 fb$^{-1}$.

\begin{figure}[htbp]
\begin{center}
\epsfig{file=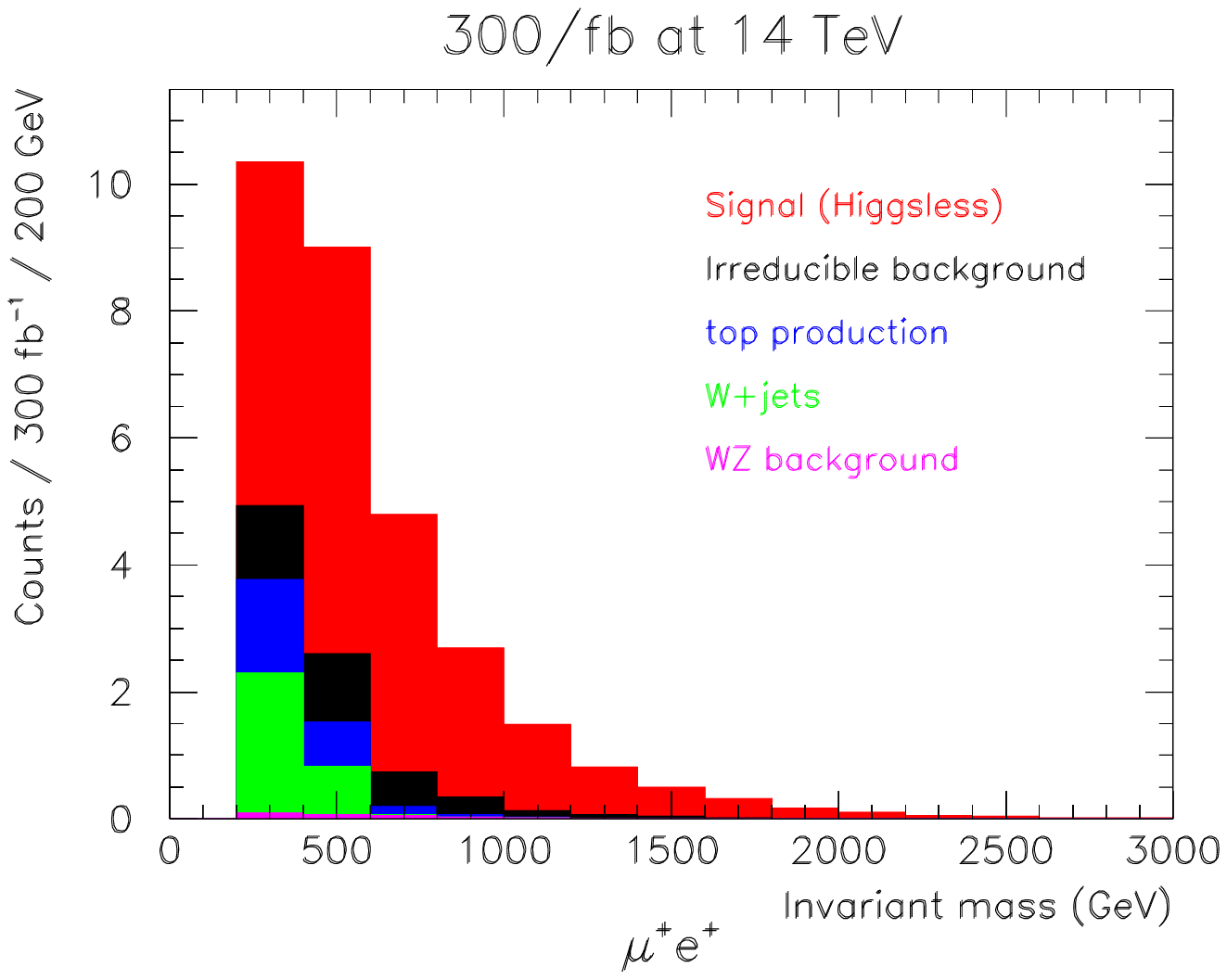,width=0.52\linewidth}\hspace{-1cm}
\epsfig{file=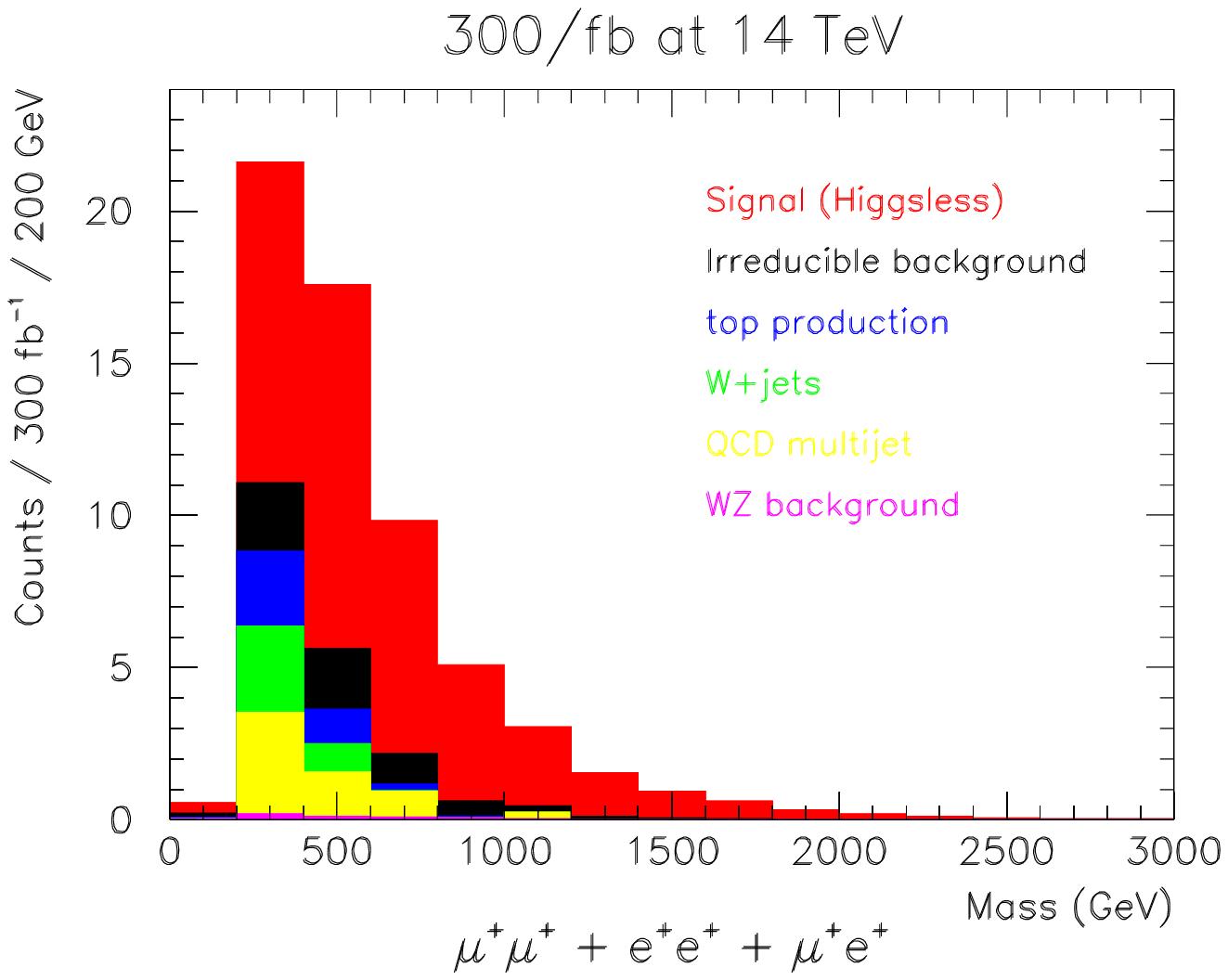,width=0.52\linewidth}
\end{center}
\caption{Invariant mass distributions of the two leptons resulting from
the process $pp \rightarrow jjW^+W^+$ at 14 TeV,
with one $W$ decaying into
a muon and another decaying into an electron (left) and with each $W$
decaying either into a muon or an electron (right).
Shown are the signal calculated under the Higgsless hypothesis and various
contributions to the background, normalized to 300/fb of data.  For the left plot,
applied were all the signal selection criteria foreseen for the
same-sign mixed muon+electron channel (cuts 1-5 listed in Fig.~\ref{cuts1}).
The $W$+jets background was deduced from dedicated $jjjW^+$ and $jjW^+\gamma$ samples,
where either any of the jets or the photon was assumed to be misidentified as an
electron.
The electron charge mis-ID probability
for the evaluation of the top background was assumed to be 1\%.
The final $\mu^+e^+$ kinematics was deduced by averaging out the distributions
obtained in $\mu^+\mu^+$ and the $e^+e^+$ channels, which differed only
due to different detector resolution effects assumed during processing by PGS 4.
All the remaining procedures and assumptions were identical as described in the
caption of Fig.~\ref{jjw+w+1}.
The right plot was obtained by summing up the individual $W$ decay channels.
}
\label{jjw+w+2}
\end{figure}

\begin{figure}[htbp]
\begin{center}
\epsfig{file=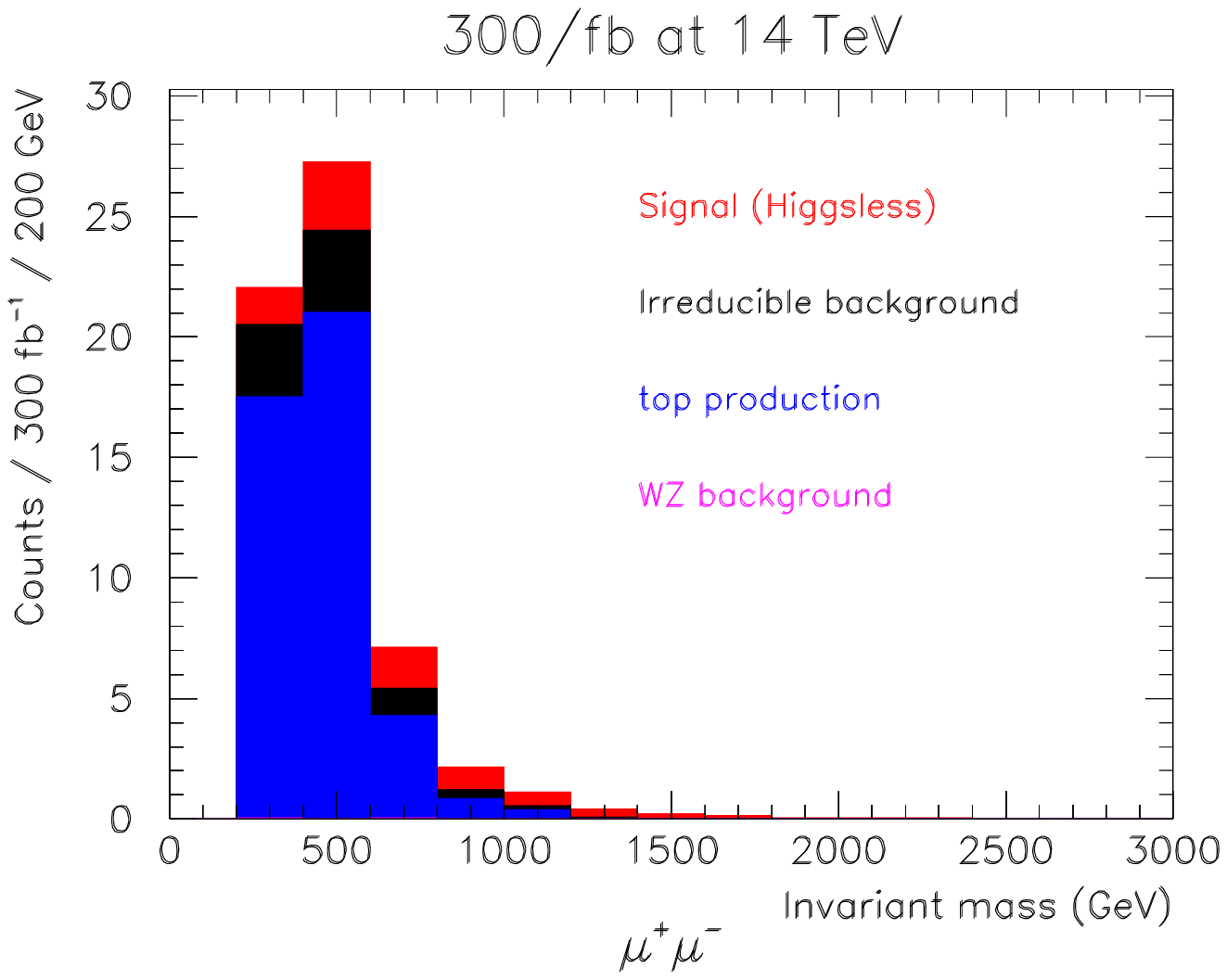,width=0.52\linewidth}\hspace{-1cm}
\epsfig{file=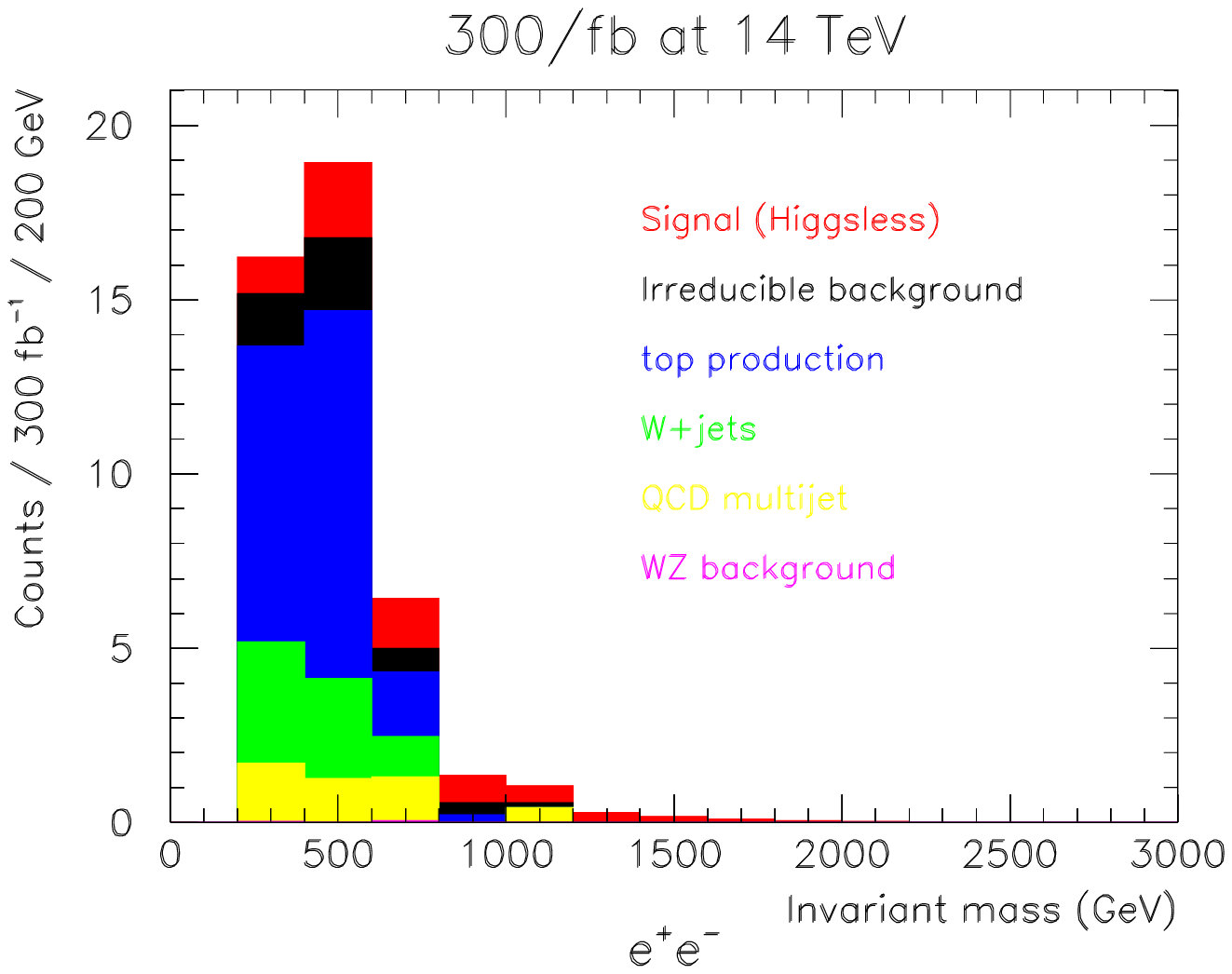,width=0.52\linewidth}
\end{center}
\caption{Invariant mass distributions of the two leptons resulting from
the process $pp \rightarrow jjW^+W^-$ at 14 TeV, with $W^\pm \rightarrow \mu^\pm\nu$
(left) and $W^\pm \rightarrow e^\pm\nu$ (right).
Shown are the signal calculated under the Higgsless hypothesis and various
contributions to the background, normalized to 300/fb of data.
Applied were respectively all the signal selection criteria foreseen for the
opposite-sign muon
channel (cuts 1-5 from Fig.~\ref{cuts1}, up to and including CJV) and
for the opposite-sign electron channel (all cuts 1-6 listed in Fig.~\ref{cuts1}).
Signal was calculated by subtracting a SM $jjW_L^+W_L^-$ sample
from a Higgsless $jjW_L^+W_L^-$ sample.  Irreducible background was calculated
from a SM $jjW^+W^-$ sample.  The top production background
was simulated as described in section~\ref{modttbar}.  The $WZ$ background was
obtained from a dedicated $jjW^+Z$ sample with subsequent $W$ and $Z$
decays into muons and an additional factor 1.5 was assumed to account for
$jjW^-Z$ (not simulated).
The $b$-tagging efficiency was assumed 50\% for a single $b$ quark,
all other efficiencies and purities were assumed 100\%.
The $W$+jets background was deduced from a dedicated $jjjW^+$ sample,
where any of the jets was assumed to be misidentified as an
electron.  An additional factor 1.5 was assumed to account for $jjjW^-$
(not simulated).
The QCD multijet background was deduced from a dedicated $jjjj$ sample
where any pair of jets was assumed to be simultaneously misidentified as electrons.
The probability of a jet faking an electron was assumed $1.1\cdot 10^{-4}$
with a further 73\% probability of sign matching.
For a more detailed explanation of the procedure see sections~\ref{modwjets} and
\ref{modqcd}.
Results of MadGraph simulations, processed by PYTHIA 6 for
$W$ decay into leptons (for signal, irreducible background and top production),
the effects of parton showering, hadronization and jet reconstruction
and further processed by PGS 4 for the effects of finite resolution in
the measurement of jet and lepton $p_T$ in a CMS-like detector.
The original PYTHIA 6 source code was modified to account for the
correct, polarization-dependent, angular distributions for the leptonic $W$ decays.
}
\label{jjw+w-1}
\end{figure}

\begin{figure}[htbp]
\begin{center}
\epsfig{file=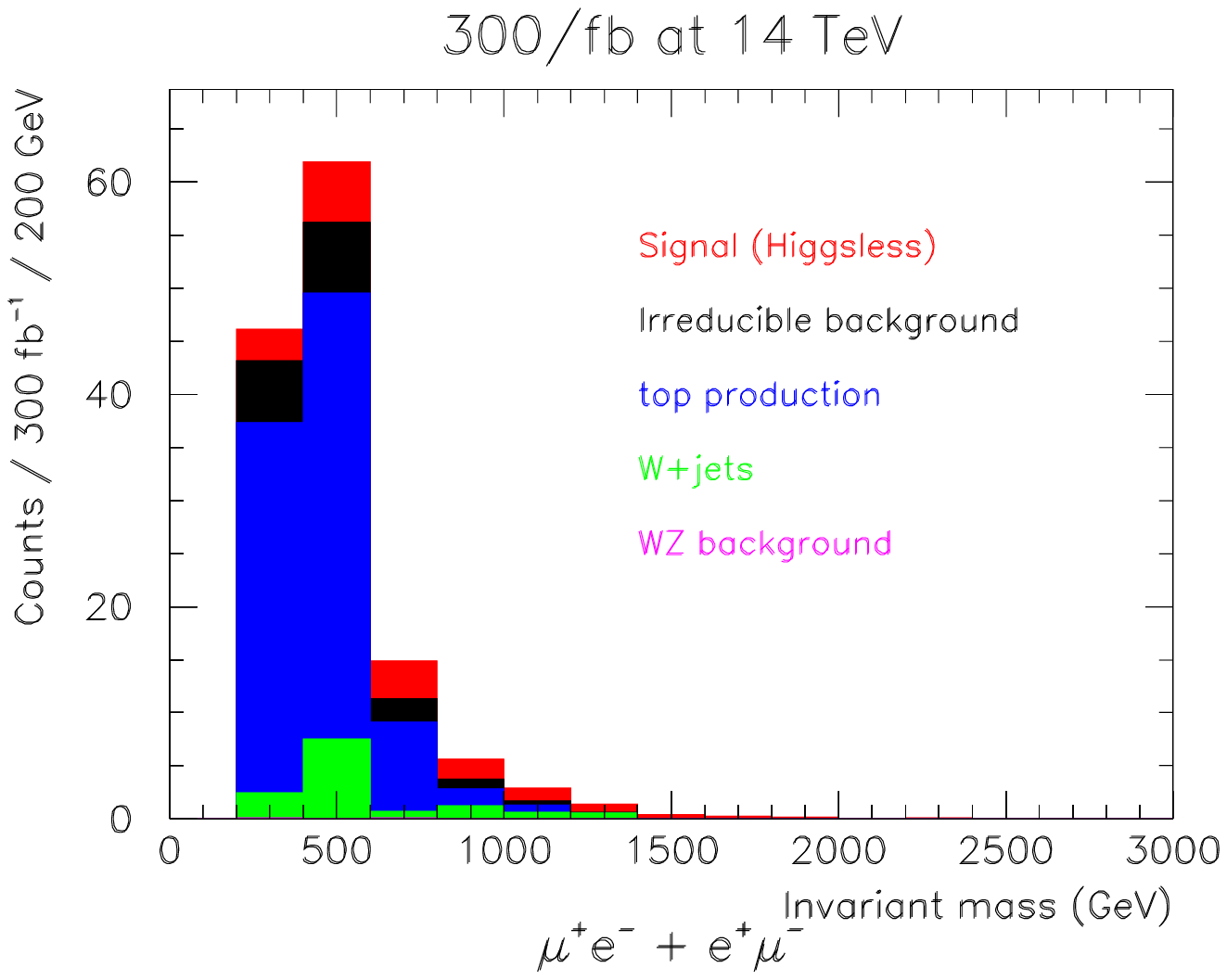,width=0.52\linewidth}\hspace{-1cm}
\epsfig{file=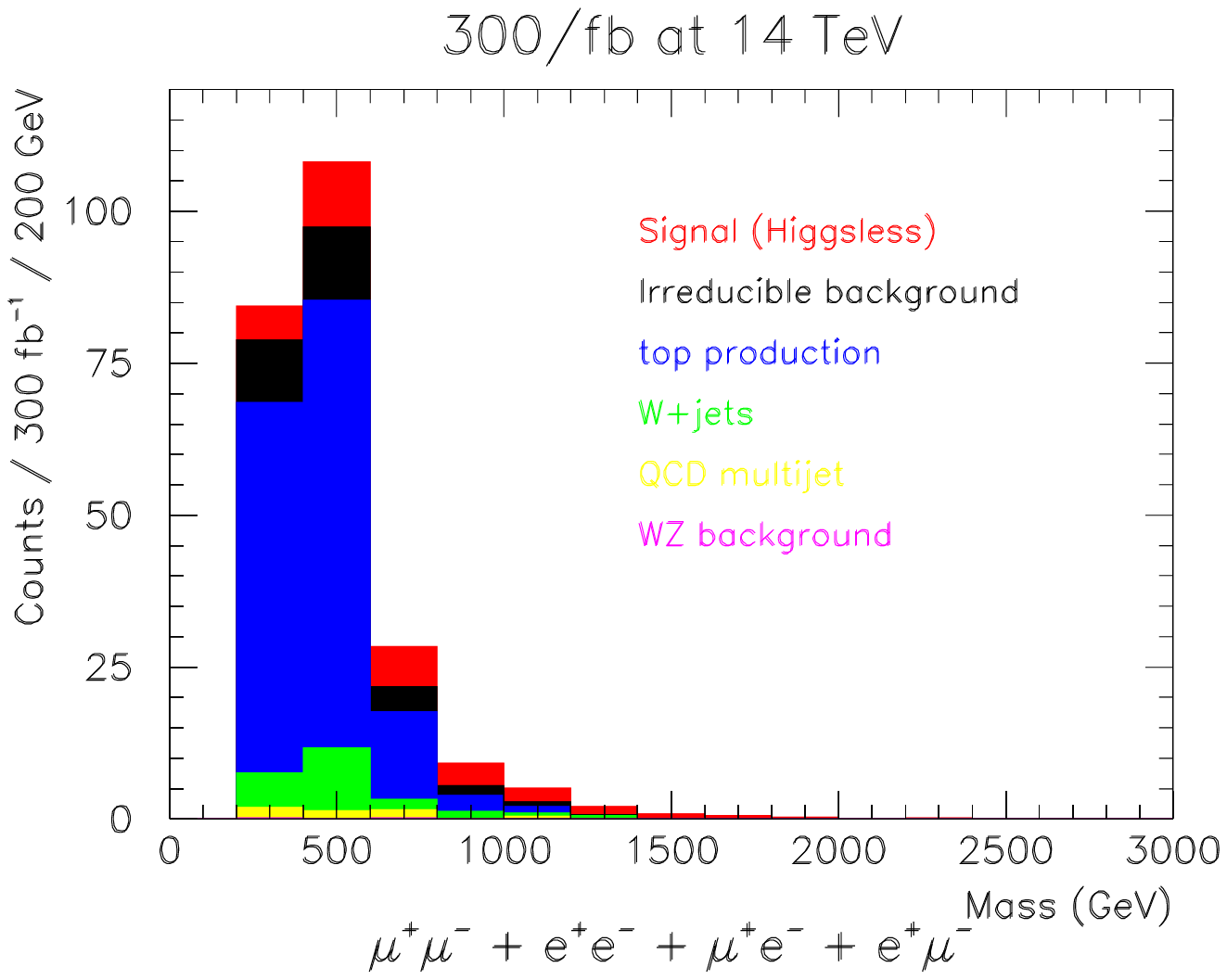,width=0.52\linewidth}
\end{center}
\caption{Invariant mass distributions of the two leptons resulting from
the process $pp \rightarrow jjW^+W^-$ at 14 TeV, with one $W$ decaying into
a muon and another decaying into an electron (left) and with each $W$ decaying
into either a muon or an electron (right).
Shown are the signal calculated under the Higgsless hypothesis and various
contributions to the background, normalized to 300/fb of data.  For the left plot,
applied were all the signal selection criteria foreseen for the opposite-sign mixed
muon+electron channel, i.e., cuts 1-5 from Fig.~\ref{cuts1} (up to and including
CJV).
The final $\mu e$ kinematics was deduced by averaging out the distributions
obtained in $\mu^+\mu^-$ and the $e^+e^-$ channels, which differed only
due to different detector resolution effects assumed during processing by PGS 4.
All the remaining procedures and assumptions were identical as described in the
caption of Fig.~\ref{jjw+w-1}.
The right plot
was obtained by summing up the individual $W$ decay channels.
}
\label{jjw+w-3}
\end{figure}

\begin{figure}[htbp]
\begin{center}
\epsfig{file=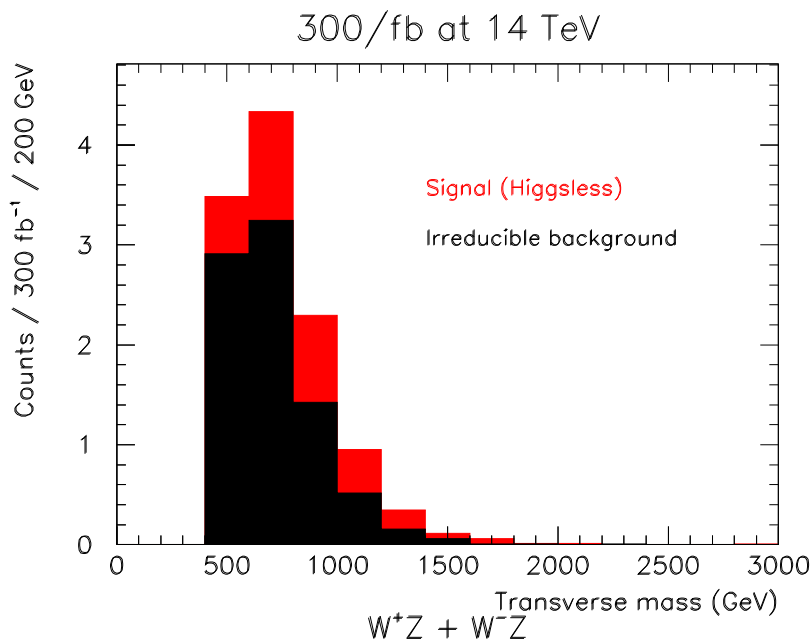,width=0.75\linewidth}
\end{center}
\caption{Distribution of the transverse mass (defined in detail in section \ref{wzzzsig})
calculated for
the process $pp \rightarrow jjW^\pm Z$ at 14 TeV, with subsequent leptonic
decays of the gauge bosons.
Shown are the signal calculated under the Higgsless hypothesis and the
irreducible background, normalized to 300/fb of data.
Applied were all the signal selection criteria foreseen for the $WZ$ leptonic
channel, no distinction was made between lepton flavors (e or $\mu$).
Signal was calculated by subtracting
the SM $jjW^+Z$ sample (by definition also identical with irreducible background)
from the Higgsless $jjW^+Z$ sample.
The cross sections were obtained by scaling the simulated $jjW^+Z$ sample by a
factor of 1.5 to account for $jjW^-Z$ (not simulated).
All the relevant efficiencies and purities were assumed 100\%.
Results of MadGraph simulations, processed by PYTHIA 6 for
parton showering, hadronization and jet reconstruction
and further processed by PGS 4 for the effects of finite resolution in
the measurement of jet and lepton $p_T$ in a CMS-like detector.
}
\label{jjw+z}
\end{figure}

It is clear that the application of MVA's enhances the possibilities
to carry an optimal analysis and isolate the signal.  The final sensitivity depends
on the kinematic separation
of signal and background in the multidimensional phase space and on the overall
signal statistics, to a lesser degree on the amount of background.  As long as
the full kinematic information on each event is taken into account,
quantitative results should in principle not depend on the applied
preselection of events, unless the latter suppresses too much of the proper signal.
It is therefore preferrable to reduce
signal losses to minimum.  A cut-based analysis, such as outlined in this work,
represents in fact the current {\it lower limit} in the achievable sensitivity.
In any case, what we have learned from the present studies, and will emphasize
this point once again, is that a conventional analysis that consists of
applying polarization-blind VBF selection criteria plus a shape analysis of
the lepton-lepton
invariant mass spectrum is suboptimal and should be replaced by a more
sophisticated analysis that explores the full kinematics of the final state
to deliver the best final result.  A lot of valuable
information sits in particular
in the jet spectra, far more than whether the process was VBF or not.
Correlated variables like $R_{p_T}$ can be thought of as a first
effective step towards fuller exploration of the entire kinematic phase of
the four particles in the final state.  Given that BSM effects in $WW$ scattering
may sit at the edge of statistical significance for the LHC, application of the
best analysis techniques may play a vital role.

\section{Anomalous triple gauge couplings}

Whether or not we will be able to observe {\it any} BSM signal is one question.
Another one is whether we will be able to correctly interpret the result in an
independent and standalone way, that is, not having to rely on other concurrent
measurements and assume consistency within a given physics scenario.

New physics may manifest itself e.g.~in anomalous triple gauge couplings which may
likewise show up as an enhancement of $WW$ scattering at high invariant mass.
Details of the shapes of kinematic distributions of the final state particles depend
on the physics scenario, but in general will also depend on specific values of the
anomalous couplings, giving rise to annoying interpretative ambiguities.
It is vital to study the dependencies of the individual kinematic variables
and single out those of them that are mostly sensitive to the scenario but not
to numerical values and vice-versa.

Updated 90\% CL limits on the dimension-6 operators that lead to anomalous triple gauge
couplings have been recently
calculated \cite{corbett}.  We take the following values:

\vspace{3mm}

\hspace{4.0cm} $c_{WWW}/\Lambda^2 ~ \epsilon ~ [-15, 3.9]$ TeV$^{-2}$,

\vspace{2mm}
\hspace{4.0cm} $c_W/\Lambda^2 ~ \epsilon ~ [-5.6, 9.6]$ TeV$^{-2}$,

\vspace{2mm}
\hspace{4.0cm} $c_B/\Lambda^2 ~ \epsilon ~ [-29, 8.9]$ TeV$^{-2}$.

\vspace{3mm}

\noindent
The above limits come from a combination of LEP, the TeVatron and the LHC Run 1 data.
They are asymmetric because in each case the
central values of the relevant couplings were determined.

It happens that anomalous couplings of roughly this size can produce BSM signals in
$WW$ scattering of
the same order of magnitude as would be produced by a $HWW$ coupling set to, e.g., 0.8
of its SM value.  Assuming that the scale of new physics, $\Lambda$, is beyond direct
LHC reach and therefore that no cutoff is necessary in evaluating signal rates,
the expected amount of signal for $W^+W^+$ in the purely leptonic decay modes
is close to 0.050 fb for $c_{WWW}/\Lambda^2 = -10/$TeV$^2$, 0.016 fb for
$c_W/\Lambda^2 = -10/$TeV$^2$ (this includes 0.011 fb of $W_LW_L$ and 0.005 fb of
$W_TW_X$ signal) and 0.003 fb for $c_B/\Lambda^2 = -10/$TeV$^2$.  In deriving these
numbers we have applied exactly the same signal selection criteria as we used
before, although $W_TW_X$ signals may be possible to further improve with
dedicated optimizations.
As it was with $HWW$, the LHC sensitivity to triple gauge couplings in $W^+W^+$
scattering is again at the very limit of present experimental bounds.

\begin{figure}[htbp]
\begin{center}
\epsfig{file=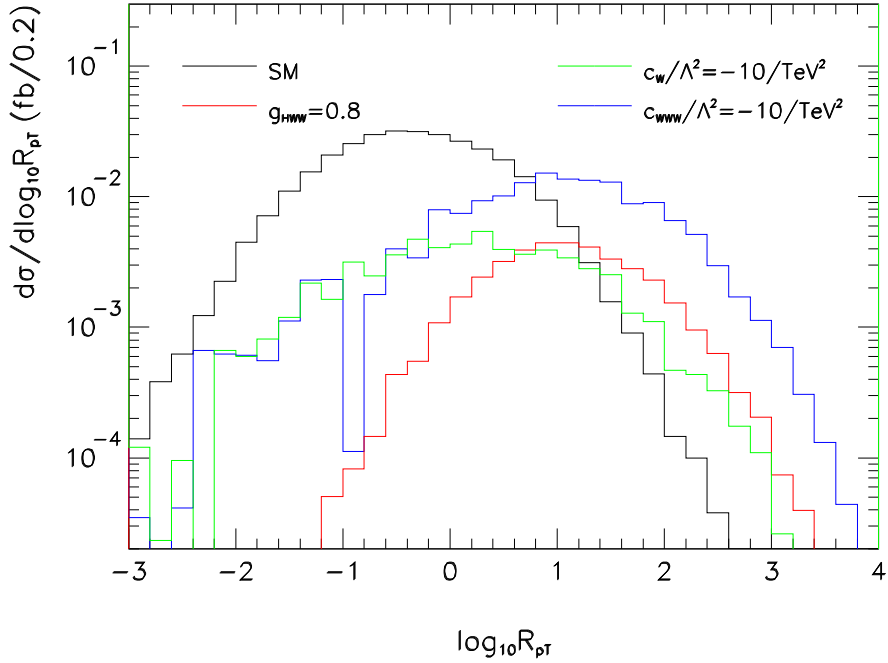,width=0.85\linewidth}
\end{center}
\caption{Distributions of log$_{10}R_{p_T}$ in the process $pp \to jjW^+W^+$ at 14 TeV,
with leptonic $W^+$ decay ($l = e, \mu$) in different physics scenarios:
Standard Model (black histo) and BSM signals for $g_{HWW} = 0.8$ (red histo),
$c_W/\Lambda^2 = -10/$TeV$^2$ (green histo) and $c_{WWW}\Lambda^2 = -10/$TeV$^2$
(blue histo).
Applied were VBF topological cuts, including $\Delta\varphi_{ll} >$ 2.5.
Result of MadGraph 5 simulations, processed by PYTHIA 6 for
$W^+$ decay, parton showering, hadronization and jet reconstruction;
no detector effects were included.
The original PYTHIA 6 source code was modified to account for the
correct polarization-dependent angular distributions for the $W$ decays.
Signals were calculated by subtracting
the SM sample from the corresponding BSM sample.  In calculating the
BSM distributions it was assumed that the scale of new physics, $\Lambda$,
is higher than the accessible energies and hence no cutoff was applied.}
\label{rpt14}
\end{figure}

\begin{figure}[htbp]
\begin{center}
\epsfig{file=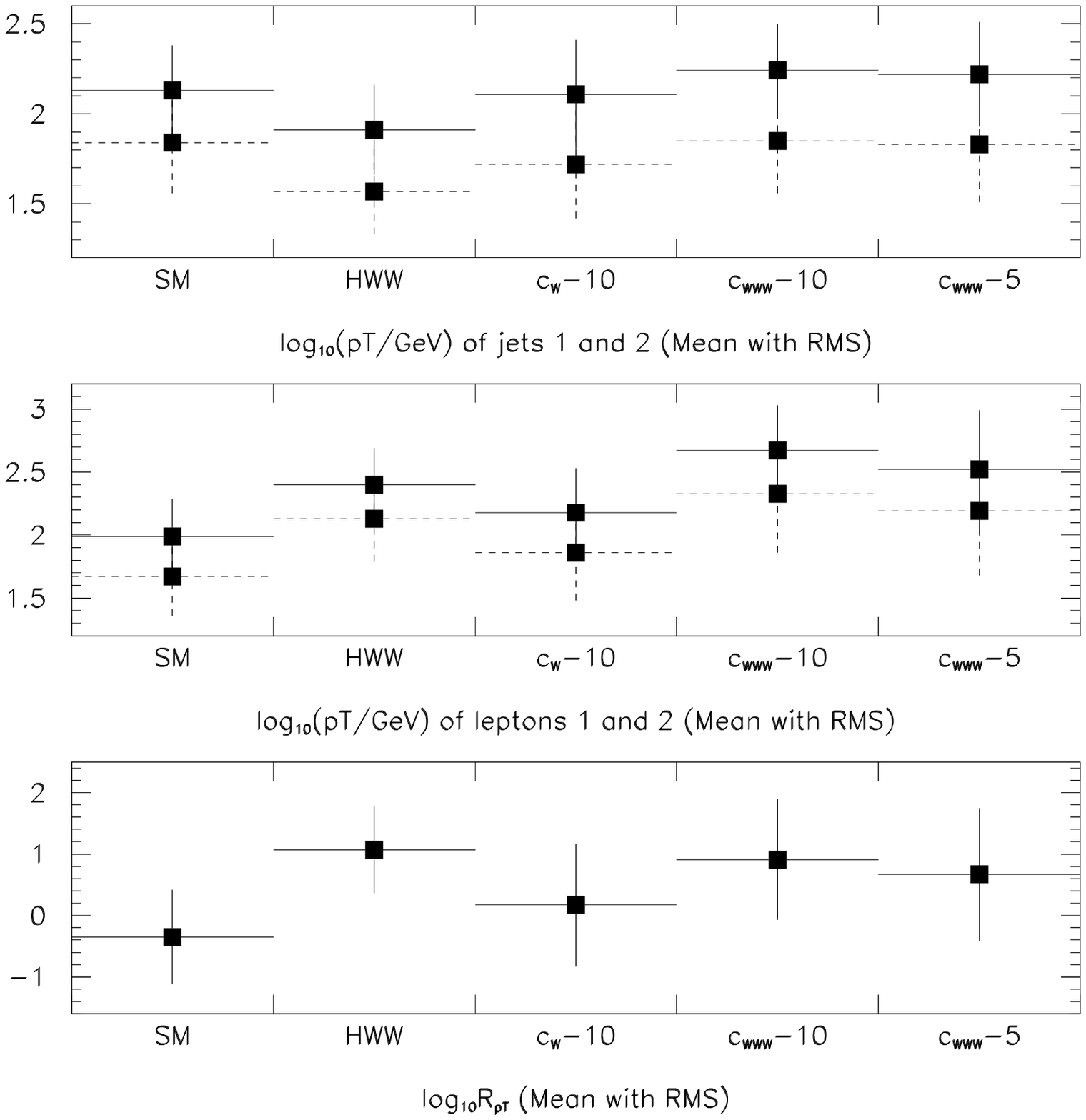,width=0.85\linewidth}
\end{center}
\caption{Mean values and RMS of the individual transverse momenta of the two
jets (upper plot) and the two leptons (middle plot), and of the resulting $R_{p_T}$
distribution (bottom plot)
in the process $pp \to jjW^+W^+$ at 14 TeV,
with leptonic $W^+$ decay ($l = e, \mu$) in different physics scenarios.
On the two upper plots, solid lines represent the leading jet and lepton,
respectively, and dashed lines represent the sub-leading jet and lepton,
respectively.  Vertical error bars represent the RMS.
Each bin on the horizontal axis represents a physics scenario; from left to right:
the Standard Model and BSM signals for $g_{HWW} = 0.8$,
$c_W/\Lambda^2 = -10/$TeV$^2$, $c_{WWW}/\Lambda^2 = -10/$TeV$^2$ and
$c_{WWW}/\Lambda^2 = -5/$TeV$^2$.
Results of MadGraph 5 simulations, all conditions and assumptions as for
Fig.~\ref{rpt14}.}
\label{pt14}
\end{figure}

The most important point is that
signal in general, understood as a sum of all possible BSM effects, may be as
much manifest in
$W_LW_L$ pairs alone (if $c_B \neq 0$), as in $W_TW_X$ alone
(if $c_{WWW} \neq 0$), both $W_LW_L$ and $W_TW_X$ in roughly similar amounts
(if $c_W \neq 0$), or any combination of the above cases.
Any of the above scenarios ultimately manifests in an
enhancement of $WW$ scattering at high invariant mass, the latter being correlated
with high lepton transverse momenta.  Therefore, the $R_{p_T}$ variable is still a
good BSM probe owing to its numerator.  Because of the denominator, however, it always
favors $W_LW_L$ pairs over $W_TW_X$ pairs.  Without prior knowledge of the
helicity composition of the signal,
an enhancement in the numerator may
be just compensated by a larger denominator.
For a complete understanding of a future experimental result we need to examine
the individual transverse momenta of the four final state objects.

What follows is a ``quick and dirty" demonstration of principles how to extract
the most physics information based on no more than four transverse momenta.
Expressed in a logarithmic scale, individual $p_T$ distributions in any given
physics scenario, as well as the corresponding
$R_{p_T}$ distributions, can be to a first rough approximation described in terms
of two parameters: the mean value and the RMS, see Fig.~\ref{rpt14}.
This allows first simple studies of
kinematic separation between different scenarios, before more detailed analyses
become available.  Since the expected signal statistics is anyway bound to be small,
any more detailed shape spectrum may not be even plausible in practice.
From such comparisons (see Fig.~\ref{pt14} bottom) we infer that $R_{p_T}$ by itself will
be enough to distinguish the SM scenario from BSM with a handful of events,
as long as the overall signal size is significant enough.
But it does not suffice to identify a BSM scenario.
In particular, a pure $c_{WWW}$ scenario can produce an $R_{p_T}$ spectrum
indistinguishable from the one produced by an anomalous $HWW$ scenario,
despite the two signal samples consisting of different $W$ polarizations.
Moreover, outgoing lepton spectra are sensitive both to the physics scenario and to the
numerical values of the anomalous coefficients and this interplay is very
sophisticated (see Fig.~\ref{pt14} middle).
Consequently, e.g., numerical variations within the $c_{WWW}$ scenario may be
larger than differences between scenarios (cf., e.g., ``$HWW$" with
``$c_{WWW}-10$" and ``$c_{WWW}-5$" in Fig.~\ref{pt14} bottom).

Fortunately, jet transverse momenta are a direct measure of helicity
composition of the $WW$ sample and nothing else.  They do not depend on
values of the anomalous coefficients (see Fig.~\ref{pt14} top).  The trend
is clear: the more $W_T$ the higher jet $p_T$.  A pure $W_LW_L$ (``$HWW$")
signal gives

\vspace{3mm}

\hspace{4.5cm} log$_{10}p_T^{j1} = 1.91 \pm 0.25$ (RMS),

\vspace{2mm}

\hspace{4.5cm} log$_{10}p_T^{j2} = 1.57 \pm 0.24$ (RMS),

\vspace{3mm}

\noindent
while for a pure $W_TW_X$ signal (e.g.~``$c_{WWW}-10$") it is

\vspace{3mm}

\hspace{4.5cm} log$_{10}p_T^{j1} = 2.24 \pm 0.26$ (RMS),

\vspace{2mm}

\hspace{4.5cm} log$_{10}p_T^{j2} = 1.85 \pm 0.29$ (RMS).

\vspace{3mm}

Combined the information from the two jets, the two extreme cases can be distinguished
at a 3$\sigma$ level with as few as 7 isolated signal events and at 5$\sigma$
with 20 events.  This will be possible with 3000 fb$^{-1}$, unfortunately not
with 300 fb$^{-1}$.  Approximately four times this statistics is required to
perform similar with a mixed helicity signal like in the case of $c_W \neq 0$.
From a back of the envelope
calculation it follows that with 20 isolated signal
events (e.g., with $g_{HWW}=0.9$ and 3000 fb$^{-1}$), the helicity composition of
the signal can be
determined to better than 20\%.  With 60 events (e.g., $g_{HWW}=0.8$, 3000 fb$^{-1}$),
it can be known to $\sim$ 10\%.  It should be kept in mind that average jet $p_T$
does increase with $M_{WW}$ for transverse polarization.  This is why the leading
jet $p_T$ of the $c_W$ signal is hardly different from the one in the SM, although
the former has a relatively larger $W_LW_L$ component.  But for any BSM scenario that
manifests as a steady enhancement at high $M_{WW}$ up to the kinematic limit of
available phase space, differences in the jet $p_T$ spectra for $W_TW_X$ are
effectively a second order effect.
On the other hand, jet $p_T$ spectra for $W_LW_L$ are very much independent of
anything and are a unique signature of longitudinal polarization.

Once we have measured the helicity compisition
and hence settled which BSM effect plays the dominant role (at least within the
limited scope of scenarios considered here), the numerical value of the leading
anomalous coefficient can be deduced by studying the lepton transverse momenta
(or alternatively the lepton-lepton invariant mass, which is quite the same thing).

In the above studies we have considered explicitly only negative values of the anomalous
coefficients.  Positive values of $c_{WWW}$ or $c_W$ happen to be more bound
experimentally and in addition
the interference with SM diagrams is in this case destructive.  The signal
will then consist of a slight depletion followed by very little enhancement
within the allowed phase space, likely beyond LHC sensitivity.

Variations of $c_B$ within the presently allowed range produce too little
signal to be detected at the LHC.  As a matter of principle, $c_B$ affecting
only $W_LW_L$ pairs
cannot be distinguished from pure $HWW$ scaling 
by means of the methodology proposed here.

\section{Anomalous quartic couplings}

Some of the dimension-6 operators discussed above modify also the gauge quartic
couplings.  It is conceivable that their values will be eventaully determined by non-VBS
processes at the LHC and applied as background in more dedicated VBS analyses.
On the other hand, quartic couplings can best de determined via VBS processes,
along with triboson production.
Since sensitivity of VBS processes to anomalous
triple couplings, including Higgs to gauge couplings, within their present bounds
is rather slim, a large deviation from SM predictions could in fact signal
non-trivial contributions from physics related to operators of yet higher
dimension than 6, namely dimension 8.  For this reason it makes sense to go
directly to the presently unbounded dimension-8 operators
and study their potential consequences.  These operators may affect VBS and
triboson production, but not other processes.  Such studies are currently in
progress, some have already been shown.

In section \ref{tqcp} we have already discussed the LHC sensitivities to
anomalous quartic couplings based on an early study of $W^\pm W^\pm$
carried at the phenomenological level.  
Most recently, sensitivity to quartic couplings in VBS processes 
has been studied by the Snowmass 2013 study group \cite{snowmass}.
Results were presented in the language of higher-dimension operators in
Effective Field Theory.  Studied were the respective sensitivities of the $ZZ$
process to parameters $f_{T,8}/\Lambda^4$ and $f_{T,9}/\Lambda^4$, of $WZ$ to
$f_{T,1}/\Lambda^4$ and of $W^\pm W^\pm$ to $f_{T,1}/\Lambda^4$.  In Effective
Field Theory, these coefficients scale dimension-8 operators ${\mathscr L_{T,8}}$,
${\mathscr L_{T,9}}$ and ${\mathscr L_{T,1}}$, respectively.
In the same work, total cross sections for $W^\pm W^\pm$ $WZ$ and $ZZ$ were
calculated varying many anomalous coefficients one at a time.
Reportedly, $f_{T,1}/\Lambda^4$ and $f_{T,0}/\Lambda^4$ were the parameters
that all the total cross sections
were found most sensitive to.  The choice of
$f_{T,8}/\Lambda^4$ and $f_{T,9}/\Lambda^4$ for $ZZ$ was motivated by the fact
that these parameters are built uniquely from the neutral field strengths $B_{\mu\nu}$
and so they can only be probed in this process.
Note that all these operators are built from field
strengths only ($B_{\mu\nu}$ or $W_{\mu\nu}$) rather than Higgs field derivatives,
and consequently affect directly only the vertices involving
transversely polarized states.  The various
anomalous coefficients were implemented in MadGraph matrix element calculations.
Simulations included typical detector resolutions parameterized in the
DELPHES program.  Only irreducible backgrounds were considered, except for
the $W^\pm W^\pm$ process for which the analysis included the $WZ$ background
scaled by an additional factor 2 to account for other backgrounds.
Applied were minimal selection criteria which consisted in principle of
a jet-jet invariant mass cut, $M_{jj}~>$ 1 TeV, in addition to the requirement
of the proper final state for each scattering process, assuming purely leptonic
decays.
Signal, understood as the total electroweak production rate, was calculated
using the alternative hypotheses of the SM and of non-zero value of the
relevant anomalous coefficients
(hence it is the difference between the two that defines signal in the
understanding used throughout this work).  
The final results were obtained by evaluating the one-dimensional spectrum
of the reconstructed $VV$ mass (for $ZZ$ and $WZ$) or of the 4-body invariant
mass, $M_{jjll}$ (for $W^\pm W^\pm$).  Different pile-up conditions were simulated
for the $W^\pm W^\pm$ process, but results reportedly did not vary much.
The authors calculated also for each process the corresponding unitarity bounds
as a function of the respective coefficient values.  Results were presented
with and without applying a sharp unitarity violation cutoff.
Snowmass study results for 14 TeV
are summarized in Table \ref{tab:snow13}.  One notices in particular that
an order of magnitude increase in integrated luminosity, between 300 and 3000 fb$^{-1}$,
translates into an increase of merely a factor $\sim$2 in the expected
sensitivities.

\vspace{5mm}

\begin{table}[htbp]
\begin{center}
\begin{tabular}{|c||c|c|c|}
\hline
Process, parameter & Luminosity & 5$\sigma$ & 95\% CL \\
\hline\hline
$ZZ$, $f_{T,8}/\Lambda^4$ & 300 fb$^{-1}$ & 5.5 (8.4) TeV$^{-4}$ & 3.2 (5.3) TeV$^{-4}$
\\
$ZZ$, $f_{T,8}/\Lambda^4$ & 3000 fb$^{-1}$ & 2.9 (4.7) TeV$^{-4}$ & 1.7 (2.4) TeV$^{-4}$
\\ \hline
$ZZ$, $f_{T,9}/\Lambda^4$ & 300 fb$^{-1}$ & 8.7 (9.0) TeV$^{-4}$ & 6.2 (6.7) TeV$^{-4}$
\\
$ZZ$, $f_{T,9}/\Lambda^4$ & 3000 fb$^{-1}$ & 5.7 (6.3) TeV$^{-4}$ & 3.9 (4.6) TeV$^{-4}$
\\ \hline
$WZ$, $f_{T,1}/\Lambda^4$ & 300 fb$^{-1}$ & 1.1 (1.6) TeV$^{-4}$ & 0.7 (1.0) TeV$^{-4}$
\\
$WZ$, $f_{T,1}/\Lambda^4$ & 3000 fb$^{-1}$ & 0.6 (0.9) TeV$^{-4}$ & 0.4 (0.5) TeV$^{-4}$
\\ \hline
$W^\pm W^\pm$, $f_{T,1}/\Lambda^4$ & 300 fb$^{-1}$ & 0.2 (0.4) TeV$^{-4}$
& 0.1 (0.2) TeV$^{-4}$ \\
$W^\pm W^\pm$, $f_{T,1}/\Lambda^4$ & 3000 fb$^{-1}$ & 0.1 (0.2) TeV$^{-4}$ 
& 0.06 (0.1) TeV$^{-4}$ \\ \hline
\end{tabular}
\end{center}
\caption{The 5$\sigma$ significance discovery values and 95\% CL limits for
coefficienits of dimension-8 operators with 300 and 3000 fb$^{-1}$ of data at
14 TeV using different VBS process.  Numbers in brackets correspond to imposing
a unitarity violation cutoff.  Results of the Snowmass13 study \cite{snowmass}.
}
\label{tab:snow13}
\end{table}

Some other studies exist that include full detector simulation and improved
background evaluation.
The ATLAS collaboration presented a new set of simulation-based
studies for 14 TeV, reformulated and updated after Higgs discovery \cite{atlasquartic}.
Under the assumption that each scattering process will be mainly
sensitive to new physics arising from just one of those higher order operators,
sensitivities to new physics have been evaluated for $ZZ$, $WZ$ and $W^\pm W^\pm$
in the purely leptonic decay modes.
These studies
were based on fully simulated events, including detector effects related to jet
clustering, pile-up, as well as parameterized reconstruction efficiencies and
resolutions for the different physics objects.  

Early work was based on the EWChL approach
and amplitude unitarization
according to the model of Dobado et al.~\cite{dobado}
to evaluate gauge boson scattering signals,
and results were given in terms of the EWChL parameters $a_4$ and $a_5$.
More recently, a newer set of analyses was presented with all the results
translated into the language of Effective
Field Theory.

The quartic $WWWW$ coupling was studied in terms of the $f_{S,0}/\Lambda^4$ coefficient
that scales the effective operator $\mathscr{O}_{S,0}$ and is best probed
via the same sign $W^\pm W^\pm$ channel.  Otherwise, the analysis was akin to
the one carried by the Snowmass study.
Various reducible backgrounds were estimated using
a combination of simulation work with existing experimental data.
The most important of them were reportedly total $jjWZ$ production, $jjW^\pm W^\pm$
production via gluon exchange (QCD) graphs and several
detector-dependent backgrounds generally termed ``mis-ID's".  The latter class included
photon conversion, jets faking leptons and lepton charge flips.

Results were presented
in terms of the 5$\sigma$ discovery reach and 95\% confidence level exclusion
limits for expected luminosities of 300/fb and
3000/fb.  With 300/fb, the 5$\sigma$ discovery limit was obtained at 10 TeV$^{-4}$,
while the expected 95\% CL exclusion limit is 6.8 TeV$^{-4}$.
An order of magnitude increase in luminosity was found to
translate into nearly an order of magnitude improvement of the exclusion limit,
but only slightly more than a factor 2 in the discovery reach.
This is due to a non-trivial relation between $f_{S,0}/\Lambda^4$ and the
signal size.
However, in the event of BSM observation with 300 fb$^{-1}$, the anomalous
coefficient could be measured with a precision better than 5\% with 3000 fb$^{-1}$,
which fully qualifies for the term precision measurement.

Based on earlier studies at the phenomenological level, similar limits can be also
expected on $f_{S,1}/\Lambda^4$.  It is important to realize that all these studies
assume just one non-vanishing
anomalous parameter at a time.  It was also shown that $f_{S,0}$ and $f_{S,1}$
produce similar signal and so
are in fact anticorrelated.  This anticorrelation is especially strong for
$W^\pm W^\pm$.  Combination of different scattering processes, in particular
$W^\pm W^\pm$ and $W^+W^-$, is instrumental in restricting the allowed ranges
of both parameters at a time so to be comparable with limits
obtained from considering just one parameter at a time.

\begin{figure}[htbp]
\vspace{5mm}
\begin{center}
\epsfig{file=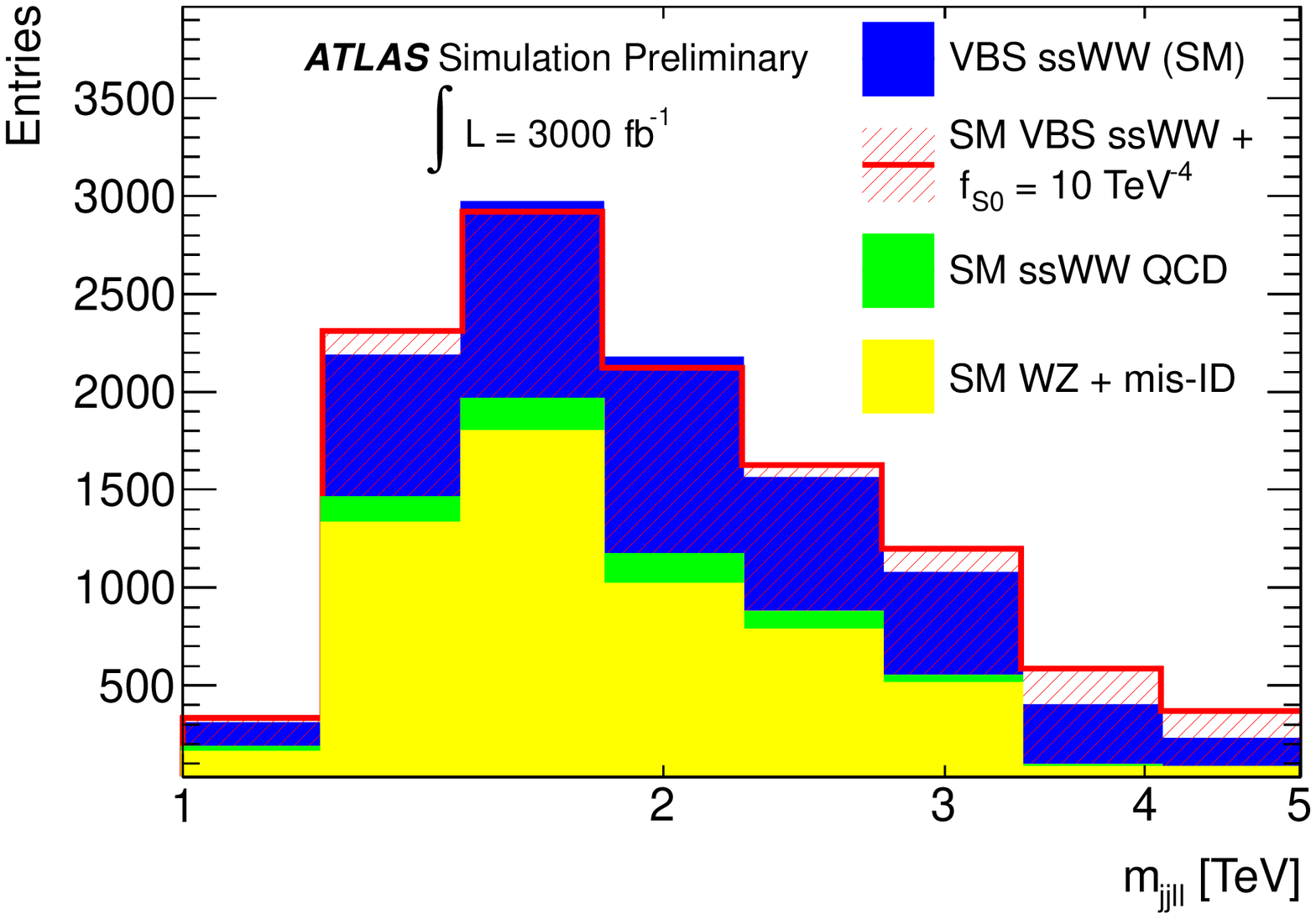,width=0.52\linewidth}
\epsfig{file=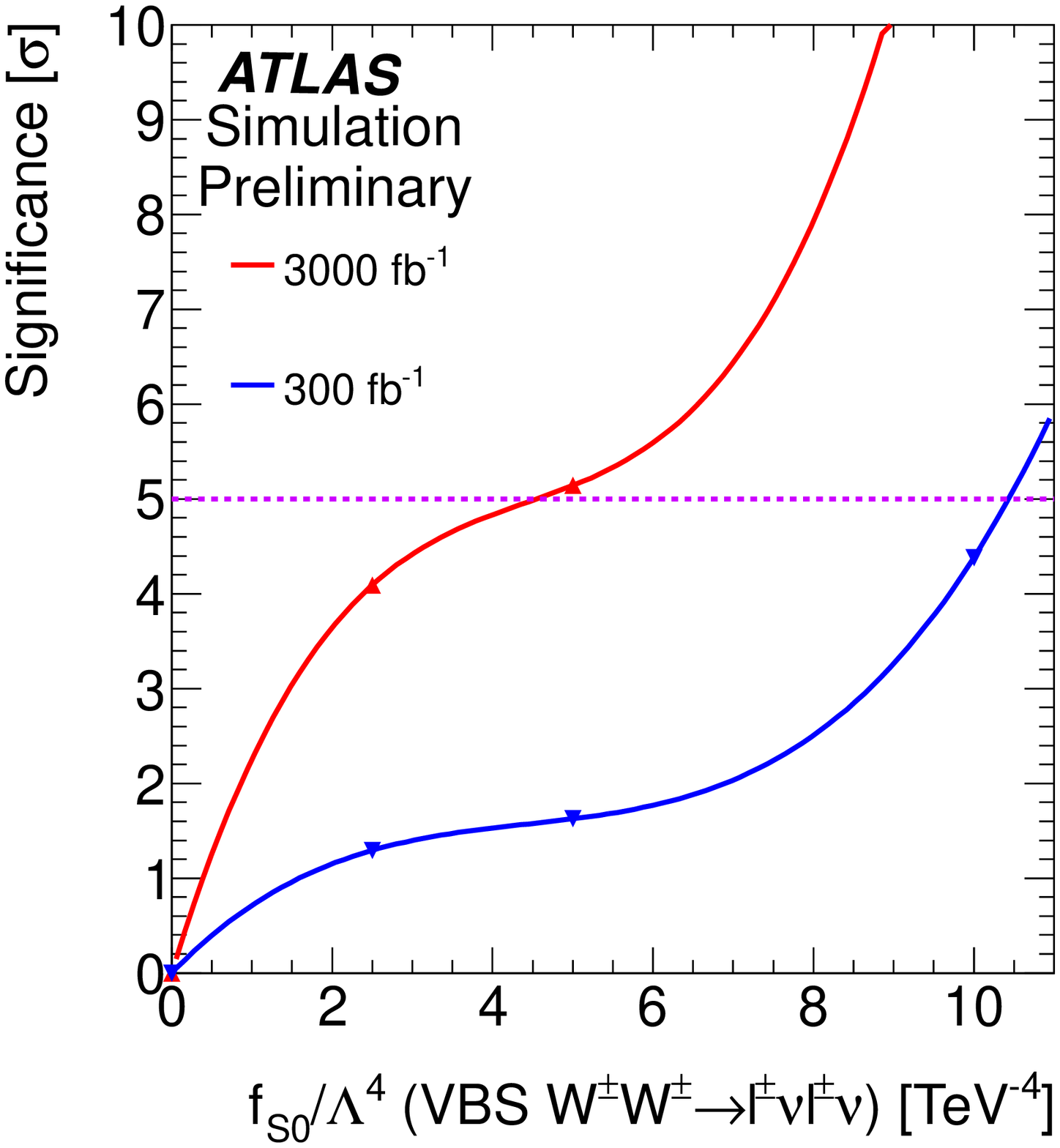,width=0.45\linewidth}
\end{center}
\caption{Reconstructed 4-body mass spectrum in the SM and in the scenario
with $f_{S,0}/\Lambda^4 =$ 10 TeV$^{-4}$ (left) and BSM signal significance in
standard deviations as a function of $f_{S,0}/\Lambda^4$ (right)
from the $pp \to jjl^\pm l^\pm\nu\nu$ process at 14 TeV.
Results of simulations done by the ATLAS experiment, images reproduced
from Ref.~\cite{atlasquartic}.}
\vspace{5mm}
\label{fig:atlasww1}
\end{figure}

\begin{figure}[htbp]
\vspace{5mm}
\begin{center}
\epsfig{file=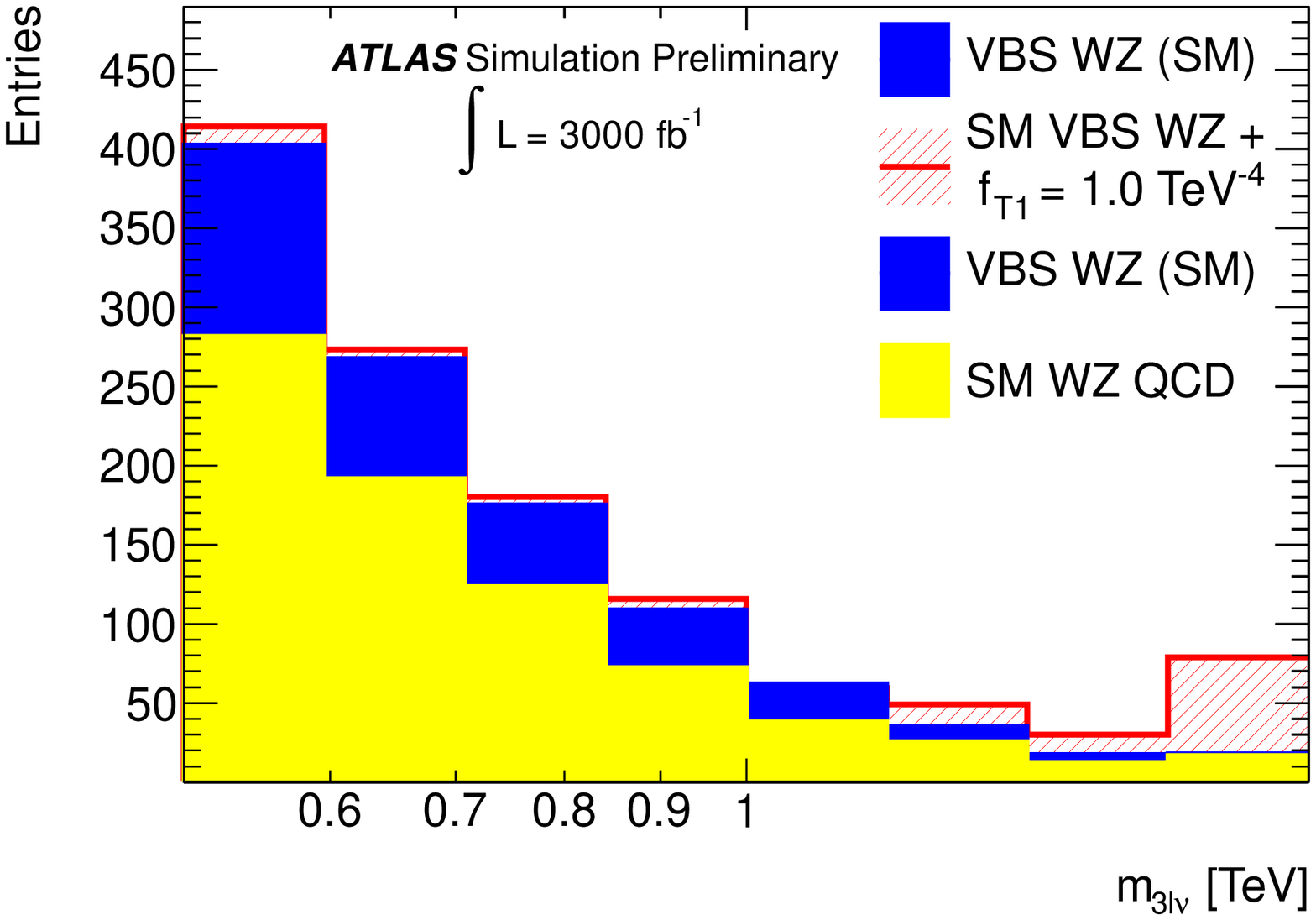,width=0.52\linewidth}
\epsfig{file=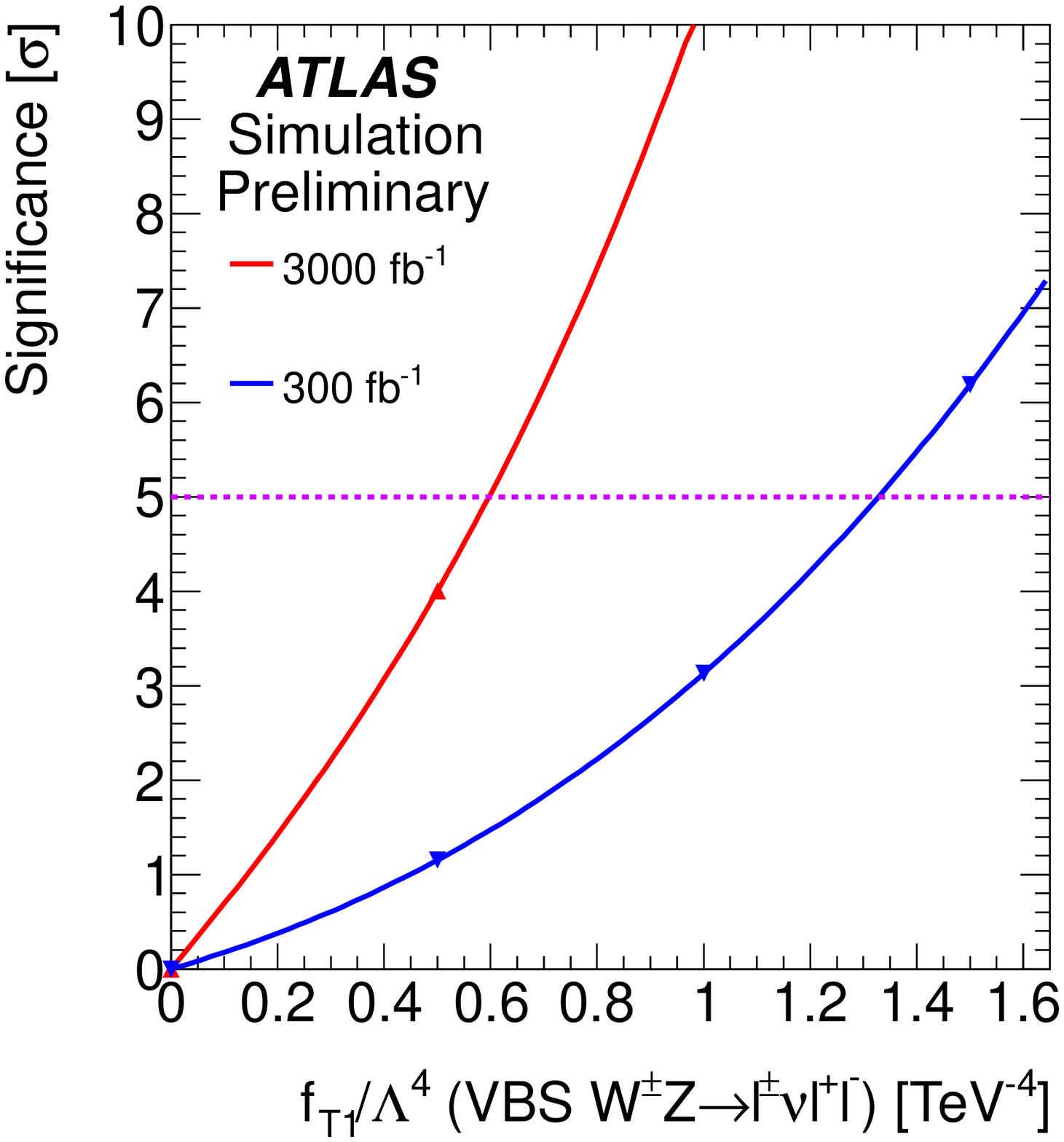,width=0.45\linewidth}
\end{center}
\caption{Reconstructed $WZ$ mass spectrum in the SM and in the scenario
with $f_{T,1}/\Lambda^4 =$ 1 TeV$^{-4}$ (left) and BSM signal significance in
standard deviations as a function of $f_{T,1}/\Lambda^4$ (right)
from the $pp \to jjl^\pm l^+l^-\nu$ process at 14 TeV.
Results of simulations done by the ATLAS experiment, images reproduced
from Ref.~\cite{atlasquartic}.}
\vspace{5mm}
\label{fig:atlasww2}
\end{figure}

The ATLAS study is the first complete detector-specific study of the unique physics
capabilities
of $W^\pm W^\pm$ scattering after Higgs discovery (unique in the sense that the same
cannot be measured with possibly better precision in any other processes).
Since a sizeable fraction of the background is comprised by the
various detector-specific ``mis-ID's", ATLAS results cannot be directly transferred
to CMS, but one can safely assume that more important at this stage is further
detector-independent analysis optimization.
The published study most certainly keeps much room for improvements.  
First and foremost, it is polarization-blind.
Moreover, a
high $p_T$ threshold of 50 GeV for both tag jets was used to protect against pile-up
jets.  This however at 14 TeV means automatical loss of two thirds of the $W_LW_L$
sample.  
It should be noted that $f_{S,0}$ produces BSM effects only in $W_LW_L$.
Any lowering of the jet $p_T$ thresholds will reflect in improved
sensitivities.  For example, with the $p_T$ threshold applied on only one tag jet
the $W_LW_L$ scattering statistics increases by a factor 2.
A one-dimensional
evaluation of the 4-body invariant mass does not exploit the full relevant details of
the final state kinematics.  Lack of explicit consideration of $\Delta\varphi_{ll}$
allows a lot of unnecessary non-VBS contributions, in particular $jjWZ$, without
practically any gain in terms of signal.
Finally, since $f_{S,0}$ enhances high $WW$ masses for $W_LW_L$ pairs, it is plain
to see that
$R_{p_T}$ will be as efficient a criterion to extract the BSM signal as it
was for the case of a scaled $HWW$ coupling.  The above remarks become even more
important when one notices that the Snowmass study \cite{snowmass} revealed a
much better sensitivity of the total $W^\pm W^\pm$ rate to $f_{T,1}/\Lambda^4$
than to $f_{S,0}/\Lambda^4$ or $f_{S,1}/\Lambda^4$.  There may be no other way
than to go to lower jet $p_T$ in order to conclude anything about 
$f_{S,0}/\Lambda^4$ or $f_{S,1}/\Lambda^4$ proper.

Discovery reaches and expected exclusion limits for other dimension-8 operators were
also obtained from $W^\pm Z \to W^\pm Z$ and $Z\gamma\gamma$ production.
From a study of $WZ$ scattering expected
limits and discovery reaches were derived on $f_{T,1}/\Lambda^4$, for which
this process is supposed to be most suited.  
It must be stressed however that
the Snowmass study revealed $W^\pm W^\pm$ be actually much more senisitive to this
parameter than $WZ$.
Again here, selection criteria were reduced to
a suitable combination of 3 leptons and a high jet-jet invariant mass.
The expected 95\% CL limits were found to be 0.7 TeV$^{-4}$ and 0.3 TeV$^{-4}$
for 300 fb${^-1}$ and 3000 fb${^-1}$, respectively, while 5$\sigma$ discovery reaches
were 1.3 TeV$^{-4}$ and 0.6 TeV$^{-4}$, respectively.  The results are in fact
in full agreement with those of the Snowmass study.
Because $f_{T,1}$ directly affects only transversely polarized pairs,
high jet $p_T$ threshold may not be a problem here.  However, drawing the result from
a one-dimensional analysis of the reconstructed $WZ$ mass spectrum
is beyond doubt suboptimal.

A similar study for the $WZ$ channel was recently presented by the CMS collaboration
\cite{cmsquartic}.
This analysis included fast simulation of detector response at high luminosity,
with dedicated CMS-specific packages
developed for the planned low luminosity and high luminosity phases of the LHC.
More restrictive selection criteria than applied in the ATLAS study
included most of the typical VBF criteria for $WZ$, namely
$\Delta\eta_{jj} >$ 4,
minimum MET, high jet-jet inavriant mass and a reconstructed $Z$ mass.
In addition, cuts were applied on the lepton-lepton and jet-lepton separation.
The final result was obtained from a shape analysis of the reconstructed
one-dimensional $WZ$ transverse mass spectrum.
This produced similar, if only slightly better,
sensitivities than those reported by ATLAS.  A 5$\sigma$ discovery is expected
for $f_{T,1}/\Lambda^4$ down to 1 TeV$^{-4}$ with 300 fb$^{-1}$ and to
0.55 TeV$^{-4}$ with 3000 fb$^{-1}$.
This study too used a conservative jet $p_T$ threshold
and made no specific selection as to separate $W/Z$ polarizations.
Another interesting result is that the Standard Model process of $WZ$ scattering
after the proposed preselection can be visible at 5$\sigma$ after collecting
185/fb of data.  Hence, consistency with the SM can be precisely tested in the
event of absence of new physics.

Once again it is important to realize the arbitrariness of such analyses in terms
of the assumed choice of the appropriate parameter or parameters to be studied in
relation to a particular scattering process.
Strictly speaking, $f_{S,0}$ and $f_{S,1}$ modify also the $WWZZ$ vertex.
Likewise, both vertices are sensitive to
other dimension-8 operators, constructed either from field strength
tensors and Higgs field derivatives or from field strength
tensors alone.  Most of these dependencies have not been explicitly studied so far.
They may not even add up coherently, but involve non-trivial interference effects.
The task of disentangling the different contributions will be a long and complicated
one.  It will certainly require a combination of all scattering processes.
We are only at the beginning of the real work.
The good news is that we can at least partly help this task with a technique
to separate different polarization states via their respective
jet transverse momentum spectra.

Finally, even if we understand the VBS measurements as purely Standard Model
measurements, aimed solely at setting limits on various anomalous contributions
(for the LHC such scenario cannot be disregarded), application of dedicated
techniques to separate longitudinal from transverse gauge bosons will certainly
result in better experimental limits - at least for those operators which
affect mainly $V_LV_L$ pairs.

\chapter{Beyond the LHC}

According to the European Strategy for Particle Physics, update of 2013,
{\it Europe's top priority should be the exploitation of the full potential of the LHC,
including the high-luminosity upgrade of the machine and detectors with a view to
collecting ten times more data than in the initial design, by around 2030}.
The ultimate goal of the LHC is to deliver 3000 fb$^{-1}$ of proton-proton data
within the next 15-20 years, but until then beam energy will stay at 13/14 TeV.
For beyond the High Luminosity LHC timescale, the next priority outlined by
the European Strategy for Particle Physics is pushing the energy frontier.
A ``High Energy LHC" option has been studied as a possible next step after 2035.
The replacement of the NbTi dipole magnets with 20 T dipoles based on the
novel High Temperature Superconductor (HTS) technology would allow reaching
as high as 33 TeV in the very same LHC ring.  Meanwhile, a yet more ambitious project
has been gaining momentum and attracting the attention of particle physicists.
On February 12-15, 2014, a kick-off meeting took place of the Future Circular
Collider (FCC) Study group in Geneva \cite{fcckickoff}.
A total number of 341 physicists from all over the globe met to
discuss the rationale and perspectives to build a new, more powerful collider
that one day may possibly become the LHC successor.
The main aim of the project is building a new ring in the area of Geneva
that will eventually collide
protons at a center of mass energy of 100 TeV.
The process of setting up a new international
collaboration was initiated.

\section{General features of $VV$ scattering at the FCC}

The subject of Vector Boson Scattering is widely considered as one of the key 
physics topics for the FCC (here and in what follows, we will refer as FCC
to specifically the proton-proton option, more exactly known as FCC-hh - there exist also
electron-electron and electron-hadron options that are being considered in parallel,
possibly to be realized some day in the very same FCC tunnel).

Producing any realistic simulation-based predictions for $WW$ scattering
in proton-proton collisions at 100 TeV is connected to several theoretical
issues.  Event topology changes as all the outgoing products of a collision get
generally boosted more forward than they are
in the LHC.  This means in particular that the typical signature of a VBS event
is now modified so to extend to higher pseudorapidity of the two tagging jets.
Early studies indicate that the minimum jet pseudorapidity range to be covered 
extends at least up to $|\eta| <$ 6.  From the experimental point of view it
is unlikely that useful jet reconstruction can be extended to yet higher $|\eta|$
in a real
detector, and so we will assume this minimum coverage as a reasonable compromise
between physics needs and technical possibilities.
Leading order calculations of the process, say, $pp \to jjW^+W^+$ with the
required kinematic coverage may not be of enough accuracy for 100 TeV and use of NLO
generators is officially encouraged and recommended by the FCC-hh group.
However, because of relatively little QCD contamination, the same-sign channel
in the purely leptonic decay mode is the least affected by this uncertainty.
Preliminary studies with the VBFNLO generator indicate that LO versus NLO
differences amount here to less than 10\% \cite{slawek}.
The applicability of currently available PDFs constitute a source of additional
uncertainty.
All these issues need to be aggressively addressed.
Furthermore, little is known at the present moment of the particle detectors 
and their performances for the FCC, although
some first intelligent guesses as to what these potential detectors may (are
bound to?) look like have already been presented.
In any case, development of complete, realistic simulations for the FCC is a task
for many years and many people.  Nonetheless, it is already possible to get a
rough glimpse of the possibilities and main advantages over the LHC.
And once again here, we will make the claim that a traditional analysis
consisting of selecting VBF-topology events and studying the lepton-lepton
invariant mass spectrum is by far a suboptimal strategy.

The total cross section for $pp \to jjW^+W^+$ with two forward ``tagging" jets
is over 40 times larger at 100 TeV than at 14 TeV.
A hint of the FCC physics capabilities can be grasped by simply repeating quite the
same analysis we have outlined in previous sections for the LHC, with the slight
alteration of the basic topological cuts which now will read:

\vspace{3mm}

\hspace{5.8cm} 2.5 $< |\eta_j| <$ 6,

\hspace{5.8cm} $\eta_{j1} \eta_{j2} < 0$,

\hspace{5.8cm} $|\eta_l| <$ 2.1.

\vspace{3mm}

Detailed simulation of inclusive $t\bar{t}$ background for 100 TeV
is currently missing; there is however much to suppose that this background
can be kept at a manageable level here too.  The kinematic bounds from the
top quark mass, which we have previously quantified as $m_{j1l2} < 200$ GeV
and $m_{j2l1} < 200$ GeV are valid here as well, while the signal region
in 100 TeV collisions starts typically at significantly higher values.  
Stringent cuts like
$m_{jl} > 400$ or 500 GeV can be readily imposed if necessary to suppress the
$t\bar{t}$ background to the desired level without being too costly to the signal.
In all the following considerations we will only
discuss the signal and irreducible background, the latter being defined, as usual,
as the Standard Model total $WW$ production.

It is plain to see that all the kinematic features that distinguish the BSM
signal from the SM background are qualitatively still the same as we saw for 14 TeV.
In particular, $W_LW_L$ signal clearly populates a region of lower jet transverse momenta
than $W_TW_X$ background.  The median of the leading jet $p_T$ distribution of
the signal is found
around 100 GeV which poses no problems from the detector point of view.
The sub-leading jet distribution, however, has a median around 50 GeV.
This means that requiring two tagging jets with $p_T >$ 50 GeV, as is
assumed by default in some studies, automatically means an unacceptable
50\% signal loss in every study where we are interested in $W_LW_L$.  
Special effort must be dedicated in order to keep the
machine pile-up under reasonable control and make the low $p_T$ jets accessible to
physics analysis.  A machine operating in a 5 ns mode (option considered as a
possible backup solution), and therefore having a 5 times lower pile-up,
would be clearly advantageous from this point of view.  Ideally, reaching the
$p_T >$ 20 GeV level would be the goal that best corresponds to the physics needs.
Alternatively, one should
reconsider the concept of tagging only one forward jet and setting an algorithm
to find the second jet off line.  Such studies however have not been seriously
started to the present moment.  The leading lepton $p_T$ for the
signal ranges virtually from around 100 GeV above, hence triggering on a single
lepton will not be a problem \footnote{If any triggering will still be used at all -
some authors predict, based on the so called Moore's law, that the need to have
a first level trigger will disappear altogether by the time FCC-hh is starting.}.

\section{Higgs to gauge couplings at the FCC}

\begin{figure}[htbp]
\begin{center}
\epsfig{file=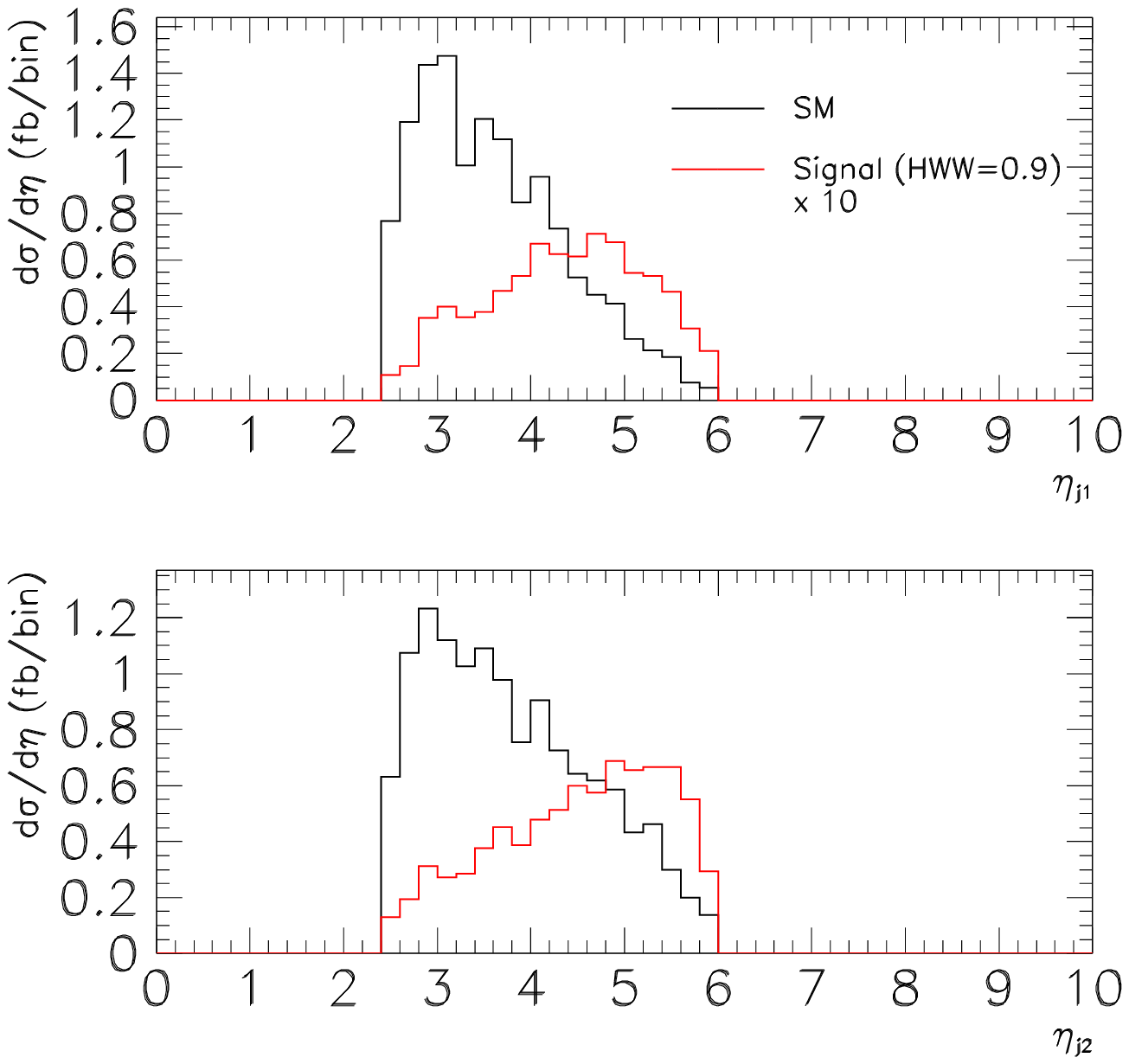,width=0.60\linewidth}
\epsfig{file=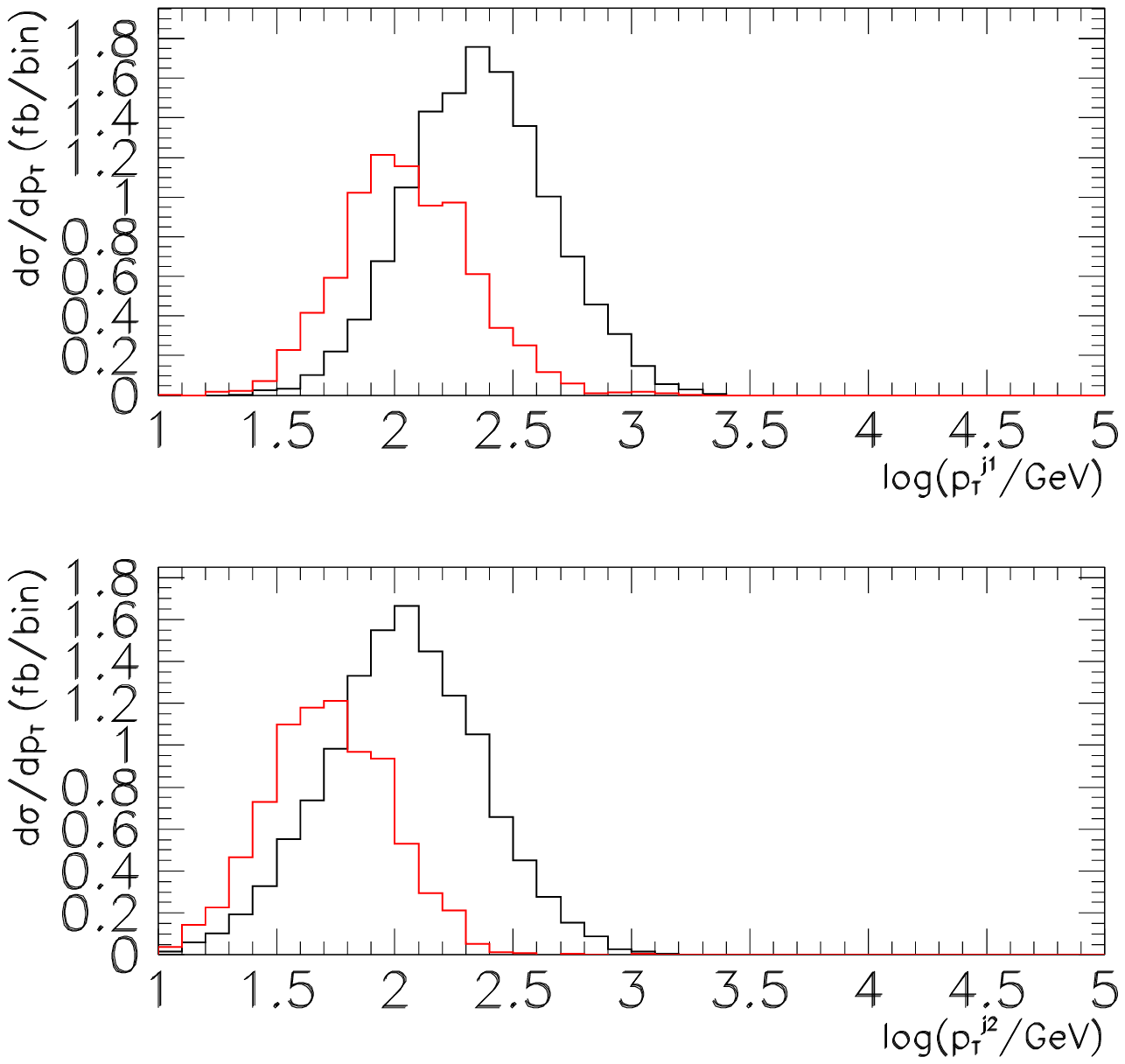,width=0.60\linewidth}
\end{center}
\vspace{-5mm}
\caption{Distributions of pseudorapidities (two upper plots) and
transverse momenta (two lower plots) of the leading
and sub-leading jets from the $pp \to jjW^+W^+$ process at
100 GeV, with leptonic $W^+$ decay ($l = e, \mu$).  Shown are the distributions for
the Standard Model irreducible background
(black histos) and the BSM signal (red histos).  BSM was defined in terms
of the $HWW$ coupling set to 0.9 of its SM value.  Signal was calculated by
subtracting the SM sample from the BSM sample and multiplied by a factor 10 for
better visibility.  Only basic topological cuts were applied (see text).
Result of MadGraph simulations, processed by PYTHIA 6 for
parton showering, hadronization and jet reconstruction.
No detector effects were taken into account.}
\label{fcc4hww1}
\end{figure}

At the level of basic topological cuts, the two kinematic variables that offer the
best sensitivity to the $HWW$ coupling are still $\Delta\varphi_{ll}$ and $R_{p_T}$.
The former, as usual, selects hard scattering events, the latter separates
BSM from SM contributions, the more effectively if BSM manifests in $W_LW_L$ pairs.
Since we are working here with energy-independent variables, the respective
signal and background regions can be taken
to a first approximation the same as we had before, before more detailed,
dedicated optimization is done.  Simple cuts like
$\Delta\varphi_{ll} > 2.5$ and $R_{p_T} > 3.5$ already suffice to isolate
significant amounts of signal in a scenario with the $HWW$ coupling being as close
to the SM one as to a few per cent.  The amount of irreducible background left
will be close to 0.66 fb, while signal ranges from 2.54 fb to 0.69 fb, to 0.26 fb
and to 0.06 fb for the scenario of the $HWW$ coupling being equal to 0.8, 0.9, 0.95
and 0.98 times its Standard Model value, respectively.  For details, see
Figs.~\ref{fcc4hww1} thru \ref{fcc4hww7}.
Assuming an integrated luminosity of
1000 fb$^{-1}$ (the order of magnitude that is usually considered for the FCC-hh),
a 3-4\% deviation from the Standard Model coupling will be measurable with a 5$\sigma$
significance.  Assuming 3000 fb$^{-1}$, we reach a 2\% sensitivity.
A combined shape analysis of the two respective distributions will ultimately
produce even more accurate results.  As long as we can restrict our analysis
to the case of $g_{HWW} < 1$, shape analysis in the
lepton-lepton invariant mass distribution gives little improvement in this
measurement and is
not much more efficient than a simple counting experiment as we have just done.
Nonetheless, an interpretative ambiguity may still exist between $g_{HWW} < 1$ and
$g_{HWW} > 1$ cases.  The lepton-lepton invariant mass spectrum, or equivalently
lepton transverse momenta, solve this puzzle unambiguously, since they directly
reflect the interference pattern between the Higgs exchange graph on the one hand
and the sum of the $Z/\gamma$ exchange and the 4-$W$ contact graphs on the other.
To be more explicit, the spectrum montonously rises with respect to the SM
if $g_{HWW} < 1$, while it initially falls and ultimately rises if $g_{HWW} > 1$.
The turning point lies well within the allowed kinematic phase space of the FCC if
only the $HWW$ coupling differs from unity enough to produce a statistically
significant signal.
Past it the $g_{HWW} > 1$ spectrum tends to catch up with its mirror
$g_{HWW} < 1$ spectrum and the two become asymptotically identical.

\begin{figure}[htbp]
\begin{center}
\epsfig{file=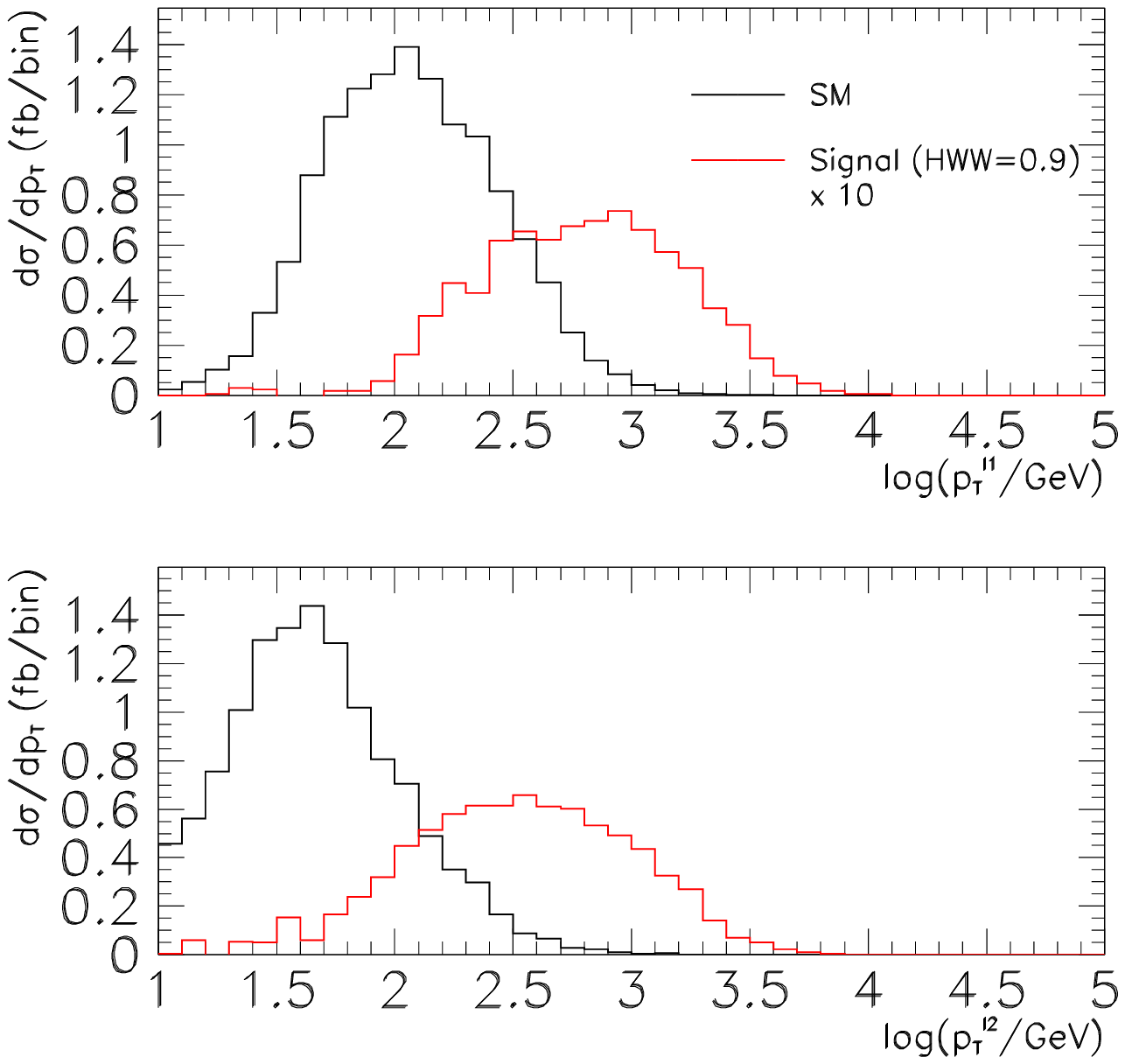,width=0.60\linewidth}
\epsfig{file=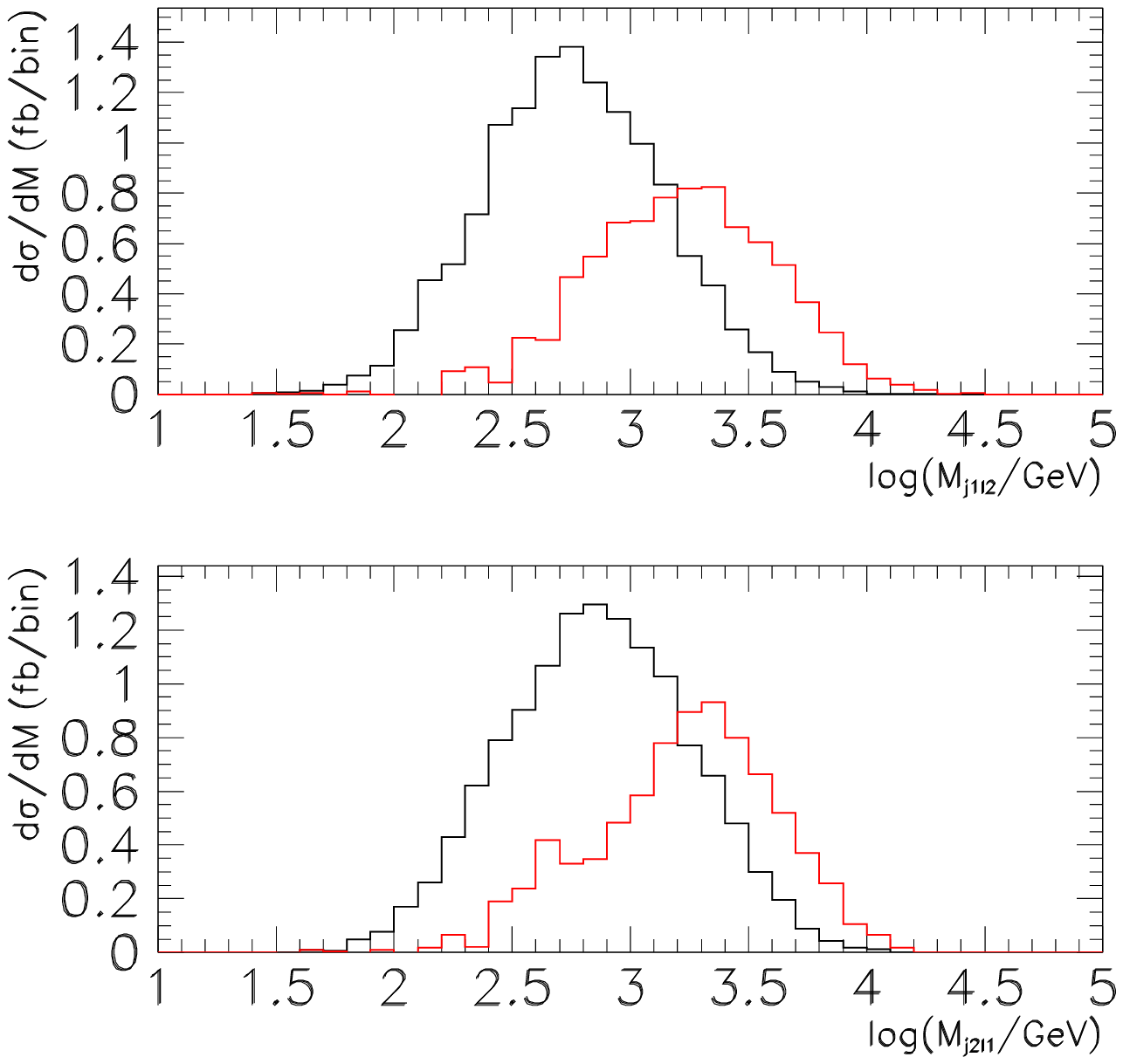,width=0.60\linewidth}
\end{center}
\vspace{-5mm}
\caption{{\small Distributions of transverse momenta of the leading
and sub-leading (two upper plots) leptons, and
invariant mass distributions of combinations of jets and leptons (two lower plots)
 from the $pp \to jjW^+W^+$ process at
100 GeV, with leptonic $W^+$ decay ($l = e, \mu$).  
Shown are the distributions for
the Standard Model irreducible background
(black histos) and the BSM signal (red histos).  BSM was defined in terms
of the $HWW$ coupling set to 0.9 of its SM value.  Signal was calculated by
subtracting the SM sample from the BSM sample and multiplied by a factor 10 for
better visibility.  Only basic topological cuts were applied.
Result of MadGraph simulations, processed by PYTHIA 6 for $W$ decay into
leptons, parton showering, hadronization and jet reconstruction.
The original PYTHIA 6 source code was modified to account for the
correct, polarization-dependent, angular distributions for the $W$ decays.
No detector effects were taken into account.
}}
\label{fcc4hww4}
\end{figure}

\begin{figure}[htbp]
\begin{center}
\epsfig{file=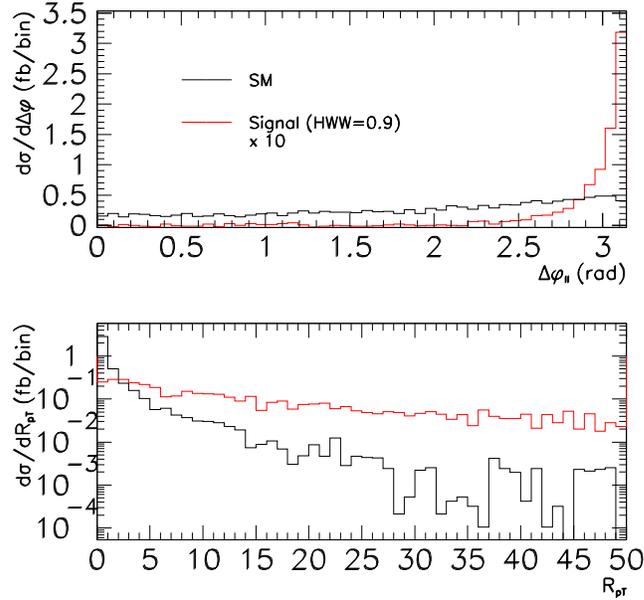,width=0.60\linewidth}
\end{center}
\vspace{-1cm}
\caption{Top: distribution of the lepton-lepton azimuthal separation, $\Delta\varphi$,
and bottom: distribution of the ratio $p_T^{l1} p_T^{l2} / (p_T^{j1} p_T^{j2})$,
from the $pp \to jjW^+W^+$ process at
100 GeV, with leptonic $W^+$ decay ($l = e, \mu$).
Shown are the distributions for the Standard Model irreducible background
(black histos) and the BSM signal (red histos).  All assumptions and conditions
as in Fig.~\ref{fcc4hww4}.  For the lower plot, an additional cut on
$\Delta\varphi > 2.5$ was applied.}
\vspace{-1cm}
\label{fcc4hww6}
\end{figure}

\begin{figure}[htbp]
\begin{center}
\epsfig{file=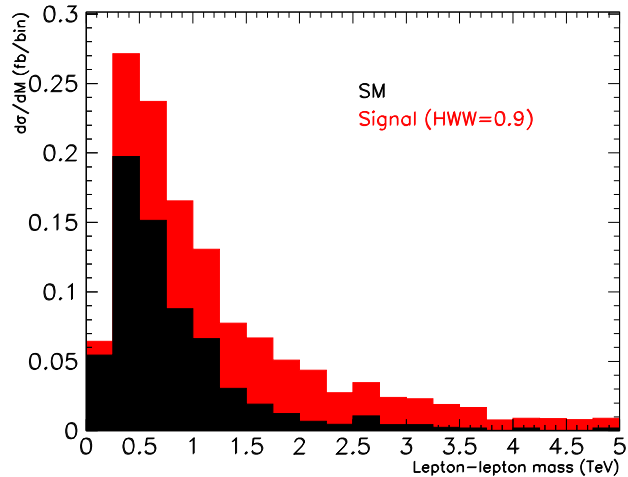,width=0.57\linewidth}
\end{center}
\vspace{-1cm}
\caption{Lepton-lepton invariant mass in the $pp \to jjW^+W^+$ process at
100 GeV, with leptonic $W^+$ decay ($l = e, \mu$).
Shown are the BSM signal (red histo) stacked on the Standard Model irreducible background
(black histo).  
In addition to basic topological cuts, required was $\Delta\varphi > 2.5$ and
$R_{p_T} >$ 3.5.  All other assumptions and conditions as in Fig.~\ref{fcc4hww4}.}
\label{fcc4hww7}
\end{figure}

\begin{figure}[htbp]
\begin{center}
\epsfig{file=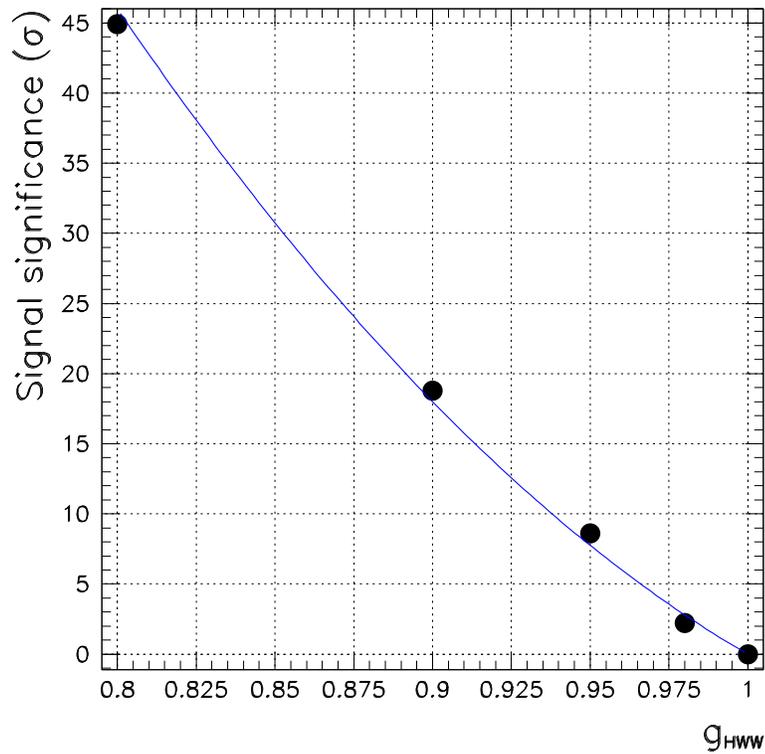,width=0.75\linewidth}
\end{center}
\caption{Signal significance, expressed in terms of the ratio $S/\sqrt{S+B}$
($S$ - BSM signal, $B$ - SM irreducible background) as a function of the actual
value of the $HWW$ coupling relative to its SM value, simulated in
the $pp \to jjW^+W^+$ process at
100 TeV, with leptonic $W^+$ decay ($l = e, \mu$), and
assuming an integrated luminosity of 1000 fb$^{-1}$.
All assumptions and conditions as in Fig.~\ref{fcc4hww7}.
In signal evaluation
no unitarity cutoff was applied and so the leftmost points
may be slightly overestimated.
}
\label{fcc4hww8}
\end{figure}
\vspace{5mm}

\section{Anomalous gauge couplings at the FCC}

As mentioned before, the most sensitive probes of triple gauge couplings will
come from measurements of total diboson production.  Preliminary simulation work
indicates that the relevant VBS modes will be able to independently cross check
these results and push the
sensitivity to well within the present limits in terms of anomalous operators
$\mathscr{O}_{WWW}$, $\mathscr{O}_W$ and even $\mathscr{O}_B$.  Of course, these
limits will be still improved by the LHC.
More or less detailed quantitative estimates of such
sensitivities are currently being worked out by many people, but perhaps are not
the most urgent
question at the present moment.  In fact a far more important issue needs to be
tackled.  The analyses carried so far 
(like the one we have just reported on in the previous section!)
usually focus on a single BSM effect or a single anomalous operator at a time.
This is acceptable for LHC energies where the main question is whether we can
observe {\it any} BSM effect given present bounds on anomalous couplings, but
our ability to identify a physics scenario wuthout relying on other measurements
is limited.  The aim of the FCC is however to identify a physics scenario.
A single anomalous operator
is unlikely what we will eventually observe in an experiment.
The key question to address is whether we can disentangle the different effects
for a correct interpretation of the experimental result.  This requires a careful
comparative study of the phenomenology associated to the possible different scenarios.

In the scenario with non-SM Higgs to gauge couplings, signal manifests solely
in $W_LW_L$ pairs rising with energy above SM prediction.  
It happens that a quartic $WWWW$ coupling scaled by a constant factor
will also be mostly observable in this way.
The entire
contribution of the quartic vertex to the dominant $W_TW_T$ scattering cross section
is rather minute, so the total rate varies very little with it.  Mixed $W_TW_L$ pairs
get some energy dependence in addition to an overall normalization shift, but this
last effect is even
less appreciable than in $W_TW_T$ because of lower absolute rates, and the first
effect is dwarfed by a much steeper energy dependence coming from $W_LW_L$ pairs.
Consequently, it is the functional form of the $W_LW_L$ energy dependence through
which one must
distinguish a scaled $HWW$ coupling from a scaled $WWWW$ coupling.
Various anomalous contributions to the quartic coupling may however affect
$W_TW_X$ as well as $W_LW_L$.

There is one result published from the Snowmass 2013 study \cite{snowmass}
that is of direct interest for the FCC.  The sensitivity of the $W^\pm W^\pm$
scattering process to the coefficient $f_{T,1}/\Lambda^4$, measured in terms
of the expected 5$\sigma$ discovery reach and 95\% CL limit increases by an
impressive factor of 100 between 14 and 100 TeV, assuming the same integrated
luminosity and excatly the same data analysis.  By comparison, sensitivities to 
anomalous coefficients studied in
the $WZ$ and $ZZ$ processes were compared at beam energies of 14 and 33 TeV
and revealed improvement by merely a factor 1.2-1.8, depending on the analysis.
This already gives a glimpse of the superb physics capabilities of
the FCC, but does not answer the question of being able to identify the scenario.

A yet different story is the one with triple gauge couplings $WWZ$ and $WW\gamma$.
They affect $W_LW_L$, $W_TW_T$ and $W_TW_L$ pairs in different ways, as well
as they affect both VBS and non-VBS processes.  Since the VBS sample is a
fraction of the non-VBS sample in absolute counts and because the kinematics
of their respective final states partly overlap, VBS signals
can only be studied on their own right once stringent criteria are predefined to
suppress the unavoidable non-VBS contamination to a negligible level.  For a
correct evaluation of pure VBS signals, the non-VBS contribution must be
negligible not only in the SM scenario, but in
any arbitrary scenario with anomalous couplings within their present experimental
bounds.  In case
of $W^\pm W^\pm$ this can be effectively achieved by tightening the lepton back-to-back
requirement to $\Delta\varphi_{ll} > 2.8$.  This is because of two classes of
non-scattering processes.  One involves a $u-u$ quark collision with one of the quarks
interacting after $W^+$ emission, the other is $u-\bar{d}$ annihilation.
They are negligible in the SM, but become part of the non-VBS signal with
anomalous $WWZ$ and $WW\gamma$ couplings. 
Luckily, tightening $\Delta\varphi_{ll}$ does not significantly reduce the VBS
signal at 100 TeV.
For the other $VV$ scattering processes, the large number of diagrams potentially
contributing to the non-VBS signal may prove this much more complicated.

In any BSM scenario, new physics ultimately ends up enhancing the $VV$ scattering cross
section at a sufficiently high invariant mass.  It remains true regardless of whether
or not this cross section gets depleted at some intermediate scale, depending on
the signs of the anomalous coefficients and therefore the pattern of interference
between the individual scattering diagrams.  It is also true regardless of whether
it is $W_LW_L$ or $W_TW_X$ the primary source of signal.  It should be noted
that even if $W_TW_X$ contribute to the VBS signal, it does not make $W_LW_L$
any less important.  Quite the contrary, the $\mathscr{O}_W$ operator produces a similar
amount of VBS signal for both helicity combinations, in clear contrast with
$\mathscr{O}_{WWW}$ on one side and $\mathscr{O}_B$, $\mathscr{O}_{\Phi d}$,
$\mathscr{O}_{\Phi W}$ and the relevant
dimension-8 operators on the other.  It makes the ability to separate the two samples
experimentally a bonus of special interest.

\begin{figure}[htbp]
\begin{center}
\epsfig{file=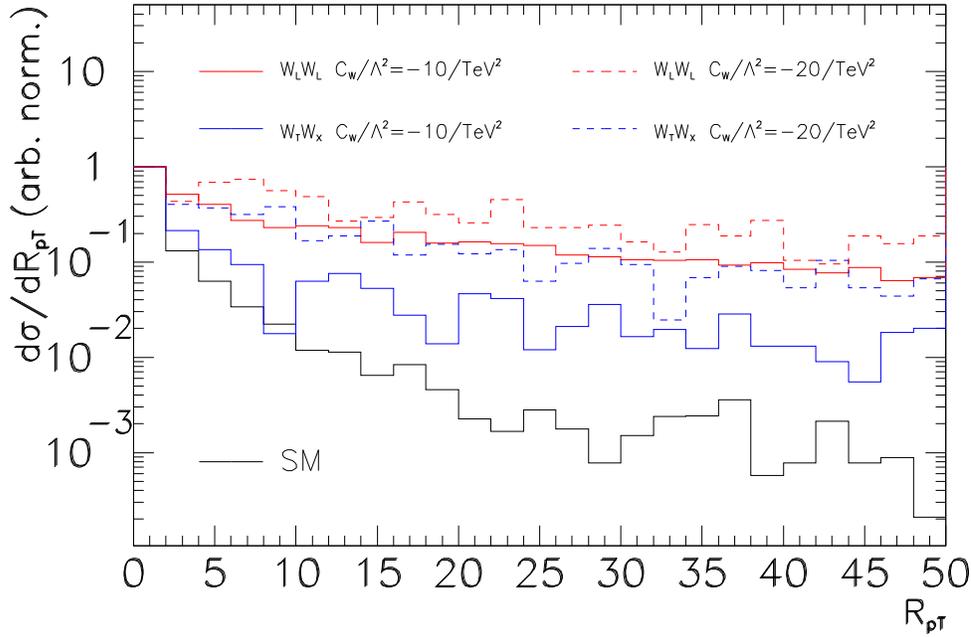,width=0.9\linewidth}
\end{center}
\caption{The shapes of $R_{p_T}$ distributions resulting from 
the $pp \to jjW^+W^+$ process at 100 TeV with
leptonic $W^+$ decay ($l = e, \mu$).  Shown are:
the Standard Model scenario (black histo, all helicity combinations summed up),
the $C_W/\Lambda^2 =$ -10/TeV$^2$ additive signals
(blue solid histo - $W_TW_X$ pairs, red solid histo - $W_LW_L$ pairs) and
the $C_W/\Lambda^2 =$ -20/TeV$^2$ additive signals 
(blue dashed histo - $W_TW_X$ pairs, red dashed histo - $W_LW_L$ pairs).
For the sake of a convenient comparison, each distribution was
individually scaled to the contents of its first bin,
$R_{p_T} < 2$.  VBF topological cuts (see text) and a cut on $\Delta\varphi_{ll} > 2.8$
were applied.  Signals were evaluated without applying any $\Lambda$ cutoff.
Results of MadGraph simulations,
all assumptions and conditions as in Fig.~\ref{fcc4hww7}.}
\label{fcc4rpt}
\end{figure}
\vspace{5mm}

\begin{figure}[htbp]
\begin{center}
\epsfig{file=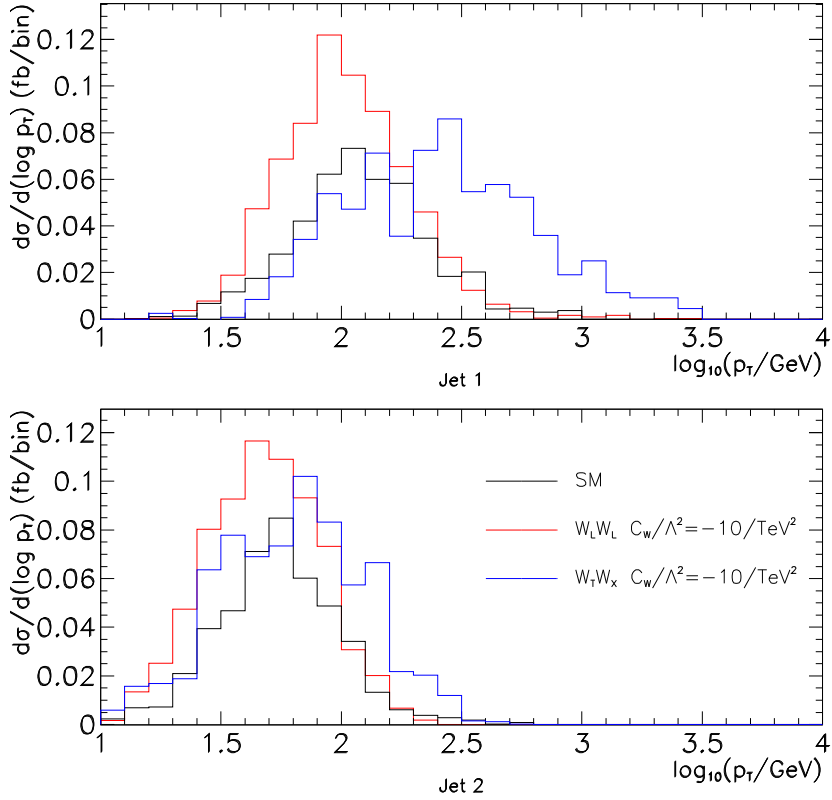,width=0.75\linewidth}
\end{center}
\caption{Transverse momentum distributions of the leading and subleading jets in
the $pp \to jjW^+W^+$ process at 100 TeV with
leptonic $W^+$ decay ($l = e, \mu$).  Shown are the
Standard Model scenario (black histo, all helicity combinations summed up),
and the additive signals of $C_W/\Lambda^2 =$ -10/TeV$^2$
(blue histo - $W_TW_X$ pairs, red histo - $W_LW_L$ pairs).
Applied were all signal selection criteria discussed in the text. 
Results of MadGraph simulations,
all assumptions and conditions as in Fig.~\ref{fcc4rpt}.}
\label{fcc4ptj}
\end{figure}
\vspace{5mm}

\begin{figure}[htbp]
\begin{center}
\epsfig{file=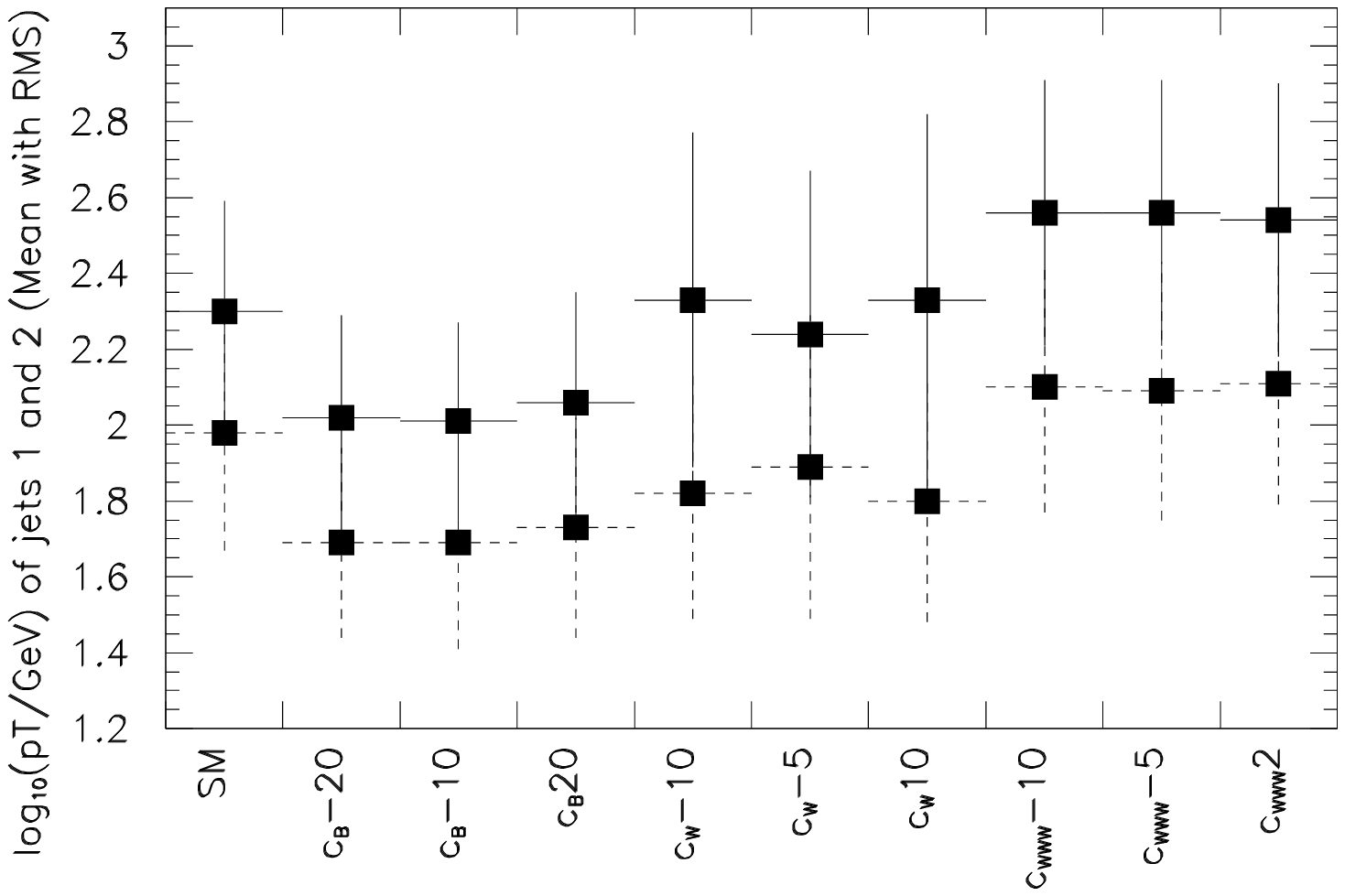,width=0.95\linewidth}
\end{center}
\caption{Mean values and RMS of the individual transverse momenta of the two jets
in the process $pp \to jjW^+W^+$ at 100 TeV,
with leptonic $W^+$ decay ($l = e, \mu$) in different physics scenarios.
Solid lines represent the leading jet and
dashed lines represent the sub-leading jet.
Vertical error bars represent the RMS.
Each bin on the horizontal axis represents a physics scenario; from left to right:
the Standard Model and BSM signals for $c_B/\Lambda^2 = -20/$TeV$^2$,
$c_B/\Lambda^2 = -10/$TeV$^2$, $c_B/\Lambda^2 = 10/$TeV$^2$,
$c_W/\Lambda^2 = -10/$TeV$^2$, $c_W/\Lambda^2 = -5/$TeV$^2$,
$c_W/\Lambda^2 = 10/$TeV$^2$, $c_{WWW}/\Lambda^2 = -10/$TeV$^2$
$c_{WWW}/\Lambda^2 = -5/$TeV$^2$ and $c_{WWW}/\Lambda^2 = 2/$TeV$^2$.
Applied were VBF selection criteria, including $\Delta\varphi_{ll} > 2.8$.
Results of MadGraph 5 simulations, all conditions and assumptions as for
Fig.~\ref{fcc4rpt}.}
\label{pt100}
\end{figure}
\vspace{5mm}

As was the case for the LHC,
use of the $R_{p_T}$ variable, the way we did just above in the scenario with modified
Higgs to
gauge couplings, is effective for any BSM scenario that enhances high $WW$ invariant
masses.  This is because of the still holding strong correlation between $M_{WW}$
and the lepton transverse momenta.
It is also always automatically more effective for $W_LW_L$ signals
than for $W_TW_X$ signals because of the jet transverse momenta in its denominator.
However, if we allow both $W_LW_L$ and $W_TW_X$ signals, there is no way to
separate these by looking at $R_{p_T}$ alone without a priori knowledge of the
physics scenario.  A study of
respective signals associated to, e.g., the $\mathscr{O}_W$ operator clearly shows this
interpretative ambiguity: the entire shape of the $R_{p_T}$ distribution for a pure
$W_LW_L$ sample with, say, $C_W/\Lambda^2 =$ -10/TeV$^2$ almost exactly coincides
with that of a pure $W_TW_X$ sample with $C_W/\Lambda^2 =$ -20/TeV$^2$
(see Fig.~\ref{fcc4rpt}).  
The ambiguity is solved
by looking at the individual jet transverse momenta, and chiefly the $p_T$ of
the leading jet.  The separation of the two helicity sub-samples is much better
for 100 TeV than for 14 TeV.
The maximum of the leading jet $p_T$ distribution is clearly shifted
with respect to the SM to lower values for $W_LW_L$ signals and to higher values
for $W_TW_X$ signals.
A shape analysis of the leading jet $p_T$ distribution,
measured from the events that pass the standard $R_{p_T}$ cut, should suffice
to determine the helicity composition to a satisfactory accuracy, enough to
resolve whether the signal is indeed $W_LW_L$-driven (via $\mathscr{O}_B$, pure $HWW$
or $WWWW$)
or $W_TW_X$-driven (via $\mathscr{O}_{WWW}$) or mixed (via $\mathscr{O}_W$ or
any suitable combination).
A peak at around 100 GeV of the measured excess over the SM (itself having a median
around 200 GeV) is an unequivocal sign of $W_LW_L$.  A broader peak at around 300 GeV
signals $W_TW_X$.  The signal peak positions link directly to helicity and
hardly vary with the actual physics scenario or specific values of the anomalous
coefficients.  The fact that the peak position slightly shifts for different
values of $c_W/\Lambda^2$ (see Fig.~\ref{pt100}), is because the proportion of
the selected $W_LW_L$ and $W_TW_X$ pairs changes likewise.
The RMS of the log$_{10}$($p_T^{j1}$) distributions are close to
0.3 and 0.4 for $W_LW_L$ and $W_TW_X$, respectively, making a clear distinction
possible whenever signal itself becomes statistically significant.
Larger widths are already a clear indication that signal is in fact a mixture of
$W_LW_L$ and $W_TW_X$, with two distinct sub-samples vaguely emerging from the spectrum.
The sub-leading jet is a less powerful discriminator on its own because it receives a
substantial $W_L$ contribution from $W_TW_L$ pairs.  Nonetheless it can be used as
an additional consistency cross check.  Once combined the information from the two
jets, it turns out that as few as 10 events suffice to distinguish a pure $W_LW_L$
from a pure $W_TW_X$ signal at the 5$\sigma$ level.  In other words, with an isolated
signal sample of
$N$ events, the helicity composition can be deduced to a precision of 
$\sim1/\sqrt{2.5~N}$.  That makes, e.g., for $HWW=0.95$ and no other anomalous
couplings (260 signal events in
1000 fb$^{-1}$) a 4\% measurement.

\begin{figure}[htbp]
\begin{center}
\epsfig{file=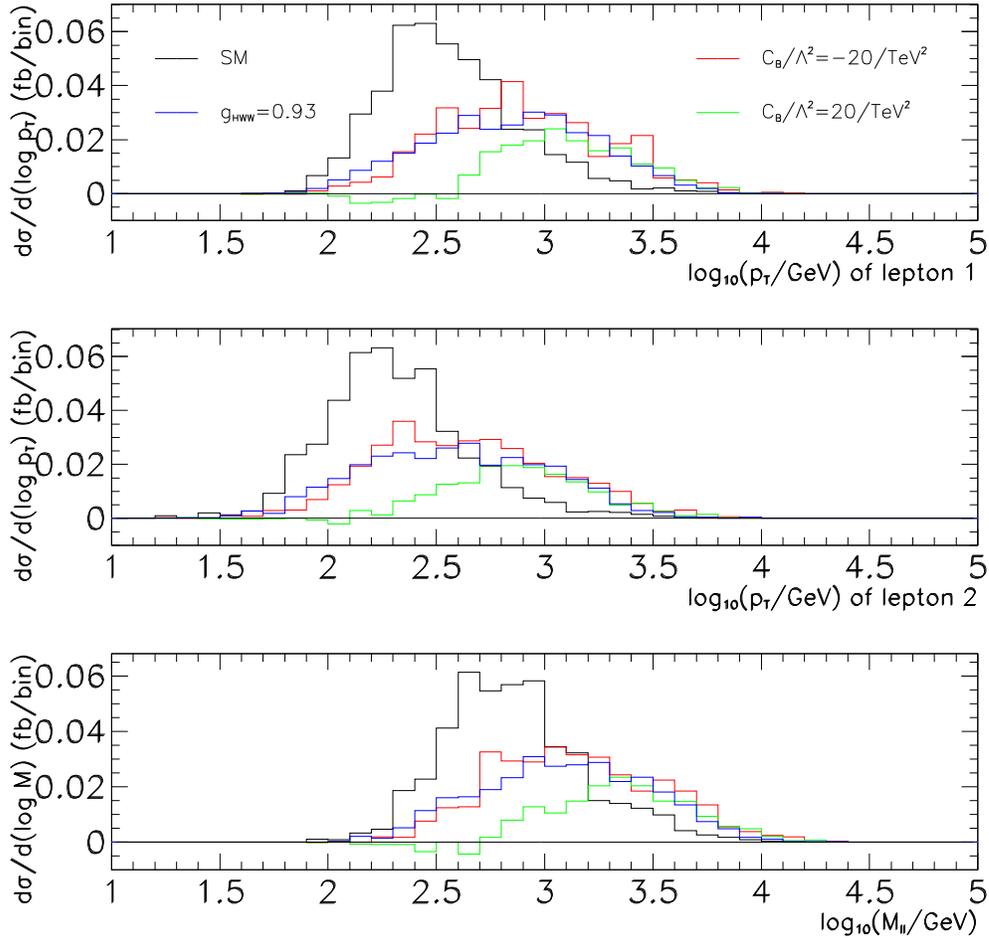,width=0.9\linewidth}
\end{center}
\caption{Transverse momentum distributions of the leading and subleading leptons 
(upper and middle plots) and lepton-lepton invariant mass distribution (lower plot)
in the $pp \to jjW^+W^+$ process at 100 TeV with
leptonic $W^+$ decay ($l = e, \mu$).  Shown are the
Standard Model scenario (black histo, all helicity combinations summed up),
and the additive $W_LW_L$ signals of $C_B/\Lambda^2 =$ -20/TeV$^2$ (red histo),
$C_B/\Lambda^2 =$ 20/TeV$^2$ (green histo) and $g_{HWW}=0.93$ (blue histo).
Applied were all signal selection criteria discussed in the text.  
Results of MadGraph simulations,
all assumptions and conditions as in Fig.~\ref{fcc4rpt}.}
\label{fcc4ptl}
\end{figure}
\vspace{5mm}

In a given physics scenario, $W_LW_L$ and $W_TW_X$ signals do not differ significantly
in terms of the outgoing lepton kinematics.  Somewhat different widths of the
respective $p_T$
distributions (larger for $W_TW_X$ than for $W_LW_L$) are expected
as a simple consequence of the angular distributions in $W$ decay.
These differences alone may however easily prove not significant enough or too entangled
with other effects to be of much practical use unless we know beforehand the
helicity composition of the selected sample from other sources.
But once we have independently established the helicity composition, leptons in
the final state help resolve the remaining ambiguities concerning
the physical scenario and the sign of the anomalous coefficients.  The 
$W_LW_L$-driven scenarios clearly differ in the lepton transverse momenta and/or 
invariant mass distributions.  For example, the median of the leading lepton $p_T$
distribution is around 600 GeV for a purely $HWW$-driven signal, but may become
larger with new physics manifesting itself in modified $WWWW$ or $WWZ$ couplings.
These numbers are a direct consequence
of the $s$-dependences of the relevant amplitudes plus a common phase space factor,
and so they have also very little sensitivity
to the actual values of the parameters in question.  It is the total signal rate
that determines the coefficient values.
The lepton-lepton invariant mass also unambiguously fixes the sign of
the relevant anomalous parameter in case an ambiguity exists on measuring the
signal rate alone.  Because of the sign of the interference terms between the
three basic graphs contributing to $W^+W^+ \to W^+W^+$, any of the following
scenarios: $g_{HWW} < 1$, $C_W < 0$ and $C_B < 0$, produces a steady enhancement
in the $M_{ll}$ spectrum, ultimately suppressed by phase space.
By contrast, $g_{HWW} > 1$, $C_W > 0$ and $C_B > 0$ are bound to produce a moderate
depletion at intermediate masses, followed by a rise at larger values, eventually
catching up asymptotically with the former.  For $C_{WWW}$, the sign may be
more difficult to determine.  In any case, there is a strong correspondence
between invariant mass distrubutions and transverse momenta distrubutions of the
outgoing leptons and
detailed simulation work will ultimately have to tell which approach is preferrable.

There are however also ambiguities that would take more effort to resolve.
For example, the signal of $C_B < 0$ coincides with that of scaling the 
Higgs to $WW$ coupling by an appropriately chosen constant, although the former
does not modify the $HWW$ coupling at all.  Such ambiguities
may be ultimately solvable only in the context of combining data from 
different processes.

Another source of complication in this type of analyses arises from the unknown
scale of new physics, $\Lambda$.
In the Effective Field Theory approach, values of the anomalous coefficients are
intimately connected to the value of $\Lambda$.  Theoretical
predictions depend only on the ratio $c/\Lambda^2$ or $c/\Lambda^4$, but one
cannot separate the coefficient from the energy.  However, if the data fit the
theoretical curve in the entire kinematic phase space covered by the experiment,
then $\Lambda$ must be
at least equal to the $WW$ invariant mass of the highest data point.  Otherwise,
data would indicate the appropriate cutoff value.  On the other end, the value
of $\Lambda$ must be lower than a calculable upper limit defined by the unitarity
condition.  Hence one could at least bound $\Lambda$ from above and below.
Nevertheless, for practical purposes the unitarity condition may be
completely irrelevant in
terms of evaluating the expected sensitivity limits, because the FCC sensitivity
reaches to anomalous coefficient values that do not lead to unitarity violation
within the available energy scale \cite{snowmass}.

All in all, we have emphasized the primary importance of studying same-sign
$WW$ interactions at the FCC.  Moreover, it is
the full event kinematics studied from a clean $W^+W^+$ scattering 
sample with leptonic
$W$ decay, and most of all the transverse momenta of
all four final state particles, that carry the bulk of the necessary information
in order to disentagle
the underlying physics scenario and correctly interpret the results.

\chapter{Summary}

The Higgs boson is an empirical fact.  Moreover, based on all the data collected
at Run 1 of the LHC, it looks by all means
consistent with the Standard Model one.  In particular, Higgs couplings to vector
bosons are consistent with SM ones to an accuracy of roughly $\sim$20\%.
No other hints of physics BSM have been observed so far, either.
This does not preclude that the dynamics of electroweak symmetry breaking may still
be partially strong.  Whether or not we observe Higgs couplings deviate from
their SM predictions in future measurements, only direct observation of $VV$
scattering at high energies will ultimately tell if this is indeed the case.

The full phenomenology of $VV$ scattering at high energy depends on the Higgs
mass, Higgs to
gauge couplings, gauge boson triple couplings and gauge boson quartic couplings.
With the present experimental bounds on these inputs, only the Higgs mass can
be considered definitely fixed for VBS studies.  Effects from non-SM Higgs
couplings and triple couplings can still be observed at the LHC with $\sqrt{s}$ =
13 TeV with hard work and some luck and consistency checks can be done with
new, more precise measurements of all the relevant quantities that will come
directly from Higgs physics on one side and total diboson production on the other.
Agreement between these three classes of measurements can be translated into
the first real experimental limits on anomalous quartic gauge couplings.
Alternatively, disagreement may signal existence of the latter.  In the event
of absence of direct observation of new resonances, the best process to
study VBS-related physics is same-sign $WW$ scattering in the purely leptonic
decay mode, but with further improvements the semi-leptonic decay modes may
prove equally important.
However, with 300 fb$^{-1}$ VBS processes on their own offer little
possibilities to interpret the results in a standalone way, i.e., without
relying on concurrent measurements.  This is beacuse
the BSM effects are bound to be tiny and statistics too low to carry more
precise studies.  The 300 fb$^{-1}$ program is likely to end up as a
Standard Model measurement, of a similar philosophy as the ones already carried
by ATLAS and CMS from 8 TeV data.  However, the focus for the High Luminosity
LHC program should be BSM and it is time now to plan an analysis strategy
different from a Standard Model analysis.
The High Luminosity program has chances to provide enough
data for at least some physics scenarios be distinguished from others based
on studies of VBS processes alone.  Among other things, this can be done by
applying novel techniques to separate different helicity combinations in the
selected samples of $VV$ pairs that we advocate in this work.  In particular,
should an excess over SM
predictions be observed, it should be possible to tell whether this excess
is related to the mechanism of electroweak symmetry breaking or to other physics.
Improvement in the sensitivity to new physics, and especially to those effects
that affect mainly $V_LV_L$ pairs, should be sought at low transverse momenta and 
large pseudorapidities of the tagging jets.  This should be taken into account
in planning future machine and detector upgrade activities for the HL-LHC phase.
Not least, even in the absence of new physics, application of analysis
techniques that fully exploit vector boson helicities will result in better
exclusion limits, at least for those scenarios that do not modify the
dominant transverse polarizations.

A qualitative improvement in sensitivity to BSM effects in VBS processes
can only be achieved via further increase in beam energy.  The FCC with its
$\sqrt{s}$ = 100 TeV has all the potential to observe many BSM effects in
$VV$ scattering and to identify the physical sources of these effects.
Consistency of VBS measurements with Higgs physics, diboson production and
triboson production measurements at the FCC will provide an ultimate closure
test of the Standard Model or the theory that will replace it.

\chapter{Acknowledgments}

The list of people I feel indebted to is long and, as always, subjective.
But there are a few names I definitely must mention here.

First and foremost,
I thank Profs.~Stefan Pokorski, Jan Kalinowski and S{\l}awek Tkaczyk
for their long collaboration, countless meetings and discussions throughout the
years, always inspiring and enlightening.  The complicated interplay between
theory and experiment,
the conceptual feedback I have received on the
theory side (SP, JK) and the support on the practical, experimental side (ST),
was a very stimulating experience and crucial to complete this work.
To Jan and S{\l}awek I am also
grateful for kindly agreeing to review parts of this work before publication
and sending me their
valuable comments.  Any remaining mistakes, typos and other shortcomings
that no doubt can still be found, are of course my fault.

I thank my CMS colleagues from the NCBJ and the University of Warsaw for their
tolerance and patience to put up with me during all this time, as well as for all
their intellectual feedback.  In particular, I am indebted to Prof.~Krzysztof Doroba
for his pioneering role in starting up the $WW$ related activities within the 
Warsaw CMS group, and to Prof.~Jan Kr{\'o}likowski, Micha{\l} Bluj, 
Artur Kalinowski, Marcin Konecki,
Piotr Zalewski and many others for sharing their
wisdom and experience in our group meetings and informal discussions.

And I thank Mayda Velasco, my former boss at Northwestern, for it is there
that my adventure with CMS has started.

\end{document}